\documentclass[a4paper]{book}

\usepackage[T1]{fontenc}
\usepackage[utf8]{inputenc}
\usepackage[english]{babel}
\usepackage{graphicx}
\usepackage{colortbl} 
\graphicspath{ {./images/} }
\usepackage{latexsym}
\usepackage{amsmath}
\usepackage{cases}
\usepackage{titlesec}
\usepackage{circuitikz}
\usepackage[toc,page]{appendix}
\usepackage{amsthm}
\usepackage{amssymb}
\usetikzlibrary{circuits.ee.IEC, circuits.logic.US, circuits.logic.IEC}
\usetikzlibrary{patterns}
\usetikzlibrary{calc}
\usepackage{graphicx}
\usepackage{enumerate}
\usepackage[toc,page]{appendix}
\usepackage{caption}
\usepackage{subcaption}
\usepackage{empheq}
\usepackage{multirow}
\usepackage[version=3]{mhchem}
\usepackage{braket}
\usepackage{centernot}
\usepackage{relsize}
\usepackage{geometry}
\usepackage{lmodern}
\usepackage[tone]{tipa}
\usepackage{hyperref}
\usepackage{array}
\usepackage{bm}
\usepackage{slashed}
\definecolor{processblue}{cmyk}{0.96,0,0,0}
\definecolor{greeen1}{cmyk}{1.0,0,57,0,42}
\definecolor{green2}{cmyk}{0,68,0,73,27}

\usetikzlibrary{decorations.markings}

\newcommand{\nn}{\bigskip\noindent}
\newtheorem{defn}{Definition}[chapter]
\newtheorem{Ese}{Example}[chapter]
\newtheorem{Pre}{Preposition}[chapter]
\newtheorem{teo}{Theorem}[chapter]
\newcommand{\V}{\mathcal{V}}
\newcommand{\B}{\mathcal{B}}
\newcommand{\A}{\mathcal{A}}
\newcommand{\Ha}{\mathcal{H}}
\newcommand{\La}{\mathcal{L}}
\newcommand{\T}{\mathcal{T}}
\newcommand{\R}{\mathbb{R}}
\newcommand{\G}{\mathcal{G}}
\newcommand{\K}{\mathbb{K}}
\newcommand{\Z}{\mathbb{Z}}
\newcommand{\C}{\mathbb{C}}
\newcommand{\D}{\mathcal{D}}
\newcommand{\M}{\mathcal{M}}
\newcommand{\z}{\bar{z}}
\newcommand{\be}{\begin{equation}}
\newcommand{\ee}{\end{equation}}
\newcommand{\f}{\varphi}

\newcommand{\m}{\mu}
\newcommand{\n}{\nu}
\newcommand{\e}{\varepsilon}
\newcommand{\bs}{\begin{split}}
\newcommand{\es}{\end{split}}
\newcommand{\x}{\hat{x}}

\newcommand{\de}{\partial}

\newcommand{\tu}{\rightarrow}
\newcommand{\tw}{\Delta\otimes u_\Delta}
\newcommand{\Lod}{\mathcal{L}^{\vee}_\omega}
\newcommand{\Lo}{\mathcal{L}_\omega}
\DeclareMathOperator*{\res}{Res}
\newcommand{\lf}[1]{\langle\f_L^{(#1)}|}
\newcommand{\rf}[1]{\f_R^{(#1)}\rangle}

\newcommand{\cycleone}{
\node at (-4,0) {$\left (S^1_{\e,i}\otimes U_{S^1_{\e,i}}\right )\cdot \left (\gamma \otimes U_{(z_i,z_{i+1})}\right )= $};
				
				 \begin{scope}[very thick,decoration={markings,mark=at position 0.5  with {\arrow{>}},mark=at position 1  with {\arrow{>}}}] 
                \draw[postaction={decorate}] (1,0) arc (0:350:1);            
                 \end{scope}	
                 \begin{scope}[very thick,decoration={markings,mark=at position 0.5  with {\arrow{>}}}]       	
                   \draw [postaction={decorate}] (0,0) -- (1.5,1);
                   \end{scope}
			  \fill (0,0) circle (2pt);
			 
	              \node [above] (0,0) {$z_i$};
	               \node at (-1,1) {$S^1_{\e,i}$};
	                \node at (1.4,0.6) {$\gamma$};
	                \node at (2,0) {$=-$};

	                	 \begin{scope}[very thick,decoration={markings,mark=at position 0.5  with {\arrow{>}},mark=at position 1  with {\arrow{>}}}] 
                \draw[postaction={decorate}] (3,0) arc (-180:170:1);            
                 \end{scope}	
                 \begin{scope}[very thick,decoration={markings,mark=at position 0.5  with {\arrow{<}}}]       	
                   \draw [postaction={decorate}] (4,0) -- (2.5,-1);
                   \end{scope}
			  \fill (4,0) circle (2pt);
			 
	              \node at (4,0.2) {$z_i$};
	               \node at (5,1) {$S^1_{\e,i}$};
	                \node at (2.4,-0.6) {$\gamma$};
	                 \node at (5.5,0) {$=-1$};
	       }

		\newcommand{\cycletwo}{	
				\node at (-4,0) {$\left (S^1_{\e,i}\otimes U_{S^1_{\e,i}}\right )\cdot \left (\gamma \otimes U_{(z_i,z_{i+1})}\right )= $};
				 \begin{scope}[very thick,decoration={markings,mark=at position 0.5  with {\arrow{>}},mark=at position 1  with {\arrow{>}}}] 
                \draw[postaction={decorate}] (1,0) arc (0:350:1);            
                 \end{scope}	
                 \begin{scope}[very thick,decoration={markings,mark=at position 0.5  with {\arrow{>}}}]       	
                   \draw [postaction={decorate}] (0,0) -- (1.5,-1);
                   \end{scope}
			  \fill (0,0) circle (2pt);
			 
	              \node [above] (0,0) {$z_i$};
	               \node at (-1,1) {$S^1_{\e,i}$};
	                \node at (1.4,-0.6) {$\gamma$};
	                  \node at (2,0) {$=-$};

	                	 \begin{scope}[very thick,decoration={markings,mark=at position 0.5  with {\arrow{>}},mark=at position 1  with {\arrow{>}}}] 
                \draw[postaction={decorate}] (3,0) arc (-180:170:1);            
                 \end{scope}	
                 \begin{scope}[very thick,decoration={markings,mark=at position 0.5  with {\arrow{<}}}]       	
                   \draw [postaction={decorate}] (4,0) -- (2.5,1);
                   \end{scope}
			  \fill (4,0) circle (2pt);
			 
	              \node at (4,0.2) {$z_i$};
	               \node at (5,1) {$S^1_{\e,i}$};
	                \node at (2.4,0.6) {$\gamma$};
	                    \node at (6,0) {$=-e^{2\pi i \alpha_i}$};}

	                    \newcommand{\cycletwoo}{	
				 \begin{scope}[very thick,decoration={markings,mark=at position 0.5  with {\arrow{>}},mark=at position 1  with {\arrow{>}}}] 
                \draw[postaction={decorate}] (1,0) arc (0:350:1);            
                 \end{scope}	
                 \begin{scope}[very thick,decoration={markings,mark=at position 0.5  with {\arrow{>}}}]       	
                   \draw [postaction={decorate}] (0,0) -- (1.5,-1);
                   \end{scope}
			  \fill (0,0) circle (2pt);
			 
	              \node [above] (0,0) {$z_i$};
	               \node at (-1,1) {$S^1_{\e,i}$};
	                \node at (1.4,-0.6) {$\gamma$};
	                \node at (1.3,0) {$\e$};
	                }

\newcommand{\twistedcycle}{					
				 \begin{scope}[very thick,decoration={markings,mark=at position 0.5  with {\arrow{>}},mark=at position 1  with {\arrow{>}}}] 
                \draw[postaction={decorate}] (1,0) arc (0:350:1);            
                 \end{scope}	
                  \begin{scope}[very thick,decoration={markings,mark=at position 0.4  with {\arrow{>}}}]   
                 \draw [postaction={decorate}] (1,0)--(3,0);    	
                   \end{scope}
			  \fill (0,0) circle (2pt);	 
	              \node [above] (0,0) {$0$};
	               \node at (-1,1) {$S^1_{\e}(0)$};
	                 
	                	 \begin{scope}[very thick,decoration={markings,mark=at position 0.5  with {\arrow{>}},mark=at position 1  with {\arrow{>}}}] 
                \draw[postaction={decorate}] (3,0) arc (-180:170:1);            
                 \end{scope}	
			  \fill (4,0) circle (2pt); 
	              \node at (4,0.2) {$1$};
	               \node at (5,1) {$S^1_{\e}(1)$};           
	                \node at (1.3,0.2) {$\e$};
	                \node at (2.7,0.2) {$\e$};
	                \node at (2,-0.3) {$\langle \e, 1-\e \rangle$};}

\newcommand{\cyclethree}{					
				 \begin{scope}[very thick,decoration={markings,mark=at position 0.5  with {\arrow{>}},mark=at position 1  with {\arrow{>}}}] 
                \draw[postaction={decorate}] (1,0) arc (0:350:1);            
                 \end{scope}	
                 \begin{scope}[very thick,decoration={markings,mark=at position 0.4  with {\arrow{>}}}]   
                 \draw [postaction={decorate}] (1,0)--(3,0);    	
                   \end{scope}
			  \fill (0,0) circle (2pt);	 
	              \node [above] (0,0) {$z_i$};
	               \node at (-1,1) {$S^1_{\e,i}$};
	                \node at (1.4,0.6) {$\gamma$};               
	                 	 \begin{scope}[very thick,decoration={markings,mark=at position 0.2  with {\arrow{>}},mark=at position 0.85  with {\arrow{>}}}] 
             		\draw [very thick,postaction={decorate}] (0,0) to[out=45,in=120] (2,0) to[out=-60,in=-135] (4,0);
                   \end{scope}
                   \node at (2,-0.7) {$\langle z_i, z_{i+1}\rangle$};          
	                	 \begin{scope}[very thick,decoration={markings,mark=at position 0.5  with {\arrow{>}},mark=at position 1  with {\arrow{>}}}] 
                \draw[postaction={decorate}] (3,0) arc (-180:170:1);            
                 \end{scope}	
			  \fill (4,0) circle (2pt); 
	              \node at (4,0.2) {$z_{i+1}$};
	               \node at (5,1) {$S^1_{\e,i+1}$};           
	                 \node at (7.5,0) {$=-\frac{1}{e^{2\pi i \alpha_i}-1}-1-\frac{1}{e^{2\pi i \alpha_{i+1}}-1}$};}

	                 \newcommand{\cyclemixone}{
	                 \begin{scope}[very thick,decoration={markings,mark=at position 0.5  with {\arrow{>}},mark=at position 1  with {\arrow{>}}}] 
                \draw[postaction={decorate}] (1,0) arc (0:350:1);            
                 \end{scope}	
                 \begin{scope}[very thick,decoration={markings,mark=at position 0.4  with {\arrow{>}}}]   
                 \draw [postaction={decorate}] (1,0)--(3,0);    	
                   \end{scope}
			  \fill (0,0) circle (2pt);	 
	              \node [above] (0,0) {$z_i$};
	               \node at (-1,1) {$S^1_{\e,i}$};
	                \node at (5.4,0.6) {$\gamma$};               
	                 	 \begin{scope}[very thick,decoration={markings,mark=at position 0.2  with {\arrow{>}},mark=at position 0.85  with {\arrow{>}}}] 
             		\draw [very thick,postaction={decorate}] (4,0) to[out=45,in=120] (6,0) to[out=-60,in=-135] (8,0);
                   \end{scope}
                   \node at (2,-0.7) {$\langle z_i, z_{i+1}\rangle$};          
	                	 \begin{scope}[very thick,decoration={markings,mark=at position 0.5  with {\arrow{>}},mark=at position 1  with {\arrow{>}}}] 
                \draw[postaction={decorate}] (3,0) arc (-180:170:1);            
                 \end{scope}	
			  \fill (4,0) circle (2pt); 
			   \fill (8,0) circle (2pt); 
	              \node at (4,0.2) {$z_{i+1}$};
	              \node at (8,0.2) {$z_{i+2}$};
	              \node at (9,0.2) {$,$};
	               \node at (5,1) {$S^1_{\e,i+1}$};   
	               \node at (-2,0){$I_{ii+1}=$};    
	               
	                 \begin{scope}[very thick,decoration={markings,mark=at position 0.5  with {\arrow{>}},mark=at position 1  with {\arrow{>}}}] 
                \draw[postaction={decorate}] (5,-4) arc (0:350:1);            
                 \end{scope}	
                 \begin{scope}[very thick,decoration={markings,mark=at position 0.4  with {\arrow{>}}}]   
                 \draw [postaction={decorate}] (5,-4)--(7,-4);    	
                   \end{scope}
			  \fill (0,-4) circle (2pt);	 
	              \node [above] (0,-4) {$z_i$};
	               \node at (3,-3) {$S^1_{\e,i}$};
	                \node at (1.4,-3.4) {$\gamma$};               
	                 	 \begin{scope}[very thick,decoration={markings,mark=at position 0.2  with {\arrow{>}},mark=at position 0.85  with {\arrow{>}}}] 
             		\draw [very thick,postaction={decorate}] (0,-4) to[out=45,in=120] (2,-4) to[out=-60,in=-135] (4,-4);
                   \end{scope}
                   \node at (6,-4.7) {$\langle z_{i+1}, z_{i+2}\rangle$};          
	                	 \begin{scope}[very thick,decoration={markings,mark=at position 0.5  with {\arrow{>}},mark=at position 1  with {\arrow{>}}}] 
                \draw[postaction={decorate}] (7,-4) arc (-180:170:1);            
                 \end{scope}	
			  \fill (4,-4) circle (2pt); 
			   \fill (8,-4) circle (2pt); 
			    \node at (0,-3.8) {$z_i$};
	              \node at (4,-3.8) {$z_{i+1}$};
	               \node at (8,-3.8) {$z_{i+2}$};
	               \node at (7,-3) {$S^1_{\e,i+1}$};           
	                 \node at (-2,-4) {$I_{i+1i}=$};     
	                }

	                 \newcommand{\cyclethreee}{					
				 \begin{scope}[very thick,decoration={markings,mark=at position 0.5  with {\arrow{>}},mark=at position 1  with {\arrow{>}}}] 
                \draw[postaction={decorate}] (1,0) arc (0:350:1);            
                 \end{scope}	
                 \begin{scope}[very thick,decoration={markings,mark=at position 0.4  with {\arrow{>}}}]   
                 \draw [postaction={decorate}] (1,0)--(3,0);    	
                   \end{scope}
			  \fill (0,0) circle (2pt);	 
	              \node [above] (0,0) {$z_i$};
	               \node at (-1,1) {$S^1_{\e,i}$};
	                \node at (1.4,0.6) {$\gamma$};               
	                 	 \begin{scope}[very thick,decoration={markings,mark=at position 0.2  with {\arrow{>}},mark=at position 0.85  with {\arrow{>}}}] 
             		\draw [very thick,postaction={decorate}] (0,0) to[out=45,in=120] (2,0) to[out=-60,in=-135] (4,0);
                   \end{scope}
                   \node at (2,-0.7) {$\langle z_i, z_{i+1}\rangle$};          
	                	 \begin{scope}[very thick,decoration={markings,mark=at position 0.5  with {\arrow{>}},mark=at position 1  with {\arrow{>}}}] 
                \draw[postaction={decorate}] (3,0) arc (-180:170:1);            
                 \end{scope}	
			  \fill (4,0) circle (2pt); 
	              \node at (4,0.2) {$z_{i+1}$};
	               \node at (5,1) {$S^1_{\e,i+1}$};           
	                 }

\newcommand{\cyclefour}{	
				 \begin{scope}[very thick,decoration={markings,mark=at position 0.5  with {\arrow{>}},mark=at position 1  with {\arrow{>}}}] 
                \draw[postaction={decorate}] (1,0) arc (0:350:1);            
                 \end{scope}	
                 \begin{scope}[very thick,decoration={markings,mark=at position 0.4  with {\arrow{>}}}]   
                 \draw [postaction={decorate}] (1,0)--(3,0);    	
                   \end{scope}
			  \fill (0,0) circle (2pt);
			 
	              \node [above] (0,0) {$z_i$};
	               \node at (-1,1) {$S^1_{\e,i}$};
	                \node at (2.4,0.6) {$\gamma$};
	                
	                 	 \begin{scope}[very thick,decoration={markings,mark=at position 0.2  with {\arrow{>}},mark=at position 0.85  with {\arrow{>}}}] 
             		\draw [very thick,postaction={decorate}] (0,0) to[out=-45,in=-120] (2,0) to[out=60,in=135] (4,0);
                   \end{scope}
                    \node at (2,-0.7) {$\langle z_i, z_{i+1}\rangle$};

	                	 \begin{scope}[very thick,decoration={markings,mark=at position 0.5  with {\arrow{>}},mark=at position 1  with {\arrow{>}}}] 
                \draw[postaction={decorate}] (3,0) arc (-180:170:1);            
                 \end{scope}	
			  \fill (4,0) circle (2pt);
			 
	              \node at (4,0.2) {$z_{i+1}$};
	               \node at (5,1) {$S^1_{\e,i+1}$};

	                 \node at (7.5,0) {$=-\frac{e^{2\pi i \alpha_i}}{e^{2\pi i \alpha_i}-1}+1-\frac{e^{2\pi i \alpha_{i+1}}}{e^{2\pi i \alpha_{i+1}}-1}$};
}

\newcommand{\cyclefive}{				
				
				 \begin{scope}[very thick,decoration={markings,mark=at position 0.5  with {\arrow{>}},mark=at position 1  with {\arrow{>}}}] 
                \draw[postaction={decorate}] (1,0) arc (0:350:1);            
                 \end{scope}	
                 \begin{scope}[very thick,decoration={markings,mark=at position 0.4  with {\arrow{>}}}]   
                 \draw [postaction={decorate}] (1,0)--(3,0);    	
                   \end{scope}
			  \fill (0,0) circle (2pt);		 
	              \node [above] (0,0) {$z_i$};
	               \node at (-1,1) {$S^1_{\e,i}$};
	                \node at (1.4,0.6) {$\gamma$};          
	                 	 \begin{scope}[very thick,decoration={markings,mark=at position 0.2  with {\arrow{>}},mark=at position 0.85  with {\arrow{>}}}] 
             		\draw [very thick,postaction={decorate}] (0,0) to[out=45,in=135] (4,0);
                   \end{scope}
                    \node at (2,-0.3) {$\langle z_i, z_{i+1}\rangle$};
   	 \begin{scope}[very thick,decoration={markings,mark=at position 0.5  with {\arrow{>}},mark=at position 1  with {\arrow{>}}}] 
                \draw[postaction={decorate}] (3,0) arc (-180:170:1);            
                 \end{scope}	
			  \fill (4,0) circle (2pt);		 
	              \node at (4,0.2) {$z_{i+1}$};
	               \node at (5,1) {$S^1_{\e,i+1}$};  
	                 \node at (7.5,0) {$=-\frac{1}{e^{2\pi i \alpha_{i}}-1}-\frac{e^{2\pi i \alpha_{i+1}}}{e^{2\pi i \alpha_{i+1}}-1}$};}

\newcommand{\cyclesix}{				
				
				 \begin{scope}[very thick,decoration={markings,mark=at position 0.5  with {\arrow{>}},mark=at position 1  with {\arrow{>}}}] 
                \draw[postaction={decorate}] (1,0) arc (0:350:1);            
                 \end{scope}	
                 \begin{scope}[very thick,decoration={markings,mark=at position 0.4  with {\arrow{>}}}]   
                 \draw [postaction={decorate}] (1,0)--(3,0);    	
                   \end{scope}
			  \fill (0,0) circle (2pt);		 
	              \node [above] (0,0) {$z_i$};
	               \node at (-1,1) {$S^1_{\e,i}$};
	                \node at (1.4,-1) {$\gamma$};                
	                 	 \begin{scope}[very thick,decoration={markings,mark=at position 0.2  with {\arrow{>}},mark=at position 0.85  with {\arrow{>}}}] 
             		\draw [very thick,postaction={decorate}] (0,0) to[out=-45,in=-135] (4,0);
                   \end{scope}
                    \node at (2,-0.3) {$\langle z_i, z_{i+1}\rangle$};          
	                	 \begin{scope}[very thick,decoration={markings,mark=at position 0.5  with {\arrow{>}},mark=at position 1  with {\arrow{>}}}] 
                \draw[postaction={decorate}] (3,0) arc (-180:170:1);            
                 \end{scope}	
			  \fill (4,0) circle (2pt);			 
	              \node at (4,0.2) {$z_{i+1}$};
	               \node at (5,1) {$S^1_{\e,i+1}$};           
	                   \node at (7.5,0) {$=-\frac{e^{2\pi i \alpha_i}}{e^{2\pi i \alpha_i}-1}-\frac{1}{e^{2\pi i \alpha_{i+1}}-1}$};
}

\newcommand{\zeroone}{
                 \begin{scope}[very thick,decoration={markings,mark=at position 0.4  with {\arrow{>}}}]   
                 \draw [postaction={decorate}] (0.5,0)--(3.5,0);    	
                   \end{scope}
			  \fill (0,0) circle (2pt);	 
	              \node [above] (0,0) {$z_i$};
	                \node at (.5,0.6) {$\gamma$};               
	                 	 \begin{scope}[very thick,decoration={markings,mark=at position 0.2  with {\arrow{>}},mark=at position 0.85  with {\arrow{>}}}] 
             		\draw [very thick,postaction={decorate}] (0,0) to[out=45,in=135] (1,0) to[out=-60,in=-135] (2,0) to[out=45,in=135] (3,0)to[out=-60,in=-135] (4,0);
                   \end{scope}
                   \node at (2,0.8) {$\langle z_i, z_{i+1}\rangle$};          
			  \fill (4,0) circle (2pt); 
	              \node at (4,0.2) {$z_{i+1}$};     
	              
\begin{scope}[very thick,decoration={markings,mark=at position 0.4  with {\arrow{>}}}]   
                 \draw [postaction={decorate}] (0.5,-1)--(3.5,-1);    	
                   \end{scope}
			  \fill (0,-1) circle (2pt);	 
	              \node [above] (0,-1) {$z_i$};
	                \node at (.5,-1.4) {$\gamma$};               
	                 	 \begin{scope}[very thick,decoration={markings,mark=at position 0.2  with {\arrow{>}},mark=at position 0.85  with {\arrow{>}}}] 
             		\draw [very thick,postaction={decorate}] (0,-1) to[out=-45,in=-135] (1,-1) to[out=60,in=135] (2,-1) to[out=-45,in=-135] (3,-1)to[out=60,in=135] (4,-1);
                   \end{scope}
                   \node at (2,-1.7) {$\langle z_i, z_{i+1}\rangle$};          
			  \fill (4,-1) circle (2pt); 
	              \node at (4,-0.7) {$z_{i+1}$};

	                 \node at (5.2,0) {$=$};
	                 \node at (-0.7,-1) {$=-$};
	                  \node at (5.5,-1) {$=-1$};
	                 \node at (-4,0) {$\left (\langle z_i,z_{i+1}\rangle\otimes U_{(z_i,z_{i+1})}\right )\cdot \left (\gamma \otimes U^{-1}_{(z_i,z_{i+1})}\right )=$};}

	\newcommand{\zeroonetwo}{
                 \begin{scope}[very thick,decoration={markings,mark=at position 0.4  with {\arrow{>}}}]   
                 \draw [postaction={decorate}] (0.5,0)--(3.5,0);    	
                   \end{scope}
			  \fill (0,0) circle (2pt);	 
	              \node [above] (0,0) {$z_i$};
	                \node at (.5,0.6) {$\gamma$};               
	                 	 \begin{scope}[very thick,decoration={markings,mark=at position 0.2  with {\arrow{>}},mark=at position 0.85  with {\arrow{>}}}] 
             		\draw [very thick,postaction={decorate}] (0,0) to[out=45,in=135] (1.34,0) to[out=-60,in=-135] (2.68,0) to[out=45,in=135] (4,0);
                   \end{scope}
                   \node at (2,0.8) {$\langle z_i, z_{i+1}\rangle$};          
			  \fill (4,0) circle (2pt); 
	              \node at (4,0.3) {$z_{i+1}$};     
	              \node at (5.5,0) {$=0$};
	              \node at (-4,0) {$\left (\langle z_i,z_{i+1}\rangle\otimes U_{(z_i,z_{i+1})}\right )\cdot \left (\gamma \otimes U^{-1}_{(z_i,z_{i+1})}\right )=$};
	              }

	                 \newcommand{\topinter}{ 
	                  \begin{scope}[very thick,decoration={markings,mark=at position 0.8  with {\arrow{>}}}]   
                 \draw [postaction={decorate}] (0,0)--(1,1);  
                 \draw [postaction={decorate}] (0,1)--(1,0);
                   \draw [postaction={decorate}] (5,0)--(6,1);  
                 \draw [postaction={decorate}] (6,0)--(5,1);         	
                   \end{scope}
                    \node at (-1.5,0.5) {$\gamma_1\cdot \gamma_2=$};
                    \node at (3.5,0.5) {$\gamma_1\cdot \gamma_2=$};
                    \node at (-0.4,1) {$\gamma_1$};
                   \node at (-0.4,0) {$\gamma_2$};
                   \node at (4.6,1) {$\gamma_1$};
                   \node at (4.6,0) {$\gamma_2$};
                   \node at (1.5,0.5) {$=+1$};
                   \node at (6.5,0.5) {$=-1$};
	                 
	                 }

\newcommand{\bumpfunction}{                        
                 \draw (1,0)[->]--(7,0); 
                  \draw [ very thick, red](2,0)--(6,0); 
                  \draw [very thick, blue](3,0)--(5,0); 
                 \fill (4,0) circle (2pt); 
                \draw[thick, scale=1, domain=1:7, smooth, samples=100, variable=\x] plot ({\x}, {0.3*exp(1-0.1*(\x-4)*(\x-4)*(\x-4)*(\x-4)*(\x-4)*(\x-4)});
                
                \fill (-3,0) circle (2pt); 
                \draw [red, thick] (-3,0) circle (2cm);
                \draw [blue, thick](-3,0) circle (1cm);
                 \node at(-3,-0.4) {$z_j$};
                     \node [blue] at (-1.8,-0.4) {$U_j$};
                      \node[red] at (-0.8,-0.4) {$V_j$};
                
                    \node at(4,-0.4) {$z_j$};
                       \node at(6,0.5) {$h_j$};
                     \node [blue] at (5.2,-0.4) {$U_j$};
                      \node[red] at (6.2,-0.4) {$V_j$};
                      
                     \draw [very thick,blue] (5,0) arc (0:10:1); 
                     \draw [very thick,blue] (5,0) arc (0:-10:1);   
                     \draw [very thick,blue](3,0) arc (10:0:1);  
                       \draw [very thick,blue](3,0) arc (-10:0:1); 
                           \draw [very thick,red](6,0) arc (0:10:1); 
                     \draw [very thick,red](6,0) arc (0:-10:1);   
                     \draw [very thick,red](2,0) arc (10:0:1);  
                       \draw [very thick,red](2,0) arc (-10:0:1); 
                   }

\newcommand{\amplitudeone}{                        
\draw (1,0.6) ellipse (0.1cm and 0.3cm);
\draw (1,-0.6) ellipse (0.1cm and 0.3cm);
\draw (3,0.6) ellipse (0.1cm and 0.3cm);
\draw (3,-0.6) ellipse (0.1cm and 0.3cm);
	\draw  (1.03,0.9) to[out=-45,in=-135] (2.97,0.9);
	\draw  (1.03,-0.9) to[out=45,in=135] (2.97,-0.9);
	
	\draw  (1.03,0.3) to[out=-25,in=90] (1.5,0) to[out=-90,in=25] (1.03,-0.3);
	\draw  (2.97,0.3) to[out=205,in=90] (2.5,0) to[out=-90,in=155] (2.97,-0.3);

	\draw (4,0.6) ellipse (0.1cm and 0.3cm);
\draw (4,-0.6) ellipse (0.1cm and 0.3cm);
\draw (6,0.6) ellipse (0.1cm and 0.3cm);
\draw (6,-0.6) ellipse (0.1cm and 0.3cm);
	\draw  (4.03,0.9) to[out=-45,in=-135] (5.97,0.9);
	\draw  (4.03,-0.9) to[out=45,in=135] (5.97,-0.9);
		\draw  (4.03,0.3) to[out=-25,in=90] (4.5,0) to[out=-90,in=25] (4.03,-0.3);
	\draw  (5.97,0.3) to[out=205,in=90] (5.5,0) to[out=-90,in=155] (5.97,-0.3);	
	\draw (5,0) ellipse (0.3cm and 0.3cm);

		\draw (7,0.6) ellipse (0.1cm and 0.3cm);
\draw (7,-0.6) ellipse (0.1cm and 0.3cm);
\draw (9,0.6) ellipse (0.1cm and 0.3cm);
\draw (9,-0.6) ellipse (0.1cm and 0.3cm);
	\draw  (7.03,0.9) to[out=-45,in=-135] (8.97,0.9);
	\draw  (7.03,-0.9) to[out=45,in=135] (8.97,-0.9);
	\draw  (7.03,0.3) to[out=-25,in=90] (7.5,0) to[out=-90,in=25] (7.03,-0.3);
	\draw  (8.97,0.3) to[out=205,in=90] (8.5,0) to[out=-90,in=155] (8.97,-0.3);	
	\draw (7.8,0) ellipse (0.18cm and 0.18cm);
	\draw (8.2,0) ellipse (0.18cm and 0.18cm);

      \node at(0,0) {$\mathcal{A}_4^{closed}=$};
           \node at(3.5,0) {$+$};
            \node at(6.5,0) {$+$};
             \node at(9.5,0) {$+......$};

                   }
                   
   \newcommand{\amplitudetwo}{                        
\draw (1,0.6) ellipse (0.1cm and 0.3cm);
\draw (1,-0.6) ellipse (0.1cm and 0.3cm);
\draw (3,0.6) ellipse (0.1cm and 0.3cm);
\draw (3,-0.6) ellipse (0.1cm and 0.3cm);
	\draw  (1.03,0.9) to[out=-45,in=-135] (2.97,0.9);
	\draw  (1.03,-0.9) to[out=45,in=135] (2.97,-0.9);
	
	\draw  (1.03,0.3) to[out=-25,in=90] (1.5,0) to[out=-90,in=25] (1.03,-0.3);
	\draw  (2.97,0.3) to[out=205,in=90] (2.5,0) to[out=-90,in=155] (2.97,-0.3);	
	\draw (3.5,0)[->]--(4,0);
	\draw (5.3,0) ellipse (1cm and 1cm);
	 \fill (5.9,0.6) circle (2pt); 
	 \fill (5.9,-0.6) circle (2pt); 
	 \fill (4.7,0.6) circle (2pt); 
	 \fill (4.7,-0.6) circle (2pt); 
	 \draw [domain=-20:20]   plot ({5.3+0.95*cos(\x)}, {0.95*sin(\x)});
	  \draw [domain=-10:10]   plot ({5.3+0.9*cos(\x)}, {0.9*sin(\x)});
	
	\draw (8.03,0.9) -- (8.03,0.3);
	\draw (8.03,-0.9) -- (8.03,-0.3);
	\draw (9.97,0.9) -- (9.97,0.3);
	\draw (9.97,-0.9) -- (9.97,-0.3);
	\draw  (8.03,0.9) to[out=-45,in=-135] (9.97,0.9);
	\draw  (8.03,-0.9) to[out=45,in=135] (9.97,-0.9);
	
	\draw  (8.03,0.3) to[out=-25,in=90] (8.5,0) to[out=-90,in=25] (8.03,-0.3);
	\draw  (9.97,0.3) to[out=205,in=90] (9.5,0) to[out=-90,in=155]  (9.97,-0.3);	
	
		\draw (10.5,0)[->]--(11,0);
	\draw (12.3,0) ellipse (1cm and 1cm);
	 \fill (13,0.7) circle (2pt); 
	 \fill (13,-0.7) circle (2pt); 
	 \fill (11.6,0.7) circle (2pt); 
	 \fill (11.6,-0.7) circle (2pt);

	}

\newcommand{\integralordering}{
\node at (0.5,1.5){$For\quad x_1<x_2<x_3$};
\node at (0.5,0){$x \in [-\infty,0]\sim$};
\node at (5.8,0){$x \in [0,1]\sim$};
\node at (10.8,0){$x \in [1,+\infty]\sim$};

\draw (3,0) ellipse (1cm and 1cm);
 \fill (3.7,0.7) circle (2pt); 
	 \fill (3.7,-0.7) circle (2pt); 
	 \fill (2.3,0.7) circle (2pt); 
	 \fill (2.3,-0.7) circle (2pt);
	  
	  \node[right] at(3.7,0.7) {$x_3$}; 
	\node[right] at(3.7,-0.7) {$x_2$}; 
	\node[left] at(2.3,0.7) {$x$}; 
	\node[left] at(2.3,-0.7) {$x_1$};

	\draw (8,0) ellipse (1cm and 1cm);
 \fill (8.7,0.7) circle (2pt); 
	 \fill (8.7,-0.7) circle (2pt); 
	 \fill (7.3,0.7) circle (2pt); 
	 \fill (7.3,-0.7) circle (2pt);
	  
	  \node[right] at(8.7,0.7) {$x_2$}; 
	\node[right] at(8.7,-0.7) {$x$}; 
	\node[left] at(7.3,0.7) {$x_3$}; 
	\node[left] at(7.3,-0.7) {$x_1$}; 

\draw (13,0) ellipse (1cm and 1cm);
 \fill (13.7,0.7) circle (2pt); 
	 \fill (13.7,-0.7) circle (2pt); 
	 \fill (12.3,0.7) circle (2pt); 
	 \fill (12.3,-0.7) circle (2pt);
	  
	  \node[right] at(13.7,0.7) {$x$}; 
	\node[right] at(13.7,-0.7) {$x_2$}; 
	\node[left] at(12.3,0.7) {$x_3$}; 
	\node[left] at(12.3,-0.7) {$x_1$};

	\node at (0.5,-1.5){$For\quad x_1<x_3<x_2$};
\node at (0.5,-3){$x \in [-\infty,0]\sim$};
\node at (5.8,-3){$x \in [0,1]\sim$};
\node at (10.8,-3){$x \in [1,+\infty]\sim$};

\draw (3,-3) ellipse (1cm and 1cm);
 \fill (3.7,-2.3) circle (2pt); 
	 \fill (3.7,-3.7) circle (2pt); 
	 \fill (2.3,-2.3) circle (2pt); 
	 \fill (2.3,-3.7) circle (2pt);
	  
	  \node[right] at(3.7,-2.3) {$x_2$}; 
	\node[right] at(3.7,-3.7) {$x_3$}; 
	\node[left] at(2.3,-2.3) {$x$}; 
	\node[left] at(2.3,-3.7) {$x_1$};

	\draw (8,-3) ellipse (1cm and 1cm);
 \fill (8.7,-2.3) circle (2pt); 
	 \fill (8.7,-3.7) circle (2pt); 
	 \fill (7.3,-2.3) circle (2pt); 
	 \fill (7.3,-3.7) circle (2pt);
	  
	  \node[right] at(8.7,-2.3) {$x_3$}; 
	\node[right] at(8.7,-3.7) {$x$}; 
	\node[left] at(7.3,-2.3) {$x_2$}; 
	\node[left] at(7.3,-3.7) {$x_1$}; 

\draw (13,-3) ellipse (1cm and 1cm);
 \fill (13.7,-2.3) circle (2pt); 
	 \fill (13.7,-3.7) circle (2pt); 
	 \fill (12.3,-2.3) circle (2pt); 
	 \fill (12.3,-3.7) circle (2pt);
	  
	  \node[right] at(13.7,-2.3) {$x$}; 
	\node[right] at(13.7,-3.7) {$x_3$}; 
	\node[left] at(12.3,-2.3) {$x_2$}; 
	\node[left] at(12.3,-3.7) {$x_1$};

}     

\newcommand{\riemann}{\draw [red,line width=0.4mm](7,0) [->]-- (10,0);
				\draw [blue,line width=0.4mm](7,0) [->]-- (7,-3);
				\draw [red,line width=0.4mm](7,-3) [->]-- (10,-3);
				\draw [blue,line width=0.4mm](10,0) [->]-- (10,-3);
				
				 \draw [greeen1] (9,0) -- (8,-1.5);
				 	 \draw [green] (9,-3) -- (8,-1.5);
		   	 
			   	 	 \fill ((8,-1.5) circle (2pt);
			   	 	 \fill ((9,0) circle (2pt);
			   	 	 \fill ((9,-3) circle (2pt);
			   	 
			   	 \node [red] at (8,0.3) {$a_1$};
			   	 \node [red] at (8,-3.3) {$a_1^{-1}$};
			   	 \node  [blue] at (6.7,-2) {$b_1$};
			   	 \node [blue] at (10.3,-2) {$b_1^{-1}$};
		
				\node  at (8,-1.8) {$P$};
					\node  at (9,0.3) {$x$};
						\node  at (9,-3.3) {$y(x)$};

	                 \draw [greeen1] (16.75,-1.5) to[out=90,in=-30] (16.5,-0.8);
	                 \draw [green] (16.75,-1.5) to[out=-80,in=40] (16.42,-2.5);
	                     \draw [green] (16.75,-1.5) to[out=-80,in=40] (16.6,-1.4);
	                 \draw [dashed,green] (16.2,-1.25) to[out=250,in=150] (16.42,-2.5);
	             
	              \draw[line width=0.4mm,rotate=-50] ($(10,11)+ (0:1.5cm and 0.5cm)$) arc   (0:360:0.5cm and 1cm);
	              \draw[line width=0.4mm,rotate=-50] ($(11,11)+ (0:1.5cm and 2cm)$) arc   (0:360:1.4cm and 2.8cm);
	                 \draw[red,line width=0.4mm,rotate=-50] ($(11,11)+ (0:1cm and 2cm)$) arc   (0:360:1cm and 2cm);
	              \draw[blue,line width=0.4mm] ($(13.3,-2.1)+ (0:1.5cm and 1cm)$) arc   (90:205:1cm and 1cm);
	              \draw[blue,line width=0.4mm,dashed,rotate=-30] ($(12.4,5.6)+ (0:1.5cm and 1cm)$) arc   (90:-90:0.5cm and 0.8cm);
	              
	               \fill ((16.75,-1.5) circle (2pt);
	                \fill ((16.5,-0.8) circle (2pt);
	               \node [right]  at (16.8,-1.5) {$x=y(x)$};
	                 \node at (16.5,-0.6) {$P$};}

	                  \newcommand{\sfera}{  \begin{scope}[very thick,decoration={markings,mark=at position 0.5  with {\arrow{>}},mark=at position 1  with {\arrow{>}}}] 
                \draw[postaction={decorate},red] (-1.6,0) arc (0:350:0.4);            
                 \end{scope}	
                  \begin{scope}[very thick,decoration={markings,mark=at position 0.4  with {\arrow{>}}}]   
                 \draw [postaction={decorate},red] (-1.6,0) to[out=-25,in=190] (-0.4,-0.3);    	
                   \end{scope}

	                	 \begin{scope}[very thick,decoration={markings,mark=at position 0.5  with {\arrow{>}},mark=at position 1  with {\arrow{>}}}] 
                \draw[postaction={decorate},red] (-0.4,-0.3) arc (-180:170:0.4);            
                 \end{scope}	                     

	\draw (0,0) ellipse (3cm and 3cm);
	 \fill (0,-0.3) circle (2pt); 
	 \fill (-2,0) circle (2pt); 
	 \fill (2,0) circle (2pt); 
	 \node [below] at (-2,0)  {$z_1$};
	 \node [below] at (0,-0.3)  {$z_2$};
	 \node [below] at (2,0)  {$z_3$};
	 \draw [domain=-20:20]   plot ({2.95*cos(\x)}, {2.95*sin(\x)});
	  \draw [domain=-10:10]   plot ({2.9*cos(\x)}, {2.9*sin(\x)});}

	    \newcommand{\multinterese}{   
	    \fill[processblue!20] (1.5,0)--(0,1.5)--(0,-0.65)--(1.5,0);
	 \fill[green!20] (0,-1.5)--(-0.3,-0.7)--(0,-0.65)--(0,-1.5);
	 \fill[orange!20] (-1.06,1.06)--(-0.3,-0.7)--(-1.06,-1.06)--(-1.06,1.06);

	 \draw [red,thick] (0,1.5) --(0,-1.5);
	  \draw [red,thick] (-1.06,1.06) --(0,-1.5);
	   \draw [red,thick] (-1.06,1.06) --(-1.06,-1.06);
	    \draw [red,thick] (-1.06,-1.06) --(1.5,0);
	     \draw [red,thick] (1.5,0) --(0,1.5);
	\draw (0,0) ellipse (1.5cm and 1.5cm);
	 \fill (0,1.5) circle (2pt); 
	 \fill (1.5,0) circle (2pt); 
	 \fill (0,-1.5) circle (2pt); 
	  \fill (-1.06,-1.06) circle (2pt); 
	 \fill (-1.06,1.06) circle (2pt); 
	 \node [above] at (0,1.5)  {$5$};
	 \node [right] at (1.5,0)  {$1$};
	 \node [below] at (0,-1.5)  {$2$};
	 \node [left] at (-1.06,-1.06)  {$3$};
	 \node [left] at (-1.06,1.06)  {$4$};
	 \draw (2.5,0) [->]--(3.5,0);
	 \fill [blue] (-0.3,-0.7) circle (2pt); 
	 \fill [blue](0,-0.65) circle (2pt); 
	 
	 \draw (6,0) ellipse (1.5cm and 1.5cm);
	 \fill (6,1.5) circle (2pt); 
	 \fill (7.5,0) circle (2pt); 
	 \fill (6,-1.5) circle (2pt); 
	  \fill (4.94,-1.06) circle (2pt); 
	 \fill (4.94,1.06) circle (2pt); 

	 \node [above] at (6,1.5)  {$5$};
	 \node [right] at (7.5,0)  {$1$};
	 \node [below] at (6,-1.5)  {$2$};
	 \node [left] at (4.94,-1.06)  {$3$};
	 \node [left] at (4.94,1.06)  {$4$};
	 \draw [red,thick] (6,1.5)--(7.5,0);
	 \draw [red,thick] (4.94,1.06)--(4.94,-1.06);
	 \draw [red,thick] (6,-1.5)--(6,-0.8);
	 
	 \draw [blue,thick] (4.94,0)--(6,-0.8);
	 \draw [blue,thick] (6,-0.8)--(6.6,0.7);
	 
	 \draw [red, thick, fill=processblue!20] (6.6,0.7) ellipse (0.3cm and 0.3cm);
	 \draw [red, thick, fill=orange!20] (4.94,0) ellipse (0.3cm and 0.3cm);
	 \draw [red, thick, fill=green!20] (6,-0.8) ellipse (0.3cm and 0.3cm);
	 \node at (9,0) {$= \frac{1}{\sin{s_{15}}}\frac{1}{\sin{s_{34}}}.$};
	 }

	  \newcommand{\winding}{
	  \node at (-3,0) {$w(12345|13542)=$};
	  \node at (2.5,0) {$=3$};
	  \draw (0,0) ellipse (1.5cm and 1.5cm);
	  \node [above] at (0,1.5)  {$5$};
	 \node [right] at (1.5,0)  {$1$};
	 \node [below] at (0,-1.5)  {$2$};
	 \node [left] at (-1.06,-1.06)  {$3$};
	 \node [left] at (-1.06,1.06)  {$4$}; 
	\fill (0,1.5) circle (2pt); 
	 \fill (1.5,0) circle (2pt); 
	 \fill (0,-1.5) circle (2pt); 
	  \fill (-1.06,-1.06) circle (2pt); 
	 \fill (-1.06,1.06) circle (2pt); 
	 
	 \begin{scope}[very thick,decoration={markings,mark=at position 1  with {\arrow{>}}}] 
               
                \draw [postaction={decorate},red,domain=0:-130]   plot ({1.3*cos(\x)}, {1.3*sin(\x)});
                \draw [postaction={decorate},red,domain=220:90]   plot ({1.3*cos(\x)}, {1.3*sin(\x)});
                \draw [postaction={decorate},red,domain=85:-220]   plot ({1.15*cos(\x)}, {1.15*sin(\x)});
                \draw [postaction={decorate},red,domain=135:-90]   plot ({1*cos(\x)}, {1*sin(\x)});
                 \draw [postaction={decorate},red,domain=-90:-360]   plot ({0.85*cos(\x)}, {0.85*sin(\x)});
                
                 \end{scope}}

 \newcommand{\multinteresedue}{   
	    \fill[processblue!20] (1.06,1.06)--(1.5,0)--(1.06,-1.06)--(-0.4,0.4)--(1.06,1.06);
	 \fill[green!20] (-1.06,1.06)--(-0.4,0.4)--(-1.06,0.2)--(-1.06,1.06);
	 \fill[orange!20] (-1.5,0)--(-1.06,-1.06)--(-1.06,0.2)--(-1.5,0);

	 \draw [red,thick] (1.06,1.06) --(1.5,0);
	  \draw [red,thick] (1.06,-1.06) --(1.5,0);
	   \draw [red,thick] (1.06,-1.06) --(-1.06,1.06);
	    \draw [red,thick] (-1.06,1.06) --(-1.06,-1.06);
	     \draw [red,thick] (-1.06,-1.06) --(-1.5,0);
	      \draw [red,thick] (1.06,1.06) --(-1.5,0);
	      \fill [blue] (-0.4,0.4) circle (2pt); 
	 \fill [blue](-1.06,0.2) circle (2pt);

	\draw (0,0) ellipse (1.5cm and 1.5cm);
	 \fill (1.06,1.06) circle (2pt); 
	 \fill (1.5,0) circle (2pt); 
	 \fill (1.06,-1.06) circle (2pt); 
	  \fill (-1.06,-1.06) circle (2pt); 
	 \fill (-1.06,1.06) circle (2pt);
	 \fill (-1.5,0) circle (2pt);

	 \node [above] at (1.06,1.06)  {$1$};
	 \node [right] at (1.5,0)  {$2$};
	 \node [below] at (1.06,-1.06)  {$3$};
	 \node [below] at (-1.06,-1.06)  {$4$};
	 \node [above] at (-1.06,1.06)  {$6$};
	 \node [left] at (-1.5,0)  {$5$};
	 
	 \draw (2.5,0) [->]--(3.5,0);

\draw (6,0) ellipse (1.5cm and 1.5cm);
	 \fill (7.06,1.06) circle (2pt); 
	 \fill (7.5,0) circle (2pt); 
	 \fill (7.06,-1.06) circle (2pt); 
	  \fill (4.94,-1.06) circle (2pt); 
	 \fill (4.94,1.06) circle (2pt);
	 \fill (4.5,0) circle (2pt);

	 \node [above] at (7.06,1.06)  {$1$};
	 \node [right] at (7.5,0)  {$2$};
	 \node [below] at (7.06,-1.06)  {$3$};
	 \node [below] at (4.94,-1.06)  {$4$};
	 \node [above] at (4.94,1.06)  {$6$};
	 \node [left] at (4.5,0)  {$5$};

	 \draw [blue,thick] (6.6,0)--(5.5,0.8);
	 \draw [blue,thick] (5.2,-0.5)--(5.5,0.8);
	 
	 \draw [red,thick] (7.06,1.06) --(6.6,0);
	  \draw [red,thick] (7.06,-1.06) --(6.6,0);
	   \draw [red,thick] (7.5,0) --(6.6,0);
	    \draw [red,thick] (4.94,-1.06) --(5.2,-0.5);
	     \draw [red,thick] (4.5,0) --(5.2,-0.5);
	      \draw [red,thick] (4.94,1.06) --(5.5,0.8);

	 \draw [red, thick, fill=processblue!20] (6.6,0) ellipse (0.3cm and 0.3cm);
	 \draw [red, thick, fill=orange!20] (5.2,-0.5) ellipse (0.3cm and 0.3cm);
	 \draw [red, thick, fill=green!20] (5.5,0.8) ellipse (0.3cm and 0.3cm);}

	 \newcommand{\multiselfinter}{   
	    \fill[processblue!20] (-1.06,1.06)--(0,1.5)--(1.06,1.06)--(1.5,0)--(1.06,-1.06)--(0,-1.5)--(-1.06,-1.06)--(0,0)--(-1.06,1.06);
	    
	   \draw[dashed, processblue!20, pattern=north east lines,pattern color=processblue!20] (-1.06,1.06)--(0,0)--(-1.25,0.5)--(-1.06,1.06);
	    \draw[dashed, processblue!20, pattern=north east lines,pattern color=processblue!20] (-1.06,-1.06)--(0,0)--(-1.25,-0.5)--(-1.06,-1.06);
	    
	    \draw [domain=-150:130]   plot ({1.5*cos(\x)}, {1.5*sin(\x)});
	    \draw [dashed, domain=90:210]   plot ({1.5*cos(\x)}, {1.5*sin(\x)});

\draw [red,thick] (0,1.5) --(1.06,1.06);
	 \draw [red,thick] (1.06,1.06) --(1.5,0);
	  \draw [red,thick] (1.06,-1.06) --(1.5,0);
	   \draw [red,thick] (1.06,-1.06) --(0,-1.5);
	   \draw [red,thick] (-1.06,-1.06) --(0,-1.5);
	   \draw [red,thick] (0,1.5) --(-1.06,1.06);
	    \draw [red,thick,dashed] (-1.06,-1.06) --(-1.25,-0.5);
	    \draw [red,thick,dashed] (-1.06,1.06) --(-1.25,0.5);

	     \fill (0,1.5) circle (2pt); 
	 \fill (1.06,1.06) circle (2pt); 
	 \fill (1.5,0) circle (2pt); 
	 \fill (1.06,-1.06) circle (2pt); 
	  \fill (-1.06,-1.06) circle (2pt); 
	 \fill (0,-1.5) circle (2pt); 
	 \fill (-1.06,1.06) circle (2pt);
	 
	 \node [above] at (0,1.5)  {$1$};
	 \node [above] at (1.06,1.06)  {$2$};
	 \node [right] at (1.5,0)  {$3$};
	 \node [below] at (1.06,-1.06)  {$4$};
	 \node [below] at (-1.06,-1.06)  {$6$};
	 \node [below] at (0,-1.5)  {$5$};
	 \node [above] at (-1.06,1.06)  {$n$};

	 \draw (2.5,0) [->]--(3.5,0);

	  \draw [domain=-150:130]   plot ({6+1.5*cos(\x)}, {1.5*sin(\x)});
	    \draw [dashed, domain=90:210]   plot ({6+1.5*cos(\x)}, {1.5*sin(\x)});
	 
     \draw [red,thick] (6,1.5) --(6,0);
     
	 \draw [red,thick] (7.06,1.06) --(6,0);
	  \draw [red,thick] (7.06,-1.06) --(6,0);
	   \draw [red,thick] (7.5,0) --(6,0);
	    \draw [red,thick] (4.94,-1.06) --(6,0);
	       \draw [red,thick] (4.94,1.06) --(6,0);
	     \draw [red,thick] (6,-1.5) --(6,0);

	         \fill (6,1.5) circle (2pt); 
	 \fill (7.06,1.06) circle (2pt); 
	 \fill (7.5,0) circle (2pt); 
	 \fill (7.06,-1.06) circle (2pt); 
	  \fill (4.94,-1.06) circle (2pt); 
	 \fill (6,-1.5) circle (2pt); 
	 \fill (4.94,1.06) circle (2pt);
	 
	     \node [above] at (6,1.5)  {$1$};
	 \node [above] at (7.06,1.06)  {$2$};
	 \node [right] at (7.5,0)  {$3$};
	 \node [below] at (7.06,-1.06)  {$4$};
	 \node [below] at (4.94,-1.06)  {$6$};
	 \node [below] at (6,-1.5)  {$5$};
	 \node [above] at (4.94,1.06)  {$n$};

	  \draw [thick,red,domain=-40:140, fill=processblue!20]   plot ({6+0.5*cos(\x)}, {0.5*sin(\x)});
	  \draw [thick, red,domain=-135:45, fill=processblue!20]   plot ({6+0.5*cos(\x)}, {0.5*sin(\x)});
	    \draw [red,thick, dashed, domain=0:360,pattern=north east lines,pattern color=processblue!20]   plot ({6+0.5*cos(\x)}, {0.5*sin(\x)});
	     \draw [thick,red,domain=-40:140, fill=processblue!20]   plot ({6+0.5*cos(\x)}, {0.5*sin(\x)});
	  \draw [thick, red,domain=-140:45, fill=processblue!20]   plot ({6+0.5*cos(\x)}, {0.5*sin(\x)});
	  \node at (8,0) {$+$};
	   \node at (12,0) {$+$};
	  
	  \draw [domain=-150:130]   plot ({10+1.5*cos(\x)}, {1.5*sin(\x)});
	    \draw [dashed, domain=90:210]   plot ({10+1.5*cos(\x)}, {1.5*sin(\x)});

	         \fill (10,1.5) circle (2pt); 
	 \fill (11.06,1.06) circle (2pt); 
	 \fill (11.5,0) circle (2pt); 
	 \fill (11.06,-1.06) circle (2pt); 
	  \fill (8.94,-1.06) circle (2pt); 
	 \fill (10,-1.5) circle (2pt); 
	 \fill (8.94,1.06) circle (2pt);
	     
	     \node [above] at (10,1.5)  {$1$};
	 \node [above] at (11.06,1.06)  {$2$};
	 \node [right] at (11.5,0)  {$3$};
	 \node [below] at (11.06,-1.06)  {$4$};
	 \node [below] at (8.94,-1.06)  {$6$};
	 \node [below] at (10,-1.5)  {$5$};
	 \node [above] at (8.94,1.06)  {$n$};
	 \draw[red,thick] (9.5,0.4)--(10.5,-0.4);
	 \draw[red,thick] (11,-0.2)--(10,-0.6);
	 
	  \draw [thick,red,domain=0:360, fill=processblue!20]   plot ({10.5+0.3*cos(\x)}, {-0.4+0.3*sin(\x)});
	  \draw [thick,red,domain=-40:140, fill=processblue!20]   plot ({9.5+0.3*cos(\x)}, {0.4+0.3*sin(\x)});
	  \draw [thick, red,domain=-135:45, fill=processblue!20]   plot ({9.5+0.3*cos(\x)}, {0.4+0.3*sin(\x)});
	    \draw [red,thick, dashed, domain=0:360,pattern=north east lines,pattern color=processblue!20]   plot ({9.5+0.3*cos(\x)}, {0.4+0.3*sin(\x)});
	     \draw [thick,red,domain=-40:140, fill=processblue!20]   plot ({9.5+0.3*cos(\x)}, {0.4+0.3*sin(\x)});
	  \draw [thick, red,domain=-140:45, fill=processblue!20]   plot ({9.5+0.3*cos(\x)}, {0.4+0.3*sin(\x)});
	  
	 \node at (9.5,0.8) {$n-2$};
	 
	 \draw [domain=-150:130]   plot ({6+1.5*cos(\x)}, {-4+1.5*sin(\x)});
	    \draw [dashed, domain=90:210]   plot ({6+1.5*cos(\x)}, {-4+1.5*sin(\x)});
	 
      \fill (6,-2.5) circle (2pt); 
	 \fill (7.06,-2.94) circle (2pt); 
	 \fill (7.5,-4) circle (2pt); 
	 \fill (7.06,-5.06) circle (2pt); 
	  \fill (4.94,-5.06) circle (2pt); 
	 \fill (6,-5.5) circle (2pt); 
	 \fill (4.94,-2.94) circle (2pt);
	 
	     \node [above] at (6,-2.5)  {$1$};
	 \node [above] at (7.06,-2.94)  {$2$};
	 \node [right] at (7.5,-4)  {$3$};
	 \node [below] at (7.06,-5.06)  {$4$};
	 \node [below] at (4.94,-5.06)  {$6$};
	 \node [below] at (6,-5.5)  {$5$};
	 \node [above] at (4.94,-2.94)  {$n$};
	 
	 \draw[red,thick] (5.4,-3.5)--(6.5,-4.5);
	 \draw[red,thick] (6.5,-4.5)--(7,-4);
	 \draw[red,thick] (6.5,-4.5)--(6,-5);

	  \draw [thick,red,domain=-40:140, fill=processblue!20]    plot ({5.4+0.3*cos(\x)}, {-3.5+0.3*sin(\x)});
	  \draw [thick, red,domain=-135:45, fill=processblue!20]    plot ({5.4+0.3*cos(\x)}, {-3.5+0.3*sin(\x)});
	    \draw [red,thick, dashed, domain=0:360,pattern=north east lines,pattern color=processblue!20]    plot ({5.4+0.3*cos(\x)}, {-3.5+0.3*sin(\x)});
	     \draw [thick,red,domain=-40:140, fill=processblue!20]    plot ({5.4+0.3*cos(\x)}, {-3.5+0.3*sin(\x)});
	  \draw [thick, red,domain=-140:45, fill=processblue!20]    plot ({5.4+0.3*cos(\x)}, {-3.5+0.3*sin(\x)});
	  \draw [thick,red,domain=0:360, fill=processblue!20]   plot ({6.5+0.3*cos(\x)}, {-4.5+0.3*sin(\x)});
	  \node at (4,-4) {$+$};
	 \node at (5.5,-3) {$n-4$};
	  \node at (8.5,-4) {$+...$};
	  
	  }

 \newcommand{\fourselfinter}{
 \draw[red, thick, fill=processblue!20] (1.06,1.06)--(1.06,-1.06)--(-1.06,-1.06)--(-1.06,1.06)--(1.06,1.06);
 
\draw (0,0) ellipse (1.5cm and 1.5cm);
	 \fill (1.06,1.06) circle (2pt); 
	 \fill (1.06,-1.06) circle (2pt); 
	 \fill (-1.06,-1.06) circle (2pt); 
	  \fill (-1.06,1.06) circle (2pt); 
	 
	 \node [above] at (1.06,1.06)  {$1$};
	 \node [right] at (1.06,-1.06)  {$2$};
	 \node [below] at (-1.06,-1.06)  {$3$};
	 \node [above] at (-1.06,1.06)  {$4$};
     \draw (2.5,0) [->]--(3.5,0);

     \draw (6,0) ellipse (1.5cm and 1.5cm);
	 \fill (7.06,1.06) circle (2pt); 
	 \fill (7.06,-1.06) circle (2pt); 
	 \fill (4.94,-1.06) circle (2pt); 
	  \fill (4.94,1.06) circle (2pt); 
	 
	 \node [above] at (7.06,1.06)  {$1$};
	 \node [right] at (7.06,-1.06)  {$2$};
	 \node [below] at (4.94,-1.06)  {$3$};
	 \node [above] at (4.94,1.06)  {$4$};
	 
	 \draw [red, thick] (7.06,1.06)--(6.7,0);
	 \draw [red, thick] (7.06,-1.06)--(6.7,0);
	 \draw [red, thick] (4.94,1.06)--(5.3,0);
	 \draw [red, thick] (4.94,-1.06)--(5.3,0);
	 
	 \draw [red, thick] (5.3,0)--(6.7,0);
	 
	 \draw [red, thick, fill=processblue!20] (5.4,0) ellipse (0.3cm and 0.3cm);
	 \draw [red, thick, fill=processblue!20] (6.6,0) ellipse (0.3cm and 0.3cm);

	 \node at (8,0) {$+$};
	 
	 \draw (10,0) ellipse (1.5cm and 1.5cm);
	 \fill (11.06,1.06) circle (2pt); 
	 \fill (11.06,-1.06) circle (2pt); 
	 \fill (8.94,-1.06) circle (2pt); 
	  \fill (8.94,1.06) circle (2pt); 
	 
	 \node [above] at (11.06,1.06)  {$1$};
	 \node [right] at (11.06,-1.06)  {$2$};
	 \node [below] at (8.94,-1.06)  {$3$};
	 \node [above] at (8.94,1.06)  {$4$};
	 
	 \draw [red, thick] (11.06,1.06)--(10,0.6);
	 \draw [red, thick] (8.94,1.06)--(10,0.6);
	 \draw [red, thick] (11.06,-1.06)--(10,-0.6);
	 \draw [red, thick] (8.94,-1.06)--(10,-0.6);
	 
	 \draw [red, thick] (10,0.6)--(10,-0.6);
	 
	 \draw [red, thick, fill=processblue!20] (10,0.6) ellipse (0.3cm and 0.3cm);
	 \draw [red, thick, fill=processblue!20] (10,-0.6) ellipse (0.3cm and 0.3cm);
 
 }

  \newcommand{\fiveselfinter}{
 \draw[red, thick, fill=processblue!20] (0,1.5)--(1.43,0.46)--(1.06,-1.06)--(-1.06,-1.06)--(-1.43,0.46)--(0,1.5);
 
\draw (0,0) ellipse (1.5cm and 1.5cm);
	 \fill (0,1.5) circle (2pt); 
	 \fill (1.43,0.46) circle (2pt); 
	 \fill (1.06,-1.06) circle (2pt); 
	 \fill (-1.06,-1.06) circle (2pt);
	  \fill (-1.43,0.46) circle (2pt); 
	 \node [above] at (0,1.5)  {$1$};
	 \node [right] at (1.43,0.46)  {$2$};
	 \node [below] at (1.06,-1.06)  {$3$};
	 \node [below] at (-1.06,-1.06)  {$4$};
	 \node [above] at (-1.43,0.46)  {$5$};
     \draw (2.5,0) [->]--(3.5,0);

     \draw (6,0) ellipse (1.5cm and 1.5cm);
	 \fill (6,1.5) circle (2pt); 
	 \fill (7.43,0.46) circle (2pt); 
	 \fill (7.06,-1.06) circle (2pt); 
	 \fill (4.94,-1.06) circle (2pt);
	  \fill (4.57,0.46) circle (2pt); 
	 
	\node [above] at (6,1.5)  {$1$};
	 \node [right] at (7.43,0.46)  {$2$};
	 \node [below] at (7.06,-1.06)  {$3$};
	 \node [below] at (4.94,-1.06)  {$4$};
	 \node [above] at (4.57,0.46)  {$5$};

	 \draw [red, thick] (6,1.5)--(6,0.46);
	 \draw [red, thick] (7.43,0.46)--(6,0.46);
	 
	 \draw [red, thick] (7.06,-1.06)--(6.6,-0.4);
	 
	 \draw [red, thick] (4.94,-1.06)--(5.3,-0.4);
	 \draw [red, thick] (4.57,0.46)--(5.3,-0.4);
	 
	 \draw [red, thick] (5.4,-0.4)--(6.6,-0.4);
	 \draw [red, thick] (6.6,-0.4)--(6,0.5);
	 
	 \draw [red, thick, fill=processblue!20] (5.4,-0.4) ellipse (0.3cm and 0.3cm);
	 \draw [red, thick, fill=processblue!20] (6.6,-0.4) ellipse (0.3cm and 0.3cm);
	 \draw [red, thick, fill=processblue!20] (6,0.5) ellipse (0.3cm and 0.3cm);

	 \node at (8,0) {$+$};
	 
	      \draw (10,0) ellipse (1.5cm and 1.5cm);
	 \fill (10,1.5) circle (2pt); 
	 \fill (11.43,0.46) circle (2pt); 
	 \fill (11.06,-1.06) circle (2pt); 
	 \fill (8.94,-1.06) circle (2pt);
	  \fill (8.57,0.46) circle (2pt); 
	 
	\node [above] at (10,1.5)  {$1$};
	 \node [right] at (11.43,0.46)  {$2$};
	 \node [below] at (11.06,-1.06)  {$3$};
	 \node [below] at (8.94,-1.06)  {$4$};
	 \node [above] at (8.57,0.46)  {$5$};

	 \draw [red, thick] (10,1.5)--(10,0.46);
	 \draw [red, thick] (11.43,0.46)--(10.6,-0.4);
	 
	 \draw [red, thick] (11.06,-1.06)--(10.6,-0.4);
	 
	 \draw [red, thick] (8.94,-1.06)--(9.3,-0.4);
	 \draw [red, thick] (8.57,0.46)--(9.3,-0.4);
	 
	 \draw [red, thick] (9.4,-0.4)--(10,0.5);
	 \draw [red, thick] (10.6,-0.4)--(10,0.5);
	 
	 \draw [red, thick, fill=processblue!20] (9.4,-0.4) ellipse (0.3cm and 0.3cm);
	 \draw [red, thick, fill=processblue!20] (10.6,-0.4) ellipse (0.3cm and 0.3cm);
	 \draw [red, thick, fill=processblue!20] (10,0.5) ellipse (0.3cm and 0.3cm);
	 
	 \node at (12,0) {$+$};
 
 \draw (6,-4) ellipse (1.5cm and 1.5cm);
	 \fill (6,-2.5) circle (2pt); 
	 \fill (7.43,-3.54) circle (2pt); 
	 \fill (7.06,-5.06) circle (2pt); 
	 \fill (4.94,-5.06) circle (2pt);
	  \fill (4.57,-3.54) circle (2pt); 
	 
	\node [above] at (6,-2.5) {$1$};
	 \node [right] at (7.43,-3.54)  {$2$};
	 \node [below] at (7.06,-5.06)  {$3$};
	 \node [below] at (4.94,-5.06)  {$4$};
	 \node [above] at (4.57,-3.54)  {$5$};

	 \draw [red, thick] (6,-2.5)--(6,-3.54);
	  \draw [red, thick]  (4.57,-3.54)--(6,-3.54);
	  \draw [red, thick] (4.94,-5.06)-- (5.4,-4.4);
	  \draw [red, thick] (7.43,-3.54)--(6.6,-4.4);
	  	  \draw [red, thick] (7.06,-5.06)--(6.6,-4.4);

	 \draw [red, thick] (5.4,-4.4)--(6.6,-4.4);
	 \draw [red, thick] (5.4,-4.4)--(6,-3.5);
	 
	 \draw [red, thick, fill=processblue!20] (5.4,-4.4) ellipse (0.3cm and 0.3cm);
	 \draw [red, thick, fill=processblue!20] (6.6,-4.4) ellipse (0.3cm and 0.3cm);
	 \draw [red, thick, fill=processblue!20] (6,-3.5) ellipse (0.3cm and 0.3cm);
	 
  \node at (12,-4) {$+$};
  \node at (8,-4) {$+$};
  \node at (8,-8) {$+$};
 
 \draw (10,-4) ellipse (1.5cm and 1.5cm);
	 \fill (10,-2.5) circle (2pt); 
	 \fill (11.43,-3.54) circle (2pt); 
	 \fill (11.06,-5.06) circle (2pt); 
	 \fill (8.94,-5.06) circle (2pt);
	  \fill (8.57,-3.54) circle (2pt); 
	 
	\node [above] at (10,-2.5) {$1$};
	 \node [right] at (11.43,-3.54)  {$2$};
	 \node [below] at (11.06,-5.06)  {$3$};
	 \node [below] at (8.94,-5.06)  {$4$};
	 \node [above] at (8.57,-3.54)  {$5$};

	 \draw [red, thick] (10,-2.5)--(10,-3.54);
	  \draw [red, thick]  (8.57,-3.54)--(10,-3.54);
	  \draw [red, thick] (8.94,-5.06)-- (9.4,-4.4);
	  \draw [red, thick] (11.43,-3.54)--(10.6,-4.4);
	  	  \draw [red, thick] (11.06,-5.06)--(9.4,-4.4);

	 \draw [red, thick] (9.4,-4.4)--(10.6,-4.4);
	 \draw [red, thick] (10.6,-4.4)--(10,-3.5);
	 
	 \draw [red, thick, fill=processblue!20] (9.4,-4.4) ellipse (0.3cm and 0.3cm);
	 \draw [red, thick, fill=processblue!20] (10.6,-4.4) ellipse (0.3cm and 0.3cm);
	 \draw [red, thick, fill=processblue!20] (10,-3.5) ellipse (0.3cm and 0.3cm);
 
 \draw (6,-8) ellipse (1.5cm and 1.5cm);
	 \fill (6,-6.5) circle (2pt); 
	 \fill (7.43,-7.54) circle (2pt); 
	 \fill (7.06,-9.06) circle (2pt); 
	 \fill (4.94,-9.06) circle (2pt);
	  \fill (4.57,-7.54) circle (2pt); 
	 
	\node [above] at (6,-6.5) {$1$};
	 \node [right] at (7.43,-7.54)  {$2$};
	 \node [below] at (7.06,-9.06)  {$3$};
	 \node [below] at (4.94,-9.06)  {$4$};
	 \node [above] at (4.57,-7.54)  {$5$};

	 \draw [red, thick] (6,-6.5)--(6,-7.54);
	  \draw [red, thick]  (4.57,-7.54)--(5.4,-8.4);
	  \draw [red, thick] (4.94,-9.06)-- (6.6,-8.4);
	  \draw [red, thick] (7.43,-7.54)--(6,-7.54);
	  	  \draw [red, thick] (7.06,-9.06)--(6.6,-8.4);

	 \draw [red, thick] (5.4,-8.4)--(6.6,-8.4);
	 \draw [red, thick] (5.4,-8.4)--(6,-7.5);
	 
	 \draw [red, thick, fill=processblue!20] (5.4,-8.4) ellipse (0.3cm and 0.3cm);
	 \draw [red, thick, fill=processblue!20] (6.6,-8.4) ellipse (0.3cm and 0.3cm);
	 \draw [red, thick, fill=processblue!20] (6,-7.5) ellipse (0.3cm and 0.3cm);
	 
	  \draw (10,-8) ellipse (1.5cm and 1.5cm);
	 \fill (10,-6.5) circle (2pt); 
	 \fill (11.43,-7.54) circle (2pt); 
	 \fill (11.06,-9.06) circle (2pt); 
	 \fill (8.94,-9.06) circle (2pt);
	  \fill (8.57,-7.54) circle (2pt); 
	 
	\node [above] at (10,-6.5) {$1$};
	 \node [right] at (11.43,-7.54)  {$2$};
	 \node [below] at (11.06,-9.06)  {$3$};
	 \node [below] at (8.94,-9.06)  {$4$};
	 \node [above] at (8.57,-7.54)  {$5$};

	 \draw [red, thick] (10,-6.5)--(10,-8);
	  \draw [red, thick]  (8.57,-7.54)--(10,-8);
	  \draw [red, thick] (8.94,-9.06)-- (10,-8);
	  \draw [red, thick] (11.43,-7.54)--(10,-8);
	  	  \draw [red, thick] (11.06,-9.06)--(10,-8);

	 \draw [red, thick, fill=processblue!20] (10,-8) ellipse (0.3cm and 0.3cm);

 }  
 
  \newcommand{\vanishingone}{
 
\draw (0,0) ellipse (1.5cm and 1.5cm);
\fill[orange] (-0.45,0.46)--(0.45,0.46)--(0.65,0)--(0,-0.4)--(-0.65,0)--(-0.45,0.46);
 \draw[red,thick] (0,1.5)-- (-1.06,-1.06)--(1.43,0.46)--(-1.43,0.46)--(1.06,-1.06)--(0,1.5);
	 \fill (0,1.5) circle (2pt); 
	 \fill (1.43,0.46) circle (2pt); 
	 \fill (1.06,-1.06) circle (2pt); 
	 \fill (-1.06,-1.06) circle (2pt);
	  \fill (-1.43,0.46) circle (2pt); 
	 \node [above] at (0,1.5)  {$1$};
	 \node [right] at (1.43,0.46)  {$2$};
	 \node [below] at (1.06,-1.06)  {$3$};
	 \node [below] at (-1.06,-1.06)  {$4$};
	 \node [above] at (-1.43,0.46)  {$5$};
	 \node at (-3.5,0){$m(12345|13524)=$};
	  \node at (2,0){$=0.$};}
     
  \newcommand{\vanishingtwo}{
 
\draw (0,0) ellipse (1.5cm and 1.5cm);
\draw (6.5,0) ellipse (1.5cm and 1.5cm);
\fill[orange] (-0.45,0.46)--(0.45,0.46)--(0.65,0)--(0,-0.4)--(-0.65,0)--(-0.45,0.46);
\fill[orange] (6.05,0.46)--(6.95,0.46)--(7.15,0)--(6.5,-0.4)--(5.85,0)--(6.05,0.46);
 \draw[red,thick](1.06,-1.06)--(0,1.5)--(-1.06,-1.06)--(1.43,0.46)--(-1.43,0.46)--(1.06,-1.06) ;
 \draw[red,thick](7.56,-1.06)--(6.5,1.5)--(5.44,-1.06)--(7.93,0.46)--(5.07,0.46)--(7.56,-1.06);
	 \fill (0,1.5) circle (2pt); 
	 \fill (1.43,0.46) circle (2pt); 
	 \fill (1.06,-1.06) circle (2pt); 
	 \fill (-1.06,-1.06) circle (2pt);
	  \fill (-1.43,0.46) circle (2pt); 
	 \node [above] at (0,1.5)  {$1$};
	 \node [right] at (1.43,0.46)  {$2$};
	 \node [below] at (1.06,-1.06)  {$3$};
	 \node [below] at (-1.06,-1.06)  {$4$};
	 \node [above] at (-1.43,0.46)  {$5$};
	 \fill (6.5,1.5) circle (2pt); 
	 \fill (7.93,0.46) circle (2pt); 
	 \fill (7.56,-1.06) circle (2pt); 
	 \fill (5.44,-1.06) circle (2pt);
	  \fill (5.07,0.46) circle (2pt); 
	 \node [above] at (6.6,1.5)  {$1$};
	 \node [right] at (7.93,0.46)  {$3$};
	 \node [below] at (7.56,-1.06)  {$2$};
	 \node [below] at (5.44,-1.06)  {$4$};
	 \node [above] at (5.07,0.46)  {$5$};
	 \node at (-3.5,0){$m(12345|31425)=$};
	  \node at (3.2,0){$=m(13245|21435)=$};
	    \node at (8.5,0){$=0,$};}

  	    \newcommand{\multintereseone}{   
  	    \node at (2.5,0) {$m(12345|21435)=$};
  	    \node at (10,0) {$=\frac{1}{\sin{\pi\alpha'k_1k_2}\sin{\pi\alpha'k_3k_4}},$};
	 \draw (6,0) ellipse (1.5cm and 1.5cm);
	 \fill (6,1.5) circle (2pt); 
	 \fill (7.5,0) circle (2pt); 
	 \fill (6,-1.5) circle (2pt); 
	  \fill (4.94,-1.06) circle (2pt); 
	 \fill (4.94,1.06) circle (2pt); 

	 \node [above] at (6,1.5)  {$1$};
	 \node [right] at (7.5,0)  {$2$};
	 \node [below] at (6,-1.5)  {$3$};
	 \node [left] at (4.94,-1.06)  {$4$};
	 \node [left] at (4.94,1.06)  {$5$};
	 \draw [red,thick] (6,1.5)--(7.5,0);
	 \draw [red,thick] (6,-0.8)--(4.94,-1.06);
	 \draw [red,thick] (5.2,0.5)--(4.94,1.06);
	 \draw [red,thick] (6,-1.5)--(6,-0.8);
	 
	 \draw [blue,thick] (5.2,0.5)--(6,-0.8);
	 \draw [blue,thick] (5.2,0.5)--(6.6,0.7);
	 
	 \draw [red, thick, fill=processblue!20] (6.6,0.7) ellipse (0.3cm and 0.3cm);
	 \draw [red, thick, fill=orange!20] (5.2,0.5) ellipse (0.3cm and 0.3cm);
	 \draw [red, thick, fill=green!20] (6,-0.8) ellipse (0.3cm and 0.3cm);
	 }

  	    \newcommand{\multinteresetwo}{   
  	    \node at (2.5,0) {$m(13245|31425)=$};
  	    \node at (10,0) {$=\frac{1}{\sin{\pi\alpha'k_1k_3}\sin{\pi\alpha'k_2k_4}}.$};
	 \draw (6,0) ellipse (1.5cm and 1.5cm);
	 \fill (6,1.5) circle (2pt); 
	 \fill (7.5,0) circle (2pt); 
	 \fill (6,-1.5) circle (2pt); 
	  \fill (4.94,-1.06) circle (2pt); 
	 \fill (4.94,1.06) circle (2pt); 

	 \node [above] at (6,1.5)  {$1$};
	 \node [right] at (7.5,0)  {$3$};
	 \node [below] at (6,-1.5)  {$2$};
	 \node [left] at (4.94,-1.06)  {$4$};
	 \node [left] at (4.94,1.06)  {$5$};
	 \draw [red,thick] (6,1.5)--(7.5,0);
	 \draw [red,thick] (6,-0.8)--(4.94,-1.06);
	 \draw [red,thick] (5.2,0.5)--(4.94,1.06);
	 \draw [red,thick] (6,-1.5)--(6,-0.8);
	 
	 \draw [blue,thick] (5.2,0.5)--(6,-0.8);
	 \draw [blue,thick] (5.2,0.5)--(6.6,0.7);
	 
	 \draw [red, thick, fill=processblue!20] (6.6,0.7) ellipse (0.3cm and 0.3cm);
	 \draw [red, thick, fill=orange!20] (5.2,0.5) ellipse (0.3cm and 0.3cm);
	 \draw [red, thick, fill=green!20] (6,-0.8) ellipse (0.3cm and 0.3cm);
	 }

	  \newcommand{\complex}{   
	  
  	    \draw [thick, fill=processblue!20] (0,0) ellipse (1.5cm and 3cm);
  	     \draw [thick, fill=gray!20] (8,0) ellipse (1.5cm and 3cm);
  	    \draw [thick, fill=greeen1!20] (4,0) ellipse (1.5cm and 3cm);
  	    \draw [thick, fill=red!20] (4,0) ellipse (1cm and 2cm);
  	    \draw [thick, fill=processblue!20] (4,0) ellipse (0.5cm and 1cm);
  	     \draw [thick, fill=greeen1!20] (8,0) ellipse (0.5cm and 1cm);
  	     \draw[fill=red!20] (8,0) circle (2pt); 
  	      \node at (0,3.3) {$\Omega^{k-1}(\M)$};
  	      \node at (4,3.3) {$\Omega^{k}(\M)$};
  	      \node at (8,3.3) {$\Omega^{k+1}(\M)$};
  	      \node at (2,-2.5) {$d$};
  	      \node at (6,-2.5) {$d$};
  	       \node at (4,0) {$Im(d)$};
  	        \node at (8,-0.5) {$Im(d)$};
  	       \node at (4,1.4) {$Ker(d)$};
  	      
				 \begin{scope}[very thick,decoration={markings,mark=at position 0.5  with {\arrow{>}}}] 
                \draw[postaction={decorate}] (0,3)--(4,1);
                \draw[postaction={decorate}] (0,-3)--(4,-1);
	  \draw[postaction={decorate}] (4,2)--(8,0.1);
	  \draw[postaction={decorate}] (4,-2)--(8,-0.1);
	  \draw[postaction={decorate}] (4,3)--(8,1);
	  \draw[postaction={decorate}] (4,-3)--(8,-1);
                 \end{scope}	
  	      
  	    }
\usepackage[sorting=none]{biblatex}
\usepackage{csquotes}
\begin{document}
\let\tempmargin\oddsidemargin
\let\oddsidemargin\evensidemargin
\let\evensidemargin\tempmargin
\begin{titlepage}
   \begin{center}

	\begin{center}
	\begin{figure}\centering 
	\includegraphics[scale=0.4]{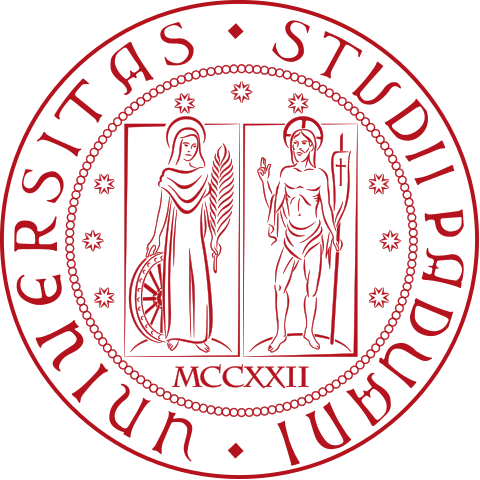}
	\end{figure}
	
    {\Large{\textbf{UNIVERSIT\`A DEGLI STUDI DI PADOVA}}}\\
    \vspace{5mm}
    \large{Dipartimento di Fisica e Astronomia "Galileo Galilei"}\\
    \large{Laurea Magistrale in Fisica}\\
    \vspace{8mm}
    \Large{\textit{\textbf{A modern approach to String Amplitudes \\ and Intersection Theory}}}
\end{center}

       \vfill
 \vspace{1cm}
  \begin{minipage}{2.4in}
\large{Supervisor:} \\
\large{\textbf{Prof. Pierpaolo Mastrolia}}\\\\
\large{Co-Supervisor:} \\
\large{\textbf{Prof. Sergio Luigi Cacciatori}}
\end{minipage}
\hfill
\begin{minipage}{1.6in}
\large{Candidate:} \\
\large{\textbf{Anthony Massidda}}\\\\
\end{minipage}
    \vfill
       
       \large{Academic year 2020/2021}
            
   \end{center}
\end{titlepage}
\newpage\null\thispagestyle{empty}\newpage
\begin{center}
    \textbf{Abstract}\\
    \vspace{1cm}

 \begin{minipage}{5in}
In this thesis, we study the properties of \textit{String theory amplitudes} within the framework of \textit{Intersection Theory for twisted (co)homology},
which, as recently proposed, offered a novel approach to analyze relations between \textit{scattering amplitudes}, in string theory as well as in Quantum Field Theory. As only recently pointed out, thanks to intersection theory, the analytic properties of scattering amplitudes can be related to the topological properties of the manifolds characterizing their integral representation.
 
Tree-level string amplitudes, as well as Feynman integrals, obey both linear and quadratic relations governed by intersection numbers, which act as scalar products between vector spaces.
We show how (co)homology with values in a local system allows to interpret closed strings tree amplitudes as intersection numbers between twisted cocycles, and open strings tree amplitudes as parings between a twisted cocycle and a twisted cycle.
We present different algorithms to evaluate univariate and multivariate \textit{intersection numbers} between both logarithmic and non-logarithmic twisted cocycles. We explore a diagrammatic method for the computation of intersection number between twisted cycles of the moduli space of the $n-$ punctured Riemann sphere. 
We use intersection theory to rederive \textit{Kawai-Lewellen-Tye relations}, naturally emerging as a twisted version of Riemann period relations. We compute intersection matrix between two-dimensional twisted cycles to explicitly obtain the  Kawai-Lewellen-Tye decomposition of five closed tachyons tree amplitudes into partial five open tachyons tree amplitudes. 
We implement a Mathematica code for multivariate logarithmic twisted cocycles intersection number, and we use it to explicitly determine the intersection matrix between two-dimensional Parke-Taylor forms. We use a recursive algorithm for generic $n$ cocycle intersection numbers to project tachyon amplitudes integrand into a Parke-Taylor basis, and we apply it to the scattering of four and five tachyons.  
The methods discussed in the thesis can be broadly applied to problems involving Aomoto-Gel'fand integrals.  
\end{minipage}

\end{center}
\tableofcontents
\addcontentsline{toc}{chapter}{Introduction}

\chapter*{Introduction}

High energy particle collisions actually are the most efficient tool for accessing new information on matter constituents and forces of Nature. The higher the energy of the colliding particles, the richer the landscape of the produced ones. The discovery of new Physics interactions is definitely associated to the discovery of new particles emerging from collisions of ever increasing energy. Therefore, advances in High Energy Particle Physics necessarily depend on our ability to describe the scattering processes at very high accuracy.\\
The unification of electric and magnetic phenomena under the same framework, encoded in the Maxwell equations, triggered physicists toward the effort of describing all forces of Nature by the same language. Electromagnetism and Weak interaction were indeed unified in the Electroweak Theory, developed by Glashow, Salam and Weinberg, and, together with Quantum Chromodynamics, that describe the Strong nuclear processes, 
they constitute the Standard Model (SM) of Elementary Particles. With the inclusion of the Higgs mechanism and of the Yukawa interaction, 
the SM is a consistent Quantum Field Theory (QFT) of Fundamental Interactions, compatible with the gauge invariance principle associated to the $SU(3) \times SU(2) \times U(1)$ symmetry.

The distribution of matter in the Universe, as well as experimental data on particle properties at very high-accuracy (such as the anomalous magnetic moment of the muon, for instance), point towards some tension on the completeness of the SM, which, although passing many tests, is indeed considered an effective model of a more fundamental theory (maybe related to some wider symmetry group), yet to be discovered. The search for New Physics is related to the discovery of new interactions, hence of new particles which may manifest themselves in scattering processes, by inducing deviations between theoretical predictions and experimental data.

In QFT, scattering processes are described in terms of \textit{scattering amplitudes}, complex numbers whose absolute squared value expresses the transition probabilities from a given asymptotic incoming state to an asymptotic outcoming one. In the perturbative approach, the scattering amplitudes can be series expanded in powers of the coupling constant, hence generating an infinite number of terms, which give rise to the so-called Dyson series. Feynman rules allow to establish a diagrammatic correspondence for each term of the expansion, hence providing a graphical representation of the scattering amplitudes. Feynman diagrams represent the various ways the particles involved in the scattering process, identified for a fixed incoming and outgoing configuration, can interact.
Therefore, in order to compute a scattering amplitude, at any given order in perturbation theory, it is sufficient to draw all the possible diagrams contributing at that given order, namely containing the necessary number of interaction vertices.
Leading order contributions, known as tree level, turn to be rational functions of kinematic variables (energy, momenta and masses of the particles) and, therefore, they do not represent any computational obstruction. Higher order terms, instead, involve loop-integrals, whose evaluation is, in general, very challenging: the higher the number of loops, the more challenging is the integral. Moreover, the number of diagrams itself increases exponentially, therefore it becomes prohibitive when considering higher order contributions. 
Dimensional regularization, introduced by 't Hooft and Veltman to regulate otherwise divergent multi-loop integrals \cite{tHooft:1972tcz}, can be exploited to show that, for any given process, \textit{Feynman integrals} are not independent. This implies that the evaluation of cumbersome amplitudes, beyond tree-level, can be simplified by minimizing the number of integrals which need to be actually computed.
In particular, in the last four decades, the theoretical progress in high-energy precision physics relied on the exploitation of linear relations for Feynman integrals known as integration by part identities (IBPs), which are generated from the vanishing contribution of surface terms \cite{TKACHOV198165,Chetyrkin1981IntegrationBP} - a generalization of Gauss theorem in arbitrary dimensions.
Beside linear relations, Feynman integrals, and more generally Scattering amplitudes, are found to obey also 
quadratic relations, which have led to the double-copy formalism and to the color-kinematic duality  \cite{2008},\cite{BCJ}. 

The existence of linear relations between Feynman integrals suggests a natural way to compute multi-loop scattering amplitudes, according to a two step procedure: i) decomposition in terms of independent integrals, named \textit{Master Integrals}; ii) evaluation of the latter.
The standard approach for the decomposition of any loop integral through integration by parts identities has been implemented in the Laporta Algorithm \cite{2000}. 
For many years, the existence of linear relations between integrals, as well as the role of the Master Integrals, has stimulated the search for an actual vector basis structure for Feynman integrals, which require the identification of two main properties: the determination of the space dimension, namely the number of Master Integrals, and of a scalar/inner product between the elements of the space.
Finding a proper basis of such space and evaluating coefficients for the decomposition onto them allows one both to drastically reduce computational effort and to investigate more deeply features of the theory structure. 
Recently, it has been found that ideas borrowed from Algebraic and Differential Geometry, and Topology, provide the correct framework to prove the existence of a vector space structure of Feynman integral, and more generally, of Aomoto-Gel'fand integral \cite{Mastrolia:2018uzb,Frellesvig:2019uqt,Mizera:2017rqa,Frellesvig:2019kgj,Frellesvig:2020qot}, recently reviewed in \cite{Cacciatori:2021nli}: \textit{Intersection Theory for twisted de Rham homology and cohomology} \cite{1997HypergeometricFM,Aomoto,Aomoto1975OnVO,Aomoto2,Mimachi:2002gi,Mimachi:2004ez,Matsumoto1998IntersectionNF,cho_matsumoto_1995}
allow to establish an inner product between integrals, and to derive linear as well as quadratic relations among them.
Differently from the currently adopted techniques, intersection theory yields a direct integral decomposition, bypassing the needs of generating and solving huge systems of integral relations.\\
An integral can be seen as a pairing between a form, namely the integrand, and a path, namely the contour of integration; by Stokes' theorem we know that there are several forms that, paired with a given contour, give the same integral, as well as many contours that, paired with a given form, leave the integral unchanged; this means that integrals are indeed pairings among equivalence classes: equivalence classes of forms are known as cocycles and equivalence classes of paths are called cycles.
The spaces of cocycles and cycles of a given dimension are endowed by a group structure and they are respectively named homology group and cohomology group.\\
By defining suitable dual spaces of (co)homology spaces, one is able to introduce two non degenerate bilinear pairings called \textit{intersection numbers}. This suggestive name is due to the fact that in considering ordinary homology, the scalar product among cycles is actually just their topological intersection. The techniques needed to explicit evaluate intersection numbers clearly depend on the considered (co)homology, but, independently from that, intersection numbers and cycle-cocycle parings turn not to be independent, but related by both linear (contiguity relations) and quadratic relations (Riemann period relations). Applying Aomoto-Gelfand interpretaion of Hypergeometric integrals to Feynman integrals, these latter relations become the key to Master decompose them.
Remarkably the dimension of the (co)homology space, equivalent to the number of MIs, is a priori deducible by topological proprieties of the integration domain \cite{Milnor,matsumoto2002introduction,Lee:2013hzt}.\\\\

While QFT describes three of the four currently known interactions, Gravity is framed into a classical and completely different formalism: General Relativity. The attempt to also describe gravitational interaction into a quantum framework is one of the most fascinating task of theoretical physics.\\
\textit{String theory} \cite{zwiebach_2004,Polchinski:1998rq,kiritsis:in2p3-00714916,Green:2012oqa,Green:2012pqa,Tong:2009np,Staessens:2010vi} is, at the moment, one of the best candidates to describe nature at scales where we suppose gravity should enter into processes comparably to other forces. \\
The basic idea behind string theory is to consider the fundamental objects of nature not as pointlike particles, but as extended one-dimensional strings. 
Strings come in two types, closed and open, and just like particles moving in space-time describe a one-dimensional worldline, strings, moving in space-time, swipe a bidimensional (Riemann) surface called worldsheet. Different string excitation states are interpreted as different particles.
Bosonic string theory, only containing bosonic degrees of freedom, is the simplest string theory one can consider.
Both open and closed bosonic strings spectra contain massless states interpretable as gluons/photons and gravitons respesctively, however they turn out not to be the fundamental states as one would expect, but the first excitations of negative mass states, known as tachyons. The presence of tachyons leads to vacuum instability that, together with the absence of fermions, makes bosonic string theory hard to be applied for describing Nature. Moreover, in order for classical symmetries to be preserved at quantum level, bosonic string theory turns out to be coherently defined only in a 26 dimensional space-time, known as critical dimension. Imposing supersymmetry on the bosonic string theory is the classical way for adding matter; this procedure remarkably eliminates tachyonic states and reduces critical dimension to $D=10$.
Nevertheless, bosonic string theory turns out to be a useful toy model for accessing information about its supersymmetric counterpart, becouse they share important structural features: for instance, tree level amplitudes in bosonic and supersymmetric string theory have the some structure, due to the decoupling of fermionic degrees of freedom only contributing to an overall factor.
String amplitudes represent a bidimensional generalization to QFT amplitudes; in the perturbative approach to string theory, the QFT discrete sum over Feynman diagrams is replaced by an integral over a continuous worldsheet. Each worldsheet shape, compatible with the scattering process, represents a different way a given incoming configuration can evolve into a given outcoming one. 
In the bidimensional extension, loops become holes in the worldsheet, then the expansion of the string amplitudes in the coupling constant turns out to be an expansion over the genus of the worldsheet: higher genera worldsheet integration corresponds to higher order contributions. Tree level contributions are then associate to genus $0$ Riemann surfaces, first order contributions to genus $1$ and so on.
Thanks to conformal symmetry of string theory, the integration over a given worldsheet representing an $n$ string scattering process is equivalent to the integration over a Riemann surface of the same genus with $n$ points removed, called punctures. This implies that not all contributions to the amplitude are inequivalent: equivalence classes of $n-$ punctured Riemann surfaces of a given genus $g$ are described by the so called \textit{ Moduli space} $\M_{g,n}$.
The possible shapes of the worldsheet clearly depend on the string type considered: closed strings swipe closed worldsheets and open strings opened worldsheets.
Therefore, contributions to $n$ closed strings amplitudes come from the integration over the Moduli space of the $n-$ punctured sphere (tree level), of the $n-$ punctured torus (first order) and so on; while $n$ open strings amplitudes contributions come from the integration over the Moduli space of the $n-$ (boundary) punctured disk (tree level), of the $n-$ (boundary) punctured annulus (first order) and so on.\\
Open and closed string tree amplitudes turn out to be related by quadratic relations known as \textit{Kawai-Lewellen-Tye (KLT) relations} \cite{KAWAI19861}, expressing closed string tree amplitudes as a sum of products of suitable partial open string tree amplitudes.\\
\textit{String amplitudes}, as well as Feynman integrals, can be written as the integral of a rational form times a multivalued functions, polynomials to some non integer power: the Koba-Nielsen factor and the graph polynomial, respectively. 
In order for intersection theory to be applied one has to suitable choose homology and cohomology; different choices are possible, like relative (co)homology or intersection (co)homology, but at the moment which one should provide the most appropriate description is still an open field of research. 
In this thesis, we elaborate on the recent work of Mizera \cite{Mizera:2017cqs,Mizera:2017rqa}, who proposed the use of twisted de Rham theory in the analyses of scattering theory amplitudes. In particular,
we will explore the possibility to describe string amplitudes by means of the so called twisted (co)homology, defined on a proper line bundle on the integration domain, called Local system  \cite{10.2307/1969099,10.1007/978-3-0346-0209-9_5}. We will study how to compute intersection numbers between twisted cycles and between twisted cocycles, and we will see how a suitable choice of the dual (co)homology space allows to interpret closed string amplitudes as twisted cocycle-cocycle intersections and open string amplitudes as twisted cycle-cocycle intersections, showing how KLT relations naturally emerge as a twisted version of Riemann period relation. 
The key objects to be evaluated are then twisted  cycle-cycle and cocycle-cocycle intersection numbers.
The evaluation of intersection numbers for twisted cycles is in general a very challenging task of algebraic geometry \cite{Kita1994IntersectionTF,Kita2,Kita3,Mimachi:2002gi,Mimachi:2004ez,Mimachi}. It has been recently showed \cite{Mizera:2016jhj,Mizera:2017cqs,Mizera:2019gea} that the diagrammatic method proposed by Cachazo, He and Yaun (CHY) \cite{Cachazo:2013iea} for the computation of the $\alpha'$ bi-adjoint scalar can be used to evaluate intersection numbers among twisted multidimensional cycles of the punctured Riemann sphere Moduli space.
We will use this method to explicitly KLT decompose a five closed string tree amplitudes into partial open string tree amplitudes.\\
The evaluation of intersection numbers for univariate twisted cocycles requires the solution of differential equations and the application of Cauchy’s residue theorem \cite{cho_matsumoto_1995}.
For the case of meromorphic n-forms, an iterative method for the determination of intersection numbers was proposed in \cite{Mizera:2019gea} and successively refined in \cite{Frellesvig:2019uqt,Frellesvig:2020qot,Weinzierl:2020xyy}. In the special case of logarithmic differential forms (\emph{dlog} forms) a remarkable semplification occurs and the above mentioned iterative method can be completely bypassed: intersection numbers completely localize and they can be expressed in terms of a multivariate residue \cite{Matsumoto1998IntersectionNF,Mizera:2017rqa}.
We implement a Mathematica code to evaluate intersection numbers for \emph{dlog} forms and we will apply it for computing intersection numbers between two-variables Parke-Taylor forms \cite{Parke:1986gb}, analytically verifying the agreement with the result obtained by the recursive algorithm. Using then the latter, we will compute the projections onto the Parke-Taylor basis, obtaining the explicit Master integral decomposition of a five partial open tachyon tree amplitudes.\\
The study of string amplitudes turns to be a very useful tool to investigate quantum field amplitudes structure: duality relations among QFT amplitudes, often conjectured, can be elegantly obtained by strings dualities in the field theory limit; for instance double-copy duality, expressing quantum gravity amplitudes as two Yang-Mills amplitudes, can be seen as a KLT decomposition of closed massless string amplitudes (graviton) into two partial open massless strings amplitudes (gluons). This fact suggests the possibility to find new relations among QFT amplitudes by better undestanding string amplitudes structure and by exploring new duality relations among them.

\nn
\textbf{Outline}\\
In chapter \ref{one}, we will use \textit{Morse Theory} \cite{Milnor,matsumoto2002introduction} to cell decompose a Manifold and naturally introduce cellular homology groups. After discussing simplicial and singular homology \cite{friedman2021elementary}, we will study cohomology groups, with special emphazis to De Rham cohomology, the cohomology of differential forms, proving the De Rham theorem. Finally, we will discuss Mayer-Vietoris sequence and Riemann period relations.\\
In chapter \ref{two}, we will focus on (co)homology with values in a local system, known as twisted (co)homology, defining and discussing intersection numbers. We will present different algorithms for computing intersection numbers between twisted forms, and a diagrammatic method to compute intersections of one-dimensional twisted cycles. We will also derive the expressions of linear and quadratic relations by means of intersection numbers.  
Chapter \ref{Chapter3}, constitutes a general introduction to the basic notions of bosonic string theory, discussing different method of quantization, and introducing the string spectrum. 
In chapter \ref{four}, we will define string amplitudes, discussing vertex operators for both fundamental and excited states, and exploring the role of the Moduli space emerging from Faddeev-Popov quantization procedure. We will obtain the expressions for tree level amplitudes, for both open and closed strings. We explicitly compute $3$ strings scattering to show the role of on-shellness in order to preserve conformal invariance, and we will compute $4$ strings scattering, recovering Veneziano \cite{Veneziano:1968yb} and Shapiro-Virasoro amplitudes.\\
Finally, in chapter \ref{five}, we will discuss the KLT relations, and we will rederive them by means of (co)homology theory, interpreting them as twisted Riemann period relations. We will then discuss a diagrammatic method to compute multidimensional intersection numbers between twisted cycles in the moduli space of the punctured Riemann sphere. We will explicitly compute a KLT decomposition for five closed strings tree amplitude and we will Master decompose a five open strings tree amplitude onto a Parke-Taylor basis.

g\chapter{Homology and Cohomology Theory }\label{one}
Homology and cohomology groups represents the fundamental mathematical objects of this work. In the first part of this chapter we are going to introduce Homology group by means of cellular decomposition, achieved by Morse theory. Subsequently we will generalize the notion of homology in order to introduce cohomology in a more rigorous way. Then we will study De Rham cohomology deriving its main result: De Rham Theorem. Finally we will present Mayer-Vietoris sequence and Riemann Period Relations. Main refernces are \cite{Milnor,matsumoto2002introduction,friedman2021elementary,Bott}. See \cite{Cacciatori:2021nli} for review.

\section{Morse theory and Homology}\label{sectionone}

Let $\M$ be an m-manifold without boundaries and $f: \M \rightarrow \R$ a smooth function defined on it. 
\begin{defn}[Critical points] A point $p_0$ on $\M$ is said to be a \textbf{critical point} of $f$ if 

\begin{equation}
\de_i f(p_0)=0 \quad \quad \forall i=1,2...m \quad \quad \mbox{(no sum on i)}
\end{equation}

\nn
with respect to a local coordinate system $\{x_i\}$ about $p_0$. Moreover we said the real number $c$ is a critical value of $f$ if it exists a critical point $p_0$ of $f$ such that $f(p_0)=c$. 

\end{defn}

\nn
This definition is independent on the coordinate choice.

\begin{defn}Let $H_{ij}^{(f)}=\de_i\de_j f$ the Hessian of $f$ respect to the coordinate system $\{x_i\}$; a critical point $p_0$ of $f$ is said to be \textbf{non-degenerate} if the determinant of the Hessian matrix of $f$ at $p_0$ is not null, otherwise we say $p_0$ is \textbf{degenerate}

\begin{equation}
\begin{cases} \det H_{ij}^{(f)}(p_0)\neq 0 \quad \quad  \mbox{ (non-degenerate)}\\ \det H_{ij}^{(f)}(p_0)= 0 \quad \quad \mbox{ (degenerate)} \end{cases}
\end{equation}
\end{defn}

\nn
Degeneracy propriety of $p_0$ does not depend on the coordinate system. Indeed, if we consider another coordinate system $\{x'_i\}$ about $p_0$, the two Hessian matrices $H^{(f)'}(p_0)$ and $H^{(f)}(p_0)$ are related as 

\begin{equation}
H^{(f)'}(p_0)=J^T(p_0)H^{(f)}(p_0) J(p_0),
\end{equation}

\nn
where $J$ is the Jacobian of the coordinate transformation $J_{ij}=\de_i' x_j$. Since the Jacobian matrix at $p_0$ has a non-zero determinant, the above relation implies $H^{(f)'}(p_0)\neq 0$ iff $H^{(f)}(p_0)\neq 0$.

\begin{defn}[Morse function] A function $f: \M \rightarrow \R$ is called a \textbf{Morse function} if every critical point of $f$ is non-degenerate.

\end{defn}

\begin{teo} Let $p_0$ be a critical point of $f:\M\rightarrow \R$ and $c=f(p_0)$; we can always choose a local coordinate system $\{X_i\}$ about $p_0$ such that $p_0$ corresponds to the origin and $f$ can be written in the standard form

\begin{equation}
f=-X_1^2-X_2^2-...-X_\lambda^2+X_{\lambda+1}^2+...+X_m^2+c.
\end{equation}

\nn
The integer number $\lambda \in [0,m]$, representing the number of minus signs in the standard form, is called index of $p_0$.
 \end{teo}
 
 \nn
As a direct consequence of this Theorem, non-degenerate critical points are always isolated and a Morse function defined on a compact manifold admits only a finite number of critical points.\\
To be more concrete let's consider a simple example. Let $f=xy+1$, the point $p=(0,0)$ is a non-degenerate critical point of $f$, in fact 

\begin{equation}
\de_x f(0,0)=\de_y f(0,0)=0 \quad \quad H^{(f)}(0,0)=\begin{pmatrix}
0&2\\2&0
\end{pmatrix}.
\end{equation}

\nn
If we perform the coordinate transformation 

\begin{equation}
\begin{cases} x=X-Y\\y=X+Y \end{cases},
\end{equation}

\nn
we obtain the standard form 

\begin{equation}
f=X^2-Y^2+1.
\end{equation}

\begin{teo}Let $\M$ be a closed m-manifold and let $g: \M \rightarrow \R$ a smooth function defined on it. There exists a Morse function $f:\M \tu\R$ arbitrarily close to $g$.\end{teo}

\nn
Where by closed manifold we mean a compact manifold without boundary, while saying arbitrarily close we mean $C^2$ close. Notice that this means that Morse functions are dense on $\M$.

\nn
\begin{teo}Let $f: \M \rightarrow \R $ be a Morse function on $\M$ and $p_i$ its critical values. Then there exists a Morse function $\tilde{f}:\M \tu\R$ whose critical points are the same as $f$ and such that 

\begin{equation}
\tilde{f}(p_i)=\tilde{f}(p_j) \leftrightarrow p_i=p_j.
\end{equation} \end{teo}\label{differentcriticalvalues}

\nn
Let $B^n$ be the n-dimensional ball. A topological space homeomorphic to the interior of the ball, i.e. $int(B^n)=B^n-\de B^n$, is called \textit{i-cell} and denoted by $e^i$. An i-cell endowed with the boundary, that is a topological space homeomorphic to the whole $B^n$, is called a \textit{close i-cell} and denoted by $\bar{e}^i$. In the 0 dimensional case both $e^0$ and $\bar{e}^0$ are a single point, so $e^0=\bar{e}^0$.

\begin{teo}Let $f:\M\tu\R$ be a Morse function and set  
\begin{equation}
\M_s(t) = \{p\in \M | s\leq f(p)\leq t\}
\end{equation}
if $f$ has no critical values in $[a,b]$, then $\M_s(a)$ and $\M_s(b)$ are diffeomorphic. \label{morsediffeomorme}
 \end{teo}
 
 \nn
 It is clear from the definition that $s<t$. 
According to Theorem \ref{differentcriticalvalues} we can always assume that $f$ takes different critical values at distinct critical points $p_i$, moreover we know the number of critical points is finite, say $n+1$, so we can order them setting

 \begin{equation}
f(p_i)\equiv c_i<c_j \quad \quad \mbox{for}\quad  i<j.
\end{equation}

\nn
If follows from Theorem \ref{morsediffeomorme} that

\begin{equation}
\M \cong\bigsqcup_{i=0}^{n} \M_{c_i}(c_{i}+\epsilon),
\end{equation}

\nn
with $\epsilon$ a fairly small positive number. Roughtly speaking, if we cut $\M$ in correspondence of $f$ critical points we obtain $n-1$ manifolds whose disjoint union, obviously, gives back the starting manifold, but, less obviously, each of those submanifolds is diffeomorphic to the one obtained cutting a thin slice around the corresponding critical point, no matter its thickness; this will turn to be a great advantage in understanding their topology, as we are going to see in the next. \\
We now analyse the changes of $\M_s(t)$ around $f$ critical points, starting from those corresponding to the minimum and maximum critical values. If $p_0$ is the $f$ critical point with minimum critical value $c_0$, it's clear there exists no point $p$ in $\M$ such that $f(p)<c_0$, so that $M_s(t)=\emptyset$ if $t<c_0$. For the same reason we can take $s\geq c_0$ and now on, for notation clarity, when we omit $s$ we mean $s=c_0$. 
Furthermore, we have that the standard form of $f$ around $p_0$ has index 0

\begin{equation}
f=x_1^2+x_2^2+...x_m^2+c_0 \quad \quad (\mbox{around} \quad p_0),
\end{equation}

\nn
so we can write

\begin{equation}
\M_{c_0}(c_0+\epsilon)\equiv \M(c_0+\epsilon)=\{(x_1..x_m)| x_1^2+x_2^2+..x_m^2\leq \epsilon\},
\end{equation} 

 \nn
 that is, $\M(c_0+\epsilon)$ is diffeomorphic to the m-dimensional ball $B^m$. As you can see, this propriety does not closely depend on the fact that $c_0$ is the minimum critical values, but just on its index, so, in general, for any critical value $c$ of index 0, we have 
 
 \begin{equation}
 \M(c+\epsilon)\cong \M(c-\epsilon)\sqcup B^m.
\end{equation}  
 
 \nn
 For reasons that will be clear soon, the $B^m$ ball appearing here is called \textit{m-dimensional 0-handle}.\\
 Consider now the point $p_n$ corresponding to the maximum critical value $c_n$. Evidently $f$ cannot take values larger then $c_n$, hence $\M_t=\M$ for $t>c_n$ and

\begin{equation}
f=-x_1^2-x_2^2+...x_m^2-c_n \quad \quad (\mbox{around} \quad p_n),
\end{equation}

\nn
that is, $p_n$'s index is $m$. Using this expansion of $f$, we have

\begin{equation}
\M_{c_n-\epsilon}(c_n)=\{x_1^2+x_2^2+..x_m^2\leq \epsilon\}.
\end{equation}

\nn
This manifold again corresponds to $B^m$. As $\epsilon$ approach to zero, the boundary of $\M_{c_n-\epsilon}$ is covered by a m-dimensional ball, resulting in a manifold without boundary. Such m-ball is said an \textit{m-dimensional m-handle}. As before, we can extend this result to general case : whenever $t$ passes a critical value $c_i$ with index $m$, the m-handle caps a connected component of the boundary of $\M_{c_i-\epsilon}$. \\
Let's finally consider the changes of $\M_t$ around a generic critical point $p_i$ with index $\lambda \in[0,m]$ as $t$ crosses the corresponding critical values $c_i$. Taking as usual a local coordinate system around $p_i$ such that $f$ has the standard form 

\begin{equation}
f=-x_1^2-x_2^2-...-x_\lambda^2+x_{\lambda+1}^2+..+x_m^2 + c_i,
\end{equation}

\nn
we can write 

\begin{equation}
\M_{c_i+\delta}(c_i+\epsilon)=\{p\in \M | \begin{cases}x_1^2+x_2^2-...+x_\lambda^2-x_{\lambda+1}^2-..-x_m^2\leq \delta \\ x_{\lambda+}^2+..+x_m^2\leq \epsilon\end{cases}\}.
\end{equation}

\nn
with $\delta$ a positive number smaller than $\epsilon$. This manifold is called an \textit{m-dimensional $\lambda$ handle} and it can easily proved it is diffeomorphic to the direct product of an $\lambda$-dimensional ball and a $m-\lambda$ one:

 \begin{equation}
\M_{c_i+\delta}(c_i+\epsilon)\cong B^\lambda\times B^{m-\lambda}.
\end{equation}
 
 \nn
 The manifold 
 
 \begin{equation}
 B^\lambda\times 0 ={x_1,..,x_\lambda,0,..,0|x_1^2+x_2^2+..x_\lambda^2\leq \delta}
 \end{equation}
 
 \nn
 is know as the \textit{core} of the $\lambda$-handle, and $0\times B^{m-\lambda}$ the \textit{co-core} of the $\lambda$-handle. \\
 So, in general, we have 
 
 \begin{equation}
 \M(c_i+\epsilon)\cong\M(c_i-\epsilon) \sqcup  B^m\times B^{m-\lambda}.
 \end{equation}
 
 \begin{defn}A manifold obtained from $B^m$ attaching handle of various indices one after another
 
 \begin{equation}
 H^m \cong B^m\sqcup B^{\lambda_1}\times B^{m-\lambda_1}\sqcup...\sqcup B^{\lambda_n}\times B^{m-\lambda_n} 
\end{equation}  
 
 \nn
 is called an $m$-dimensional \textbf{handlebody}.
 \end{defn}
 
 \begin{center}
     \begin{figure}[h!]
         \centering
         \includegraphics[scale=0.4]{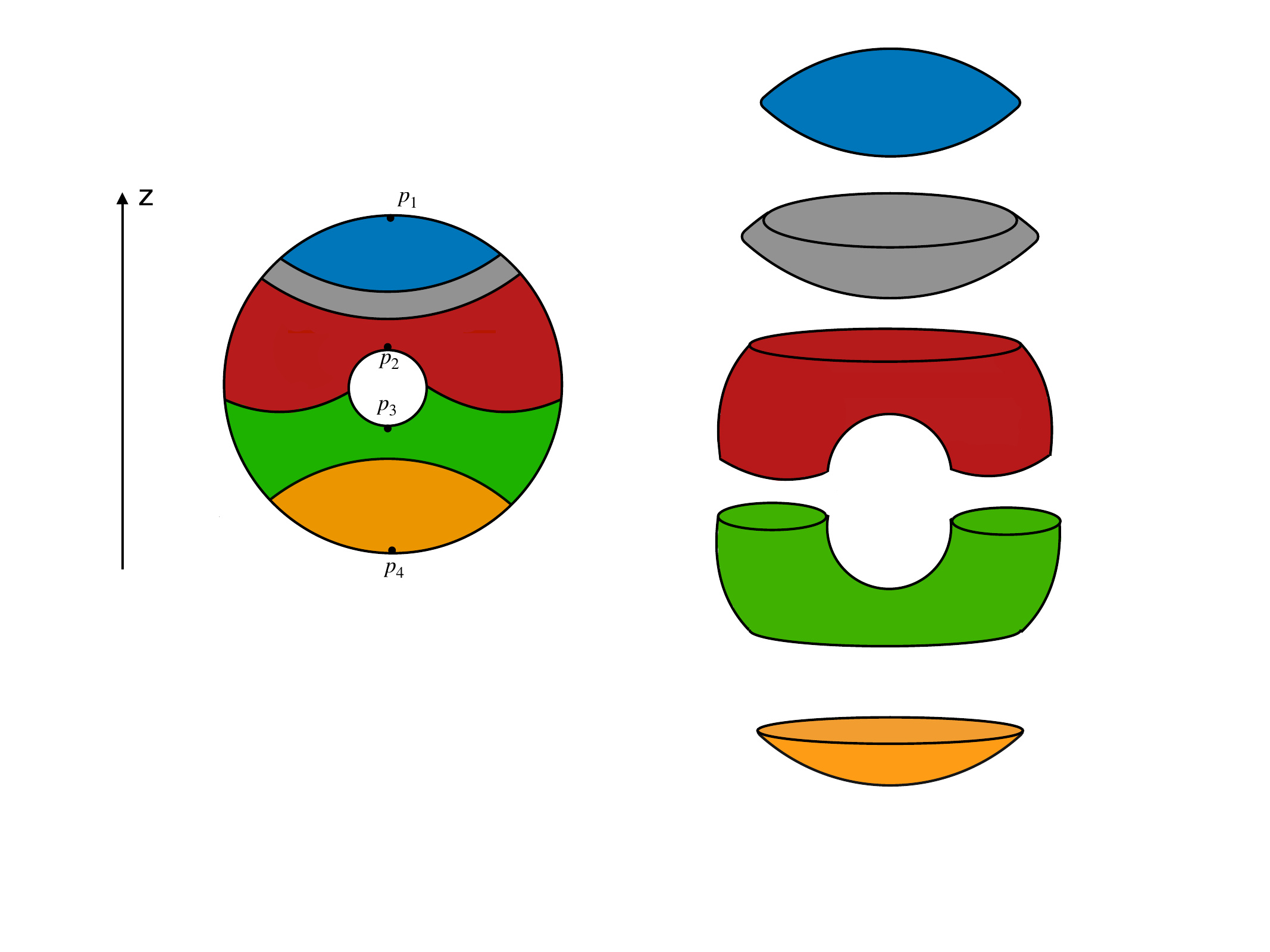}
         \caption{Torus handle decomposition.}
     \end{figure}
 \end{center}

 \nn
 We are now ready to enunciate the following theorem 
 
  \begin{teo} Let $\M$ be a closed manifold and $f:\M\tu\R$ a Morse function defined on it, then the structure of a handlebody on $\M$ is determined by $f$. The handles of this handlebody correspond to the critical points of $f$ and their indices coincide with the indices of the corresponding critical point. \end{teo}
  
\nn
To be more concrete let's see an example.
\begin{Ese}
 Let $S^m$ be the $m$ dimensional sphere and define $f:\M\tu \R$ by 

\begin{equation}
f(x_1,x_2,...x_{m+1})=x_{m+1}= \pm\sqrt{1-x_1^2-x_2-...-x_m^2}.
\end{equation}
 
\nn
Taking the derivative of $f$ and solving for $\de_i f=0$ we see there are only two critical points, $p_+=(0,..0,1)$ and $p_-=(0,..,0,-1)$, and Hessian of $f$ in those points is $H_+=-H_-=diag(-2,-2...-2)$, so they are non-degenerate and $f$ is a Morse function. Expanding $f$ around the two critical points we find 

\begin{equation}
f_{\pm}=\pm1\mp \frac{1}{2}x_1^2\mp\frac{1}{2}x_2^2\mp...\mp \frac{1}{2}x_m^2=\pm 1\mp X_1^2\mp X_2^2\mp..\mp X_m^2,
\end{equation}
 
 \nn
 we then conclude $p_+$ is a critical point of index $m$ and $p_-$ a critical point of index $0$. Therefore, the handle decomposition of $S^m$ consists of a $0$-handle and $m$-handle:
 
 \begin{equation}
 S^m \cong B^m \sqcup B^m.
\end{equation}  
  \end{Ese}\label{Esempiosfera1}
 \nn
 For the bidimensional sphere this means it is equivalent to the union of two $B^2$: we can take two circles, continuously deform them into two hemispheres and then obtain the sphere attaching them together.
 This example introduce as to the following theorem 
 
 \begin{teo}Let $\M$ be a closed manifold and $f:\M\tu\R$ a Morse function defined on it, if $f$ has only two critical points, then $\M$ is homeomorphic to $S^m$. Furthermore, if $m\leq 6$, then $\M$ is diffeomorphic to $S^m$.
 
 \end{teo}
 
 \begin{defn}A topological space homeomorphic to the interior $int(B^m)=B^m-\de B^m$ of the $m$-dimensional ball is called an $m$-cell adn denoted by $e^m$ A topological space homeomorphic to $ B^m$ is called a close $m$-cell and denoted by $\bar{e}^m$.
 \end{defn}
 
 \nn
 In the zero dimensional case both $e^0$ and $\bar{e}^0$ correspond to a single point.
 
  \begin{defn}{(cell complex)}A space obtained from $0$-cells by attaching closed cells one after another is called a \textbf{cell complex}
 \end{defn}
 
 \nn
Before to continue in our discussion about cells, we need to introduce the concept of orientation of a manifold. \\
Consider a point $p$ of an $m$dimensional manifold $\M$ and a neighborhood $U$ of $p$. Because, for definition, $U\cong V\subset\R^m$, we can define a basis on it. Two different basis are related by an isomorphism, whose determinant is a non-zero real number, so we can recognize two equivalence classes of basis: two basis are equivalent if they are related by a positive determinant transformation. Each class defines an orientation on $U$.

\begin{defn}
Let $\M$ be an $m$-dimensional manifold and $U_i$ a family of open sets covering $\M$. If we can simultaneously define a basis for each $U_i$ such that in the intersection $U_i \cap U_j$ of any pair of neighborhoods their orientation coincide, then $\M$ is said to be \textbf{orientable}, and giving  such orientation is called \textbf{orienting} $\M$. An oriented manifold $\M$ is denoted $\langle\M\rangle$.
\end{defn}
 
 \nn
 If $\M$ is orientable and connected, then there are only two orientations on it, one opposite to the other.\\
Consider now a cell complex $X$ having $k_n$ oriented $n$-cells, the linear combination 

\begin{equation}
c_n = \sum_i a_i \langle e_i^n\rangle \quad \quad \mbox{with} \quad a_i\in \mathbb{Z},
\end{equation}
 
 \nn
 is called an $n$-\textit{chain} of $X$. The set of all $n$-chains of $X$, with the operation of addition defined as 

 \begin{equation}
 c_n+c'_n = \sum_i (a_i+a'_i) \langle e_i^n\rangle ,
 \end{equation}

 \nn
  forms an abelian group, denoted by $C_n(X)$, called $n$-\textit{dimensional chain group of}$X$. Since $C_n(X)$ is the free abelian group generated by $k_q$ $n$-cells, we have
  
  \begin{equation}
  C_n(X)\cong  \mathbb{Z}\oplus\mathbb{Z}\oplus...\oplus\mathbb{Z} \quad\quad (k_n copies).
  \end{equation}
 
 \nn
 Now we define the \textit{boundary homomorphism} $\de_q : C_q(X)\rightarrow C_{q-1}(X)$ as
 
 \begin{equation}
\de_q \langle e_k^q\rangle = \sum_{i=1}^{q-1} a_{ki}\langle e_k^{q-1}\rangle,
 \end{equation}
 
 \nn
 when the orientation of the $q-1$-cells is naturally induced from the $q$-cell orientation. Obviously, if $X$ contains no $q-1$-cells, all the coefficients $a_{ki}$ are null, and the imagine of the boundary homomorphism is the empty set. A sequence consisting of chain groups of a cell complex $X$ and boundary homomorphism is called a \textit{chain complex} of $X$
 
 \begin{equation}
...\rightarrow C_q(X)\xrightarrow{\de_{q}}C_{q-1}(X)\xrightarrow{\de_{q-1}}C_{q-2}(X)\rightarrow ..\rightarrow C_1(X)\xrightarrow{\de_{1}}\{0\}.
 \end{equation}
 
 \nn
 We are now ready to introduce homology group: let $Z_q(X)\equiv Ker \de_q$ the $C_q(X)$ subgroup of $q$-chains without boundary 
 
 \begin{equation}
 Z_q(X)=\{c\in C_q(X)|\de_q(c)=0\},
 \end{equation}
 
 \nn
 called $q$-dimensional cycle group and whose elements are known as $q$-cycles; let $B_q(X)\equiv Im \de_{q+1}$ the $C_q(X)$ subgroup of  $q$-chains that are themselves boundary of some $q+1$-chain
 
  \begin{equation}
 B_q(X)=\{c\in C_q(X)|c=\de_{q+1}(c_{q+1})\},
 \end{equation}
 
 \nn
 called $q$-dimensional boundary group and whose elements are said $q$-boundaries. 
 Because of the identity $\de_q\de_{q+1}=0$, we have $B_q(X)\subset Z_q(X)$, and we can therefore define their quotient group:
 
 \begin{defn}The quotient group 
 
 \begin{equation}
 H_q(X)\equiv \frac{Z_q(X)}{B_q(X)},
 \end{equation}
 
 \nn
 defines the \textbf{$q$-dimensional (cellular)homology group of $X$}. It's elements are called homology classes.
 
 \end{defn}
 \nn
Two q-cycles $c, c' \in Z_q(X)$ belongs to the same homology class if they differ by a boundary $q-$chain.\\
Notice that the first homology group $H_1(X)$ contains the equivalence classes of $1-$ chains, that are nothing else that inequivalent loops on $X$, this fact suggests a close relation between $H_1(X)$ and the first homotopy group (fundamental group) $\pi_1(X)$; despite one could at first guess they are isomorphic, this is in general not true,  since the fundamental group is not abelian, in general, while the homology group is, by definition. In particular it turns out that 

\begin{equation}
H_1(X)\cong \pi (X)^{ab}, \label{gruppofondamentale1}
\end{equation}

\nn
where $\pi(X)^{ab}$ is the abelianization of $\pi(X)$. This relation could be vary useful: the isomorphism between them evidently implies they have the same dimension, but the dimension of the fundamental group is related to the genus of the manifold. Remember the (topological) genus $g$ of a manifold $\M$ is defined as the maximum number of cuts along non-intersecting closed simple curves that may be performed without disconnect $\M$, for instance in the case of a surface $g$ corresponds to the number of "holes"; it turns out that\footnote{Notice that, being $\pi_1$ homotopic invariant, $\pi(\M)=\pi(X)$, for every cell complex $X$ on $\M$.}

\begin{equation}
 dim \,\pi_1(M)= 2g(\M).\label{gruppofondamentale2}
\end{equation}

\nn
An important property of the homology group is homotopic invariance: if two cell complexes $X$ and $Y$ are homotopic equivalent, i.e. there exist two continuous maps $f:X\rightarrow Y$ and $g:Y\rightarrow X$ such that $f\circ g= id_Y$ and $g\circ f=id_X$, then the respective $q-$th homology groups are isomorphic:
 
 \begin{equation}
 H_q(X)\cong H_q(Y).
\end{equation} 

\nn
This means that homology groups are related to topological properties of the manifold. 
The homology groups $H_q(X)$ of a finite cell complex $X$ have, in general the following form:

\begin{equation}
H_q(X)\cong \bigoplus \Z \oplus T,
\end{equation}

\nn where $T$ is an abelian group called torsion part of $H_q(X)$ while the $\Z\oplus\Z\oplus...\oplus \Z$ represents the free part\footnote{A group $G$ is said to be free its elements can be uniquely expressed as the product of a finite number of element of a subset $S \subset G$} and, as we will see in a moment, it plays an important role in our discussion. The dimension of the the largest free abelian subgroup of a group $G$ is called \textbf{rank} of $G$, then in our case the number of copies of $\Z$ in the free part of $H_q(x)$ is precisely the rank of the homology group

  \begin{defn} (Betti numbers) The rank of the $q-$th homology group $H_q(X)$ is called $q-$dimensional Betti number of $X$ and it is denoted by $b_q(X)$:
  
  \begin{equation}
  b_q(X) \equiv rank H_q(X).
  \end{equation}

 \end{defn}
  \nn
  From \eqref{gruppofondamentale1} and \eqref{gruppofondamentale2} immediately follows $b_1(X)=2g(X)$.\\
 The importance of Betti numbers is related to the following theorems: 
 
  \begin{teo}(Euler-Poincaré formula)Let $X$ be an $m-$dimensional cell complex and $k_q$ the number of $q-$cells contained in $X$. Then we have 
  
  \begin{equation}
  \chi(X) \equiv \sum_{q=0}^m (-1)^q k_q = \sum_{q=0}^m (-1)^q b_q,
  \label{chi_1}
  \end{equation}
  
  \nn
  where $\chi(X)$ is the Euler characteristic of $X$;
  \end{teo}
 \nn
  and 
   \begin{teo}(Morse inequality) Let $\M$ be a closed $m-$manifold, $f:\M\rightarrow \R$ a Morse function on $\M$ and $k_\lambda$ the number of its critical points of order $\lambda$. Betti numbers are constrained by the following inequality:
   
   \begin{equation}
   k_\lambda\geq b_\lambda(\M).
   \end{equation}
  \end{teo}
  
  \nn
  The connection between the metric and the Euler characteristic is given by the following theorem:
  
  \begin{teo}Gauss-Bonnet Theorem 
 Let $\M$ be a compact Riemann Manifold with boundary $\de\M$, $K$ the gaussian curvature of $M$ and $k$ the geodesic curvature of $\de \M$, then 
 
 \begin{equation}
 \int_\M K dS + \int_{\de\M} k dl = 2\pi \chi (\M).
 \end{equation}
 \label{GBtheorem}
 \end{teo}
  
  \nn
  \begin{Ese} In example \ref{Esempiosfera1} we saw the $m-$sphere can be handle-decomposed as $S^m\cong e^0\sqcup e^m$, then 
  
  \begin{equation}
  \chi (S^m) = 1+(-1)^m
  \end{equation}
  \end{Ese}
  
  \nn
  The first Homology group $H_0(\M)$ contains the inequivalent classes of 0-chains, i.e. points; if the manifold is path-connected all points are equivalent, because for any two points on the manifold a path can be found having them as extrema, so $H_0={P}$, with $P\in \M$, contains just one element and $b_0=1$. Of course, if $\M$ it's not path connected but composed by the union of $n$ path-connected components, the first homology group would counts $n$ elements, a point per connected component.
  
  \subsection{Simplicial and singular homology}\label{simplicialhomology}
  Consider an euclidian space $\R^n$; via the map 
  
  \[(x_1,...,x_n) \rightarrow (x_1,...,x_n,0),\] 
  
	\nn
	it can be naturally embedded in $\R^{n+1}$. Let $\{e^i\}$ be the standard basis of $R^{n+1}$ with $e^0$ the origin of $\R^n$, we define a \textbf{standard p-simplex} $\Delta_p$ to be the set
  
  \begin{equation}
  \Delta_p= \left\lbrace \sum_i^p \alpha_i e^i  | \sum \alpha_i =1  \quad \mbox{and} \quad \alpha_i\geq 0 \forall i \right\rbrace,
  \end{equation}
  
  \nn
  One immediately see a $0$-simplex is just a point. Consider now a $1-$simplex, by definition we have 
  
  \begin{equation}
  \Delta_1=\left \lbrace \alpha_0 e^0+\alpha_1 e^1 | \alpha_0+\alpha_1=1\right \rbrace=\left \lbrace \alpha_0 (1,0)+\alpha_1 (0,1) | \alpha_0+\alpha_1=1\right \rbrace,
  \end{equation}
  
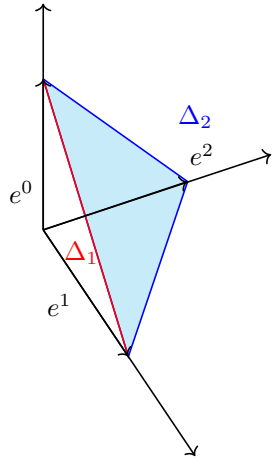
\begin{figure}[h!]
 \centering
			\begin{tikzpicture}[circuit ee IEC,semithick,small circuit symbols,set resistor graphic= var resistor IEC graphic] [h!]

				\fill [processblue!20](0,2)--(1.9,0.64)--(1.12,-1.68)--(0,2);
				\draw[blue](0,2)--(1.9,0.64)--(1.12,-1.68)--(0,2);
				\draw[red](0,2)--(1.12,-1.68);
			   	\draw  [->](0,0)--(0,3);
			   	\draw  [->](0,0)--(3,1);
			   	\draw  [->](0,0)--(2,-3);
			   	\draw [->](0,0)--(0,2);
			   	\draw [->](0,0)--(1.12,-1.68);
			   	\draw [->](0,0)--(1.9,0.64);

					\node [left] at (0,0.5) {$e^0$};
					\node [left] at (0.5,-1) {$e^1$};
					\node [right] at (1.8,1) {$e^2$};
						\node [text=red] at (0.5,-0.3) {$\Delta_1$};
							\node [text=blue] at (2,1.5) {$\Delta_2$};

		\end{tikzpicture}
		\caption{1-simplex and 2-simplex representation.}
		\label{delta12}
		\end{figure}

  \nn
 \ref{delta12}this, as can been seen from picture \ref{delta12}, is nothing else that the segment in $\R^2$ with extrema the points $(1,0)$ and $(0,1)$. Thus a $2-$ simplex is the triangle in $\R^3$ with vertices $(1,0,0),(0,1,0)$ and $(0,0,1)$, and so on for higher dimensional simplices.
  
  \nn
  It is convenient now to introduce a more compact notation for $k$ simplices: denote with $i$ the point $P_i$ identified by the basis vector $e^i$, as we saw a $k$- simplex is the planar $k$-dimensional subset of $\R^{n+1}$ having vertices at the points $P_i$; we can endow the $k$-simplex of a direction by fixing the order of vertices, thus a standard ordered $k-$ simplex can be written as:
  
  \begin{equation}
  \Delta_p=\langle 012....p \rangle.
  \end{equation}
  
  \nn
  For instance the $1-$ simplex $\Delta_1=\langle 01\rangle$ is the segment with extrema $P_0$ and $P_1$ with direction $P_0\rightarrow P_1$.
  Notice that a permutation of two vertices in a $k$-simplex switches the direction of the respective segment, so in general
  
  \begin{equation}
  \langle 012..p \rangle = sgn(\sigma)  \langle \sigma(012..p)\rangle,
  \end{equation}
  
  \nn
  with $\sigma \in S_p$ a permutation of $p$ elements. \\
  We now define a linear combination with integer coefficient of $p-$ simplices to be a simplicial p-chain 
  
  \begin{equation}
  c_p= \alpha_i \Delta_p^i.
  \end{equation}
  
  \nn
  These $p$-chains form an abelian group know as simplicial chain group $C_p^{simpl}$. Consider now the map $\de_p^i$, called $i$-th face, given by

\begin{equation}
\de_p^i\langle 01...i..p\rangle = \langle 01..\hat{i}..p\rangle,
\end{equation}

\begin{figure}[h!]
 \centering
			\begin{tikzpicture}[circuit ee IEC,semithick,small circuit symbols,set resistor graphic= var resistor IEC graphic] [h!]

				\draw [fill=processblue!20](0,0)--(3,3)--(6,0)--(0,0);
			   	\draw (-1,1)--(2,4);
			   	\draw (7,1)--(4,4);
			   	\draw (0,-1)--(6,-1);
			   	
			   	\draw [->](3,-0.2)--(3,-0.8);
			   	\draw [->](1.2,1.5)--(0.6,2.2);
			   	 \draw [->](4.8,1.5)--(5.4,2.2);
			   	
			   	\fill (0,0) circle (2pt);
			   		\fill (3,3) circle (2pt);
			   			\fill (6,0) circle (2pt);
			   				\fill (-1,1) circle (2pt);
			   					\fill (2,4) circle (2pt);
			   						\fill (7,1) circle (2pt);
			   							\fill (4,4) circle (2pt);
                   \fill (0,-1) circle (2pt);
					\fill (6,-1) circle (2pt);

					\node [left] at (0,0) {$0$};
					\node [right] at (6,0) {$1$};
					\node at (3,3.2) {$2$};
					\node [right] at (1,2) {$\de_2^1$};
					\node [left] at (5,2) {$\de_2^0$};
					\node [right] at (3,-0.5) {$\de_2^2$};
					
					\node  at (3,1.5) {$\Delta_2 $};
					
				\node [left] at (-1,1) {$0$};
					\node [left] at (0,-1) {$0$};
					\node [right] at (6,-1) {$1$};
						\node [right] at (7,1) {$1$};
					\node  [right] at (2,4) {$2$};		
					\node  [left] at (4,4) {$2$};

		\end{tikzpicture}
		\caption{Face map $\de_2^i$ applied on a 2-simplex.}
		\label{facemapfigure}
		\end{figure}
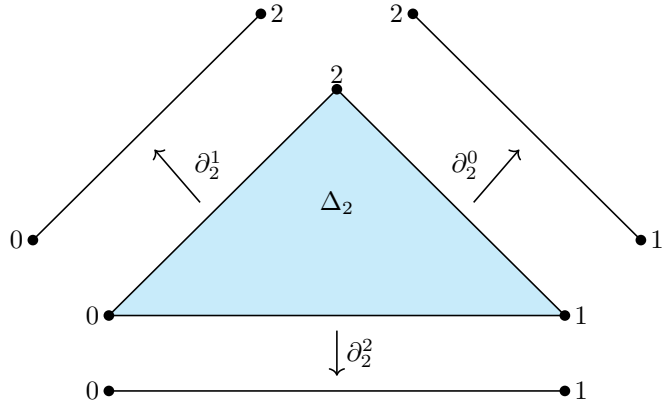

\nn
where $\hat{j}$ means $i$ has been removed. This map acts on a $p-$ simplex associating to it the $(p-1)$- dimensional face obtained removing the $i$th vertex(\ref{facemapfigure}). It is immediate to see this map allows one to define a boundary operator $\de_p$

\begin{equation}
\de_p: \Delta_p \rightarrow \Delta_{p-1},
\end{equation}

\nn
as

\begin{equation}
\de_p = \sum_{i_0}^p(-1)^i \de_p^i,
\label{simplexboundary}
\end{equation}

\nn
where the sign depends on the permutation order of the simplex left over and it's introduced to preserve the correct order. For instance, if we consider a $2$-simplex $\langle 012\rangle$, the above formula gives  

\begin{equation}
\de_2 \langle 012 \rangle = \langle 12\rangle -\langle 02\rangle + \langle 01\rangle =\langle 12\rangle +\langle 20\rangle + \langle 01\rangle,
\end{equation}

\nn
that is the correct boundary of the triangle trodden in the counter-clockwise direction. The $\de_p$ operator can be linearly extended to chains and on can trivially check that $\de_p\de_{p-1}=0$. We can therefore construct a simplicial chain complex with simplicial chain groups and the boundary operator $\de$: the $p$-th homology group of this complex is called \textbf{p-th simplicial homology group} $H_p().$

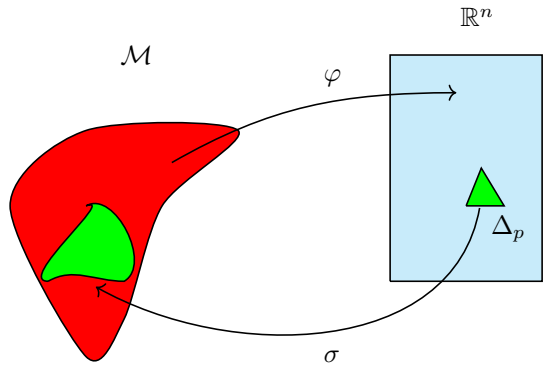
\begin{figure}[h!]
 \centering
			\begin{tikzpicture}[circuit ee IEC,semithick,small circuit symbols,set resistor graphic= var resistor IEC graphic] [h!]

				\draw [fill=red!] plot [smooth cycle] coordinates {(0,0) (1,1) (3,1) (2,0) (1.5,-1.5)(1,-2)};
				\draw [fill=processblue!20](5,-1)--(5,2)--(7,2)--(7,-1)--(5,-1);
			    \draw [fill=green](6,0)--(6.5,0)--(6.2,0.5)--(6,0);			    
			    \node (a) at (6.2, 0.1) {};
                \node (b) at (1,-1) {};
                \draw[->] (a)  to [out=-100,in=-30] (b);               
                 \node (c) at (2, 0.5) {};
                \node (d) at (6,1.5) {};
                \draw[->] (c)  to [out=30,in=180] (d);             
                \draw  [fill=green] (0.5,-1) to [out=30,in=180] (1.5,-1) to [out=30,in=30] (1,0) to [out=20,in=180] (0.5,-1);		
					\node [left] at (2,2) {$\M$};
					\node [left] at (6.5,2.5) {$\R^n $};
					\node [right] at (4,-2) {$\sigma $};
					\node [right] at (4,1.7) {$\f $};
					\node [right] at (6.2,-0.3) {$\Delta_p $};
		\end{tikzpicture}
		\caption{Singular simplex illustration}
		\label{singluarfigure}
		\end{figure}

  \nn 
 Notice that any $p$- cell is omeomorphic to a $p$-simplex, that means that a $p$- chain can be equivalenty build up starting from simplices instaed of cells.
  For a $n$-dimensional manifold $\M$, locally diffeomorphic to $\R^n$ for definition, we define a \textbf{singular q-simplex} to be a continuous map $\sigma:\Delta_p \rightarrow X$(Figure \ref{singluarfigure}). A \textbf{singular q-chain} is therefore defined as finite linear combination with integer coefficient of singular $p$-simplices. Collectively these $p$-chains form an Abelian group $S_p(\M,\Z)$. Starting from the face map defined on simplices one can introduce a boundary operator on singular chains as
  
  \begin{equation}
  \de_p=\sum_{i=0}^p (-1)^i\sigma \circ \de_p^i.
  \label{facemap}
  \end{equation}

\nn
as usual, this map turns to be nillpotent and it allows us to define singular chain complex. The homology group of this complex is called \textbf{singular homology of $\M$ with integer coefficient} $H_*^{sing}(\M,\Z)$. \\
  In order to state the fundamental theorem of this section we need the following definition
  
  \begin{defn}{Triangularizable manifold}
  A manifold $\M$ is said to be trangularizable if it can be covered by a finite number
  \end{defn}
  \nn
  Simplicial and singular homology are related by the following theorem.
  \begin{teo}
  For any triangularizable manifold $\M$ the simplicial homology is isomorphic to the singular homology:
  
  \begin{equation}
  H_p^{simpl}(\M, G) \cong H_p^{sing}(\M, G) \quad \quad, \forall p. 
  \end{equation}
  \end{teo}

	\nn
	 Let $S_p^\infty(\M)\subset S_p(\M)$ be the subset of singular $p$-chains of $\M$ generated by the smooth singular simplices. We call $p$th smooth singular homology group $H_p^\infty(\M)$ the quotient group 
  
  \begin{equation}
  H_p^\infty(\M)=\frac{Ker[\de:S_k^\infty(\M)\rightarrow S_{k-1}^\infty(\M)]}{Im[\de:S_{k+1}^\infty(\M)\rightarrow S_{k}^\infty(\M)]}.
  \end{equation}
  
  \nn
  Since the inclusion map $i:S_p^\infty(\M)\rightarrow S_p^(\M)$ commutes with the boundary operator $\de$, such a map induces a map $i_*:H_p^\infty(\M)\rightarrow H_p^(\M)$ between respective homology groups. It can be proven that the map $i_*$ is an isomorphism, so 
  
  \begin{equation}
  H_p^\infty(\M)\cong H_p^{sing}(\M).
  \end{equation}

  \section{Cohomology group}\label{cohomology section}
 An other important tool in topology, close related to Homology group, is the cohomology group. To introduce it in the most general way let's briefly revisited the homology group in a slightly different manner. We saw how to build a chain complex starting from handle decomposition and via chain group, however the notion of chain complex is a little more abstract and independent of this specific procedure; in particular a chain complex $(G_q,\delta_q)$ is a structure associable to a topological space $\mathcal{T}$ consisting of a sequence of abelian groups $G_0,G_1,..,G_n$, connected by homomorphisms $\delta_q: G_q \rightarrow G_{q-1}$ such that $\delta_q\delta_{q+1}=0$ and written as
 
 \begin{equation}
...\rightarrow G_n\xrightarrow{\delta_{n}}G_{n-1}\xrightarrow{\delta_{n-1}}G_{n-2}\rightarrow ..\rightarrow G_1\xrightarrow{\delta_{1}}....
 \end{equation}
 
 \nn
 Relation $\delta_q\delta_{q+1}=0$ implies $Im\, \delta_{q+1}\subseteq Ker\,\delta_q$; when the equality holds the chain complex is said to be exact.
 As we will see in the next a particular role is played by sequences consisting in just three non-zero terms, that is of the form 
 
 \begin{equation}
 0\rightarrow G_3 \xrightarrow{\delta_{3}}G_2 \xrightarrow{\delta_{2}}G_1 \rightarrow 0,
 \end{equation}
 
 \nn
these sequences are called \textbf{short sequences}. On the other hand, sequences with more non null terms are said long sequences. 
A short sequence turns to be exact if $\delta_3$ is injective, $\delta_2$ surjective and  $Im\, \delta_{3}\subseteq Ker\,\delta_2$.

  The $q-$dimensional Homology group is defined as 
 
 \begin{equation}
 H_q(\mathcal{T}) = \frac{Ker \,\delta_{q}}{Im\, \delta_{q+1}}.
 \end{equation}
 
\nn
 We therefore notice that homology groups of an exact chain complex are all null; on the other and we can roughly interpret homology groups as a measure of deviation of the chain complex from exactness. \\
 Notice that, if $\mathcal{T}$ is a manifold that admits a cell complex, identifying groups $G_q$ with chain groups $C_q(X)$ and homomorphism $\delta_q$ with the boundary operators $d_q$ we obtain the same definition for the Homology group given in section \ref{sectionone}.
In a similar manner let define a \textit{cochain complex} $(G^q,\delta^q)$ as a sequence of abelian groups  $G^0,G^1,..,G^n$ on a topological space $\mathcal{T}$ connected by homomorphisms $\delta^q: G^q \rightarrow G^{q+1}$ such that $\delta_q\delta_{q-1}=0$, and let's write it in dual notation as 

 \begin{equation}
...\leftarrow G^n\xleftarrow{\delta^{n}}G_{n-1}\xleftarrow{\delta^{n-1}}G^{n-2}\leftarrow ..\leftarrow G^1\xleftarrow{\delta^{1}}....
 \end{equation}
 
 \begin{defn}
 Let $(G^q,\delta^q)$ be a cochain complex, the \textbf{q-th Cohomology group} $H^q$ is defined as the quotient group 
 
 \begin{equation}
 H^q(\mathcal{T}) \equiv \frac{Ker\, \delta^q}{Im \, \delta^{q-1}}
 \end{equation}
 \end{defn}
 
 \nn
Homology and Cohomology group are related by the following theorem:

\begin{teo}(Poincaré duality) Let $\M$ be an $n-$dimensional oriented closed manifold, then 
\begin{equation}
H_q(\M)\cong H^q(\M)\cong H^{n-q}(\M), \quad\quad \mbox{for}\quad q=0,...,n .
\end{equation} 
\end{teo}

 \subsection{Simplicial and singular cohomology}
 Let $C_k^\infty(\M)$ be the smooth $k-$chain group of $\M$ and $H_k^\infty$ the smooth singular homology group defined in section \ref{simplicialhomology}.
 Consider the sequence 
 
 \begin{equation}
 ...\rightarrow C_{p+1}^\infty(\M)\xrightarrow{\de_{p+1}}C_{p}^\infty(\M)\xrightarrow{\de_p}C_{p-1}^\infty(\M)\rightarrow ...\label{chainsequence}
 \end{equation}
 
 \nn
and define the map\footnote{Contravariant functor}  $Hom(-,K)$, with $K$ a field,  acting on the sequence as follows: $Hom(-,K)$ maps each group $C_p^\infty(\M)$ to the set of morphisms $Hom(C_p^\infty(\M),K)$ and each operator $\de_p$ to the function 

\[Hom(\de_p,K)=\delta_p:(C_{p+1}^\infty(\M),K)\rightarrow Hom(C_{p}^\infty(\M),K),\]

\nn
given by $\delta(\tilde{c}_{p+1})\equiv \delta_p= \tilde{c}_{p+1}\circ \de, $ for each $ \tilde{c}_{p+1} \in Hom(C_{p+1}^\infty(\M),K) $.\\
Applying $Hom(-,K)$ on the sequence \eqref{chainsequence} we obtain the dual complex 

\begin{equation}
 ...\rightarrow Hom(C_{p-1}^\infty(\M),K)\xrightarrow{\delta_{p}}Hom(C_{p}^\infty(\M),K)\xrightarrow{\delta_{p+1}}Hom(C_{p+1}^\infty(\M,K)\rightarrow ..
\end{equation}

\nn
The $p$-th cohomology group of the previous cochain complex

\begin{equation}
H^p_\infty(\M,K)= \frac{Ker[\delta_{p+1}] }{Im[\delta_p]},
\end{equation}
 
 \nn
 is called the $p$-th \textbf{smooth singular cohomology group} of $\M$.

 \section{De Rham cohomology}

The way we build the chain complex on the topological space $\mathcal{T}$ distinguishes different cohomologies on it. A special role, for our interest, is played by the so called \textit{De Rham Cohomology}, constructed via differential forms. Before discussing it a brief introduction to differential forms is required.

 \subsection{Differential forms}
 Before giving a rigorous definition of differential forms is instructive to introduce them heuristically.
 
  \begin{defn} Let $M$ be a differentiable manifold and $T\M$ its tangent bundle, the cotangent bundle $T^*\M$ is defined as the dual bundle of $T\M$, consisting of all maps:
  
  \begin{equation}
  T^*\M: T\M\rightarrow \R.
\end{equation}   
\end{defn}
\nn
On the other side the cotangent bundle can be seen as the bundle over $\M$ having the cotangent space $T^*_P\M$ at each point $P$ as fiber. \\
 Elements of the cotangent space in a point are called (algebraic) $1-$form: $\omega_1 \in T_P^*\M$.\\
 Sections of the cotangent bundle, i.e. maps associating an algebraic  $1-$ form to each point of the manifold are known as differential $1-$ forms.
Think at the difference between a vector and vector field, a vector is an element of the tangent space in a point of the manifold, while a vector field is a section of the tangent bundle, roughly speaking a function that maps a point of the manifold into a vector, similarly, as far as concern forms, we call algebraic $1-$ forms the elements of the cotangent space in a point, namely maps mapping vectors to real numbers, and differential $1-$forms, the sections of the cotangent bundle, that is maps associating to each point of the manifold an element of the cotangent space in that point. Those objects are commonly called $1-$forms, and we will use the same word to identify both, the meaning will be clear from the context. The same ambiguity will also be obviously true for higher order forms.\\
Consider now a differentiable manifold $\M$ and a set of local coordinates $x_i$ in a neighborhood $U$ around a point $P$ of $\M$. The set ${\de^i}$ of partial derivatives operators provides a basis of the tangent space $T_P\M$, a basis of the cotangent space $T_P^*\M$ will therefore be given by dual elements of $\de^i$, called differentials $dx_i$ and defined by
 
 \begin{equation}
 dx_i \de^j = \delta_i^j.
 \end{equation}
 
 \nn
 Thanks to the vector space structure of cotangent space, a $1-$form $\omega_1$ can be expressed as a linear combination of differentials
 
 \begin{equation}
 \omega_1 = a_i(x) dx_i,
 \end{equation}
 
 \nn
 where $a_i(x)$ are elements of the space $\mathcal{E} (U)$ of differential functions on $U$.\\ 
 A $1-$form is so nothing else that a covariant vector.
In order to define higher order forms, we need to introduced the following antisymmetric product:

  \begin{defn}(wedge product) Let $\omega_1=\omega dx_1$ and $\gamma_1=\gamma dx_2$ be two $1-$forms, the \textbf{wedge product} or \textbf{exterior product} is an operation denoted by $\wedge$, antisymmetric

  \begin{equation}
      \omega_1\wedge \gamma_1=-\gamma_1 \wedge \omega_1,
  \end{equation}
 \nn
 acting as 
 
 \begin{equation}
    \omega_2= \omega_1\wedge \gamma_1 = \omega\gamma dx_1\wedge dx_2.
 \end{equation}
 
 \nn
The form $omega_2$ is called a $2$ form.
 \end{defn}
 
 \nn
 We are now ready to formally define $k-forms$.
 
  \begin{defn} A $k-$form $\omega_k$ is a smooth section of the $k-th$ external algebra of the cotangent bundle:
 
 \begin{equation}
 \omega_k : \M \rightarrow \bigwedge^k(T^*\M),
 \end{equation}
 \end{defn}

 \nn
 in other words it's a map
 
 \begin{equation}
 \omega_k(x): \underbrace{T_x(\M) \times T_x(\M) \times...\times T_x(\M)}_{k} \rightarrow \R,
 \end{equation}
 
 \nn
 where $T_x\M$ is the tangent space of $\M$ in the point $x$. Denote by $\Omega^k(\M)$ the vector space of $k-$forms on $\M$.
 The dimension of $\Omega^k(\M)$ can be immediately read from the antisymmetry of the wedge product:
 
 \begin{equation}
 dim \Omega^k(\M) = \binom{d}{k}=\frac{d!}{k!(d-k)!};
 \end{equation}
 
 \nn
 from the binomial identity 
 
 \begin{equation}
 \binom {d}{k}=\binom {d}{d-k}
 \end{equation}

\nn
 it follows that $dim \Omega^k(\M)=dim \Omega^{d-k}(\M)$, we can therefore define an isomorphism, called Hodge dual and denoted by "$\ast$" mapping each $k-$form to the$(d-k)-$form:

 \begin{equation}
 \ast \omega = \frac{1}{(d-k)!} dx^{i_{d-k}}\wedge...\wedge dx^{i_1}\tilde{\omega}_{i_1..i_{d-k}},
 \end{equation}
 
 \nn
 where 
 \begin{equation}
 \tilde{\omega}_{i_1..i_{d-k}}= \frac{1}{(k)!}\varepsilon_{i_1..i_d,j_1,j_k} \omega^{j_1..j_k},
 \end{equation}
 
 \nn
 with $\varepsilon_{i_1..i_d,j_1,j_k}$ the Levi-Civita tensor in $d$ dimensions.

  \begin{defn}
  Let $\Omega^k(\M)\ni\omega=f_Idx^I=f_{i_1..i_k}dx^{i_1}\wedge dx^{i_2}\wedge...\wedge dx^{i_k}$ be a $k-$form on an $n$ dimensional manifold. The exterior derivative $d$ is a differential operator
  
  \begin{equation}
      d: \Omega^k \rightarrow \Omega^{k+1}
  \end{equation}
  \nn
  acting as
  \begin{equation}
  d\omega=\sum_{i=1}^n \frac{\de f}{\de x^i} dx^i \wedge dx^I.
  \end{equation}
  \end{defn}
 
 \begin{Ese}
 Consider $\M=\R^3$, the only non trivial $\Omega^k(\R^3)$ are for $k=0,1,2,3$; denoting by $x,y,z$ a set local coordinates, their elements can be written as 
 
 \begin{equation}
     \begin{split}
    &\Omega^0(\R^3) \ni \omega_0 = f_0(x,y,z),\\
    &\Omega^1(\R^3) \ni \omega_1= f_1(x,y,z) dx+f_2(x,y,z) dy+f_3(x,y,z) dz,\\
    &\Omega^2(\R^3) \ni \omega_2= f_{12}(x,y,z) dx\wedge dy+f_{13}(x,y,z)dx\wedge dz +f_{23}(x,y,z)dy\wedge dz,\\
    &\Omega^3(\R^3) \ni \omega_3 = f_{123}(x,y,z)dx\wedge dy\wedge dz.
     \end{split}
 \end{equation}
 
 \nn
 with $f_I\in \C^{\infty}$. Applying the external differential operator, one obtains
 
 \begin{equation}
     \begin{split}
    &d \omega_0 = \de_x f_0 dx+\de_y f_0 dy +\de_z f_0 dz,\\
    &d\omega_1= \left (\de_xf_2-\de_yf_1 \right )dx\wedge dy-\left (\de_zf_1-\de_xf_3 \right )dx\wedge dz +  \left (\de_yf_3-\de_zf_2 \right )dy\wedge dz,\\
    &d\omega_2= \left (\de_x f_{23}+\de_y f_{13}+\de_zf_{12}\right ) dx\wedge dy\wedge dz,\\
    &d\omega_3 = 0.
     \end{split}
 \end{equation}
 
 \nn
 Therefore, the exterior derivative acting on zero, one and to two forms (of $\R^3$) respectively corresponds to ordinary gradient, curl an divergence.
 \end{Ese}
 
 \begin{Pre}\label{prepnihil}
 The exterior differential operator $d$ is nihilpotent:
 
 \begin{equation}
     d^2=0.
 \end{equation}
 It immediately follows by Schwarz Theorem.
 \end{Pre}
 
 \nn
 A form whose exterior derivative is null, as $\omega_3$ in the previous example, is said a closed form. On the other hand, a $k-$ form that can be written as the exterior derivative of a $k-1$ form is called exact. Notice that, because of preposition \ref{prepnihil}, any exact form is also closed.
 
 \begin{teo}Poincaré Lemma (for differential forms)
 Let $\omega$ be a smooth close $p-$ differential form defined on $U\subset \R^n$. If $U$ is contractible, then $\omega$ is exact.
 \end{teo}
 
 \nn
 In order to make same practice with forms, let consider the example of classical electrodynamic. 
\begin{Ese}Electromagnetism.\\
Let associate to the antisymmetric tensor $F^{\m\n}$ the $2-$ form:

\begin{equation}
F=\frac{1}{2}dx^\n\wedge dx^\m F_{\m\n},
\end{equation} 

\nn
and to the four-current $J^\m$ the $1-$form

\begin{equation}
J=dx^\m J_\m.
\end{equation}

\nn
Performing the exterior derivative of $F$

\begin{equation}
dF= d\left (\frac{1}{2} dx^\n\wedge dx^\m F_{\m\n} \right )=\frac{1}{2}dx^\n\wedge dx^\m\wedge dx^\rho \de_{[\rho}F_{\m\n]},
\end{equation}

\nn
we see that the Bianchi identity, that encodes two of the Maxwell's equations (Faraday's law and magnetic field solenoidality condition), can be expressed in terms of forms just like the closure of $F$: 

\begin{equation}
dF=0.
\end{equation}

\nn
Thanks to Poincarè lemma, the closure of $F$ implies its local exactness, it therefore exists a $1-$form $A$ such that 

\begin{equation}
F=dA.
\end{equation}

\nn
If we consider two forms $A$ and $A+d\Lambda$ that differ by an exact $1$ form, we see $F$ turns to be unchanged, we can therefore interpret gauge invariance as the equivalence of forms belonging to the same cohomology class.  
Let now see that the other Maxwell equation can be written as an equation between $3-$forms as

\begin{equation}
d\ast F=\ast J,
\label{Eq.Maxwell2}
\end{equation}

\nn
In order to obtain a relation between $1-$ forms, we can apply the hodge operator to both sides of \eqref{Eq.Maxwell2}, the right hand side, bypassing some algebra, becomes:

\begin{equation}
\ast d\ast F = \ast d\left (\frac{1}{2} dx^\n\wedge dx^\m \tilde{F}_{\m\n} \right )= dx^\m(\de^\n F_{\n\m}),
\end{equation}

\nn
so we have:

\begin{equation}
dx^\m (\de^\n F_{\n\m})=dx^\m J_\m,
\end{equation}

\nn
that is equivalent to Maxwell equation. From \eqref{Eq.Maxwell2}  we can immediately read off current conservation, it's sufficient to notice that the right hand side is a close form, so the left hand one must also be close for consistency, and therefore

\begin{equation}
d\ast J =0.
\end{equation}

\nn
Differential forms formalism allows us to reinterpret electromagnetic laws in a topological sense: consider the integral over a tridimensional manifold $V$ of the $3-$form $d\ast F$;  if $V$ does not contain any singularity (charges or currents) we can apply twice Stokes' theorem in order to obtain:

\begin{equation}
\int_V d\ast F=\int_{\de V} \ast F=\frac{1}{2}\int_{\de V}dx^\m\wedge dx^\n \epsilon_{\m\n\rho\sigma}F^{\rho\sigma}= \int_{\de V}\left [ dx^0\wedge dx^i B_i +dx^j\wedge dx^i \epsilon_{ijk}E^k \right ]=0,
\label{Max3}
\end{equation}

\nn
where in the last equality we use the fact that $S=\de V$ has no boundary. Conversely, if $V$ contains $n$ singularities localized at $x_i$, we can choose $n$ small balls $\B_i$ centered at $x_i$ and such that $\B_i$ lies in the interior of $V$ and $B_i\cap B_j =\emptyset$, for $i\neq j$, then we can write:

\begin{equation}
\int_V d\ast F = \int_{V\setminus \cup \B_i} d\ast F + \sum_i \int_{\B_i} d\ast F;
\end{equation}

\nn
the first term in the right hand side can be handled just like in equation \eqref{Max3}, while the second term, using equation \eqref{Eq.Maxwell2} becomes

\begin{equation}
\sum_i \int_{\B_i} d\ast F=\sum_i \int_{\B_i} \ast J= \frac{1}{3!}\sum_i \int_{\B_i} dx^\m\wedge dx^\n \wedge dx^\rho \epsilon_{\m\n\rho\sigma}J^\sigma,
\end{equation} 

\nn
putting all together we finally find 

\begin{equation}
\int_{\de V}\left [ dx^0\wedge dx^i B_i +dx^j\wedge dx^i \epsilon_{ijk}E^k \right ]=\frac{1}{3!}\sum_i \int dx^\m\wedge dx^\n \wedge dx^\rho \epsilon_{\m\n\rho\sigma}J^\sigma.
\label{finally1}
\end{equation}

\nn
If we choose a spacial region of integration, that is we fix the time, and we therefore focus on electrostatic, the previous equation reads 

\begin{equation}
\int_{V_{spacial}} d\ast F=\int_{\de V_{spacial}}dx^j\wedge dx^i \epsilon_{ijk}E^k = \int_{\de V_{spacial}}\vec{E}\cdot d\vec{S}= \sum_i \int d^3x J^0 = Q.
\end{equation}

\nn
We have obtained Gauss theorem.
On the other and, if we fix a spacial coordinate and we consider a space-time volume region $V$, the \eqref{finally1} becomes

\begin{equation}
\int_{\de V}\left [ dx^0\wedge dx^i B_i +dx^j\wedge dx^i \epsilon_{ijk}E^k \right ]= \int  \vec{B}\cdot dxdt + \int \vec{E}\cdot d\vec{S}= \sum_i \int \vec{J}\cdot d\vec{S} dt, 
\end{equation} 
\nn
and this is nothing else that Maxwell-Ampère law.

\end{Ese}

\nn
We are now ready to introduce De Rham cohomology. 
 
 \begin{defn}De Rham Cohomology Group\\
 Let $\M$ be a $n-$ dimensional smooth manifold, $\Omega^k(\M)$ the group of $k-$ forms defined on $\M$ and $d$ the exterior derivative operator, the $p-$th cohomology group of the complex 
 
 \begin{equation}
 0\xrightarrow{d} \Omega^0(\M)\xrightarrow{d} \Omega^1(\M)\xrightarrow{d}...\xrightarrow{d} \Omega^n(\M)\xrightarrow{d} 0,
 \end{equation}
 
 \nn
 is called $p-$th de Rham cohomology group of $\M$:
 
 \begin{equation}H_{dR}^p(\M)=\frac{Ker[d:\Omega^p(\M)\rightarrow \Omega^{p+1}(\M)]}{Im[d:\Omega^{p-1}(\M)]\rightarrow \Omega^p(\M)}].
 \end{equation}
 \end{defn}
 
\begin{figure}[h!]
  \centering
\begin{tikzpicture}
\complex{(0,0)}
\end{tikzpicture}
\caption{Pictorial illustration of a portion of the De Rham complex.}
\label{formcomplex}
 \end{figure}
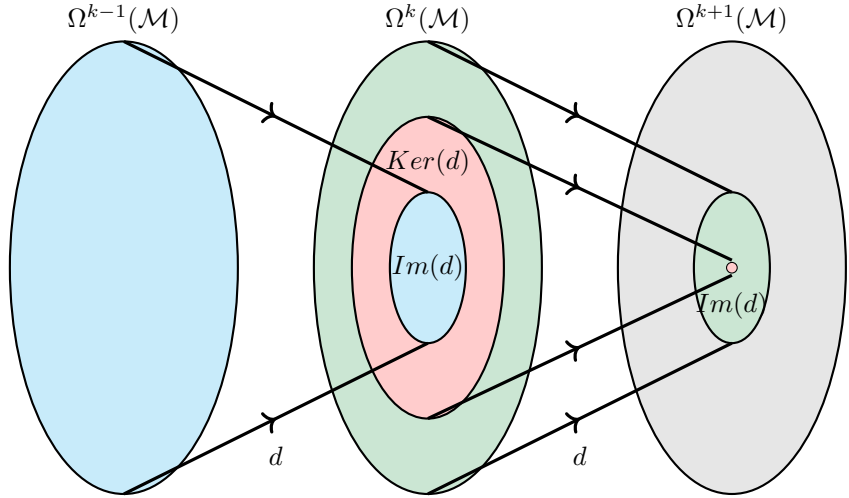

 \nn
 In figure \ref{formcomplex} a pictorial representation of the De Rham complex is shown. In the middle picture, the red subspace of $\Omega^k(\M)$ is the set of closed $k$ forms and the light blue subspace the one of exact $k-$forms.
  
   \section{Mayer-Vietoris Sequence}
 
 Mayer-Vietoris sequence provides a way to compute the cohomology of the union of two open sets. Consider $\M=U\cup V$, with $U$ and $V$ open and the following diagram:
 
 \begin{center}
			\begin{tikzpicture}[circuit ee IEC,semithick,small circuit symbols,set resistor graphic= var resistor IEC graphic] [h!]
				
				\draw (0,0)[->] -- (1,1);
				\draw (1.2,1) [->]-- (2.2,0);
				\draw (0,-0.2) [->]-- (1,-1.2);
				\draw (1.2,-1.2) [->]-- (2.2,-0.2);
		
				\node [left] at (0,0) {$U\cup V$};
				\node [above] at (1.2,1) {$U$};
				\node [below] at (1.2,-1.2) {$V$};
					\node [right] at (2.2,0) {$\M$};
					
					\node [left] at (0.5,0.7) {$\tau$};
					\node [left] at (0.5,-0.8) {$\lambda $};
					\node [right] at (1.5,0.7) {$\f $};
					\node [right] at (1.5,-0.8) {$\psi $};
		\end{tikzpicture}
		
		\end{center}
		\nn
		where $\tau,\lambda,\f$ and $\psi$ are inclusion maps. Consider then the pullback maps 
		
		\begin{equation}
		\begin{split} &\tau^*: \Omega^p(U)\rightarrow\Omega^p(U\cup V) \quad \quad \tau^*(\omega)=\omega |_{U\cup V},\\&\lambda^*: \Omega^p(V)\rightarrow\Omega^p(U\cup V)\quad \quad \lambda^*(\omega)=\omega |_{U\cup V},\\ &\f^*: \Omega^p(\M)\rightarrow\Omega^p(U)\quad \quad \,\,\,\quad \f^*(\omega)=\omega |_{U},\\&\psi^*: \Omega^p(\M)\rightarrow\Omega^p(V)\quad \quad \,\,\,\quad \psi^*(\omega)=\omega |_{V},\end{split}
		\end{equation}
 
 \nn
 then the following sequence, called Mayer-Vietoris sequence,

 \begin{equation} 0\rightarrow \Omega^p(\M) \xrightarrow{\f^*\oplus\psi^*} \Omega^p(U)\oplus \Omega^p(V) \xrightarrow{\tau^*-\lambda^*}\Omega^p(U\cap V)\rightarrow 0 
 \label{MVsequence}
 \end{equation}
 
 \nn
 where $\f^*\oplus\psi^*(\omega)=(\omega |_U, \omega |_V)$ and $(\tau^*-\lambda^*)(\omega,\omega')=\omega |_{U\cap V}-\omega' |_{U\cap V}$, is exact. 
 To proof the exactness of sequence \eqref{MVsequence} we should first notice it is a short sequence and as we said in section \ref{cohomology section} for these type of sequences the exactness is granted by the injectivity of $\f^*\oplus\psi^*$, the surjectivity of $(\tau^*-\lambda^*)$ and the condition $Ker [\tau^*-\lambda^*]=Im[ \f^*\oplus\psi^*] $, let proof them:
 
 \begin{enumerate}
 \item \textbf{Injectivity of} $\f^*\oplus\psi^*$: \\Let $\omega \in \Omega^p(U \cap V)$ then  $\f^*\oplus\psi^*(\omega)=(\omega |_U,\omega |_V)=0$ implies  $\omega |_U=\omega |_V=0$ and so $\omega=0$ on $\M$;
 
  \item \textbf{Surjectivity of} $\tau^*-\lambda^*$: \\ we should prove that $\forall \omega \in \Omega^p(U\cap V)\, \exists \, (\eta,\eta')\in \Omega^p(U)\oplus \Omega^p(V) | (\tau^*-\lambda^*)(\eta,\eta')=\omega $. Let $\{\rho_U,\rho_V\}$ a partition of the unity subordinate to the covering $\{U,V\}$\footnote{A partition of the unity subordinate to a cover $\{U_\alpha\}$ of a manifold $\M$ is a collection of non negative smooth functions $\{\rho_\alpha\}$ with support $Supp [\rho_\alpha]\subset U_\alpha$, such that $\sum \rho_\alpha=1$ and  $\sum \rho_\alpha$ is a finite sum in a neighborhood of any point of $\M$.}, then for any $\omega \in \Omega^p(U \cap V)$ we can define 
  
  \[\eta =\rho_V\cdot\omega \in \Omega^p(U) \quad\quad\mbox{and} \quad\quad \eta'=-\rho_U\cdot\omega \in \Omega^p(V),\]
  
  \nn
  and we have 
  
  \[(\tau^*-\lambda^*)(\eta,\eta')=\eta_{U\cap V}-\eta'_{U\cap V}=\rho_V\omega+\rho_U\omega=\omega.\]
  
  \item $Ker [\tau^*-\lambda^*]=Im[ \f^*\oplus\psi^*] $:\\
  Let $\omega \in  \Omega^p(\M)$, we have :
  
  \[(\tau^*-\lambda^*)\circ (\f^*\oplus\psi^*)(\omega)=(\tau^*-\lambda^*)(\omega |_U,\omega |_V)=(\omega |_U)_{U\cap V}-(\omega |_V)_{U\cap V}=0, \]

\nn
so $Im[ \f^*\oplus\psi^*]\subseteq Ker [\tau^*-\lambda^*]$. On the other hand, for any $(\omega,\omega')\in Ker[\tau^*-\lambda^*]$ we have 
$\omega |_{U\cap V}=\omega'|_{U\cap V}$, so we can define $\eta\in \Omega^p(\M)$ as

\[\eta(p) = \begin{cases} \omega \,, \mbox{for}\,\, p \in U,\\\omega' ,  \mbox{for}\,\,p \in V, \end{cases}  \] 
   
   \nn
   and we have $( \f^*\oplus\psi^*)(\eta)=(\omega,\omega')$ so $Ker [\tau^*-\lambda^*]\subseteq Im[ \f^*\oplus\psi^*]$. Combining thes two results the equality trivially follows.

 \end{enumerate}

 \nn
 The Mayer-Vietoris sequence induces a long exact sequence of cohomology groups, also called Mayer Vietoris sequence:
 
 \begin{equation}
 ...\rightarrow H^p(\M)\rightarrow H^p(U)\oplus H^p(V)\rightarrow H^p(U\cap V)\rightarrow H^{p+1}(\M)\rightarrow H^{p+1}(U)\oplus H^p{p+1}(V)\rightarrow H^{p+1}(U\cap V)\rightarrow ...
 \end{equation}
 
 the commutative diagram\footnote{A diagram is said to be commutative if all arrows paths having same start and endpoint give the same result.}
 \begin{center}
			\begin{tikzpicture}[circuit ee IEC,semithick,small circuit symbols,set resistor graphic= var resistor IEC graphic] [h!]
				
				\draw [fill=processblue!20](-2,2)--(13,2)--(13,-6)--(-2,-6)--(-2,2);
				
				\draw (0,0) [->]-- (1,0);
				\draw (3,0) [->]-- (4,0);
				\draw (7.5,0) [->]-- (8.5,0);
				\draw (11,0) [->]-- (12,0);
              	\node [left] at (0,0) {$0$};
				\node [right] at (1,0) {$\Omega^{p-1}(\M)$};
				\node [right] at (4,0) {$\Omega^{p-1}(U)\oplus\Omega^{p-1}(V)$};
			   \node [right] at (8.5,0) {$\Omega^{p-1}(U\cap V)$};
			   \node [right] at (12,0) {$0$};

			   			\draw (0,-2) [->]-- (1,-2);
				\draw (3,-2) [->]-- (4,-2);
				\draw (7.5,-2) [->]-- (8.5,-2);
				\draw (11,-2) [->]-- (12,-2);
              	\node [left] at (0,-2) {$0$};
				\node [right] at (1,-2) {$\Omega^{p}(\M)$};
				\node [right] at (4.4,-2) {$\Omega^{p}(U)\oplus\Omega^{p}(V)$};
			   \node [right] at (8.5,-2) {$\Omega^{p}(U\cap V)$};
			   \node [right] at (12,-2) {$0$};

			   			\draw (0,-4) [->]-- (1,-4);
				\draw (3,-4) [->]-- (4,-4);
				\draw (7.5,-4) [->]-- (8.5,-4);
				\draw (11,-4) [->]-- (12,-4);
              	\node [left] at (0,-4) {$0$};
				\node [right] at (1,-4) {$\Omega^{p+1}(\M)$};
				\node [right] at (4,-4) {$\Omega^{p+1}(U)\oplus\Omega^{p+1}(V)$};
			   \node [right] at (8.5,-4) {$\Omega^{p+1}(U\cap V)$};
			   \node [right] at (12,-4) {$0$};

			   	\draw (1.5,1.4) [->]-- (1.5,0.4);
			   	 \draw (5.7,1.4) [->]-- (5.7,0.4);
			   	 \draw (9.2,1.4) [->]-- (9.2,0.4);
			   	 
			   	   	\draw (1.5,-0.5) [->]-- (1.5,-1.5);
			   	 \draw (5.7,-0.5) [->]-- (5.7,-1.5);
			   	 \draw (9.2,-0.5) [->]-- (9.2,-1.5);
			   	 
			   	    	\draw (1.5,-2.5) [->]-- (1.5,-3.5);
			   	 \draw (5.7,-2.5) [->]-- (5.7,-3.5);
			   	 \draw (9.2,-2.5) [->]-- (9.2,-3.5);
			   	 
			   	 	    	\draw (1.5,-4.5) [->]-- (1.5,-5.5);
			   	 \draw (5.7,-4.5) [->]-- (5.7,-5.5);
			   	 \draw (9.2,-4.5) [->]-- (9.2,-5.5);

		\end{tikzpicture}
		
		\end{center}
		\nn
		Consider a closed $k-$form $\omega\in \Omega^k(\M)$ and a singular $k-$simplex $\sigma:\Delta_k\rightarrow M$. Define the integral of $\omega$ over $\sigma$ by 
		
		\begin{equation}
		\int_\sigma \omega \equiv \int_{\Delta_k}\sigma^*\omega,
		\label{integraloversigma}
		\end{equation}
 
 \nn
 Now we define the integral of $\omega$ over any smooth chain $c=\sum_\sigma n_\sigma \sigma \in C_p^\infty(\M)$ extending \eqref{integraloversigma} linearly:
 
 \begin{equation}
 \int_c \omega \equiv \sum_\sigma n_\sigma\int_\sigma\omega.
 \end{equation}
 
 \nn
 Integration provides a homomorphism $\Psi(\omega):C_p^\infty(\M)\rightarrow \K$, thus, for each $p$-form $\omega$ on $\M$, we have a homomorphism $\Psi: \Omega^p(\M) \rightarrow Hom (C_p^\infty(\M),\K)$ between vector spaces. We want to show now that $\Psi$ induces a homomorphism between cohomology groups. Consider the following diagram 

 \begin{center}
			\begin{tikzpicture}[circuit ee IEC,semithick,small circuit symbols,set resistor graphic= var resistor IEC graphic] [h!]
				
				\draw [fill=processblue!20](-2,1)--(14,1)--(14,-3)--(-2,-3)--(-2,1);
				
				\draw (-1,0) [->]-- (1,0);
				\draw (3,0) [->]-- (5,0);
				\draw (6.5,0) [->]-- (8.5,0);
				\draw (11,0) [->]-- (13,0);
             
				\node [right] at (1,0) {$\Omega^{p-1}(\M)$};
				\node [right] at (5,0) {$\Omega^{p}(\M)$};
			   \node [right] at (8.5,0) {$\Omega^{p+1}(\M)$};
			   		   
			   			\draw (-1,-2) [->]-- (0,-2);
				\draw (3.2,-2) [->]-- (4.2,-2);
				\draw (7.5,-2) [->]-- (8.5,-2);
				\draw (12,-2) [->]-- (13,-2);
				\node [right] at (0,-2) {$Hom(C_{p-1}^\infty(\M), \K)$};
				\node [right] at (4.4,-2) {$Hom(C_{p}^\infty(\M), \K)$};
			   \node [right] at (8.5,-2) {$Hom(C_{p+1}^\infty(\M), \K)$};

			   	   	\draw (1.5,-0.5) [->]-- (1.5,-1.5);
			   	 \draw (5.7,-0.5) [->]-- (5.7,-1.5);
			   	 \draw (9.2,-0.5) [->]-- (9.2,-1.5);

			   	 	\node at (3.7,0.2) {$d$};
			   	 	\node at (8,0.2) {$d$};
			   	 	
			\node  at (1.7,-1) {$\Psi$};
				\node  at (5.9,-1) {$\Psi$};
				\node  at (9.4,-1) {$\Psi$};
				
		   \node  at (3.7,-1.8) {$\delta$};
			     \node  at (8,-1.8) {$\delta$};

		\end{tikzpicture}
		
		\end{center} 
		
		\nn
		and let $\omega\in \Omega^{p-1}(\M)$ and $\sigma \in C_p^\infty(\M)$, using equation\eqref{facemap} and Stokes' theorem, we have
		
		\begin{equation} 	 
		\begin{split}\delta(\Psi(\omega))\sigma &=\Psi(\omega)(\de\sigma)=\int_{\de\sigma}\omega= \sum_{i=0}^p(-1)^i\int_{\Delta_{p-1}}{\de^i_p}^*\circ\sigma*\omega=\\&\sum_{i=0}^p\int_{\de^i_p(\Delta_{p-1})} \sigma^*\omega =\int_{\de\Delta_p}\sigma^*\omega\\&=\int_{\Delta_p}d(\sigma^*\omega)=\int_{\Delta_p}\sigma^*(d\omega)=\Psi(d\omega)(\sigma)\end{split},
		\end{equation}
		
		\nn
		then there is a homomorphism $\Psi^* : H^p_{DR}(\M)\rightarrow H^p_\infty(\M,\K)$. \\
		 When $\Psi^*$ is an isomorphism for any $p$, the manifold is said to be De Rham. We are now ready to state the fundamental result of De Rham theory: De Rham theorem.

 \begin{teo}\textbf{(De Rham Theorem).} 
 Every smooth manifold is De Rham.
 
 \begin{equation}
 H^p_{DR}(\M)\cong H^p_\infty(\M,K).
 \end{equation}
 
 \end{teo}

The following bilinear forms are non degenerate:
\begin{teo}
There exist the following non-degenerate bilinear forms:
\begin{equation}
\begin{split}
 (i) \quad  H_p (\mathcal{M}, \mathbb{C}) \times H^p (\M,d) \quad & \mapsto \quad \mathbb{C} \\
\langle \psi |\gamma ] \quad & \mapsto \quad \int_{\gamma} \psi \\
(ii) \quad H^{2n-p} (\M,d)\times H^p (\mathcal{M}),d) \quad & \mapsto \quad \mathbb{C} \\
\langle \psi |\f\rangle  \quad & \mapsto \quad \int_{\mathcal{M}} \psi \wedge \f \\
\end{split}
\label{BilineariDeRham}
\end{equation}
\end{teo}
 
\section{Riemann period relations}
\begin{figure}[h!]
 \centering
		\begin{tikzpicture}[circuit ee IEC,semithick,small circuit symbols,set resistor graphic= var resistor IEC graphic] [h!]
				
		\begin{scope}[very thick,decoration={markings,mark=at position 0.5 with {\arrow{>}}}] 
     \draw[postaction={decorate}] (0,0)--(-0.5,-1);
    \draw[red,postaction={decorate}] (0,0)--(0.5,0.8);
    \draw[blue,postaction={decorate}] (0.5,0.8)--(2,1);
    \draw[red,postaction={decorate}] (3,0)--(2,1);
    \draw[blue,postaction={decorate}] (3.5,-1)--(3,0);
     \draw[green,postaction={decorate}] (3.5,-1)--(4.5,-1);
\end{scope}	
	\fill (3.5,-1) circle (2pt);
  	\fill (0,0) circle (2pt);
  	 \fill (0.5,0.8) circle (2pt);
  	 \fill (2,1) circle (2pt);
  	 \fill (3,0) circle (2pt);
  	 \coordinate (P) at ($(0, 0) + (30:3cm and 2cm)$);
  	 
		\draw (4,0.5)--(5,0.5);
		\draw (4,0.4)--(5,0.4);
		\draw (5,0.3)--(5,0.4);
		\draw (5,0.3)--(5.5,0.45);
		\draw (5,0.6)--(5.5,0.45);
		\draw (5,0.6)--(5,0.5);

  	 \draw[processblue] plot [smooth, tension=2] coordinates { (5.5,-1) (6,0) (6.24,1.65)};
  	  \draw[processblue] plot [smooth, tension=2] coordinates { (9,-1.2) (7.8,-0.8)  (8.95,0.32)};
  	 \begin{scope}[very thick,decoration={markings,mark=at position 0.5 with {\arrow{<}}}] 
     \draw[red,postaction={decorate}] (6.9,0.95) -- (6,-1);
    \draw[red,postaction={decorate}] (7.6,0.55) -- (6.7,-1.1);  
    \draw[blue,postaction={decorate},rotate=-30] ($(5,4)+ (30:1.5cm and 0.5cm)$) arc   (-80:250:1.5cm and 0.5cm);
     \draw[blue,postaction={decorate}] (6.7,-1.1) -- (8,-1.5);
      \draw[postaction={decorate}] (5.5,-1.5) -- (6,-1);
       \draw[green,postaction={decorate}] (9,-1.5)--(8,-1.5);
     
   \end{scope}	
   
			 \fill ((8,-1.5) circle (2pt);
			 	\fill (6,-1) circle (2pt);
			 	\fill (6.9,0.95) circle (2pt);
			 	 	\fill (7.6,0.55) circle (2pt);
			 	 	\fill (6.7,-1.1) circle (2pt);

		\end{tikzpicture}	
		\caption{Riemann surface construction from $4$g-gone }
\label{4ggone}
 \end{figure}
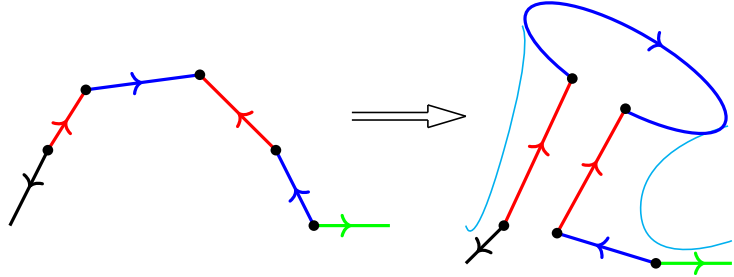

In this section we would like to give a geometric interpretation of the intersection between cocycles on De Rham cohomology and to compute-them through \textit{Riemann bilinear relation}.  \\
Each closed oriented Riemann surface $\Sigma$  of genus $g$ can be realized starting to a plane figure $\tilde{\Sigma}$ with $4g$  edges ($4g$-gone ) in the order: $a_1 b_1 a_1^{-1} b_1^{-1}....a_gb_ga_g^{-1}b_g^{-1}$., where $a_i$ and $b_i$ are clockwise oriented,  while $a_i^{-1}$ and $b_i^{-1}$ are counterclockwise oriented.  Gluing each edge with his "inverse edge",  according to the orientation,  we obtain the desired surface $\Sigma$, as showed in \ref{4ggone}.
Notice that if we change the orientation, the  construction with the same number of edges carries out different surfaces.  \\

\begin{Ese}
Consider the 4-gone, i.e. the parallelogram. The four possible orientations of the parallelogram edges, shown in figure \ref{gone_1}, generate four different topologies of bidimensional surfaces: two orientable surfaces, the sphere and the torus, and two unorientable ones, the projective plane and the Klein bottle.

\begin{figure}[h!]
 \begin{center}

			\begin{tikzpicture}[circuit ee IEC,semithick,small circuit symbols,set resistor graphic= var resistor IEC graphic] [h!]

		      \draw [red,line width=0.4mm](0,0) [->]-- (1,0);
				\draw [blue,line width=0.4mm](0,-1) [->]-- (1,-1);
				\draw [red,line width=0.4mm](0,0) [->]-- (0,-1);
				\draw [blue,line width=0.4mm](1,0) [->]-- (1,-1);

		      \draw [red,line width=0.4mm](3,0) [->]-- (4,0);
				\draw [blue,line width=0.4mm](3,-1) [->]-- (3,0);
				\draw [red,line width=0.4mm](4,-1) [->]-- (3,-1);
				\draw [blue,line width=0.4mm](4,0) [->]-- (4,-1);
				
				   \draw [red,line width=0.4mm](6,0) [->]-- (7,0);
				\draw [blue,line width=0.4mm](6,0) [->]-- (6,-1);
				\draw [red,line width=0.4mm](7,-1) [->]-- (6,-1);
				\draw [blue,line width=0.4mm](7,0) [->]-- (7,-1);
				
				  \draw [red,line width=0.4mm](9,0) [->]-- (10,0);
				\draw [blue,line width=0.4mm](9,0) [->]-- (9,-1);
				\draw [red,line width=0.4mm](9,-1) [->]-- (10,-1);
				\draw [blue,line width=0.4mm](10,0) [->]-- (10,-1);

			   	 	\node at (0.5,-1.5) {$Sphere$};
			   	 	\node at (3.5,-1.5) {$RP^1$};
			   	 	
			\node  at (6.5,-1.5) {$Klein\, bottle$};
				\node  at (9.5,-1.5) {$Torus$};

		\end{tikzpicture}
	\end{center}
		\caption{$4$-gone }
		 \label{gone_1}
	
 \end{figure}

\nn
Each curve $a_i$ or $b_i$ is a closed curve on $\Sigma$ that satisfy to following intersection relations:
\begin{equation}
\begin{split}
& a_i \cdot a_k = b_i \cdot b_k =0 \\
& a_j \cdot b_k = - b_k \cdot a_j = \delta_{jk} \\
\end{split}
\end{equation}
These are homotopically distinct curves that form a \textit{canonical basis} for the first homology group of $\Sigma_g$, that, as we saw in the first chapter, is isomorphic to the fundamental group: 
\begin{equation}
H_1 (\Sigma_g , \mathbb{Z}) \simeq \mathbb{Z}^{2g}.
\end{equation}

\end{Ese}

\noindent
Let's call \textit{period} the first bilinear form of the \eqref{BilineariDeRham} between closed cycles and closed forms:
\begin{equation}
\begin{split}
\pi : \quad H_1 ( \Sigma_g , \mathbb{R}) \quad & \mapsto \quad H^1 (\Sigma_g , \mathbb{R}) \\
([c]; [\omega]) \quad & \mapsto \quad \pi (c , \omega) = \oint_{c} \omega \\
\end{split}
\end{equation}
In particular, choosing the canonical basis cycles we can define the following periods:
\begin{equation}
\pi^a_j (\omega) = \oint_{a_j} \omega \quad \quad \quad \quad \pi^b_j = \oint_{b_j} \omega,
\end{equation}
by which we can compute every intersection number between closed forms. \\
Consider two de Rham closed 1-forms $\omega_1 , \omega_2 \in H^1_{DR} (\Sigma_g ; \mathbb{R})$ on $\Sigma_g$.  The integration on $\Sigma_g$ of their wedge product satisfies the following relation: 
\begin{equation}
\int_{\Sigma} \omega_1 \wedge \omega_2 = \sum_{j=1}^g \left[ \pi^a_j (\omega_1) \pi^b_j (\omega_2) - \pi^b_j (\omega_1) \pi^a_j (\omega_2) \right].
\label{PeriodRiemann1}
\end{equation}

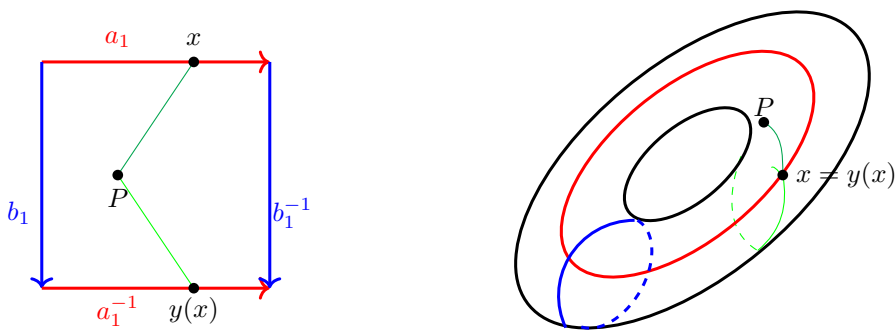
\begin{figure}[h!]
\centering
\begin{tikzpicture}
\riemann{(0,0)}
\end{tikzpicture}
		 \caption{The torus.}
		 \label{gone_2}
 \end{figure}
 \nn
\textbf{Proof.} \\
If we cut the $\Sigma_g$ surface along the representative cycles $a_i , b_i$, we obtain the $4g-$gone $\tilde{\Sigma}$ with boundary:
\begin{equation}
\tilde{\partial \Sigma} = a_1 b_1 a^{-1}_1 b^{-1}_1 .....a_g b_g a^{-1}_g b^{-1}_g.
\end{equation}
This plane figure is simply connected,  and,  by Poincarè Lemma, any closed form on it is exact.  In particular, for the form $\omega_2$, there exist on $\tilde{\Sigma}$ a well defined function $\phi_2$ such that $\omega_2 = d \phi_2$.  \\
Fix now a point $P$ in $\tilde{\Sigma}$, since $\tilde{\Sigma}$ is convex and simply connected, we can connect each point $x \in \tilde{\Sigma}$ to $P$ by a segment, and we define the function $\phi_2$ as a potential through the integration along that segment as follows:

\begin{equation}
\phi_2 (x) = \int_P^x \omega_2.
\end{equation} 

\nn
Since $\omega_1$ and $\omega_2$ are respectively closed and exact on $\tilde{\Sigma}$,the 2-form $\omega_1 \wedge \omega_2$ is exact too on the same domain, it is straightforward indeed to see that exists a well-defined 1-form $\Omega=- \phi_2 \omega_1$ on $\tilde{\Sigma}$, allowing one to write $d \Omega = \omega_1 \wedge \omega_2$.\\ 
Moreover, as one can easily convince him/herself with the help of figure \ref{4ggone}, the following equality holds:
\begin{equation}
\Sigma - \cup_{j=1}^g \left( a_j \cup b_j \right) = \tilde{\Sigma} - \tilde{\partial \Sigma},
\end{equation}

\nn
where the subtracting terms have vanishing measure on, respectively,  $\Sigma$ and $\tilde{\Sigma}$. Manipulating the integration domain in this way one can rewrite the integral in the lhs of \eqref{PeriodRiemann1} over the exactness region of the integrand, in order for Stokes' theorem to be applied; thus:
\begin{equation}
\begin{split}
 \int_{\Sigma} \omega_1 \wedge \omega_2 &= \int_{\Sigma - \cup_{j=1}^g \left( a_j \cup b_j \right)}  \omega_1 \wedge \omega_2 =
\int_{\tilde{\Sigma}} \omega_1 \wedge \omega_2 - \int_{\tilde{\partial \Sigma}} \omega_1 \wedge \omega_2 =\\& =\int_{\tilde{\Sigma}} d \Omega = \int_{\tilde{\partial \Sigma}} \Omega = \int_{\tilde{\partial \Sigma}} \left( - \phi_2 \omega_1 \right) = -  \int_{\tilde{\partial \Sigma}} \omega_1 (x) \int_P^x \omega_2.   \\
\end{split}
\end{equation}

\nn 
Since the boundary $\tilde{\partial \Sigma}$ is just the oriented union of the $4g$-gone edges, we have:
\begin{equation}
\int_{\Sigma} \omega_1 \wedge \omega_2 = - \sum_{j=1}^g \left[ \int_{a_j} \omega_1 (x) \int_0^x \omega_2 + \int_{b_j} \omega_1 (x) \int_0^x \omega_2  +\int_{a^{-1}_j} \omega_1 (x) \int_0^x \omega_2  +\int_{b^{-1}_j} \omega_1 (x) \int_0^x \omega_2 \right].
\end{equation}

\nn
The remaining step to conclude our proof consists in noticing, as shown in the torus example of figure \ref{gone_2}, that if $y(x)$ is the point on $a_j^{-1}$ identified with $x$ under the gluing then:

\begin{equation}
    \int_0^x \omega_2-\int_0^{y(x)}\omega_2=\int_{b_j}\omega_2.
    \end{equation}

 \chapter{Twisted De Rham Theory}\label{two}
  In this chapter we want to extend De Rham theory in order to treat integrals involving multi-valued functions.

\section{Introduction}

Consider a finite set of $m$ polynomials $P_j(z):\C^n \rightarrow \C$ and the algebraic variety $D_j=\{P_j(z)=0\}$ generated by the zeros of the $j$-th polynomial, assume then $\M$ to be the $n-$dimensional complex manifold obtained removing those points from $\C^n$:

\begin{equation}
\M \equiv \C^n \setminus D, \quad\quad \mbox{with}\quad\quad D=\bigcup_1^m D_j.
\end{equation}

\nn
Consider the multi-valued meromorphic function $u$

\begin{equation}
u(z)=\prod_{j=1}^m P_j(z)^{\alpha_j}, \quad \alpha_j \in \C\setminus \Z, 
\end{equation}

\nn
the holomorphic domain of $u$ is $\M$, because $D_j$ are either branch points or poles (eventually both), depending on the sign of the correspondent exponent; so restrict $u$ on $\M$. 
The simplest way to deal with multivalued functions is to establish a prescription that restricts its image by hand, selecting only one solution among the all possible ones, this procedure eliminate the multivaluedness of $u$, but the resulting single-valued function is not holomorphic any more: a discontinuity has arose along the segments (cuts) connecting branch points, and there is no way to avoid it.
The formal way to remove polidromy preserving holomorphicity is to defined $u$ on a suitable covering $\tilde{\M}$ of $\M$ such that it becomes single-valued; for instance, if our function was just the $n-$th root of $z$, the cover $\tilde{M}$ would be the space obtained pasting together $n$ Riemann sheets along the cut connecting the two branch points $0$ and $\infty$. However the covering map is in general too much complicated in order for this method to be useful for concrete calculations; that is why we want to develop a more handy formalism.
What we are interesting in is the integral of such a multivalued function, i.e. 

\begin{equation}
I_C = \int_C u(z)\f(z),
\end{equation}

\nn
where $\f(z)=\hat{\f}(z)d^pz \in \Omega^p(\M)$ is a differential $p-$form. 
Suppose $\M$ admits a triangulation $K$ and let $\Delta \in K$ be a $p-$ simplex; to determine the integral $I_\Delta$ we have to fix a branch of $U$ on $\Delta$; denote by $\Delta\otimes U_\Delta$ the pair of the $p$-simplex and the branch $U_\Delta$ fixed on it, so we have 

\begin{equation}
\int u_\Delta \f \equiv \int_{\tw} u\f.\label{twistedef}
\end{equation}

\nn
Since the integrand is now a single-valued function, Stokes theorem can be applied and we can write

\begin{equation}
\int_\Delta d(u_\Delta \f)= \int_{\de\Delta} u_\Delta \f,\label{stoke}
\end{equation}

\nn
on the other and we have 

\begin{equation}
d(u_\Delta \f)=du\wedge \f + u_\Delta d\f = u_\Delta \left (d\f +\frac{du_\Delta}{u_\Delta} \wedge \f \right ),
\end{equation}

\nn
setting $\omega = d\ln u$ and defining the covariant derivative (twisted derivative)

\begin{equation}
\nabla_\omega \equiv d +\omega\wedge,
\label{covariant1}
\end{equation} 

\nn
the relation \eqref{stoke} becomes

\begin{equation}
\int_{\tw} u\nabla_\omega \f = \int_{\de\Delta} u_\Delta \f.
\end{equation}

\nn
Notice the covariant derivative \eqref{covariant1} is nihilpotent 

\begin{equation}
\nabla_\omega^2 =0.
\label{nihilpotent1}
\end{equation}

\nn
Focus now on the right-hand side of the later equation. The boundary of the $p$-simplex is given by \eqref{simplexboundary}

\begin{equation}
\de \Delta \equiv \de \langle 01...p\rangle = \sum_{j=0}^{p} (-1)^j \langle 01..\hat{j}...p \rangle,
\end{equation}

\nn
then 

\begin{equation}
\int_{\de\Delta} u_\Delta \f = \sum_{j=0}^{p} (-1)^j \int _{ \langle 01..\hat{j}...p\rangle} u_{\langle 01..\hat{j}...p\rangle} \f =
 \sum_{j=0}^{p} (-1)^j \int _{ \langle 01..\hat{j}...p\rangle\otimes u_{\langle 01..\hat{j}...p\rangle}} u \f,
\end{equation}

\nn
where in the second equality we used the definition \eqref{twistedef}. Since we want to right down an expression analogous to Stokes theorem, we require right-hand side integration contour to be the boundary of $\tw$, i.e. we introduce a twisted boundary operator as

\begin{equation}
\de_\omega (\tw) \equiv \sum_{j=0}^{p} (-1)^j \langle 01..\hat{j}...p\rangle\otimes u_{\langle 01..\hat{j}...p\rangle}.\label{twistboundary}
\end{equation}

\nn
Notice the twisted boundary operator $\de\omega$ is nihilpotent:

\begin{equation}
\de_\omega^2(\tw)=0.
\label{nihilpotent2}
\end{equation}

\nn
We have obtained a twisted version of Stokes' theorem by introducing a twisted boundary operator and a twisted derivative, that both turn to be nihilpotent. This fact suggests one to introduce a twisted version of (co)-homology groups. 

\section{Twisted Homology}
\subsection{Homology with local coefficients}

Let $X$ be an arcwise connected topological space, $x,y \in X$ any two points of $X$ and $\gamma^i_{xy}$ the classes of homotopically equivalent curves on $X$ having extrema $x$ and $y$. Denoting then with $\gamma^i_x \in \Pi_x$ the classes of loops with base point on $x$. Each class $\gamma^{i}_{xy}$ determines an isomorphism $\f^i_{xy}:\Pi_x \rightarrow \Pi_y$ between fundamental groups on $x$ and $y$ defined by $\f^i_{xy}(\gamma^j_x)=\gamma^i_{yx}\gamma^j_x\gamma^i_{xy}$, (one can easily convince him-herself with the help of a drawing).
We say we have a local system in a space $X$ and we denote it by $\{G_x\}$ if we can associating to each point $x \in X$ a group (ring) $G_x$ and to each class of paths $\gamma^i_{xy}$ a group (ring) isomorphism $\f[\gamma^i_{xy}]:G_x \rightarrow G_y$ such that 

\begin{equation}
\f[\gamma^i_{xy}]\circ \f[\gamma^j_{yx}]=\f[\gamma^i_{xy}\gamma^j_{yx}].\label{trnsitivity}
\end{equation}

\nn
It follows from \eqref{trnsitivity} that the identity path from $x$ to $x$ is the identity transformation in $G_x$ and that inverse of the isomorphism 
is the one associated to the inverse path. Moreover each loop on $x$ determines an automorphism of $G_x$, so $\Pi_x$ is the group of automorphisms of $G_x$. If $\Pi_x$ acts as the identity on $G_x$, and if this is true in $x$ it must be true in every point, then the isomorphism $\f[\gamma^i_{xy}]$ does not depend on the path from $x$ to $y$ and, as a consequence, all $G_x$ are isomorphic among themselves, i.e. the local system consists of one group $G$ and a copy of it for each point of $X$. Let call such a local system a simple local system. It can be proven that if $G_x$ is an abelian group than $\{G_x\}$ is isomorphic to a simple local system. We are not interested in non abelian local systems or ring local systems, so in the next we will just use the term local system having in mind a simple local system of abelian groups. 
For us a local system is then just a fibre bundle with trivial transition function, that is having the same fibre at every point.\\
Let now $\{G_x\}$ be a local system in the space $X$ admitting a cell complex decomposition  $K$, denoting a $q-$cell by $\langle e_q\rangle$, we define a $q-$ chain with values in the local system, or twisted chain, the linear combination

\begin{equation}
c^q = \sum_i \langle e^q_i\rangle\otimes G_x,
\end{equation}

\nn
for some $x\in \langle e_q\rangle $. Twisted chains form a group, the twisted chain group $C_p (X, \{G_x\})$.
We can now introduce a twisted boundary operator as

\begin{equation}
\de_\omega : C_p (X,\{G_x\}) \tu C_{p-1} (X, \{G_x\}).
\end{equation}

\nn
and define the twisted homology group as 

\begin{defn}
\begin{equation}
H_p(X, \{G_x\}) = \frac{Ker \de_\omega}{Im \de_\omega}
\end{equation}

\nn
The elements of $H_k (X,\{G_x\})$, are called twisted cycles and denoted by $|C]$.
\end{defn}

\subsection{Constructing a local system}\label{constrlocsys}

Coming back to our initial issue, we want to explicitly construct a local system on the manifold $\M$.  Consider the differential equation on $\M$

\begin{equation}
\nabla_\omega h = dh + \sum_{j=1}^m \alpha_j \frac{dP_j}{P_j}h=0,\label{eq.deff1}
\end{equation}

\nn
it admits a general solution that can be formally written as

\begin{equation}
h=c\prod_{j=1}^m P_j^{-\alpha_j}, \quad c\in \C.
\end{equation}

\nn
and the space generated by the local solutions of \eqref{eq.deff1} has dimension $1$. Cover now $\M$ by a finite set of open subsets $U_i$ and fix a single-valued non-zero solution $h_i$ on each $U_i$; since, for $i\neq j$, $h_i$ and $h_j$ are solutions of the same differential equation we have 

\begin{equation}
h_i(x)=g_{ij}(x)h_j(x), \quad\quad\quad x\in U_i \cap U_j,
\end{equation}

\nn
on the other the solution $h(x)$ in the intersection set can be written both as a linear combination of the solution on $U_i$ and $U_j$, so

\begin{equation}
h(x)= a_ih_i = a_j h_j,
\end{equation}

\nn
with $a_i,a_j \in \C$. Comparing the last two equations we see that $g_{ij}$ is just a constant. As a consequence, the set of local solutions of \eqref{eq.deff1} defines a local system with fiber $\C$. Denote by $\mathcal{L}_\omega(\gamma)$ the local system with transition function $g^{-1}_{ij}$ and by $\Lod$ the local system, said dual to $\Lo$, with transition function $g_{ij}$.
The function $u(z)$ turns to be a local section of the local system, this fact concretely provides the connection between the previous formal approach and intuitive explanation given at the beginning.

\begin{Ese} 
Consider the multivalued function $u(z)=z^{\alpha_1}(z-1)^{\alpha_2}$, with $\alpha_j\notin \Z$, on $M=\C \backslash \{0,1\}$. In order to find the elements of twisted homology groups, consider the path showed in figure \ref{Twistedcycle2}, composed by two small circles $S^1_\e(0)$ and $S^1_\e(1)$ of radius $\e$ around the branch points $0$ and $1$ respectively, and a segment $\langle \e,1-\e\rangle$.  The boundaries of the three components can be obtained applying \eqref{twistboundary}:

\begin{equation}
\begin{split}
&\de_\omega S^1_\e (0)= \langle\e\rangle \otimes u(\e e^{2\pi i})-\langle\e\rangle \otimes u(\e)=\e(e^{2\pi i \alpha_1}-1)\otimes\langle\e\rangle, \\&\de_\omega S^1_\e (1)= \langle\e\rangle \otimes u(1-\e e^{2\pi i})-\langle\e\rangle \otimes u(1-\e)=\e(e^{2\pi i \alpha_2}-1)\otimes\langle\e\rangle, \\& \de_\omega \langle \e,1-\e\rangle=\langle 1-\e\rangle -\langle\e\rangle.\end{split}
\end{equation}

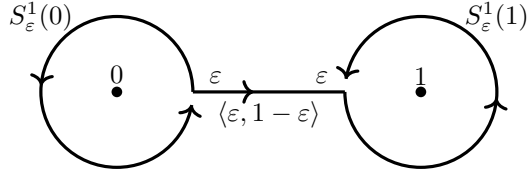
\begin{figure}[h!]
\centering
\begin{tikzpicture}
\twistedcycle{(0,0)}
\end{tikzpicture}
\caption{Twisted cycle between $0$ and $1$.}
\label{Twistedcycle2}
 \end{figure}

\nn
Omitting the irrelevant overall coefficient $\e$ \footnote{The important part of the coefficient is the one expressing the changing from one fiber to another, a constant prefactor is completely irrelevant}, we see that the twisted cycle 

\begin{equation}
\Delta(\omega)= \frac{1}{e^{2\pi\alpha_1}-1}S^1_\e (0) \otimes u_{S^1_\e(0)}+\langle \e,1-\e\rangle \otimes u_{(0,1)}- \frac{1}{e^{2\pi\alpha_2}-1}S^1_\e (1)\otimes u_{S^1_\e(1)}
\label{twistedcycle}
\end{equation}

\nn
satisfies $\de_\omega\Delta(\omega)=0$, i.e. it is closed in the twisted sense, so the class defined by $\Delta(\omega)$ up to exact twisted cycles belongs to the first homology group with coefficient in $\Lod$:

\begin{equation}
[\Delta(\omega)]\in H_1(\M,\Lod).
\end{equation}

\nn
This twisted cycle is called regularization of the open $(0,1)$ interval, and it coincides to Pochhammer contour. 
\end{Ese}

\nn
This twisted cycle turn to be the fundamental block to write down the twisted homology group of more complicated spaces. Consider indeed a function $U$ of the form 

\begin{equation}
u(z)=\prod_{j=0}^{n-1} (z-z_j)^{\alpha_j} \quad\quad\quad \alpha_i \in \C\setminus \Z,
\end{equation}

\nn
defined on $\mathbb{P}^1\setminus \{z_0,z_1,...z_n\}$, with $z_n=\{\infty\}$. Suppose for future convenience $z_j \in \R$. The zero-th homology group is trivially null due to the connectedness of $\M$, $H_0(\M,\Lod)=0$, i.e. any two twisted zero chains are boundary of a twisted one chain. The 2-th homology group is also zero, now for the non compactness of $\M$. Finnally being  $\M$ a two (real) dimensional space, the higher homology group are zero too. The only non trivial homology group is the first. Applying \eqref{chi_1} we have

\begin{equation}
\chi(\M)=-dim H_1,
\end{equation}

\nn
on the other 

\begin{equation}
\chi (\M)=\chi (\mathbb{P}^1\setminus\{z_1,z_2,...z_n\} )=\chi (\mathbb{P}^1)-\chi (\{z_0,z_1,...z_n\} )=2-(n+1)=-n+1,
\end{equation}

\nn
thus 

\begin{equation}
dim H_1 = n-1.
\end{equation}

\nn
A basis of $n-1$ independent twisted cycles is provided by the twisted cycles $\Delta_j(\omega)$ constructed, in the same manner then above, between $z_j$ and $z_{j+1}$:

\begin{equation}
\Delta_j(\omega)= \frac{1}{c_j-1}S^1_\e (z_j) \otimes u_{S^1_\e(z_j)}+\langle z_j+\e,z_{j+1}-\e\rangle \otimes u_{(z_j,z_{j+1})}- \frac{1}{c_{j+1}-1}S^1_\e (z_{j+1})\otimes u_{S^1_\e(z_{j+1})},
\label{twistedcycle3}
\end{equation}

\nn
where we have defined $c_i=exp{2\pi i \alpha_i}$. We therefore have:

\begin{equation}
H_1(\M,\Lod)\simeq\bigoplus_{j=1}^{n-1}\C \cdot \Delta_j(\omega).
\end{equation}

\section{Twisted cohomology}
In order to define a twisted version of cohomology groups, we proceed analogously to what we did for homology group. The first step is to consider differential forms with values in the line bundle defined by the local system $\Lo$

\begin{equation}
\Omega^k(\M,\Lo)=\Omega^k(\M,\C)\otimes \Lo.
\end{equation}

\nn
The elements of this space are called $k-$ left forms and denoted by $\f_k^L$. Later on when we will not need to explicitly write the order we will simply denote it by $\f_L$. \\Notice the twisted derivative defined in \eqref{covariant1} acts on a $k$ left form as 

\begin{equation}
\nabla_\omega \f_L^k = \nabla (\f_k u )= d\f_k u = \f_{k+1} u = \f_{k+1}^L
\end{equation}

\nn
that is 

\begin{equation}
\nabla_\omega :\Omega^k(\M,\Lo) \rightarrow \Omega^{k+1}(\M,\Lo).
\end{equation}

\nn
Then, because the twisted derivative is nihilpotent, as one can easily verify, we can write the cochain complex: 

\begin{equation}
0\xrightarrow{\nabla_\omega^0} \Omega^0(\M, \Lo)\xrightarrow{\nabla_\omega^1} \Omega^1(\M,\Lo)\xrightarrow{\nabla_\omega^2}...\xrightarrow{\nabla_\omega^n} \Omega^n(\M,\Lo)\xrightarrow{d} 0,
\label{cochaincomplextwisted}
\end{equation}

\nn
The cohomology associated to this complex is precisely what we call twisted cohomology group:

\begin{defn}
Let $\{\Omega^k(\M,\Lo),\nabla_\omega^k\}$ be the cochain complex of differential forms groups with coefficient in a local system $\Lo$ and $\nabla_\omega^k$ the covariant derivative defined by the connection $\omega$. We call $k-th$ twisted cohomology group and we denoted it by $H^k(\M,\nabla_\omega)$, or just $H^k_\omega$, the quotient group:

\begin{equation}
H^k(\M,\nabla_\omega) \equiv \frac{ker [\nabla_\omega^k]}{Im[\nabla_\omega^{k+1}]}.
\end{equation}

\nn
The elements of the $H^k_\omega$ are classes of $k$ left forms under the equivalent relation induced by the twisted derivative, that is are covariantly closed $k$ left forms up to exact ones; we denote them by $\langle\f_k^L |$. 
\end{defn}

\nn
As it emerged during the discussion of twisted homology, one can define dual quantities to the ones above considering the dual local system $\Lod$;  introduce then the group of $k-$ forms with values in $\Lod$ 

\begin{equation}
 H^k(\M, \Lod)=H^k(\M, \C)\otimes \Lod,
\end{equation} 

\nn
whose elements are called right $k$ forms and denoted by $\f^R_k$. The connection associated to the dual local system is just $-\omega$, so the covariant derivative is $\nabla_{-\omega}$; we therefore have

\begin{equation}
H^k(\M,\nabla_{-\omega}) \equiv \frac{ker [\nabla_{-\omega^k}]}{Im[\nabla_{-\omega^{k+1}}]}.
\end{equation}

\section{Vectorial structure of (Co)homology spaces}

Vectorial structure of $\Omega^n(\M,\C)$ induces a vectorial structure to the twisted cohomology group either, this allows one to deduce important result by purely algebraical considerations. This fact was firstly noticed by Mastrolia and Mizera in \cite{Mastrolia:2018uzb} and subsequently succesfully applied to Feynman integral and formalized by Frellesvig et all in \cite{Frellesvig:2019kgj},\cite{Frellesvig:2019uqt} and \cite{Frellesvig:2020qot}.\\
Let $\{\langle e_i|\}$ be a basis of $H^n_\omega$, any left form $\langle \f_L |\in | $ can be written as 

\begin{equation}
\langle \f_L| = \sum_{i=1}^\nu c_i \langle e_i|,
\label{decomposition1}
\end{equation}

\nn
where $\n=dim (H^n_\omega)$. Analogously, for any right form $\f_R\rangle \in H^n_{-\omega}$ we can write

\begin{equation}
|\f_R\rangle = \sum_{i=1}^\nu \tilde{c}_i | h_i\rangle,
\label{decomposition2}
\end{equation}

\nn
with $\{ | h_i\rangle\}$ a basis of $H^n_{-\omega}$.\\
Consider now the paring between a left form and twisted cycle $C$, one has

\begin{equation}
\langle \f_L | C] = \sum_{i=1}^\nu c_i \langle e_i|C],
\label{Masterintegraldecomposition}
\end{equation}

\nn
this means that any integral can be decomposed in the sum of "fundamental" integrals, such integrals are called Master integrals. The problem of evaluating the integral in the left-hand side is then transferred into computing the coefficient and the master integrals appearing in the right-hand one. This decomposition is known from long time and more or less successfully applied in the computation of Feynman integrals, however in its original version it was obtained using integration by parts identities; the language of twisted co-homology theory provides a more deeply understanding of its nature and a vary powerful tool to compute its coefficients.

\nn
We call intersection matrix between the basis, the matrix:
\begin{equation}
C_{ij}\equiv\langle e_i| h_j\rangle.
\end{equation}

\nn
Let $\{\langle h_i |\}$ be a basis of $H^n_{\omega}$ such that

\begin{equation}
\langle h_i | h_j \rangle = \delta_{ij}, \quad\quad | h_i \rangle\langle h_i |= \mathbb{I},
\label{dual1}
\end{equation}

\nn
and $\{|e_i \rangle\}$ a basis of $H^n_{-\omega}$ such that 

\begin{equation}
\langle e_i | e_j \rangle = \delta_{ij}, \quad\quad | e_i \rangle\langle e_i |= \mathbb{I}.
\label{dual2}
\end{equation}

\nn
Than we can formally write:

\begin{equation}
\sum_{ij}\langle \f_L| h_j\rangle \langle h_j | e_i\rangle \langle e_i | = \mathbb{I}.
\end{equation}

\nn
Now, the matrix $\Lambda_{ji}=\langle h_j | e_i\rangle$ is, by construction, the inverse of the matrix $C$, in fact:

\begin{equation}
C_{ij}\Lambda_{jk}=\langle e_i| h_j\rangle\langle h_j | e_k\rangle=\langle e_i| e_k\rangle = \delta_{ik}\quad  \rightarrow \quad \Lambda=C^{-1},
\end{equation}

\nn
thus we obtain a representation of the identity operator as 

\begin{equation}
\sum_{ij}\langle | h_j\rangle C^{-1}_{ji} \langle e_i | = \mathbb{I}.
\label{identityoper}
\end{equation}

\nn
The advantage of the \eqref{identityoper} is that it is fully expressed in terms of the original basis $|h_i\rangle$ and $\langle e_i|$, and it does not require the knowledge of the dual basis defined in \eqref{dual1} and \eqref{dual2}, actually this is more then an advantage because finding them explicitly is in general a vary challenging task; on the other hand, if we were we could have directly chosen $|h_i\rangle = |e_i\rangle$, and one of our biggest problem would have been solved. By means of \eqref{identityoper} we can instead represent the identity operator on the cohomology space picking up any two basis for $H^n_\omega$ and $H^n_{-\omega}$. Once we have the identity operator we can use it to project left and right forms into their respective basis:

\begin{equation}
\begin{split}
&\langle \f_L| = \sum_{ij}\langle \f_L| h_j\rangle C^{-1}_{ji}\langle e_i | ,\\& |\f_R\rangle = \sum_{ij} | h_j\rangle C^{-1}_{ji}\langle e_i |\f_R\rangle.
\end{split}
\label{decomposition3}
\end{equation}

\nn
Comparing the two equations in \eqref{decomposition3} with \eqref{decomposition1} and \eqref{decomposition2}respectively, one can immediately read the expression for the coefficients 

\begin{equation}
\begin{split}
&c_i = \sum_{j} \langle \f_L|  h_j\rangle C^{-1}_{ji},\\&\tilde{c_i}=\sum_{j}C^{-1}_{ij}\langle e_j |\f_R\rangle.
\end{split}
\end{equation}

\nn
The latter is know as \textit{master decomposition formula}.

\nn
Also the twisted cohomology space is endowed by a vectorial structure induced by the chain group one, thus one can introduce a basis and a dual basis, and decompose any left and right twisted cycles as:

\begin{equation}
\begin{split}
 &H_n(\M,\Lo) \ni |C_R]= \sum_{j=1}^\n \alpha_j |C_{Rj}],\\& H_n(\M,\Lod) \ni [C_L| = \sum_{j=1}^\n \tilde\alpha_j [C_{Lj}|.
 \end{split}
\end{equation}

\nn
Proceeding in the same way as above, one can construct a representation of the identity operator of the homology

\begin{equation}
\mathbb{I} = \sum_{ij} |C_{Ri}](H^{-1})_{ij}[C_{Lj}|,
\end{equation}
 
 \nn
 where $H_{ij}=[C_{Li}|C_{Rj}]$ is the intersection matrix for the twisted cycles of the basis.\\
 Finally, inserting the identity operator between 
 
 \begin{equation}
 \begin{split}
&\langle \f_L | \f_R\rangle = \sum_{ij}\langle \f_L|C_{Ri}](H^{-1})_{ij}[C_{Lj}|\f_R\rangle \\& [C_{L}|C_{R}]=\sum_{ij}[C_{L}| h_j\rangle C^{-1}_{ji} \langle e_i |C_{R}].
 \end{split}
 \label{TwistedRiemannperiod}
\end{equation}
 
 \nn
These quadratic expressions are the twisted version of \eqref{PeriodRiemann1} and they are known as \textit{Twisted Riemann period relations}.

\subsection{The dimension of the co-homology space}
The number $\nu$ of master integrals appearing in the master decomposition formula, clearly corresponding to the dimension of the cohomology space, turns to be related to the topology of the manifold by the following theorem.
Under the assumptions given in \cite{Aomoto} (see also \cite{Aomoto2} and \cite{Aomoto1975OnVO}) one can prove the following theorem.
\begin{teo}\label{vanishing} Vanishing theorem.\\
Let $\M = \mathbb{P}^n \backslash \cup D$ be the $n-$ dimensional complex manifold complementary to the projective variety $D$ defined by $m$ polynomials $P_j: \C^n \rightarrow \C$, and $\nabla_\omega$ the covariant derivative on $\M$ defined by the connection $\omega$ then (under some assumptions) the only non vanishing twisted cohomology group is the one of maximal dimension: 

\begin{equation}
H^k(\M,\nabla_\omega)=0 \quad \quad for \, k\neq n.
\end{equation}
\end{teo}

\nn
By Euler-Poincaré formula \eqref{chi_1} we have 

\begin{equation}
\dim H^n(\M,\nabla_\omega)= (-1)^n\chi(\M).
\end{equation}

\nn
and thus:

\begin{equation}
\nu= (-1)^n (n+1-\chi(D)).  
\end{equation}

\nn
The computation of the Euler characteristic for the projective variety $D$ is actually not trivial.
Thanks to Poincaré duality we can obtain $\nu$ by considering the Homology group. 
Let $f$ be a Morse function for $\M$, and $k_p$ the number of its critical points with Morse index $p$. Because of the vanishing theorem there are no critical points with index $k\neq n$ and thus all critical points must have the same, i.e. the highest, Morse index; $k_n$ is therefore the total number of critical points of $f$, and it must equal the $n-$th Betti number $b_n$:

\begin{equation}
\nu = b_n = k_n.
\end{equation}

\nn
In \cite{Lee:2013hzt} Lee and Pomeransky show that $\log{u(z)}$ is a Morse function for $\M$. Its critical points are nothing but the solutions of $\omega=0$:

\begin{equation}
\omega = d\log{u(z)} = \sum \de_j \log{u(z)} dz_j =\sum \omega_j dz_j;
\end{equation}

\nn
and then the number of master integrals is given by:

\begin{equation}
\nu = \# I \ni i \,|\,\omega_j(z_i)=0\,\,\, \forall\, j=1..n. 
\end{equation}

\subsection{Differential equation for Master integrals}
We have seen co-homology theory, throuth \eqref{Masterintegraldecomposition}, provides a decomposition of integral into $\n$ Master integrals. Sometimes one is interested in explicitly compute them
The elements of the basis $e_i$ and, eventually, the twisted $u$ generally depend on more variables then the integrated ones, as momenta or masses in the Feynman integral case. Let $x$ denote one of such variables, thanks to integration-derivation commutativity, one has:

\begin{equation}
\begin{split}
\de_x \langle e_i | C]&=\de_x \int_C  e_i u = \int_C u \left ( \de_x+\frac{\de_x u}{u}\right )e_i =\\&= \int_C u(\nabla_{\sigma_x}e_i)= \langle \nabla_{\sigma_x} e_i | C],
\label{Diffeqmasterintegral}
\end{split}\end{equation}

\nn
where we have introduced the covariant derivative 

\begin{equation}
\nabla_{\sigma_x}\equiv \de_x + \sigma_x \wedge,
\end{equation}

\nn
with $ \sigma_x = \de_x\log {u}$. From \eqref{Diffeqmasterintegral} one immediately reads:

\begin{equation}
\de_x \langle e_i | = \langle \nabla_{\sigma_x} e_i | = \Omega_{ij} \langle e_j |,
\label{Diffeqmasterintegral2}
\end{equation}

\nn
where in the last equality we used the fact that $\langle \nabla_{\sigma_x} e_i | \in H^n_{\omega}$.\\
Paring \eqref{Diffeqmasterintegral2} again with the right cycle we obtain a differential equation for the master integral:

\begin{equation}
\de_x \langle e_i |C] = \Omega_{ij}\langle e_i |C].
\end{equation}

\section{Intersection number between twisted cycles}\label{Intc}
Since $\M$ is not compact, it cannot be cover by a finite number of open subsets. Actually we can introduce a weaker condition then compactness, called paracompactness, 
For instance if we consider the open interval $(0,1)$ this is not a chain, because it cannot be written as a finite sum of simplices. However, 
We say a curve $\gamma$ on $\M$ is a locally finite 1-chain if it can be expressed as (infinite) sum of 1-simplices 

\begin{equation}
\gamma= \sum a_j \Delta^j_1
\end{equation}

\nn
such that for any point $x \in \M$ it exists an open subset $V(x)\subset \M$ such that the number of simplices in $\gamma$ intersecting $V$ is finite:

\begin{equation}
\forall x \in \M\, \exists\, V(x) \subset \M,\, \#\{j | \Delta_1^j \cap V(x) \neq \emptyset\}< \infty.
\label{locallyfinitedef}
\end{equation}

\nn
The open $(0,1)$ interval can be for instance expressed as 

\begin{equation}
(0,1)= \bigcup_2^\infty \left [\frac{1}{n+1},\frac{1}{n}\right ] +\bigcup_2^\infty \left [\frac{1}{n},1-\frac{1}{n+1}\right ],  
\end{equation}

\nn
consider now a point $x \in \M$. If $x \notin (0,1)$ definition \eqref{locallyfinitedef} is trivially satisfied, if $\x \in (0,1)$, let $V(x)=(x-\delta,x+\delta)$, supposing for convenience $0<x<\frac{1}{2}$ \footnote{If $\frac{1}{2}<x<0$ a similar computation will bring to the same result}, we have 

\begin{equation}
\begin{split}V(x) \cap \left \{\bigcup_2^\infty \left [\frac{1}{n+1},\frac{1}{n}\right ] +\bigcup_2^\infty \left [\frac{1}{n},1-\frac{1}{n+1}\right ]\right \}&=(x-\delta,x+\delta)\cap \bigcup _2^\infty\left [\frac{1}{n+1},\frac{1}{n}\right ]=\\&=\bigcup _2^\infty (x-\delta,x+\delta)\cap \left [\frac{1}{n+1},\frac{1}{n}\right ], \end{split}
\end{equation}

\nn
now, the set intersection appearing above is clearly empty for

\begin{equation}
\begin{split}
&\frac{1}{n+1}\geq  x+\delta\quad\quad \rightarrow\quad\quad n\leq \frac{1-x-\delta}{x+\delta},  \\& \frac{1}{n}\leq x-\delta \quad \quad \quad \quad\rightarrow \quad \quad  n \geq \frac{1}{x-\delta} .\end{split}
\end{equation}

\nn
introducing  

\begin{equation}
n_i \equiv \left \lceil\frac{1}{x-\delta}\right \rceil \quad\quad \mbox{and} \quad\quad n_f\equiv \left \lfloor  \frac{1-x-\delta}{x+\delta} \right\rfloor, 
\end{equation}

\nn
where $\lceil \cdot\rceil$ and $\lfloor \cdot \rfloor$ denote ceiling and floor function respectively. \footnote{The integer part plus one and the integer part respectively.}
One therefore has:

\begin{equation}
\bigcup _2^\infty (x-\delta,x+\delta)\cap \left [\frac{1}{n+1},\frac{1}{n}\right ]=\bigcup _{n_i}^{n_f} (x-\delta,x+\delta)\cap \left [\frac{1}{n+1},\frac{1}{n}\right ], 
\end{equation}

\nn
and clearly 

\begin{equation}
\#\{n_i,n_{i+1},...n_f\}<\infty, 
\end{equation}

\nn
thus  condition \eqref{locallyfinitedef} is satisfied. The interval $(0,1)$, is therefore a locally finite chain. 
Let $C^{lf}_k(\M, \Lo)$ be the locally finite twisted k-chain group, defined as 

\begin{equation}
C^{lf}_k(\M, \Lo)=\{a_j c_k^j \otimes U\},
\label{locallyfinitecycle}
\end{equation}

\nn
with $c_k^j \in C^{lf}_k(\M)$ locally finite k-chains of $\M$; one can easily show that a boundary a operator can be defined and, as usual an homology group introduced: we call locally finite twisted k-th homology group the group 

\begin{equation}
H_k^{lf}(\M,\Lo)\equiv \frac{Ker [\de_\omega: C_k^{lf}(\M,\Lo)\rightarrow  C_{k-1}^{lf}(\M,\Lo) ]}{Im [\de_\omega: C_{k+1}^{lf}(\M,\Lo)\rightarrow  C_{k}^{lf}(\M,\Lo)]}.
\end{equation}

\nn
If $\M$ is paracompact, one can always associate a locally finite chain to any chain, just as we saw above for the $(0,1)$ interval, it exists therefore a map, said natural map

\begin{equation}
N: C_k \rightarrow C_k^{lf},
\end{equation}

\nn
this map induces a map between homology groups $H_k \rightarrow H_k^{lf}$. It is remarkable that, in general, this last map turns to be an isomorphism. Finally, we call regularization map the inverse map 

\begin{equation}
reg_h: H_k^{lf} \rightarrow H_k,
\end{equation}

\nn
where the subscript $h$ stands for "homology". \\
Thanks to the isomorphism mentioned above and Poincarè duality, one can define the following non degenerate bilinear form, called intersection number between twisted homologies:

\begin{equation}
H_p(\M, \Lod) \times H_{2n-p}^{lf}(\M,\Lo) \rightarrow \C.
\end{equation}

\nn
Let $\{\Delta_j(\omega)\}$ be a basis of $H_1(\M, \Lod)$ and $\{\Delta_j\}$ a basis of $H_1^{lf}(\M,\Lo)$, we call intersection matrix the matrix whose elements are the intersection number between basis elements:

\begin{equation}
I_{ij}=\Delta_i(\omega) \cdot \Delta_j.
\end{equation}

\nn
 Using the shorthand notation 
 
 \begin{equation}
 S^1_\varepsilon(z_i)\equiv S^1_{\e,i} \quad\quad \langle z_i+\e,z_{j}-\e\rangle\equiv \langle i,j\rangle,
\end{equation}  
 
 \nn
 and the expressions \eqref{twistedcycle3} and \eqref{locallyfinitecycle} for the twisted cycle $\Delta_j(\omega)$ and the locally finite one respectively, the intersection number becomes:
 
 \begin{equation}
 \begin{split}
 \Delta_i(\omega)\cdot \Delta_j&=\\&=\left [\frac{1}{c_i-1}S^1_{\e,i} \otimes U_{S^1_{\e,i}}+\langle i,i+1\rangle \otimes U_{(z_i,z_{i+1})}- \frac{1}{c_{i+1}-1}S^1_{\e, i+1}\otimes U_{S^1_{\e,i+1}}  \right ]\cdot \\&\cdot \gamma_j\otimes U^{-1}_{(z_j,z_{j+1})}=\\&=\frac{1}{c_i-1}S^1_{\e,i}\cdot \gamma_j M(S^1_{\varepsilon,i}, \langle z_j,z_{j+1} \rangle)+\langle i,i+1\rangle \cdot\gamma_j + \frac{1}{c_{i+1}-1} S^1_{\e, i+1} \cdot \gamma_j M(S^1_{\varepsilon,i+1}, \langle z_j,z_{j+1} \rangle),\end{split}
 \label{intersection_2}
 \end{equation}

\nn
Where we have introduced the monodromy factors   

\begin{equation}
M(\gamma_1,\gamma_2)\equiv U_{\gamma_1}U^{-1}_{\gamma_2}, 
\end{equation}

\nn
and the topological intersection $\gamma_1 \cdot \gamma_2$.

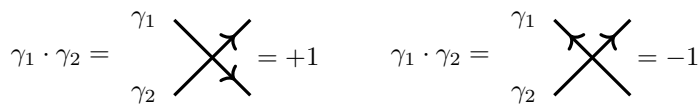
\begin{figure}[h!]
\centering
\begin{tikzpicture}
\topinter{(0,0)}
\end{tikzpicture}
\caption{Topological intersection}
\label{topologicalintersection}
 \end{figure}

\paragraph{Topological intersection.} Consider two oriented paths $\gamma_1$ and $\gamma_2$ intersecting in a point $P$, the respectively tangent vectors $v_1$ and $v_2$ at $P$ span a basis $B$ of $\R^2$, the topological intersection $\gamma_1\cdot\gamma_2$ is defined as the sing of the determinant of the matrix $V=(v_1|v_2)$ whose columns are the basis vectors \ref{topologicalintersection}. Roughly speaking, the topological intersection is obtainable applying the right hand rule: if, pointing the right index finger along the first path and the middle one along the second, the thumb turns to be upward the topological intersection is $+1$, viceversa, if the thumb points downwards, it is $-1$. 
If the two paths intersect in more points their topological intersection is just the sum of the intersection in each point, while if they do not intersect their intersection is clearly null.

\nn
With this in mind we see that the intersection \eqref{intersection_2} is different from zero only for $i=j$ and $|i-j|=1$, i.e. for twisted cycles sharing at least one extremum. Consider first the case $i=j$, at which we will refer as self-intersection. To calculate the topological intersection appearing in \eqref{intersection_2} we have to choose a deformation $\gamma$ of the oriented $(0,1)$ segment, for example the one depicted in figure \ref{examplecycle}. Whatever $\gamma$ is we see from the picture it must intersect the circles $S^1_{\e,i}$ and $S^1_{\e,i+1}$ in just a point respectively, because otherwise it was not locally finite, and the topological intersections are always $-1$ in the first case and $+1$ in the second one. The topological intersection number between $\gamma$ and the $\langle z_i,z_{i+1}\rangle$ segment, which could intersect in any number of points $n$, only depends on whether $n$ is even or odd: due the the fact that two consecutive intersections have opposite topological intersection, if $n$ is even (0 included) $\gamma\cdot \langle z_i,z_{i+1}\rangle=0$, while if $n$ is odd $\gamma\cdot \langle z_i,z_{i+1}\rangle=\pm 1$ depending on the topological intersection of the first intersection point. 

\begin{figure}[h!]
\centering
\begin{tikzpicture}
\cyclethreee{(0,0)}
\end{tikzpicture}
\caption{in}
\label{examplecycle}
 \end{figure}

\nn
\paragraph{Monodromy factors.} Given two loaded paths $\gamma_1$ and $\gamma_2$, we call monodromy factor the term

\begin{equation}
M(\gamma_1,\gamma_2)=  U_{\gamma_1}U^{-1}_{\gamma_2}.
\end{equation} 

\nn
Despite the previous quantity could seems simply to coincide with the identity, at first glance, that is not actually always true, and as we will see in a moment the key detail lies in whether $\gamma_1$ and $\gamma_2$ belong to the same branch of $U$ or they do not. 
Let $z^p_i$ be the coordinate of the intersection point $P$ obtained approaching to $P$ along the path $\gamma_i$, we can practically calculate the monodromy factor as 

\begin{equation}
M(\gamma_1,\gamma_2) = U(z^p_1) U^{-1}(z^p_2).
\end{equation}

\nn
We in general have 

\begin{equation}
U(z)=\prod_{j=1}^m P_j(z)^{\alpha_j},
\end{equation}

\nn
where $P_j(z)=(z-z_j)$. Thus we have 

\begin{equation}
M(\gamma_1,\gamma_2) =\prod_{j=1}^m P_j(z^p_1)^{\alpha_j}P_j(z^p_2)^{.-\alpha_j}=\prod_{j=1}^m(z^p_1-z_j)^{\alpha_j}(z^p_2-z_j)^{.-\alpha_j}
\label{monodromyfactors2}
\end{equation}

\nn
If $P$ does not belong to a neighborhood of a branch point, its coordinates do not depend on the direction one approaches to it, so $z^p_1=z^p_2$, all the products appearing in \eqref{monodromyfactors2} are simply $1$, and the monodromy factor reduces to the identity. If instead $P$ belongs to a neighborhood of the branch point $z_i$, the $i-th$ term of the previous equation is not necessary $1$, and one in general has:

\begin{equation}
M(\gamma_1,\gamma_2) =(z^p_1-z_i)^{\alpha_j}(z^p_2-z_i)^{.-\alpha_i}.
\label{monodromyfactors3}
\end{equation}

\nn
Since $P$ is near $z_i$, its coordinate respect to the paths $\gamma_1$ and $\gamma_2$, could in general be written as $z^p_1=z_i+\e e^{\beta_1 i}$ and $z^p_2=z_i+\e e^{\beta_2 i}$, then the equation \ref{monodromyfactors3} becomes 

\begin{equation}
M(\gamma_1,\gamma_2)= e^{i(\beta_1-\beta_2)\alpha_i}.
\end{equation}

\nn
The phases $\beta_1$ and $\beta_2$ clearly depends on where one choose to fix brunch cuts. 
Consider, for instance the paths $S_{i,\e}$ and $\gamma$ showed in figure \ref{monodromyfactors4}. The coordinate $z^p_2$ of the intersection point $P$ respect to the branch of the path $\gamma$ is just $z_i+\e$, since $\gamma$ is chosen  on the main branch of $U$ by definition. However the coordinate $z^p_1$ of $P$ respect to the branch of the cycle $S_{i,\e}$ is $z^p_1=z_i+\e e^{2\pi i}$, because $S_{i,\e}$ wraps once around the branch point $z_i$ before reaching $P$.

 \begin{figure}[h!]
 \centering
\begin{tikzpicture}
\cycletwoo{(0,0)}
\end{tikzpicture}
\caption{Example}
\label{monodromyfactors4}
 \end{figure}

\nn
Thus we have 
\begin{equation}
M(S_{i,\e},\gamma)=\left (\e e^{2\pi i}\right )^{\alpha_i}\left ( \e \right )^{-\alpha_i}=e^{2\pi i \alpha_i}.
\end{equation}

\nn
Putting together the topological intersection contribute and the one due to monodromy factors we can derive the following diagrammatic rules:

\begin{figure}[h!]
\begin{tikzpicture}
\cycleone{(0,0)}
\end{tikzpicture}
 \end{figure}
 
 \begin{figure}[h!]
\begin{tikzpicture}
\cycletwo{(0,0)}
\end{tikzpicture}
 \end{figure}

  \begin{figure}[h!]
\begin{tikzpicture}
\zeroone{(0,0)}
\end{tikzpicture}
 \end{figure}

  \begin{figure}[h!]
\begin{tikzpicture}
\zeroonetwo{(0,0)}
\end{tikzpicture}
 \end{figure}

\nn
Combining these diagrams in such a way that $\gamma$ is a locally finite deformation of the $(0,1)$ interval, that is it must not turn around poles and it must be continuous, one can perform four different representations of the intersection number: 

\begin{figure}[h!]
\centering
\begin{tikzpicture}
\cyclethree{(0,0)}
\end{tikzpicture}
 \end{figure}
 
 \begin{figure}[h!]
\centering
\begin{tikzpicture}
\cyclefour{(0,0)}
\end{tikzpicture}
 \end{figure}
 
  \begin{figure}[h!]
\centering
\begin{tikzpicture}
\cyclefive{(0,0)}
\end{tikzpicture}
 \end{figure}
 
  \begin{figure}[h!]
\centering
\begin{tikzpicture}
\cyclesix{(0,0)}
\end{tikzpicture}
 \end{figure}

\nn
The four results above, as one can easily prove with few algebra, actually coincides, thus one obtains that, whatever $\gamma$ is, the selfintersection number can be written as:
\begin{equation}
I_{ii}=\Delta_i(\omega)\cdot \Delta_i= -\frac{c_ic_{i+1}-1}{(c_i-1)(c_{i+1}-1)}=\frac{i}{2}\left ( \frac{1}{\tan \pi \alpha_i}+\frac{1}{\tan \pi \alpha_{i+1}}\right).
\label{cycleselfintersection}
\end{equation}

\nn
Consider now the intersection number between adjacent cycles, i.e. for $|i-j|=1$; diagrammatically represented below. In both cases the two cycles intersect only once, near the common extremum. 

  \begin{figure}[h!]
\centering
\begin{tikzpicture}
\cyclemixone{(0,0)}
\end{tikzpicture}
 \end{figure}
 
\nn
 We choose the right cycle to be in the main branch of $U$ ($arg(z)=0$), while the left cycle branch is obtained by analytic continuation along the lower half plane ($arg(z)=-\pi$). Finally branch cuts are chosen to be $(z_i, i\infty)$. With this convention, setting $d_i=c_i-1$ and $d_{ij}=c_ic_j-1$, the intersection matrix becomes:

\begin{equation}
I_{ij}=-\frac{d_{i,i+1}}{d_id_{i+1}}\delta_{ij}+\frac{c_i}{d_i}\delta_{ij+1}-\frac{1}{d_j}\delta_{i+1j}.
\end{equation}

\nn
It is important to notice and underlying that the intersection matrix do depend on the choose of branch and branch cuts. The latter convention we introduce here is the one chosen by Cho and Matsumoto in \cite{cho_matsumoto_1995}.\\
Intersection between $1-$ cycles turns to be the building block to compute higher dimensional $n-$cycles intersection; the latter, in fact, can be in general decomposed in intersection number between $n-1$ cycles, and then recursively brought back to one dimensional cycles intersection.  
This decomposition unfortunately is in general not trivial and the computation of higher dimension cycles intersection become a quite advanced topic in algebraic geometry. In \ref{Appendice1} we will discuss a diagrammatic way to compute them in the specific case of a Koba-Nielsen twist in the Moduli space of the punctured Riemann sphere; relevant case for string amplitude. Here we just want to give some hint on the complexity of computing high dimensional cycles intersection.
The first important thick to realize is that in considering  $n-$dimensional cycles, loci of intersections are not just points as in the bidimensional one, but objects of dimension going from $1$ to $n-1$. Consider for instance two triangles, according to their disposition they can intersect in five different ways:

\begin{itemize}
    \item Nowhere, if they do not intersect at all; 
    \item In a $0-$ cycle, if they share a vertex;
    \item In two $0-$cycles and one $1-$cycle, if they share a side;
    \item In a $0-$ cycle and a $1-$cycle, if they share a vertex and part of a side;
    \item In three $0-$cycles and three $1-$ cycles if they overlap. \footnote{The $2-$ dimensional face of one of the triangles should be homotopically deformed in order not to overlap, just like we did for the self-intersection of $1-$cycles.}
    
\end{itemize}

\nn
Clearly higher the dimension is, greater the possibilities.
In considering a multivariate twist, also poles of $\omega$ are not isolated points but more complicated varieties, that can overlap in same region; if more than two divisors intersect in the same region,the degenerancy must be eliminated by applying some suitable procedure, known as blow up.

\section{Intersection number between twisted cocycles}

Let $\Omega_c^k(\M)$ be the group of smooth forms on $\M$ having compact support, we define the k-th twisted cohomolgy group with compact support the group 

\begin{equation}
H_c^k(\M, \Lo)\equiv \frac{ker [\nabla_\omega:\Omega_c^k \rightarrow \Omega_c^{k+1}]}{Im [\nabla_\omega:\Omega_c^{k-1}\rightarrow \Omega_c^k]}.
\end{equation}

\nn
Because $\Omega_c^k(\M)\subset \Omega^k(\M)$ one can naturally define the an inclusion map $\Omega_c^k(\M)\hookrightarrow \Omega^k(\M)$, this map induces a homomorphism between cohomology groups 

\begin{equation}
N: H_c^k (\M,\nabla_\omega) \rightarrow H^k(\M,\nabla_\omega),
\end{equation}

\nn
called natural map. Such a map, just as it happened for the homology case, turns in general to be an isomorphism, then luckily 

\begin{equation}
H_c^k(\M,\nabla_\omega) \simeq H^k(\M,\nabla_\omega).
\end{equation} 

\nn
Inverting the natural map one obtains the so called (cohomology) regularization map:

\begin{equation}
reg_c : H^k(\M,\nabla_\omega) \rightarrow H_c^k(\M,\nabla_\omega).
\label{regularization_map}
\end{equation}

\nn
\begin{teo}
The following bilinear form, known as intersection between forms, is non degenerate:

\begin{equation}
H_c^k(\M,\nabla_\omega) \times H^k(\M,\nabla_{-\omega}) \rightarrow \C.
\end{equation}

\end{teo}

\nn
In order to find and explicit expression for the pairing above, let first explicitly construct the regularization map for one forms.\\
Let $U_j \subset V_j \subset \M$ be two neighborhoods $z_j$, and let $h_j: \M \rightarrow \C$ be a $\C^\infty$ function on $\M$ such that 

\begin{equation}
h_j(z)=\begin{cases} &1 \quad\quad z \in U_j, \\ &0 \quad\quad z\notin V_j .\end{cases}
\end{equation}  

\begin{figure}[h!]
\centering
\begin{tikzpicture}
\bumpfunction{(0,0)}
\end{tikzpicture}
\caption{Pictorial representation of $U_j, V_j$ and $h_j(z)$.}
\label{bumpfunction}
 \end{figure}
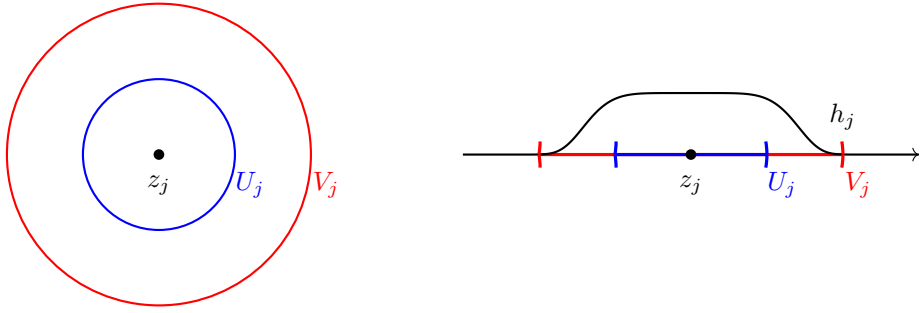

\nn
as shown in figure \ref{bumpfunction}.

 \nn
Let now $\f_L(z) \in \Omega^1_{\omega}$ be a one form and $\psi_j(z) \in \Omega^0(V_j)$ a function such that
 
 \begin{equation}
 \nabla_\omega \psi_j(z) = \f_L(z)  \quad\quad \mbox{for} z\in V_j, 
 \label{Diffeq1}
 \end{equation}
 
 \nn
 then the one form
 
 \begin{equation}
 \f'_L \equiv \f_L - \nabla_\omega (h_j\psi_j) = \begin{cases} &0 \quad \mbox{if}\, z \in U_j \\& \f_L  \quad \mbox{if}\, z \notin V_j,\end{cases}
 \label{regularizedmaprealization}
 \end{equation}
 
 \nn
where the some over repeated $j$ is understood, has support 

\begin{equation}
supp[\f'_L] \subset \left ( \bigcup_j U_j\right )^c,
\end{equation}  

\nn
where the superscript $c$ here stands for the complementary set in $\mathbb{P}^1$. Because $U_j$ are open, their complementary is closed, and then $\psi'$ has compact support; we have found the explicit action of the regularization map

\begin{equation}
reg_c(\f_L)= \f_L - \nabla_\omega (h_j\psi_j).
\end{equation}

\nn
The function $\psi_j$ introduced here will play an important role in the next, let for now notice that it can be formally written as:

\begin{equation}
\psi_j(t)=\frac{1}{(c_j-1)u(t)}\int_{C(t)} u(z)\f_L \quad\quad \mbox{for}\, t \in V_j, 
\label{function1}
\end{equation}

\nn
where $C(t)=\gamma(t)\otimes u(t)$ with $\gamma$ any loop of $V_j$ with base point $t$ and turning once around $z_j$. Clearly, $\psi_j(t)$ does not depend on the choice of $\gamma$, as once can trivially convince itself. Let now prove \eqref{function1} actually satisfies \eqref{Diffeq1}, one has:

\begin{equation}
\begin{split}
\nabla_\omega \psi_j(t)&= d\psi_j +\omega\wedge \psi_j=\left ( -\frac{du}{(c_j-1)u^2}\int u \f_L + \frac{1}{(c_j-1)u} (u\f_L) \Big |_{\de (C(s)}\right ) +\frac{du}{u}\wedge \psi_j =\\&= \frac{1}{(c_j-1)u}(c_j u\psi - u\psi)=\psi \end{split}.
\end{equation}

\nn
We define the intersection between $1$-forms as 

\begin{equation}
\langle \f_L | \f_R \rangle \equiv \frac{1}{2\pi i}\int_\M reg_c (\f_L)\wedge \f_R.
\end{equation}

\nn
We have 

\begin{equation}
\begin{split}
2\pi i \langle \f_L | \f_R \rangle =& \int_\M \left (\f_L-\nabla_\omega (h_j\psi_j)\right )\wedge \f_R = \\&\int_{\bigcup U_j} \left (\f_L-\nabla_\omega (h_j\psi_j)\right )\wedge \f_R +\int_{\bigcup (V_j-U_j)} \left (\f_L-\nabla_\omega (h_j\psi_j)\right )\wedge \f_R +\int_{\left (\bigcup V_j\right )^c} \left (\f_L-\nabla_\omega (h_j\psi_j)\right )\wedge \f_R 
\end{split}
\end{equation}

\nn
where we have just decomposed the integration region. Now, the first integral is zero because $reg_c(\f_L)=0$ in $U_j$ by definition \ref{regularizedmaprealization} and the third integral is also zero, due to the same definition, because $\f_L\wedge \f_R =0$, being the wedge product of two 1 forms in one variable. Then one has:

\begin{equation}
\begin{split}
2\pi i \langle \f_L | \f_R \rangle =& \int_{\bigcup (V_j-U_j)} \left (\f_L-\nabla_\omega (h_j\psi_j)\right )\wedge \f_R= -\sum_j \int _{V_j-U_j}\nabla_\omega (h_j\psi_j)\wedge \f_R =\\&=-\sum_j \int _{V_j-U_j} \left (d(h_j\psi_j)+\omega h_j\psi_j \right ) \wedge \f_R =  -\sum_j \int _{V_j-U_j} d(h_j\psi_j)\wedge \f_R =\\&=  -\sum_j \int _{V_j-U_j} d(h_j\psi_j \f_R)= -\sum_j \int _{\de (V_j-U_j)} (h_j\psi_j \f_R)
\end{split}
\end{equation}

\nn
where in the last equality Stokes' theorem has been applied and in the second last one we used $d^2\f_R=0$. This passage is crucial: this equality holds here because we are considering just one variable, if more variables were involved it would not be true, unless $\f_R=df$ for some $f$. Now, the boundary of the annulus $V_j-U_j$ is obviously given by the boundaries of the two neighborhoods, and reminding $h_j$ is identically zero on the boundary of $V_j$ and $1$ on the boundary of $U_j$, we have:

\begin{equation}
\begin{split}
2\pi i \langle \f_L | \f_R \rangle = -\sum_j \int _{\de (V_j} (h_j\psi_j \f_R)+\sum_j \int _{\de (U_j)} (h_j\psi_j \f_R)=\int _{\de (U_j)} \psi_j \f_R.
\end{split}
\end{equation}

\nn
Finally we can apply residue theorem and obtain 

\begin{equation}
\langle \f_L | \f_R \rangle =\sum_j Res_{z=z_j} (\psi_j \f_R).
\label{intersectionnumber1}
\end{equation}

\nn
Therefore we have found that the computation of the intersection number between two $1-$ forms $\f_L$ and $\f_R$ reduces to: (1) solve a differential equation around each pole in order to find the local potential $\psi_j$ of $\f_L$ and (2) compute the residue of the function $\psi_j\f_R$ at each pole. Expression \eqref{intersectionnumber1} assumes a further simplified form when $\f_L$ and $\f_R$ have only simple poles. Let 

\begin{equation}
\f_{jL}(z) = \sum_{n=-1}^\infty \Phi^L_{jn} (z-z_j)^n,
\label{leftformexpansion}
\end{equation}

\nn
be the Laurent expansion of $\f_L$ around the pole $z_j$, and 

\begin{equation}
\psi_j = \sum_{n=0}^{\infty} \Psi_{jn} (z-z_j)^n,
\end{equation}

\nn
the one of $\psi_j$. Reminding 

\begin{equation}
\omega=dlog(u) =\sum_{i} \frac{\alpha_i}{z-z_i} dz,
\end{equation}

\nn
we have

\begin{equation}
\begin{split}
\nabla_\omega \psi_j &= \sum_{n=1}^\infty n \Psi_{jn} (z-z_j)^{n-1} dz + \sum_{n=0}^\infty \Psi_{jn}(z-z_j)^n \left ( \frac{\alpha_j}{z-z_j} +\sum_{i\neq j}\frac{\alpha_i}{z-z_j} \right)dz =\\&=\alpha_j \Psi_{j0} (z-z_j)^{-1}+ \sum_{n=1}^{\infty}\tilde{\Psi}_{jn} (z-z_j)^n,
\end{split}
\label{expansione}
\end{equation}

\nn
where $\tilde{\Psi}_{jn}$ have been defined. Comparing \eqref{leftformexpansion} and \eqref{expansione} term by term (in order for \eqref{Diffeq1} to be satisfied) one can fully determine $\phi_j$; actually we do not need all the terms, but just the one related to the pole of $\f_{jL}$:

\begin{equation}
\Phi^L_{j,-1}=\alpha_j\Psi_{j0}.
\label{coeffequaliti}
\end{equation}

\nn
Thus, Laurent expanding also $\f_R$, the intersection number becomes:

\begin{equation}
\begin{split}
\langle \f_L | \f_R \rangle &= \sum_j Res \left [ \sum_{n=0}^\infty \Psi_{jn}(z-z_j)^n \sum_{n=-1}^\infty \Phi^R_{jn}(z-z_j)^n \right ]=\\&=\sum_j Res [\Psi_{j0}\Phi^R_{j,-1} (z-z_j)^{-1}]=\sum_j \Psi_{j0}\Phi^R_{j,-1} = \sum_j \frac{\Phi^L_{j,-1}\Phi^R_{j,-1}}{\alpha_j};
\label{simplepoleintersection}
\end{split}
\end{equation}

\nn
where in second equality we kept the only one term of the product non holomorphic in $z_j$,  and in the last equality we used \eqref{coeffequaliti}.
Clearly if $\f_L$ and $\f_R$ had higher order poles one could always repeat the previous procedure, but the computation of residues would involve more Laurent series coefficients.\\ 
Consider a $1-$ form having, pardon the pun, the following form

\begin{equation}
\f_{j_1,j_2}= d\log{ \left ( \frac{z-z_{j_1}}{z-z_{j_2}}\right )}=\left (\frac{1}{z-z_{j_1}}-\frac{1}{z-z_{j_2}}\right ) dz;
\end{equation}

\nn
such a form is called logarithmic $1-$ form, for obviously reasons. Suppose now both $\f_L$ and $\f_R$ are logarithmic forms and consider their intersection number

\begin{equation}
\langle \f_{Lj_1,j_2} |\f_{Rj_3j_4}\rangle = \sum_j \frac{\Phi^L_{j,-1}\Phi^R_{j,-1}}{\alpha_j},
\label{logintersection1}
\end{equation}

\nn
from the definition of logarithmic form one can immediately read the coefficients of the expansions: 

\begin{equation}
\begin{split}
&\Phi_{Lj_1,-1}=1, \quad  \Phi_{Lj_2,-1}=-1, \quad\Phi_{Lj_i,-1}=0 \quad i\neq 1,2  \\ &\Phi_{Rj_3,-1}=1,\quad \Phi_{Rj_4,-1}=-1 , \quad \Phi_{Rj_i,-1}=0 \quad i\neq 3,4, 
\end{split}
\end{equation}

\nn
or more compactly 

\begin{equation}
\Phi_{Lj,-1}=\delta_{jj_1}-\delta_{jj_2} \quad\quad \mbox{and} \quad \quad \Phi_{Rj,-1}=\delta_{jj_3}-\delta_{jj_4}.
\end{equation}

\nn
Putting the previous expression into \eqref{logintersection1} and summing over $j$ one finds 

\begin{equation}
\langle \f_{Lj_1,j_2} |\f_{Rj_3j_4}\rangle =\frac{1}{\alpha_{j_1}}(\delta_{j_1j_3}-\delta_{j_1j_4})+\frac{1}{\alpha_{j_2}}(\delta_{j_2j_4}-\delta_{j_2j_3}).
\label{logintersection2}
\end{equation}

\nn
Summarizing the above results, we have found a vary nice and easy expression to compute the intersection number between logarithmic $1-$ forms in terms of the exponents $\alpha_j$; despite this result, being so specific, could seems quite useless till now, its fundamental importance is due to the following preposition.

\begin{Pre} Basis of Logarithmic forms.\\
Let $\f_{j}\equiv \f_{jj+1}$ be a logarithmic $1-$ form with poles in two adjacent points. The set $\{\f_j\}_{j=0}^{n}$ forms a basis of $H^1(\M,\nabla_\omega)$.
\end{Pre}

\nn
This means, obviously, that any form $\f_L \in H^1_\omega$ can be written as linear combination of logarithmic forms,

\begin{equation}
\f_L = \sum_j c_j \f_{j}
\end{equation}

\begin{Ese}
Consider for instance the form 

\begin{equation}
\begin{split}
\nabla_\omega \left ( \frac{1}{(z-z_1)^n}\right )&=d\left ( \frac{1}{(z-z_1)^n}\right )-\sum_j \frac{dz}{z-z_j}\frac{1}{(z-z_1)^n} =\\&=-\frac{-ndz}{(z-z_1)^{n+1}}- \frac{dz}{(z-z_1)^{n+1}}+\sum_{j\neq 1} \frac{dz}{(z-z_j)(z-z_1)^n}=\\&=-\frac{(n+1)dz}{(z-z_1)^{n+1}}-\sum_{j\neq 1} \frac{dz}{(z-z_j)(z-z_1)^n},
\end{split}
\end{equation}

\nn
thus one has:

\begin{equation}
\frac{(n+1)dz}{(z-z_1)^{n+1}} = -\sum_{j\neq 1} \frac{dz}{(z-z_j)(z-z_1)^n} -\nabla_\omega \left ( \frac{1}{(z-z_1)^n}\right ).
\end{equation}

that is 
\begin{equation}
\frac{dz}{(z-z_1)^{n+1}} =-(n+1)^{-1}\sum_{j\neq 1} \frac{dz}{(z-z_j)(z-z_1)^n} \quad\quad in \quad H^1_\omega.
\end{equation}

\nn
We have therefore found that a $1-$form with order $n$ pole in $z_1$ is equivalent to the sum of $1-$forms with order $n-1$ pole at $z_1$. We can obviously apply this idea recursively until obtaining only simple poles at $z_1$. Notice that with this procedure, the forms obtained have globally the same pole order, and one just moves one order from one pole to the another one, however this is enough for our purpose, because in computing intersection one just cares about the local behaviour around one specific point.
\end{Ese}

\nn
In \cite{Matsumoto1998IntersectionNF} Matsumoto showed that \eqref{logintersection2} can be generalized to the case of $n-$ forms.

\nn
In \cite{Mizera:2017rqa} Mizera shows that the previous result can be rewrite in a more computational friendly way. In order to understand his result is convenient to rewrite the intersection number for $1-$ logarithmic forms in a slight different form. Consider again the expansion \eqref{expansione} and rewrite it as 

\begin{equation}
\begin{split}
\nabla_\omega \psi_j& = \omega\Psi_{j0} -\sum_{i\neq j} \alpha_i \Psi_{j0} (z-z_i)^{-1} \sum_{n=1}^{\infty}\tilde{\Psi}_{jn} (z-z_j)^n=\\&=\omega\Psi_{j0} + Hol(z_j),
\end{split}
\end{equation}

\nn
where we have denoted by $Hol(z_j)$ all the remaining terms holomorphic in $z_j$. Again, in order for \eqref{Diffeq1} to be satisfied this must coincide to $\f_L$, then, ignoring holomorphic parts for the same reason discussed above, we have: 

\begin{equation}
\psi_{j} = \frac{1}{\omega}\f_{jL}.
\end{equation}

\nn
Notice this expression actually does not depend any more on $j$, because the dependence on $z_j$ is encoded inside $\omega$, then this is really true globally 

\begin{equation}
\psi = \frac{1}{\omega}\f_L.
\end{equation}

\nn
One could read this result directly from \eqref{Diffeq1} once one convinces himself the potential is holomorphic in $z_j$ and the only term contributing to the residue is the zero-th one, whose derivative is clearly null. With this result \eqref{intersectionnumber1} becomes 

\begin{equation}
\langle \f_L | \f_R \rangle =\sum_j \res_{z=z_j} {\left (\frac{\f_{jL}\f_R}{\omega}\right )}.
\end{equation}

\nn
Now, the function $(\f_L\f_R)/\omega$ has poles both at the poles $z_j$ of $\omega$ and at the zeros $z^*_j$ of $\omega$; because the sum over all residues, with the one at infinity taken into account, must be zero, one can write:

\begin{equation}
\res_{z=z_j}{\left ( \frac{\f_{jL}\f_R}{\omega}\right )}=-\res_{z=z^*_j}{ \left ( \frac{\f_{jL}\f_R}{\omega}\right )}
\end{equation}

\begin{equation}
\int \delta[\omega]\f_L\f_R =\sum_{z=z^*_j}\int\frac{\delta(z-z^*_j)}{|\de\omega|} \f_L\f_R = \sum_{z=z^*_j}\frac{\f_L\f_R}{|\de\omega|}\Big |_{z=z^*_j}
\label{logamizera1}
\end{equation}

\nn
In words, we have managed to write the intersection number between one logarithmic forms in terms of an expression computed in the zeros of the connection instead that in its poles; although this is not vary illuminating from a mathematical point of view, when one has to concretely find these points it turns to be easier to identify zeros respect to poles. In \cite{Mizera:2017rqa} Mizera shows that \eqref{logamizera1} can be generalized to the multivariate case, the final result is  

\begin{equation}
\langle \f_L |\f_R \rangle = \int \prod_{i=1}^n dz_i\delta[\omega_i]\f_L\f_R=\sum_{z^*_{ij}} |\de_i\omega_j|^{-1} \f_L\f_R \Big |_{z_i=z^*_{ij}}.
\end{equation}

\nn
where $z_{ij}^*$.

\subsection{A recursive method for generic n-forms}\label{recursive}
As we said, in the one dimensional case one can always find a logarithmic basis for the cohomology space, and then compute intersection numbers by means of Matsumoto or Mizera formulae; this is no long true when one considers the multivariate case: the $n-$th cohomology space of meromorphic forms seems not to be in general isomorphic to the one of logarithmic forms, moreover also in those cases where this occurs, finding a logarithmic basis and decomposing a generic $n-$ form onto it, is vary challenging. For this reason one would like to have an algorithm able to directly compute the intersection number of generic $n-$forms. In \cite{Frellesvig:2019uqt} Frellesvig, Gasparotto, Mandal, Mastrolia, Matiazzi and Mizera, presented a recursive algorithm for constructing multivariate intersection number between generic $n-$ forms. Subsequently in \cite{Weinzierl:2020xyy} Weinzierl shows that it can be written as a global residue. 
The key idea lies in the reiterative decomposition of the space of $n-$ forms into a (outer) space of $1-$forms, depending only on the $z_n$ variable, treating all the other as parameters, wedge a (inner) space of $(n-1)-$ forms depending on the $z_1,..z_{n-1}$ variables.\footnote{One can clearly perform the decomposition choosing any variable $z_i$ for the outer space, but it can be proven that the choosing order, although it influences the intermediate steps, is irrelevant for the final result.}
\paragraph{2-forms intersection.}
Consider the intersection of two $2-$forms. 
Using the same notation of \cite{Frellesvig:2019uqt}, let $\langle\f_L^{(\mathbf{2})}| \in H^2(\M, \nabla_\omega)$ and $|C_R^{(\mathbf{2})}]\in H_2(\M,\Lo)$, and consider the intersection number

\begin{equation}
    \langle \f_L^{(\mathbf{2})} |C_R^{(\mathbf{2})}]= \int_{C_R^{(\mathbf{2})}}\f_L^{(\mathbf{2})}(z_1,z_2)u(z_1,z_2).
    \label{recursiveone}
\end{equation}

\nn
The connection is given by

\begin{equation}
    \omega = d\log u = \sum_i \omega_i dz_i,
\end{equation}

\nn
and, as we know, the number of solutions $\n_2$ of the system $\omega_i=0$ gives the dimension of $H^2_\omega$. The suitable observation here is that the number of solutions of the equation $\omega_1=0$, corresponds to the dimension of $H^1(\M,\omega_1)$ obtained treated $z_2$ as a parameter. Let $\{\langle e_i^{(\mathbf{1})}|\}$ and $\{|h_i^{(\mathbf{1})}\rangle\}$ be two basis for $H^1_\omega$ and $H^1_{-\omega}$ respectively, and let 

\begin{equation}
{(C_{((\mathbf{1}))})}_{ij} \equiv  \langle e_i^{(\mathbf{1})}|  h_i^{(\mathbf{1})} \rangle
\end{equation}

\nn
be their intersection matrix. 
Now, if we assume $\M$ admits a fibration into one-dimensional spaces, we perform the decomposition 

\begin{equation}
    \langle\f_L^{(\mathbf{2})}| = \sum_{i=1}^{\nu_1} \langle e_i^{(\mathbf{1})}|\wedge  \langle\f_{L,i}^{({2})}| 
\end{equation}

\nn
actually splitting the integral \ref{recursiveone} as:

\begin{equation}
    \langle \f_L^{(\mathbf{2})} |C_R^{(\mathbf{2})}]= \sum_{i=1}^{\n_1}\int_{C_R^{(2)}}\f_{L,i}^{(2)}(z_2)\int _{C_R^{(1)}}e_i^{(1)}(z_1,z_2)u(z_1,z_2)=\sum_{i=1}^{\n_1}\int_{C_R^{(2)}}\f_{L,i}^{(2)}(z_2) \langle e_i^{(1)}|C_R^{(1)}].
    \label{recursivetwo}
\end{equation}

\nn
It is important to notice that the last integral can be interpreted as the intersection of a one form in a space where the twist is given by $\langle e_i^{(1)}|C_R^{(1)}]$.
The inner intersection number satisfies the differential equation

\begin{equation}
    d_{z_2}\langle e_i^{(1)}|C_R^{(1)}] = \Omega{ij}^{(2)}\langle e_i^{(1)}|C_R^{(1)}];
\end{equation}

\nn
where the matrix $\Omega_{ij}$, obtainable  using Stoke' theorem and master decomposition formula is

\begin{equation}
    \Omega^{(2)}_{ij}= \langle d_{z_2}+\omega_2\wedge)e_i^{(1)}|h_k^{(1)}{(C_{(1)}^{-1})}_{kj}.
\end{equation}

\nn
For generic $0-$form $\xi_i(z_2)$, one therefor has:

\begin{equation}
    \begin{split}
        0&=\int_{C_R^{(2)}} d_{z_2}\left (\xi_i(z_2) \langle e_i^{(1)}|C_R^{(1)}]\right )=\int_{C_R^{(2)}} {(\nabla_{\Omega^{(2)}})}_{ij} \xi_i(z_2)\langle e_i^{(1)}|C_R^{(1)}],
    \end{split}
\end{equation}

\nn
where

\begin{equation}
    {(\nabla_{\Omega^{(2)}})}_{ij} \equiv \delta_{ij}d_{z_2} +\Omega^{(2)}_{ij}.
\end{equation}

\nn
Consider then the differential equation around the pole $q$

\begin{equation}
    \nabla_{\Omega^{(2)}}\psi_i^{(q)}=\f_{L,i}^{(2)},
\end{equation}

\nn
proceeding as in the previous section, this differential equation allows one to construct a realization of the regularization map, i.e. explicitly obtain a compactly supported form belonging to the same cohomology class of $\f_{L,i}^{(2)}$, and therefore apply formula \ref{intersectionnumber1}, finally obtaining:

\begin{equation}
    \langle \f_L^{(\mathbf{2})}|\f_R^{(\mathbf{2})}\rangle = \sum_{i,j=1}^{\nu_1} \sum_{q=P_\Omega^{(2)}}\res_{z_2=q}\left (\psi_i^{(q)} {(C_{(1)})}_{ij}\f_{R,j}^{(2)} \right ).\label{multivariatedecomposition}
\end{equation}

\paragraph{General case.}
The last procedure can be generalized for the generic multivariate case, paying attention to the fact that now the inner space is clearly not one-dimensional but $n-1$ dimensional. Equation \ref{multivariatedecomposition} becomes

\begin{equation}
    \lf{\mathbf{n}}\rf{(\mathbf{n})} = \sum_{p\in P_\omega} \res_{z_n=p}\left ( \psi_i^{(n)}{(C_{(n-1)})}_{ij}\f_{Rj}^{(n)}\right );
\end{equation}

\nn
applying the algorithm recursively one is able to reduce the dimension of the intersection matrix step by step, finally obtaining:

\begin{equation}
    \langle \f_L^{(\mathbf{n})} | \f_R^{(\mathbf{n})} \rangle = \sum_{p_n \in P_n}...\sum_{p_1 \in P_1} \res_{z_n=p_n}... \res_{z_1=p_1}\left (\psi^{(n)}_{i_{n-1}}\psi^{(n-1)}_{i_{n-1}i_{n-2}}...\psi^{(1)}_{i_11}\f_R^{(n)} \right ),
\end{equation}

\nn
for all $i_m=1,...,\n_m$, where $\psi^{m}_{i_mi_{m-1}}$ are the local solutions of the differtial equation 

\begin{equation}
    \de_{z_m}\psi^{m}_{i_mi_{m-1}}+\psi^{m}_{j_mi_{m-1}}\Omega^{(m)}_{j_mi_{m-1}}=\hat{e}^{(m)}_{j_mi_{m-1}},
\end{equation}

\nn
and $\hat{e}^{(m)}_{j_mi_{m-1}}$defined by 

\begin{equation}
    \langle e^{(m)}_{j_mi_{m-1}}|=\hat{e}^{(m)}_{j_mi_{m-1}}dz_m.
\end{equation}
\chapter{Introduction to String theory}\label{Chapter3}
String theory is one of the most famous attempt to describe gravity into a quantum framework, achieving the so yearned unification with the other forces of nature. The basic idea behind string theory is to interpret particles not as pointlike objects but as different excited states of a onedimensional object: a string. 
The simpler string theory one can write down is the bosonic string theory, involving only bosonic degrees of freedom. Although it does not contain matter and unstable negative energy states appear in it (tachyons), bosonic string theory turns to be an interesting toy model to be studied, because it shares many structural features with its supersymmetric counterpart, containing fermions and no tachyons.
Main references are \cite{zwiebach_2004},\cite{Polchinski:1998rq},\cite{kiritsis:in2p3-00714916},\cite{Green:2012oqa},\cite{Tong:2009np},\cite{Staessens:2010vi}.

\section{The relativistic string}
As known a point-particle moving in a $d$-dimensional Minkoswki space-time describes a one-dimensional curve called \textit{world-line}. A string, instead, being a onedimensional object, moving in the same space, sweeps a two-dimensional surface, at which will refer as \textit{world-sheet}.\\
In order to introduce the classical action and the quantization of strings, it is useful to briefly review the relativistic point particle, that will be the starting point for string treatment.
Consider a point-particle with mass $m$ in a $d$-dimensional Minkowski space-time with metric $\eta_{\mu \nu}=  diag\left(-1,1,1,...,1 \right)$. Its action is simply the length of its world-line parametrized as $x^{\mu}= x^{\mu} ( \tau)$:
\begin{equation}
S= - m \int d \tau	 \sqrt{- \frac{dx^{\mu}}{d \tau} \frac{dx^{\nu}}{d \tau} \eta_{\mu \nu}}.
\label{action}
\end{equation}
\nn
The action \eqref{action} is invariant under Poincarè transformations, manifestly, and under reparametrizations of $\tau$ by any monotonic function $\tau \rightarrow \tilde{\tau} = \tilde{\tau} (\tau)$. This gauge invariance of the theory, reflects a redundancy in our description, i.e. the presence of a non-physical degree of freedom. Indeed the conjugate momenta:
\begin{equation}
p_{\mu} = \frac{\partial \mathcal{L}}{\partial \dot{x}^{\mu}} 
\end{equation}
\nn
are not all independent, as it is well known, they must satisfy the mass-shell condition:
\begin{equation}
p_{\mu}p^{\mu} + m^2 =0.
\end{equation}

\subsection{ The Nambu-Goto action} 
As we said the motion of a string describes a bidimensional surface on a $d$-dimensional Minkowski space-time, we can therefore parametrize it by two coordinates $\sigma^a = \left(  \sigma, \tau \right)$, a spatial coordinate $\sigma$ and a temporal one $\tau$.
The parameter $\sigma$ is conventionally chosen to be $\sigma \in \left[ 0; 2 \pi \right)$ for closed strings and $\sigma \in \left[ 0; \pi \right ]$ for open strings.
The string evolution can be described as maps from two-dimensional world-sheet to space-time through the coordinates:
\begin{equation}
X^{\mu} (\tau , \sigma) \quad \quad \quad \mu=0, 1, ... d-1
\end{equation} 

\nn
A first attempt to write down a string action, naturally inspired by what we did for the point particle, is to consider the area of the world-sheet. Let $\gamma_{ab}$ be the induced metric, from the $d$-dimensional Minkowski space-time, on bidimensional surface of the world-sheet:

\begin{equation}
\gamma_{\alpha \beta} = \frac{\partial X^{\mu}}{\partial \sigma^a} \frac{\partial X^{\nu}}{\partial \sigma^b} \eta_{\mu \nu},
\label{inducedmetric}
\end{equation}

then the seeked action take the form:
\begin{equation}
S_{NG} = - T \int d^2 \sigma \sqrt{-det \gamma_{ab}} = - T \int d^2 \sigma \left[ - \left( \dot{X} \right)^2 \left( X' \right)^2 + \left( \dot{X} X' \right)^2 \right]^{1/2},
\label{NambuGoto}
\end{equation}
where $\dot{X}^{\mu} \equiv \de_\tau X^{\mu}$ and $X'^{\mu} \equiv \partial_\sigma X^{\mu}$. The action \eqref{NambuGoto} is called \textit{Nambu-Goto action}. Since space-time coordinates, in natural units, have dimension $\left[X \right]= M^{-1}$, and the parameters $\sigma^a$ are dimensionless, the constant $T$ must have dimension  $\left[T \right]= M^{2}$ so the action be dimensionless. $T$ has the physical interpretation of  \textit{tension} of string, which behaves as an elastic band and its potential energy increases linearly with length. We introduce the \textit{Regge parameter} $\alpha' = \frac{1}{2 \pi T}$, through which we define a string length scale $l_s = \sqrt{\alpha'}$ and a string mass scale $M_s  \sim \frac{1}{\sqrt{\alpha'}}$. \\
Just like the point particle action, the Nambu-Goto action turns to be invariant under space-time Poincarè transformations and reparametrizations $\sigma^a \rightarrow \tilde{\sigma}^a (\sigma)$. Notice that Poincarè invariance, from world-sheet point of view, is a global symmetry because the transformation parameters do not depend on sheet coordinates. Reparemetrization invariance reflects instead the unphysicality of sheet coordinates.
In order to derive the equation of motion induced by the action \eqref{NambuGoto} we introduce the momenta:
\begin{equation}
\begin{split}
& \Pi_{\mu}^{\tau} = \frac{\partial \mathcal{L}}{\partial \dot{X}^{\mu}} = -T \frac{\left( \dot{X} \cdot X' \right) X'_{\mu} - \left( X'^2 \right) \dot{X}_{\mu}}{\sqrt{ \left( \dot{X} \cdot X' \right)^2 - \dot{X}^2 X'^2}} \\
& \Pi_{\mu}^{\sigma} = \frac{\partial \mathcal{L}}{\partial \dot{X}^{\mu}} = -T \frac{\left( \dot{X} \cdot X' \right) \dot{X}_{\mu} - \left( X'^2 \right) X'_{\mu}}{\sqrt{ \left( \dot{X} \cdot X' \right)^2 - \dot{X}^2 X'^2}} \\
\end{split}
\end{equation}

\nn
and we apply, as usual, the variational principle, obtaining:
\begin{equation}
\frac{\partial \Pi^{\tau}_{\mu}}{\partial \tau} + \frac{\partial \Pi^{\sigma}_{\mu}}{\partial \sigma} =0
\end{equation}

\nn
To solve the equation of motion we need to impose supplemented boundary conditions. They depend on the type of string we are considering: for closed strings, having no extrema, we must impose e periodicity condition 

\begin{equation}
X^{\mu} (\tau , \sigma) = X^{\mu} (\tau , \sigma +2\pi);
\label{periodicityconditions}
\end{equation}

\nn
for open strings instead, we have to impose constrains on endpoints, the most frequently used ones are Neumann conditions

\begin{equation}
\frac{\delta\La}{\delta X'^{\m}}\Big |_{0,\pi}=0,
\end{equation}

\nn 
imposing no momentum flows off the endpoints, and Derichlet conditions
and 
\begin{equation}
\frac{\delta\La}{\delta \dot{X}^{\m}}\Big |_{0,\pi}=0,
\end{equation}

\nn
implying a fixed space-time position of endpoints.

\section{The Polyakov action}\label{Polyakovactionsection}
The Nambu-Goto action, due to the presence of the square root, is not particularly suitable to be quantized by path integral procedure. 
In order to avoid this inconvenient we would like to handle with an action equivalent to \eqref{NambuGoto} but involving no square roots in it; actually it can be found, analogously to what we do for a massless point particle action, at the cost of introducing an additional field on the world-sheet. Let $h_{\alpha \beta} ( \tau , \sigma)$ a dynamical metric of the world-sheet with signature $\left( - ; + \right)$, and consider the action

\begin{equation}
S_P = - \frac{1}{4 \pi \alpha '} \int d^2 \sigma \sqrt{-h} h^{\alpha \beta} \partial_{\alpha} X^{\mu} \partial_{\beta} X^{\nu} \eta_{\mu \nu}.
\label{Polyakov}
\end{equation}

\nn
called Polyakov action.
The equations of motion for the fields $X^{\mu} ( \tau, \sigma)$ are:
\begin{equation}
\partial_{\alpha} \left( \sqrt{-h} h^{\alpha \beta} \partial_{\beta} X^{\mu} \right) =0.
\end{equation}
Varying the action with respect to the metric field 
\begin{equation}
\delta S_P = - \frac{T}{2} \int d^2 \sigma \delta h^{\alpha \beta} \left( \sqrt{-h} \partial_{\alpha} X^{\mu} \partial_{\beta} X^{\nu} - \frac{1}{2} \sqrt{-h} h_{\alpha \beta} h^{\rho \sigma} \partial_{\rho} X^{\mu} \partial_{\sigma} X^{\nu} \right) \eta_{\mu \nu},
\end{equation}

\nn
we obtain its equation of motion:
\begin{equation}
h_{\alpha \beta } = 2 f( \tau , \sigma	) \partial_{\alpha} X \cdot \partial_{\beta} X
\end{equation}

\nn
where we put $f^{-1} = h^{\rho \sigma} \partial_{\rho} X \cdot \partial_{\sigma} X$. 
Comparing of this result with the $\gamma_{\alpha \beta}$ definition \eqref{inducedmetric}, we see they are equivalent up to a factor $f$.  Within this difference, the equations of motion for the fields $X^{\mu}$ induced by Nambu-goto action and Polyakov action are identical. The presence of this, factor, called conformal factor, reflects the existence of an extra symmetry in the Polyakov action. \\
Let us summarize the symmetries of the Polyakov action. 
\begin{itemize}
\item[] \textit{Poincarè invariance}, is a global symmetry from world-sheet point of view.
\begin{equation}
X^{\mu} \quad \longmapsto \quad \Lambda^{\mu}_{\nu} X^{\nu} + c^{\mu}
\end{equation}
\item[] \textit{Reparameterization invariance} (also \textit{diffeomorphisms-invariance}), is a gauge-symmetry on the world-sheet. We define the change of coordinates on the worldsheet as $\sigma^{\alpha} \rightarrow \tilde{\sigma}^{\alpha} (\sigma)$, under which the fields $X^{\mu}$ transforms as worldsheet scalars, while $h_{\alpha \beta}$ as a 2d metric:
\begin{equation}
\begin{split}
 X^{\mu} (\sigma) \quad & \longmapsto \quad \tilde{X}^{\mu} (\tilde{\sigma}) = X^{\mu} (\sigma) \\
h_{\alpha \beta} (\sigma) \quad & \longmapsto \quad \tilde{h}_{\alpha \beta} (\tilde{\sigma}) = \frac{\partial \sigma^{\rho}}{\partial \tilde{\sigma}^{\alpha}} \frac{\partial \sigma^{\gamma}}{\partial \tilde{\sigma}^{\beta}} h_{\rho \gamma} (\sigma). \\
\end{split}
\end{equation}
If we consider an infinitesimal transformation $\sigma^{\alpha} \rightarrow \tilde{\sigma}^{\alpha} (\sigma)= \sigma^{\alpha}- \eta^{\alpha} (\sigma)$, the corresponding infinitesimal transformations of fields are:
\begin{equation}
\begin{split}
& \delta X^{\mu} (\sigma) = \eta^{\alpha}\partial_{\alpha} X^{\mu} (\sigma) \\
& \delta h_{\mu \nu} (\sigma) = \nabla_{\alpha} \eta_{\beta} + \nabla_{\beta} \eta_{\alpha}, \\
\end{split}
\end{equation}
where the covariant derivative is defined as:
\begin{equation}
\nabla_{\alpha} \eta_{\beta} = \partial_{\alpha} \eta_{\beta} - \Gamma^{\rho}_{\alpha \beta} \eta_{\rho}
\end{equation}
with Levi-Civita connection associated to worldsheet metric:
\begin{equation}
\Gamma^{\rho}_{\alpha \beta} = \frac{1}{2} h^{\rho \gamma} \left( \partial_{\alpha} h_{\gamma \beta} + \partial_{\beta} h_{\gamma \alpha} - \partial_{\gamma} h_{\alpha \beta} \right).
\end{equation}
\item[] \textit{Weyl-invariance}, is a gauge symmetry defined by following fields transformations:
\begin{equation}
\begin{split}
 X^{\mu} (\sigma) \quad & \longmapsto \quad X^{\mu} (\sigma) \\
 h_{\alpha \beta} (\sigma) \quad & \longmapsto \quad \Omega^2 (\sigma) h_{\alpha \beta} (\sigma). \\
 \end{split}
\end{equation}
Infinitesimally we can write $\Omega^2 (\sigma) = e^{2 \phi (\sigma)}$, for small $\phi (\sigma)$, thus:
\begin{equation}
\delta h_{\alpha \beta} (\sigma) = 2 \phi (\sigma) h_{\alpha \beta} (\sigma).
\end{equation} 
The Weyl symmetry in not a coordinates transformation, it represents the theory invariance under a local change of scale preserving angles. 
This propriety is special for two-dimensional theory, however it restricts the kind of interactions that can be added to the action, and these constraints becomes even more stringent in the quantum theory.
\end{itemize}
The reparametrization-invariance allows us to choose a convenient gauge for the worldsheet metric, called \textit{conformal gauge}:
\begin{equation}
h_{\alpha \beta} (\sigma) = e^{2 \phi (\sigma)} \eta_{\alpha \beta},
\end{equation} 
i.e. the two degrees of freedom in the choice of parameters allows us to fix two of the three degrees of freedom of the metric. \\
Again, using the Weyl-invariance we can remove the last independent component of the metric and set $\phi(\sigma)=0$ such that $h_{\alpha \beta} = \eta_{\alpha \beta}$. Through this choice the action \eqref{Polyakov} takes the simple form:
\begin{equation}
S_{P} = - \frac{1}{4 \pi \alpha '} \int d^2 \sigma \partial_{\alpha} X \partial_{\beta} X,
\end{equation}
and the equations of motion become:
\begin{equation}
\partial_{\alpha} \partial^{\alpha} X^{\mu} =0
\label{eqmotostringa}
\end{equation}
We define the energy-momentum tensor to be:
\begin{equation}
T_{\alpha \beta} = - \frac{2}{T} \frac{1}{\sqrt{-h}} \frac{\delta S_P}{\delta h_{\alpha \beta}}= \partial_{\alpha}  X \partial_{\beta } X - \frac{1}{2} \eta_{\alpha \beta} \eta^{\rho \gamma} \partial_{\rho} X \partial_{\gamma} X,
\end{equation}
where we performed the  choice of flat metric. \\
The \eqref{eqmotostringa} are subject to two constraints arising from the equation of motion of the metric:
\begin{equation}
T_{\alpha \beta} =0 \quad \quad \begin{cases} & T_{01} = \dot{X} \cdot X' =0 \\ & T_{00}=T_{11} = \frac{1}{2} \left( \dot{X}^2 + X'^2 \right)=0. \\ \end{cases}
\label{eqmotometrica}
\end{equation}
The first condition  tell us that we have to choose a parametrisation such that  lines $\sigma =const$ are perpendicular to lines $\tau=const$. Now we use the residual-gauge arising from Weyl transformations to introduce the \textit{statical gauge}:
\begin{equation*}
X^0 =R \tau
\end{equation*}  
with $R=const$. The previous constraints for the spatial components in this gauge becomes:
\begin{equation*}
\begin{split}
& \quad \dot{\overline{X}} \cdot \overline{X}' =0 \\
& \dot{\overline{X}}^2 + \overline{X}'^2 =R^2 \\
\end{split}
\end{equation*}
The first condition tell us that the motion of string must be perpendicular to string itself, i.e. the oscillations will be described by transverse modes. The second condition relates $R$ with the length of string when $\overline{\dot{X}}=0$, i.e. starting from a stretched string at a time with $\overline{\dot{X}}=0$, the string will contract under its own tension to later times.

Having a nice classical action describing the relativistic string, our next purpose is naturally to quantize it. As in ordinary quantum field theory the quantization procedure may be perform via operator methods or path integral techniques. In the first case we promote classical variables to operators and we build up the Hilbert space of states of the theory, this operation may be carried on in two differ manners: we can promote to operators the unconstrained classical variables and impose constrains later on on states, or we can solve constrains classically and quantize left over variables. The first approach, know as Covariant canonical quantization, preserves manifest Lorentz invariance, but it leads to a Hilbert space with negative norm states, that must be appropriately cut off by means of suitable conditions that may be untrivial to be found. The second procedure instead, called gauge quantization, because we obviously have to choose a gauge to classically solve the theory, is in general easier to be performed, but it is not manifest Lorentz invariant. All three methods of quantization bring to equivalent

\section{Mode expansion} 
Inspired, as usual, by the harmonic oscillator and in order to canonical quantize the theory, we want to expand the equation of motions in terms of oscillation modes. To do that, it is convenient to introduce more suitable coordinates on worldsheet, called lightcone coordinates, defined as

\begin{equation}
\begin{split}
& \sigma^+ = \tau + \sigma \\
& \sigma^- = \tau - \sigma. \\
\end{split}
\label{gaugeconoluce}
\end{equation}

\nn
The Minkowski metric in these coordinates takes the nice form:
\begin{equation}
\eta_{\alpha \beta} = \left( \begin{matrix} 0 & - \frac{1}{2} \\ - \frac{1}{2} & 0 \\ \end{matrix} \right) \quad \quad \quad \eta^{\alpha \beta} = \left( \begin{matrix} 0 & -2 \\ -2 & 0 \\ \end{matrix} \right).
\end{equation}

\nn
and the equation of motion reads:

\begin{equation}
\partial_{+} \partial_{-} X^{\mu} =0.
\label{eqmotionlightcoordinates}
\end{equation}

\nn
The more general solution of \eqref{eqmotionlightcoordinates} can be written in terms of left and right oscillating modes as:
\begin{equation}
X^{\mu} (\tau , \sigma) = X^{\mu}_{L} (\sigma^+) + X^{\mu}_{R} (\sigma_-).
\end{equation}

\nn
One can Fourier expand the left and right solutions as:

\begin{equation}
\begin{split}
& X^{\mu}_L ( \sigma^+) = \frac{1}{2} x^{\mu} + \frac{1}{2} \alpha' p^{\mu} \sigma^+ +i \sqrt{\frac{\alpha'}{2}} \sum_{k \neq 0} \frac{1}{k} \tilde{\alpha}^{\mu}_k e^{-i k \sigma^+} \\
& X^{\mu}_R ( \sigma^-) = \frac{1}{2} x^{\mu} + \frac{1}{2} \alpha' p^{\mu} \sigma^- +i \sqrt{\frac{\alpha'}{2}} \sum_{k \neq 0} \frac{1}{k} \tilde{\alpha}^{\mu}_k e^{-i k \sigma^-}. \\
\end{split}
\label{solstringa2}
\end{equation}

The quantities $x^{\mu}$ and $p^{\mu}$ can be respectively interpreted as the\textit{position} and the \textit{momentum} of the center of mass of the string. This can be easily proven by studying the Noether currents associated to translation invariance $X^{\mu} \rightarrow X^{\mu}+ c^{\mu}$. Since the coordinate fields must be real, the coefficients of the Fourier expansion must satisfy:
\begin{equation}
\left( \alpha_k^{\mu} \right) = \left( \alpha_{-k}^{\mu} \right)^* \quad \quad \quad \left( \tilde{\alpha}_k^{\mu} \right) = \left( \tilde{\alpha}_{-k}^{\mu} \right)^*.
\end{equation}

\nn
In order to proceed we have to impose boundary conditions; as we said they depend on string type, thus we have to treat them separately. Consider first closed strings.  The periodicity condition \eqref{periodicityconditions} implies $k$ must be a non zero integer, thus we have

\begin{equation}
\begin{split}
& X^{\mu}_L ( \sigma^+) = \frac{1}{2} x^{\mu} + \frac{1}{2} \alpha' p^{\mu} \sigma^+ +i \sqrt{\frac{\alpha'}{2}} \sum_{n \neq 0} \frac{1}{n} \tilde{\alpha}^{\mu}_n e^{-i n \sigma^+} \\
& X^{\mu}_R ( \sigma^-) = \frac{1}{2} x^{\mu} + \frac{1}{2} \alpha' p^{\mu} \sigma^- +i \sqrt{\frac{\alpha'}{2}} \sum_{n \neq 0} \frac{1}{n} \tilde{\alpha}^{\mu}_n e^{-i n \sigma^-}, \\
\end{split}
\label{solstringa}
\end{equation}

\nn
where $n \in \Z\setminus \{0\}$.
The constrains \eqref{eqmotometrica} becomes on the lightcone coordinates:
\begin{equation}
\left( \partial_+ X \right)^2 = \left( \partial_- X \right)^2 =0.
\label{constrains3}
\end{equation}

\nn
Deriving the coordinate field respecting to $\sigma^-$ we have:

\begin{equation}
\left( \partial_- X^{\mu} \right) = \partial_- X^{\mu}_R = \frac{\alpha'}{2}p^{\mu} + \sqrt{\frac{\alpha'}{2}} \sum_{n \neq 0} \alpha^{\mu}_n e^{-i n \sigma^-} =  \sqrt{\frac{\alpha'}{2}} \sum_{n \in Z} \alpha^{\mu}_n e^{-i n \sigma^-}
\end{equation} 

\nn
where we set  $\alpha_0^{\mu}\equiv \sqrt{\frac{\alpha'}{2}} p^{\mu}$. 
Equation \eqref{constrains3} then becomes:

\begin{equation}
\begin{split}
\left( \partial_- X^{\mu} \right)^2 & = \left[  \sqrt{\frac{\alpha'}{2}} \sum_{n \in Z} \alpha^{\mu}_n e^{-i n \sigma^-} \right]^2 \\
& = \frac{\alpha'}{2} \sum_{m,p} \alpha_m \cdot \alpha_p e^{-i(m+p) \sigma^-} \\
& = \frac{\alpha'}{2} \sum_{m,n} \alpha_m \cdot \alpha_{n-m} e^{-in \sigma^-} \\
&= \alpha' \sum_n L_n e^{-in \sigma^-} =0 \\
\end{split}
\end{equation}

\nn
where we defined 
\begin{equation}
L_n= \frac{1}{2} \sum_m \alpha_{n-m} \cdot \alpha_m.
\end{equation}

\nn
Proceeding analogously for left modes, one can also define the quantity
\begin{equation}
\tilde{L}_n= \frac{1}{2} \sum_m \tilde{\alpha}_{n-m} \cdot \tilde{\alpha}_m,
\end{equation}
with $\alpha_0^{\mu} = \tilde{\alpha}_{0}^{\mu}$. The equations \eqref{eqmotometrica} impose an infinite number of conditions on the classical string solutions:
\begin{equation}
L_n = \tilde{L}_{n} = 0 \quad \quad \quad \forall n \in \mathbb{Z}.
\end{equation} 
In particular, the constraints arising from $L_0$ and $\tilde{L}_0$ have a rather special interpretation because they include the square of the momentum $p^{\mu}$ that respect the mass-shell condition on the Minkowskian space-time. Thus we have two expressions for the effective mass of string, one in terms of left-oscillators and one in terms of right-oscillators, and they must be equal to each other:
\begin{equation}
M^2 = \frac{4}{\alpha'} \sum_{n>0} \alpha_n \cdot \alpha_{-n} = \frac{4}{\alpha'} \sum_{n>0} \tilde{\alpha}_n \cdot \tilde{\alpha}_{-n}.
\label{classicallevelmat}
\end{equation}
This is called \textit{level matching condition}, and it will play an important role in the string quantization.

\section{Covariant quantization}

A first attempt to quantize the theory is naturally provided by the canonical quantization procedure. As usual we proceed promoting the fields $X^{\mu}$ and their conjugate momenta $\Pi_{\mu}= \frac{1}{2 \pi \alpha'} \dot{X}_{\mu}$ to operator, and replacing their Poisson brackets by commutators $\left\lbrace , \right\rbrace \rightarrow i^{-1} \left[ , \right]$:
\begin{equation}
\begin{split}
& \left[ X^{\mu} (\sigma , \tau), \Pi_{\nu} (\sigma' , \tau) \right] = i \delta (\sigma - \sigma') \delta^{\mu}_{\nu} \\
& \left[ X^{\mu} (\sigma , \tau), X^{\nu} (\sigma' , \tau) \right] = \left[ \Pi^{\mu} (\sigma , \tau), \Pi^{\nu} (\sigma' , \tau) \right] =0 \\
\end{split}
\end{equation}

\nn
The Fourier expansion coefficients of the equation of motion solutions became operators satisfying the commutation rules:
\begin{equation}
\left[ x^{\mu} , p_{\nu} \right] = i \delta^{\mu}_{\nu} \quad \quad \left[ \alpha^{\mu}_n , \alpha^{\nu}_m \right] = \left[ \tilde{\alpha}^{\mu}_n , \tilde{\alpha}^{\nu}_m \right]= n \eta^{\mu \nu} \delta_{n+m, 0} \quad \quad \left[ \alpha^{\mu}_n , \tilde{\alpha}^{\nu}_m \right] =0.
\end{equation}

\nn
The rescaling of these operators as $a_n = \frac{\alpha_n}{\sqrt{n}}$ and $a_{-n} = \frac{\alpha_{-n}}{\sqrt{n}}$ allows as to rewrite the commutation relations in the form
\begin{equation}
\left[ a_{n} , a_{-m} \right] = \delta_{nm},
\end{equation}

\nn
This is clearly, for each $n$, the algebra of the harmonic oscillator. We can therefore interpret the solutions of the string equation of motion as a superposition of an infinite number of harmonic oscillation modes, where $\alpha_n$ acts as an annihilation operator for $n > 0$ and as a creation operator for $n < 0$. \\
In order to construct the Hilbert space of the theory, we proceed as usual defining the vacuum state $\vert \Omega \rangle$ as the one annihilated by all $\alpha_n$ and $\tilde{\alpha}_n$, for $n>0$, and we build up the full space acting on the vacuum with creation operators

\begin{equation}
\vert \lambda \rangle = \left( \alpha_{-1}^{\mu_1} \right)^{n_{\mu_1}}  \left( \alpha_{-2}^{\mu_2} \right)^{n_{\mu_2}} ...  \left( \tilde{\alpha}_{-1}^{\nu_1} \right)^{n_{\nu_1}} \left( \tilde{\alpha}_{-2}^{\nu_2} \right)^{n_{\nu_2}}... \vert\Omega\rangle
\end{equation}

\nn
Those states represent all the possible excited states of the string, and, in the space-time, each state turns to be interpreted as a different species of particle. \\
Just as in QED canonical quantization, where we are forced to introduce Gupta-Bleuer conditions in order to avoid negative norm states, also in string covariant quantization in lightcone gauge unphysical states appear and a suitable gauge fixing condition on operators must be imposed.
Since $L_n$ and $\bar{L}_n$ are defined in terms of product of operators, in the quantum theory we have to choose an order. In particular for the zero-modes $L_0$ and $\bar{L}_0$ we discover to have an ambiguity in this definition because the operators on the product do not commutate, i. e. they are not completely determined by classical theory. Since the commutators are constant quantities, the different choices of $L_0$ and $\bar{L}_0$ are related by a  constant shift $L_0 \rightarrow L_0 +a$:
\begin{equation}
\begin{split}
& L_0 = \sum_{n=1}^{\infty} \alpha_{-n} \alpha_n + \frac{1}{2} \alpha_0^2 \\
&  \bar{L}_0 = \sum_{n=1}^{\infty} \bar{\alpha}_{-n} \bar{\alpha}_n + \frac{1}{2} \bar{\alpha}_0^2. \\
\end{split}
\end{equation}
In order to fix the constants, we impose the standard condition:
\begin{equation}
\begin{split}
& \left[ L_1 , L_{-1} \right]= 2 L_0 \\
& \left[ \bar{L}_1 , \bar{L}_{-1} \right]= 2 \bar{L}_0, \\
\end{split}
\end{equation} 
which are equivalent to require that $\left\lbrace L_{-1}, L_0 , L_1 \right\rbrace$ generate a subalgebra $SL(2, \mathbb{C})$, and also their analogues bar. Under these conditions the algebra of $L_n, \bar{L}_n$ operators take the following form:
\begin{equation}
\begin{split}
& \left[ L_m , L_n \right] = (m-n)L_{m+n} + \frac{c}{12} m (m^2-1) \delta_{m+n,0} \\
& \left[ \bar{L}_m , \bar{L}_n \right] = (m-n) \tilde{L}_{m+n} + \frac{\bar{c}}{12} m (m^2-1) \delta_{m+n,0}. \\
\end{split}
\end{equation}
\nn
We find one couple of \textit{Virasoro algebra} with central charges $c$ and $\bar{c}$. Physically these terms arises as quantum effect due to the breaking of Weyl invariance at quantum level. In the representation through $d$ free bosons $c= \bar{c}=d$, where $d=\eta^{\mu}_{\mu}$ is related to the dimension of the embedding space-time. Thus we can think that each boson contributes to central charge of one unity. \\
Let us now come back to the Virasoro generators. At classical level the constraints on the Fourier components of energy-momentum tensor are $L_n = \bar{L}_n=0$ but, because their commutations rules are not trivial, the same conditions cannot be imposed at quantum level. However, we can require that all positive mode-operators annihilate any physical state:
\begin{equation}
\begin{split}
&L_n \vert \psi \rangle_{phys} = \bar{L}_n  \vert \psi \rangle_{phys}  =0 \quad \quad \quad \quad \forall n >0 \\
& \left( L_0 -a \right)  \vert \psi \rangle_{phys}  = \left( \bar{L}_0 -a \right)  \vert \psi \rangle_{phys} =0 .\\
\end{split}
\label{condquantizzazione}
\end{equation}
The first of these two conditions is called \textit{Virasoro condition}. In second condition of \eqref{condquantizzazione} the quantity $a$ is some constant.\\
As we saw, the operators $L_0$ and $\bar{L}_0$ play an important role in determining the spectrum of the string because they include a term quadratic in the momentum $\alpha_0^{\mu} = \tilde{\alpha}_0^{\mu} = \sqrt{\frac{\alpha'}{2}} p^{\mu}$.  Using the redefinitions $L_0 +a$ and $\tilde{L}_0 +a$, the second constraint of  \eqref{condquantizzazione} and the classical condition \eqref{classicallevelmat} we obtain the \textit{quantum level condition}:
\begin{equation}
M^2 = \frac{4}{\alpha'} \left( -a + \sum_{m=1}^{\infty} \alpha_{-m} \cdot \alpha_m \right) = \frac{4}{\alpha'} \left( -a + \sum_{m=1}^{\infty} \tilde{\alpha}_{-m} \cdot \tilde{\alpha}_m \right).
\label{levelquant}
\end{equation}
We notice that the undetermined constant $a$ has a direct physical effect: it changes the mass spectrum of the string. We can write the \eqref{levelquant} through the \textit{Number operators}:
\begin{equation}
N= \sum_{m=1}^{\infty} \alpha_{-m} \cdot \alpha_m  \quad \quad \quad \tilde{N}= \sum_{m=1}^{\infty} \tilde{\alpha}_{-m} \cdot \tilde{\alpha}_m 
\end{equation}
that counts the number of excited modes of string. Then the level condition tell us that the number of left-moving modes must be equal to right-moving modes. Finally we define the Hilbert space of physical states as the quotient:
\begin{equation*}
\mathcal{H} = \frac{Ker L_n (n \geq 0)}{Im L_{-n} (n>0)}.
\end{equation*}

\section{Lightcone Quantization}
This different method of quantization for the string theory solve the constraints at classical level, leaving behind only the physical degrees of freedom, and later perform the quantization of classical solutions. \\
The gauge symmetries of the theory allows us to choose the \textit{light-cone coordinates} \eqref{gaugeconoluce}, where the metric take the form:
\begin{equation*}
ds^2 = -d \sigma^+ d \sigma^-
\end{equation*}
However we still have a residual gauge, in fact transformations like $\sigma^+ \rightarrow \tilde{\sigma}^+ (\sigma^+)$ and $\sigma^- \rightarrow \tilde{\sigma}^- (\sigma^-)$ lead an overall factor on the metric that can be remove by a compensating Weyl transformation. Again, if we fixed 3 components of worldsheet metric corresponding to 3 gauge-invariance degrees of freedom, then we wonder why do we still have some gauge symmetry left? The reason is that  $\tilde{\sigma}^+ , \tilde{\sigma}^-$   are functions of just a single variable, not two. \\
The physical implication of this residual gauge is to reduce the degrees of freedom of string solutions. In fact the equation of motion solutions  \eqref{solstringa} are $2d$ functions of a single variable, the constraints \eqref{eqmotometrica} reduces the number of independent functions to $2(d-1)$. The residual choice of parameters $\tilde{\sigma}^{\pm}$ still lowers this number to $2(d-2)$: these are the degrees of freedom of \textit{transverse} fluctuations of string. \\
In order to remove the remaining invariance we want to impose the \textit{light-cone gauge}. Let's start by choosing the space-time coordinates:
\begin{equation}
\begin{split}
& X^+ = \sqrt{\frac{1}{2}} \left( X^0 + X^{d-1} \right) \\
& X^- = \sqrt{\frac{1}{2}} \left( X^0 - X^{d-1} \right). \\
\end{split}
\end{equation}
We notice that this choice is not manifestally Lorentz-invariant.The new space-time metric is:
\begin{equation*}
ds^2 = -2 dX^+ dX^- + \sum_{i=1}^{d-2} dX^i dX^i.
\end{equation*}
Now we fix the gauge by requiring that the equation of motion solution for $X^{+}$ to be:
\begin{equation}
 X^+ = X^+_L (\sigma^+) + X^+_R (\sigma^-) = \frac{1}{2} \left( x^+ + \alpha' p^+ \sigma^+ \right)  + \frac{1}{2} \left( x^+ + \alpha' p^+ \sigma^- \right) =x^+ + \alpha' p^+ \tau.
\end{equation}
This is the light-cone gauge. Notice that $X^+$ is a null space-time coordinate proportional to timelike worldsheet parameter $\tau$. \\
The constraint equations for this coordinate are trivial, thus we wonder if there are extra-constraints on the other coordinate-solutions. Let's consider the usual ansatz for $X^-$ to solve its equation of motion:
\begin{equation*}
X^- = X^-_L (\sigma^+) + X^-_R (\sigma^-).
\end{equation*} 
Now we impose the constraints \eqref{eqmotometrica} in these new coordinates for this solution:
\begin{equation*}
\left( \partial_+ X^- \right)^2 = \left( \partial_- X^- \right)^2 =0,
\end{equation*}
using the gauge-fixing we find:
\begin{equation*}
\begin{split} 
& \partial_+ X^-_L = \frac{1}{\alpha' p^+} \sum_{i=1}^{d-2} \partial_+ X^i \partial_+ X^i \\
& \partial_- X^-_R= \frac{1}{\alpha' p^+} \sum_{i=1}^{d-2} \partial_- X^i \partial_- X^i .\\
\end{split}
\end{equation*}
So, up to an integration constant, the function $X^- (\sigma^+ , \sigma^-)$ is completely determined by other fields $X^i (\sigma^+ , \sigma^-)$:
\begin{equation*}
\begin{split}
& X^-_L (\sigma^+) = \frac{1}{2} \left( x^- + \alpha' p^- \sigma^+ \right) +i \sqrt{\frac{\alpha'}{2}} \sum_{n \neq 0} \frac{1}{n} \tilde{\alpha}^-_{n} e^{-in \sigma^+} \\
& X^-_R (\sigma^-) = \frac{1}{2} \left( x^- + \alpha' p^- \sigma^- \right) +i \sqrt{\frac{\alpha'}{2}} \sum_{n \neq 0} \frac{1}{n} \alpha^-_{n} e^{-in \sigma^-} \\
\end{split}
\end{equation*} 
$x^-$ is a integration constant, while $p^- ; \bar{\alpha}^-_n , \alpha^-_n$ are fixed by previous conditions in terms of other fields:
\begin{equation*}
\alpha^-_n = \sqrt{\frac{1}{2 \alpha'}} \frac{1}{p^+} \sum_{m = - \infty}^{+ \infty} \sum_{i=1}^{d-2} \alpha^i_{n-m} \alpha^i_m
\end{equation*}
and an equivalent expression for $\bar{\alpha}^-_n$, while for $\alpha_0^- = \sqrt{\frac{\alpha'}{2}} p^-$ we find:
\begin{equation*}
\begin{split}
& \frac{\alpha' p^-}{2} = \frac{1}{2 p^+} \sum_{i=1}^{d-2} \left( \frac{1}{2} \alpha' p^i p^i + \sum_{n \neq 0} \alpha^i_n \alpha^i_{-n} \right) \\
& \frac{\alpha' p^-}{2} = \frac{1}{2 p^+} \sum_{i=1}^{d-2} \left( \frac{1}{2} \alpha' p^i p^i + \sum_{n \neq 0} \bar{\alpha}^i_n \bar{\alpha}^i_{-n} \right) ,\\
\end{split}
\end{equation*}
from which we can reconstruct the classical level matching conditions:
\begin{equation*}
M^2 = 2 p^+ p^- - \sum_{i=1}^{d-2} p^i p^i = \frac{4}{\alpha'} \sum_{i=1}^{d-2} \sum_{n \neq 0} \alpha^i_{-n} \alpha^i_n = \frac{4}{\alpha'} \sum_{i=1}^{d-2} \sum_{n \neq 0} \bar{\alpha}^i_{-n} \bar{\alpha}^i_n.
\end{equation*}
By comparison with the level condition \eqref{classicallevelmat}, we observe that here the $i$-sum is only over transverse oscillators. Thus, the general classical solution is written in terms of $2(d-2)$ transverse oscillators and the zero-modes $x^i,p^i,p^+,x^-$ describing the center of mass and momentum of the string. \\
Finally, we promote these physical degrees of freedom to operators role, and we impose the following commutation rules:
\begin{equation}
\begin{split}
& \quad \left[ x^i , p^j \right] =i \delta^{ij} \quad \quad  \left[ x^- , p^+ \right] = -i \\
& \left[ \alpha^i_n , \alpha^j_m \right] = \left[ \bar{\alpha}^i_n , \bar{\alpha}^j_m \right] = n \delta^{ij} \delta_{n+m,0}.\\
\end{split}
\end{equation}
Now we have all the ingredients to construct the Hilbert space of the theory. Let $\vert 0, p \rangle$ be the vacuum states labelled by eigenvalues of momentum operator:
\begin{equation*}
\begin{split}
& P^{\mu} \vert 0, p \rangle = p^{\mu} \vert 0, p \rangle \\
& \alpha^i_n \vert 0, p \rangle = \tilde{\alpha}^i_n \vert 0, p \rangle =0 \quad \quad n >0 \\
\end{split}
\end{equation*}
and the other excited states are given by acting with the creation operators $\alpha^i_{-n}, \tilde{\alpha}^i_{-n}$ with $(n>0)$ on these vacua. The difference with the covariant quantization is that now we are acting only with transverse oscillators, for this reason the Hilbert space is, by construction, positive definite. \\
Defining the new number operators:
\begin{equation*}
N= \sum_{i=1}^{d-2} \sum_{n>0} \alpha_{-n}^i \alpha_n^i \quad \quad \quad \bar{N}= \sum_{i=1}^{d-2} \sum_{n>0} \bar{\alpha}^i_{-n} \bar{\alpha}^i_n
\end{equation*}
and introducing the normal ordered into the level condition we find:
\begin{equation*}
M^2 = \frac{4}{\alpha'} \left( N -a \right) =  \frac{4}{\alpha'} \left( \bar{N} -a \right) ,
\end{equation*}
where $a$ is a normal ordering constant. In order to fix the $a$ value we write the sum:
\begin{equation*}
\begin{split}
\frac{1}{2} \sum_{i=1}^{d-2} \sum_{n \neq 0} \alpha_{-n}^i \alpha_n^i & = \frac{1}{2} \sum_{i=1}^{d-2}  \left[ \sum_{n>0} \alpha_{-n}^i \alpha_n^i + \sum_{n<0} \alpha_{-n}^i \alpha_n^i \right] \\
& =\frac{1}{2} \sum_{i=1}^{d-2}  \left[ 2 \sum_{n>0} \alpha_{-n}^i \alpha_n^i + \sum_{n>0} \left[ \alpha_{-n}^i ;\alpha_n^i \right] \right] \\
& = \sum_{i=1}^{d-2}   \sum_{n>0} \alpha_{-n}^i \alpha_n^i + \frac{d-2}{2} \sum_{n>0} n \\
\end{split}
\end{equation*}
where we used the oscillators's commutation rules. We notice that $a$ correspond to  latter divergent  term of the last line. So, we consider  the Riemann zeta-function:
\begin{equation}
\zeta (s) = \sum_{n=1}^{+ \infty} n^{-s} \quad \quad \quad Re(s)>1
\end{equation}
that admits a unique analytic continuation to all $s$ values, except one pole in $s=1$. In particular we find $\zeta (-1)= -1/12$, therefore the quantum level matching condition becomes:
\begin{equation}
M^2 = \frac{4}{\alpha'} \left( \widehat{N} - \frac{d-2}{24} \right) =  \frac{4}{\alpha'} \left( \widehat{\bar{N}} - \frac{d-2}{24} \right).
\label{levelquantistica}
\end{equation}

\section{The string spectrum}\label{stringspectrum}
In this section we would like to make a brief analysis of the spectrum of a single free string. \\
Let us start with the \textit{ground state} $\vert 0, p \rangle$, that has not excited oscillators. The condition \eqref{levelquantistica} tell us that this state has  a negative mass-squared:
\begin{equation*}
M^2 = - \frac{1}{6 \alpha'} (d-2).
\end{equation*}
The corresponding particles in the space-time are called \textit{tachyons} $T(X^{\mu})$. This type of particles are a problem for the bosonic string because correspond to unstable vacuum states. However, there may be a stable vacuum which is not accessible by perturbative theory. \\
Before begging the analysis of excited states, we remember the Wigner representation of the Poincarè group on a $d$-dimensional Minkowski space-time $\mathbb{R}^{1,d-1}$:
\begin{itemize}
\item[] \textit{Massive particles} are classified by representations of $SO(d-1)$, little group in which transforms internal index of momentum $p^{\mu}$  with $p_{\mu}p^{\mu}=-M^2$.
\item[] \textit{Massless particles} are classified by irreducible representations of $SO(d-2)$, little group of momentum transformations with $p_{\mu}p^{\mu}=0.$
\end{itemize}
We now look at the first excited states $N = \bar{N}=1$. Since the level matching condition tell us that the number of states left and right are equal, we act on the vacuum state with both operators $\alpha^i_{-1}$ and $\bar{\alpha}^i_{-1}$:
\begin{equation}
\bar{\alpha}^i_{-1} \alpha^i_{-1} \vert 0; p \rangle.
\label{primostatoeccitato}
\end{equation} 
We obtain $(d-2)^2$ particle states with mass:
\begin{equation*}
M^2 = \frac{4}{\alpha'} \left( 1- \frac{d-2}{24} \right).
\end{equation*}
Since there is no way to package these $(d-2)^2$ states into a representation of $SO(d-1)$, i.e. the first excited string state cannot form a massive representation of Lorentz group. However these states can provide irreducible representations of $SO(d-2)$, thus in order to respect the Lorentz-invariance these states must be massless. And this is only the case if the dimension of space-time is:
\begin{equation}
M^2 = \frac{4}{\alpha'} \left( 1- \frac{d-2}{24} \right) = 0 \quad \quad \longmapsto \quad \quad d=26.
\end{equation}
The bosonic string theory on  $d=26$ dimensions contains massless particle on its spectrum: these particles are interesting because they give rise to long range interactions. The states \eqref{primostatoeccitato} transform on the representation $\mathbf{24} \otimes \mathbf{24}$ of $SO(24)$, that can be decomposed into three irreducible representations:
\begin{equation*}
\left( d-2 \right)^2 = \underbrace{\frac{(d-2)(d-3)}{2}}_{\text{Anti-symmetric} B_{ij}} \oplus \underbrace{\frac{(d-2)(d-1)}{2}}_{\text{Traceless symmetric} G_{ij}} \oplus \underbrace{1}_{\text{trace} \Phi}
\end{equation*}
To each of these representations we associate a massless field on space-time, respectively:
\begin{equation*}
B_{\mu \nu} (x) \quad \quad \quad G_{\mu \nu} (X) \quad \quad \quad \Phi(X).
\end{equation*}
$B_{\mu \nu}$ is the \textit{Kalb-Ramond} field, and it defines a two-form on the space-time. The scalar field $\Phi (X)$ is called \textit{dilaton}. The field $G_{\mu \nu} (X)$ represents a massless particle with spin 2 on the space-time thus, pushed by the idea of Feynman and Weinberg , that any theory of interacting massless spin two particles must be equivalent to general relativity, we would like to identify this field with the space-time metric.  \\
We  analyse higher excited states of the spectrum. The string at level $N = \bar{N}=2$ has two different states both in left-sector and in right-sector, thus the total set of states at level 2 is:
\begin{equation*}
\left( \alpha^i_{-1} \alpha^j_{-1} \oplus \alpha^k_{-2} \right) \otimes \left( \bar{\alpha}^i_{-1} \bar{\alpha}^j_{-1} \oplus \bar{\alpha}^k_{-2} \right) \vert 0, p \rangle.
\end{equation*}
This states have mass $M^2= 4/ \alpha'$. In each sector we have $\frac{1}{2} d (d-1) -1$ states,  that does fit nicely into a representation of $SO(d-1)$, in particular it is the symmetric traceless tensorial representation. \\
The only consistency requirement that we need for Lorentz invariance is to fix up the first excited states: the space-time dimension must be $d=26$. In $d=26$ all higher excited states will be massive, so they will be representations of $SO(25)$.

\section{Open string}

Let us consider an open string in $d$-dimensional space-time. We impose Neumann conditions for the first $p+1$ coordinates and Dirichlet conditions for the others:
\begin{equation}
\begin{split}
 \partial_{\sigma} X^a =0 \quad \quad \quad \quad & a=0,1,...,p \\
 X^I =c^I \quad \quad \quad \quad & I=p+1 , ..., d-1. \\
\end{split}
\end{equation}

\nn
Thus, the end-points of string can move freely on a $(p+1)$-dimensional hypersurface of space-time called \textit{Dp-brane}, where $p$ is its spatial dimension. The branes are dynamical object in string theory. \\
However this boundary condition breaks space-time Poincarè invariance, i.e. the Lorentz group is broken to:
\begin{equation}
SO(1,d-1) \quad \mapsto \quad SO(1,p) \times SO(d-p-1).
\end{equation}
\nn
Let us start from the solutions  of equation of motion written in terms of Fourier modes:

\begin{equation}
\begin{split}
& X^{\mu}_L (\sigma^+) = \frac{1}{2} x^{\mu} + \alpha' p^{\mu} \sigma^+ + i \sqrt{\frac{\alpha'}{2}} \sum_{n \neq 0} \frac{1}{n} \tilde{\alpha}^{\mu}_n e^{-i n \sigma^+} \\
& X^{\mu}_R (\sigma^-) = \frac{1}{2} x^{\mu} + \alpha' p^{\mu} \sigma^- + i \sqrt{\frac{\alpha'}{2}} \sum_{n \neq 0} \frac{1}{n} \alpha^{\mu}_n e^{-i n \sigma^-}, \\
\end{split}
\end{equation}
\nn
we impose on them the boundary conditions:
\begin{equation}
\begin{split}
\text{Neumann} \quad \quad \partial_{\sigma} X^a =0 \quad \quad & \alpha^a_n = \tilde{\alpha}^a_n \\
\text{Dirichlet} \quad \quad \quad X^I =c^I \quad \quad & x^I = c^I ,\quad p^I=0 , \quad  \alpha^a_n = - \tilde{\alpha}^a_n \\
\end{split}
\end{equation}
So for both boundary conditions, we only have one set of  independent oscillators,  the others are then determined by the boundary conditions. \\
Now we want to quantize the theory, thus we promote $x^a , p^a$ and $\alpha^{\mu}_n$ to operator, where $a=0,1,...p$ (Neumann directions). So we expect that the open string quantization gives rise to states which are restricted to lie on the brane. \\
Defining the light-cone coordinates lying on the brane:
\begin{equation}
X^{\pm} = \sqrt{\frac{1}{2}} \left( X^0 \pm X^p \right)
\end{equation} 
we can proceed to quantization in the same manner as for the closed string. The mass formula for the states is:
\begin{equation}
M^2 = \frac{1}{\alpha'} \left( \sum_{i=1}^{p-1} \sum_{n>0} \alpha_{-n}^i \alpha_n^i + \sum_{i=p+1}^{d-1} \sum_{n>0} \alpha^i_{-n} \alpha^i_n -a \right).
\end{equation}
As in the closed string case we have the normal-ordering constant $a$, and the require of reduced symmetry $SO(1,p) \times SO(d-p-1)$ leads to same constraints $d=26$ and $a=1$.  The differences with the closed case are the overall factor $(4)$ and the presence of only $\alpha$-modes. \

\section{Path integral Quantization}\label{PIQ}
Consider the Euclidian \footnote{It is convenient for convergency reason, as in QFT, to perform the Path integral in Euclidean space, that means we are replacing Minkoskian metric inside Polyakov action with the Euclidean one.} string partition function

\begin{equation}
Z = \int \frac{\mathcal{D} X \mathcal{D}h}{V_{g}} e^{-S_p \left[h,X,  \right]  },
\label{pathintegral1}
\end{equation}

\nn
obtained integrating over all possible worldsheet metrics and over all coordinates fields. The factor $V_{g}$ is the volume of gauge transformations, given by diffeomorphisms and Weyl's transformations, needed to renormalized the integral that, otherwise, would be infinite due to the redundancy in integrating over equivalent configurations.  
As we saw in section \ref{Polyakovactionsection} gauge invariance allows one to fix the metric inside Polyakov action to same fiducial metric $\hat{h}$, the conformal one for instance. Then one can a priori guess the path integral in \eqref{amplitude1} actually runs over the coordinate fields $X$ only. In order to determine the correct measure of integration and to isolate the divergent part, a standard procedure that can be performed is 
 \textit{Faddev-Popov} method. Let $\xi:h(\sigma)\rightarrow h^\xi(\sigma')$ be a combined coordinate and Weyl transformation given by

\begin{equation}
h^\xi(\sigma')_{ab}=e^{2\omega(\sigma)}\frac{\de \sigma^c}{\de\sigma^{'a}}\frac{\de \sigma^d}{\de\sigma^{'b}}h(\sigma')_{cd},
\end{equation}

\nn
let then formally introduce the Faddev-Popov determinant by 

\begin{equation}
1= \Delta_{FP}(h)\int \D\xi\delta(h-\hat{h}^\xi),
\end{equation}

\nn
with $\hat{h}$ same fiducial metric, and inserting it on \eqref{pathintegral1}, one finds

\begin{equation}
Z[\hat{h}]=\int \frac{\mathcal{D} X \mathcal{D}h\D\xi}{V_{g}}\Delta_{FP}(h)\delta(h-\hat{h}^\xi) e^{-S_p \left[\hat{h},X  \right]  }.
\label{pathintegral2}
\end{equation}

\nn
Performing the integral over the metric $h$ and renaming the dummy variable $X\rightarrow X^\xi$, the previous expression becomes 

\begin{equation}
Z[\hat{h}]=\int \frac{\mathcal{D} X \D\xi}{V_{g}}\Delta_{FP}(\hat{h}^\xi) e^{-S_p \left[\hat{h}^\xi,X^\xi \right]  }.
\label{pathintegral3}
\end{equation}

\nn
Thanks to Feddev-Popov determinant, Polyakov action and integration measure gauge invariance, the integrand does not really depend on $\xi$, then the integration over $\xi$ just produces the volume of the gauge group and cancels the denominator:

\begin{equation}
Z[\hat{h}]=\int \D X\Delta_{FP}(\hat{h}) e^{-S_p \left[\hat{h},X\right]  }.
\label{pathintegral4}
\end{equation}

\nn
As usual, after some algebra we are going to skip here, the Feddev-Popov determinant can be expressed in term of a path integral over two Grassmanian fields $b_{ab}$ and $c_a$, ghosts, as:

\begin{equation}
\Delta_{FP}(\hat{h})=\int \D b\D c e^{\frac{1}{2\pi}\int d^2\sigma\sqrt{\hat{h}}b_{ab} \nabla ^a c^b},
\end{equation}

\nn
where $b_{ab}$ is traceless and $\nabla$ is the covariant derivative. The exponent in the integrand above is called ghost action 

\begin{equation}
S_g\equiv\frac{1}{2\pi}\int d^2\sigma\sqrt{\hat{h}}b_{ab} \nabla ^a c^b.
\label{ghostaction}
\end{equation}

\nn
In order to further simplify the ghost action it is convenient to fix conformal gauge, choosing $\hat{h}_{ab}=e^{2\omega}\delta_{ab}$; indeed, one can easily proof that in this case covariant derivatives are just ordinary ones (Christoffel symbols turn to be null). Choosing then complex coordinate on the wordlsheet $z$ and $\bar{z}$, one finally finds

\begin{equation}
S_g=\frac{1}{2\pi}\int d^2z \left ( b_{zz} \de _{\bar{z}} c^z+ b_{\bar{z}\bar{z}} \de _z c^{\bar{z}}\right ).
\label{ghostaction2}
\end{equation}

\nn
Putting all together the Polyakov path integral becomes:

\begin{equation}
Z[\hat{h}]=\int \D X \D b \D c  e^{-S_p \left[\hat{h},X\right] -S_g \left[\hat{h},b, c\right] }.
\label{pathintegral5}
\end{equation}

\nn
Classical symmetries of Polyakov action are preserved in \eqref{pathintegral5} by BRST invariance. We are not going to inter in this aspect in detail, but it is interested to say that computing the BRST charge one finds that its nihipotency is granted only for a 26 dimensional spacetime. This means that the path integral quantization of String Theory is consistent only for $D=26$; this fact, known as Weyl anomaly because it was originally considered a pathology of the path integral approach, is the vary real reason why String Theory requires a 26 dimensional spacetime; the same result we obtain heuristically in section \ref{stringspectrum}.

 \chapter{String amplitudes}\label{four}

String amplitude differs from ordinary quantum field theory amplitude by two important features. In a quantum field theory one can define and compute amplitudes off-shell, and subsequently impose on them on-shell conditions; in string theory, conversely, off-shell amplitudes are not well defined, because on-shellness is required in order for conformal invariance to be preserved. Moreover, in QFT interaction coupling constants appear as external parameters, the arbitrary coefficients of the vertex terms of the action, and they can eventually be set to zero obtaining the corresponding free theory; on the contrary in string theory, the interaction coupling constant is an internal variable arising from dynamic, precisely the vacuum expectation value of a particular string state, called dilaton; as a consequence string theory is intrinsically interacting and free string theory is meaningless.\\

\section{Bosonic strings amplitudes}
We can formally express the amplitude for $n$ asymptotic states as

\begin{equation}
\A_n= \int DX Dh e^{-S_p[X,h]} \prod_{i=1}^n V_i,
\label{amplitude1}
\end{equation}

\nn
where we have omitted, for the moment, the normalization gauge volume denominator for notation clarity, and where we have introduced some operators $V_i$, called Vertex operators, specifying all informations about in and out asymptotic states.\\
Remind the Polyakov action $S_p$ \eqref{Polyakov} is invariant under worldsheet diffeomorphism and Weyl rescalings, than it has to be equivalent to the most generic action invariant under such symmetries; it turns then to be that one can rewrite $S_p$ as:

\begin{equation}
S_p[S,h,\lambda] = \frac{1}{4\pi \alpha'} \int_{\Sigma_g} d^2 \sigma\sqrt{h} h^{ab}\de_a X^\m\de_b X^\n G_{\m\n}(X)+\frac{\lambda}{4\pi}\left [ \int_{\Sigma_g} d^2 \sigma\sqrt{g} R +2\int_{\de \Sigma_g} k ds\right ] ,
\end{equation} 

\nn
where $\lambda$ is a constant, $R$ and $k$ the worldsheet Ricci scalar and the worldsheet boundary geodesic curvature respectively. The two terms we add are a pure topological quantity related to the Euler characteristic $\chi$ of the worldsheet by Gauss-Bonnet Theorem \eqref{GBtheorem}.
Moreover the invariance properties of $S_p$ imply that the integral over worldsheet metric appearing in the path integral above does not depend on the particular surface shape but only on its topology, i.e. its genus $g$. After this consideration \eqref{amplitude1} takes the form

\begin{equation}
\A_n= \sum_g \int  DX Dh e^{-S_p[X,h,\lambda]} \prod_{i=1}^n V_i =\sum_g e^{-\lambda \chi(g)}\int  DX Dh e^{-S_p[X,h]} \prod_{i=1}^n V_i.
\label{amplitude2}
\end{equation}

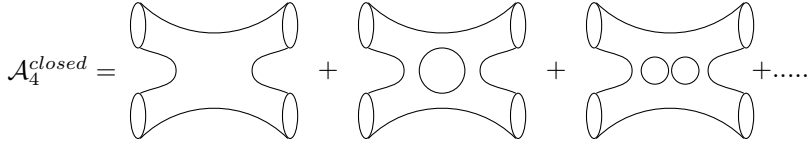
\begin{figure}[h!]
\centering
\begin{tikzpicture}
\amplitudeone{(0,0)}
\end{tikzpicture}
\caption{Diagrammatic expansion of four closed string scattering.}
\label{amplitudepicture1}
 \end{figure}

\nn
We have obtain an expansion for the S matrix in the genus $g$ of worldsheet; this expression is formally quite similar to the series expansion in QFT in terms of increasing loops contributions, indeed we can interpret it exactly in the same way: higher genera worldsheets provide higher order in the expansion. As in QFT we can interpret the Dyson series in terms of Feynman diagram, also in string theory we can provide a pictorial representation of this expansion, for instance a 4 points closed string scattering can be diagrammatically depicted as shown in picture \ref{amplitudepicture1}.
 
\nn
Where dots stand for diagrams with increasing number of holes.
An important difference with loop expansion in QFT, is that in that case to higher loops contributions correspond increasing number of diagram, in this case instead one always has just one term per each genus, no matter how high.
The factor $g_s\equiv e^{\lambda}$ appearing in the expansion \eqref{amplitude2} plays the role of a coupling constant. 
In order to explicit show the contribution of vertex operator to Euler characteristic notice that adding a closed string to the process means inserting an hole to the worldsheet, while adding an open string implies the insertion of a boundary; reminding that for a genus $g$ oriented surface with $b$ boundaries the Euler characteristic is given by

\begin{equation}
\chi=2-2g-b,
\end{equation} 

\nn
one has that for $n_c$ closed string vertex operators $V_i^c$ and $n_o$ open string vertex operator $V_i^o$, the amplitude \eqref{amplitude2} can be rewritten as

\begin{equation}
\A_n= \sum_g g_s^{-2+2g+n_c+\frac{1}{2}n_o}\int  DX Dh e^{-S_p[X,h]} \prod_{i=1}^{n_c} V_i ^{c} \prod_{i=1}^{n_o} V_i^o.
\label{amplitude3}
\end{equation}

\nn
Looking at the previous expression, it seems quite natural to absorb prefectors depending on vertex operators directly in their own definitions, as 

\begin{equation}
V^{o'} = \sqrt{g_s} V^{o} \quad\quad \mbox{and} \quad\quad V^{c'} = g_s V^{c}.
\end{equation}

\nn
Omitting the prime symbol for notation simplicity one can finally writes:

\begin{equation}
\A_n= \sum_g g_s^{2g-2}\int  DX Dh e^{-S_p[X,h]} \prod_{i=1}^{n_c} V_i ^{c} \prod_{i=1}^{n_o} V_i^o.
\label{amplitude4}
\end{equation}

\section{Vertex operator}\label{VOS}

As we saw in chapter \eqref{Chapter3} the lowest closed string energy states are tachyons, with negative mass, and graviton, dilaton and B field with no mass. 
One can guess the form of the vertex operator for each of these states just starting from taking into account Lorentz degree of freedom. The tachyon is a scalar field, then the easiest form one could guess for its vertex operator is the identity. 
In the same way one can guess for the graviton the simplest symmetric two tensor and for the B field the simplest antisymmetric one. \\
Finally, for the dilaton, that is a scalar just like the tachyon, a cleaver choice, that will be clear in a moment, is to take its vertex operator to be proportional to the Ricci scalar $R$ of the worldsheet.
These guess are summarized in the following table (\ref{VO}).

\begin{table}[h!]
\begin{center}
\begin{tabular}{|l | l|}
\hline
\textbf{State} & $\mathcal{V}(\sigma)$\\ \hline 
Tachyon & $e^{ikX}$ \\
Graviton & $g_c s_{\m\n}h^{ab}\de_a X^\m \de_b X^\n e^{ikX}$ \\
B field & $g_c a_{\m\n}\varepsilon^{ab}\de_a X^\m \de_b X^\n e^{ikX}$ \\
Dilaton & $\f \alpha' R e^{ikX}$ \\
\hline
\end{tabular} 
\caption{\label{VO}Vertex operator for lowest string states.}\end{center}
\end{table}

\nn
Where $s_{\m\n}, a_{\m\n}$ and $ \f$ are constant respectively symmetric tensor, antisymmetric tensor and real number. 
The exponential factor here, where $kX\equiv k_\m X_\m$, is added to describe the correct behaviour under spacetime translations.
In order for vertex operator to be invariant under worldsheet diffeomorphisms, it should take the form:

\begin{equation}
V=\int d^2\sigma \sqrt{h}\mathcal{V}(\sigma).
\end{equation}

\nn
One can immediately see the choice we made for the dilaton vertex operator is such that the coupling constant is nothing else that dilaton vacuum expectation value.\\
On the other and, in order to convince one the choice we made for the graviton operator vertex is the correct one, consider the path integral of the Polyakov action in a generic space time with metric $g_{\m\n}$:

\begin{equation}
Z \sim \int \mathcal{D} X \mathcal{D}h e^{-\int d^2\sigma \sqrt{h}h^{ab}\de_a X^\m \de_b X^\n g_{\m\n}},
\label{pathintegralgraviton1}
\end{equation}

\nn
suppose $g_{\m\n}=\eta_{\m\n} + \xi_{\m\n}$ to be closed to flat metric, one has

\begin{equation}
Z \sim \int \mathcal{D} X \mathcal{D}h e^{-\frac{1}{4\pi\alpha'}\int d^2\sigma \sqrt{h}h^{ab}\de_a X^\m \de_b X^\m } e^{-\frac{1}{4\pi\alpha'}\int d^2\sigma \sqrt{h}h^{ab}\de_a X^\m \de_b X^\n \xi_{\m\n}};
\label{pathintegralgraviton2}
\end{equation}

\nn
setting 

\begin{equation}
\xi_{\m\n}=-4\pi\alpha' g_c e^{ikX} s_{\m\n},
\end{equation}

\nn
the \eqref{pathintegralgraviton2} becomes

\begin{equation}
\begin{split}Z & \sim \int \mathcal{D} X \mathcal{D}h e^{-\frac{1}{4\pi\alpha'}\int d^2\sigma \sqrt{h}h^{ab}\de_a X^\m \de_b X^\m } e^{g_c\int d^2\sigma \sqrt{h}h^{ab}\de_a X^\m \de_b X^\n s_{\m\n}e^{ikX}}=\\&=\int \mathcal{D} X \mathcal{D}h e^{-S_p[h,X]} e^{\int d^2\sigma\sqrt{h} \V_g(\sigma)},\end{split}
\label{pathintegralgraviton3}
\end{equation}

\nn
where $\V_g(\sigma)$ is the graviton vertex operator defined above. Then one can see that the path integral of the Polyakov action in a curved spacetime is equivalent to the scattering amplitude of a coherent state of gravitons in flat spacetime. \footnote{A coherent state is described exponentiating the operator of the one-particle state.} 
For open strings the lowest state is a tachyon again, and its vertex operator, up to coupling constant, is the same as the closed tachyon. The first excited state is a gauge boson.

\section{Gauge fixing of Polyakov path integral}
So far we have completed ignored convergency problem of the amplitude \eqref{amplitude1}, that indeed turns to be infinite for the vary same reason we discussed in section \ref{PIQ}, the redundancy due to gauge invariance. In the next section we will gauge fix the amplitude path integral by means of Feddev-Popov procedure, that represents the most elegant and correct way to approach the problem. However in order to introduce some of concepts and quantities we will use, we try first to gauge fix it "by hand".
Consider a worldsheet with genus $g$, we saw in the introduction of this chapter the path integral of the Polyakov action consists in the integration over all possible metrics $h \in \Ha$, embedding fields $X \in \mathcal{X}$ and vertex operator positions $\{\sigma_i\}\in (\Sigma_g)^n$, with $n$ the number of inserted vertices. The full region of integration is than the space 

\begin{equation}
\mathcal{I}\equiv \Ha \times\mathcal{X}\times  (\Sigma_g)^n.
\end{equation}

\nn
As we said this space contains a lot of equivalent configurations due to gauge invariance under diffeomorphisms and Weyl transformations, in order to avoid this redundancy we considered the space 

\begin{equation}
\mathcal{I}' = \frac{\mathcal{I}}{\G_g},
\end{equation} 

\nn
with $G_g= diff \times Weyl$ is the gauge transformations space. Focus on the space 

\begin{equation}
\M_g = \frac{\Ha }{\G_g};
\end{equation}

\nn
quotienting the space of all metrics by the gauge volume one is left with only inequivalent metrics, i.e. metric uncorrelatable by conformal transformations. The space $\M_g$ is called Moduli space and its elements are said moduli. \\
Notice that among all gauge transformations there are also many that do not change the metric at all, that is $\G_g$ contains a subgroup whose  elements leave the metric invariant: the group of automorphism $Aut(\Sigma_g)$ of the worldsheet. We would like to isolate it, because as we will see the path integral gauge fixing does not eliminate this symmetry group and we will need to identify it in order to fix its redundancy "by hand". \\
Before treating the subject in a more formal way we can introduce it algebraically, an approach that could turn to be useful from a computational point a view. Under a combined infinitesimal and Weyl transformation, one can easily check the metric changes as 

\begin{equation}
\delta h_{ab}=-2(P\delta\sigma)_{ab}+ (2\delta\omega-\nabla_c\delta\sigma^c)h_{ab},
\end{equation}

\nn
where $P$ is a traceless symmetric tensor defined by 

\begin{equation}
(P\delta\sigma)_{ab} = \frac{1}{2}\left ( \nabla_a \delta\sigma_b+\nabla_b\delta\sigma_a -h_{ab}\delta_c\delta\sigma^c \right ).
\end{equation}

\nn
Consider a generic infinitesimal variation of the metric $\Delta h$, it turns that $\Delta h$ is not induced by a conformal transformation if it is orthogonal to any conformal variation $\delta h$ respect to the scalar product

\begin{equation}
\langle \Delta h | \delta h \rangle = \int d^2\sigma \sqrt{h} \Delta h_{ab} \delta h^{ab};
\end{equation}

\nn
then moduli corresponds to variations $\Delta h$ such that 

\begin{equation}
\langle \Delta h | \delta h \rangle = \int d^2\sigma \sqrt{h} \Delta h_{ab} \left[ -2(P\delta\sigma)_{ab}+ (2\delta\omega-\nabla_c\delta\sigma^c)h_{ab} \right ].
\label{moduli1}
\end{equation}

\nn
In order for \eqref{moduli1} to vanish for any $\delta\omega$ and $\delta\sigma$ one need:

\begin{equation}
h^{ab}\Delta h_{ab}=0 \quad\quad \mbox{and} \quad\quad (P^T\Delta h)_a=0;
\label{moduli2}
\end{equation}

\nn
where $(P^Th)_a \equiv -\nabla^b h_{ab}$. Notice the first equation just requires $\Delta h_{ab}$ to be traceless; then the number of moduli are the number of traceless solution of the second equation.\\ 
On the other hand transformations leaving the metric unchanged are the solutions of the equation:

\begin{equation}
\delta h_{ab}=- 2(P\delta\sigma)_{ab}+(2\delta\omega-\nabla_c\delta\sigma^c)h_{ab}=0,
\end{equation}

\nn
taking the trace of this equation one fixes the last two terms to zero, then it simplifies to 

\begin{equation}
(P\delta\sigma)_{ab}=0.
\label{CKV1}
\end{equation}

\nn
This equation is called conformal Killing equation and its solutions conformal Killing vectors (CKVs).
CKVs are by construction the automorphism of the worldsheet connected to the identity, and they span a subgroup of $Aut(\sigma_g)\subset \G_g$ known as conformal Killing group (CKG). Notice that the number of moduli corresponds to the dimension $\m$ of the kernel of $P^T$ while the number of CKVs to the dimension $k$ of the kernel of $P$; then one can use Riemann-Roch theorem to relate them to Euler characteristic, one has

\begin{equation}
\mu -k = -3\chi.
\end{equation} 

\nn
Since $\chi$ is known for closed oriented surfaces, the previous relation allows one to easily obtain the number of $\m$ or $k$ once one knows the other.
Before proceeding, because we will use it in the next, it is useful to write here that for variations around the conformal gauge and in complex coordinates, one can show that equations \eqref{moduli2} and \eqref{CKV1} further simplify into

\begin{equation}
\begin{split}
&\de_z \Delta h_{\z\z} = \de_{\z}\Delta h_{zz}=0 \\& \de_z\delta \z =\de_{\z}\delta z=0,\end{split}
\label{modulickv}
\end{equation}

\nn
so variations of the moduli correspond to holomorphic quadratic differentials and CKVs to holomorphic vector fields.

\nn
Consider the space $diff(\Sigma)$ of diffeomorphisms on the wordlsheet\footnote{Actually we should consider just the space of the ones preserving the orientation but we are going to skip this subtlety here to keep it easily readable.} and define its subgroup containing those homotopic to the identity:

\begin{equation}
diff_0 (\Sigma) \equiv \{f \in diff(\Sigma) | f \sim Id \}.
\end{equation}

\nn
Because $diff(\Sigma)$ is not in general connected, it makes sense to consider the quotient group

\begin{equation}
\Gamma_g \equiv diff(\Sigma)\slash {diff}_0(\Sigma),
\end{equation}

\nn
whose elements are right the diffeomorphisms not continuously connected to the identity. The group $\Gamma_g$ is know as \textit{Modular group} of genus $g$. Consider now the space of classes of conformally inequivalent metrics

\begin{equation}
\T_g= \frac{\Ha}{diff_0(\Sigma) \times Weyl(\Sigma)};
\end{equation}

\nn
the space $\T_g$ turns to be a finite dimensional simply connected manifold having the same dimension of the Moduli space, and it is called Teichm\"{u}ller space and its elements Teichm\"{u}ller parameters. Notice that Teichm\"{u}ller space differs from Moduli space just right by conformal Killing vectors, in other wards moduli are those Teichm\"{u}ller parameters that cannot be identified under the action of $\Gamma_g$, one therefore has

\begin{equation}
\M_g = \T_g \slash \Gamma_g.
\end{equation}

\nn
The modular group is a discrete group, then we can read the previous expression saying that Teichm\"{u}ller space is the covering group of Moduli space.\\
We have therefore see that the integration over metrics in the path integral corresponds to an integration on the moduli space and an integration over 
gauge group, up to conformal killing symmetry; this means the integration still contains $k$ unphysical degrees of freedom. Once we now CKG we could eliminate this redundancy fixing by hand the position of $k$ vertex operator, provided we have inserted enough, i.e :

\begin{equation}
\Ha\times (\Sigma_g)^n \sim  \M_g \times \G_g\times(\Sigma_g)^{n-k}
\end{equation}

\nn
Assuming functional behave just like ordinary differentials one can naively write the path integral measure in \eqref{amplitude1} as 

\begin{equation}
\D h \prod_{i=1}^n d^2\sigma_i = Jd^\m m \D \xi \prod_{i=1}^{n-k} d^2\sigma_i,
\label{cambiocoordinate1}
\end{equation} 

\nn
where $J$ is the Jacobian of the coordinate change. We are not going to perform the full $J$ computation here, because we will treat the problem in detail in the next section. Denote with $\delta a^r$ with $r={1,..2k}$ a real vector representation of infinitesimal elements of CKG and denote their action on vertex operator position by

\begin{equation}
\delta \hat{\sigma}_i = \delta a^r C^a_r (\hat{\sigma}_j),
\end{equation}

\nn
where $\hat{\sigma_j}$ are the fixed vertex operator positions.

\begin{equation}\begin{split}
\A_{j_1..j_n}(k_1..k_n) = \sum_g g_s^{2g-2}& \int d^\m m |P^\dagger P| \sum_\alpha \frac{(\psi_\alpha, \de_i \hat{h}_{ab})}{\psi_\alpha^2} |C^a_s(\hat{\sigma}_j)|\times \\&\times \int \D X e^{-S_p} \prod_{i=1}^{n-k}\int d^2 \sigma_i \sqrt{h(\sigma_i)}\prod_i \V_{j_i}(k_i,\sigma_i).\end{split}
\end{equation}

\section{Faddeev-Popov procedure}
The gauge fixing we performed in the previous section has several problems. First the measures involved in the path integral are not clearly defined, to be honest they are not defined at all. Moreover one can fix CVG redundancy only if enough vertex operators are inserted, assuming CVG is known. In fact the group of automorphism of the wordlsheet and the CVG are not in general easy to be found. In this section we are going to remedy to these inconveniences, gauge fixing the amplitude path integral proceeding by means of Faddeev-Popov procedure, analogously to section \ref{PIQ}.
Our starting point is the amplitude \eqref{amplitude1}

\begin{equation}
\A_n= \int \frac{1}{V_g}DX Dh e^{-S_p[X,h]} \prod_{i=1}^n \int d^2\sigma_i \sqrt{h}\V_i(\sigma_i),
\label{amplitude5}
\end{equation}

\nn
and the coordinate change \eqref{cambiocoordinate1} 

\begin{equation}
\D h \prod_{i=1}^n d^2\sigma_i \rightarrow d^\m m \D \xi \prod_{i=1}^{n-k} d^2\sigma_i.
\end{equation} 

\nn
We saw the gauge choice fixes $k$ of the vertex operator coordinates, denote by $\hat{\sigma}_i^a$ the $2k$ fixed real coordinates and by $F$ the set of (a,i) indices for which $\sigma_i^a=\hat{\sigma}_i^a$. In the following we will often omit the $a$ index for notation clarity, as well as other indices and summations or products extrema, restoring them when indispensable.\\
Define Faddeev-Popov determinant on the moduli space as 

\begin{equation}
1=\Delta_{FP}(h,\sigma)\int_{\M}d^m m \int_{\G_g}\D \xi \delta (h-\hat{h}(m)^\xi) \prod_{(a,i)\in F} \delta (\sigma_i^a-\hat{\sigma}_i^{a\xi}),
\label{FP2}
\end{equation}

\nn
inserting this expression in \eqref{amplitude5} and proceeding as in section \ref{PIQ} one obtain

\begin{equation}
\A_{j_1...j_n}(k_1...k_n) = \sum_g e^{\lambda \chi}\int d^\m m \Delta_{FP}(\hat{h},\hat{\sigma}) \int \D X \prod_{(a,i)\notin F} d^2\sigma_i e^{-S_p}\prod_i \sqrt{\hat{h}}\V_{j_i}(k_i,\sigma_i)
\label{amplitude6}
\end{equation}

\nn
The Faddeev-Popov determinant is non-zero only in finite number $n_R$ of points, because those points are related by symmetry and FP determinant is invariant, we can just pick one of them and multiplied by $n_R.$
Expanding the definition \eqref{FP2} around such a point one can read the inverse determinant to be:

\begin{equation}
\Delta_{FP}(\hat{h},\hat{\sigma})^{-1}=n_R \int d^\m \delta m \D \delta \xi \delta(\delta h)\prod_i \delta(\delta\sigma(\hat{\sigma}_i))
\label{FP3}
\end{equation}

\nn
The general variation of the metric is given by usual variation under conformal transformation plus the contribution due to moduli space transformations:

\begin{equation}
\delta h_{ab} = \delta m^\alpha \de_{m_\alpha} \hat{h}_{ab}- 2(P\delta\sigma)_{ab}+(2\delta\omega-\Delta_c \delta\sigma^c)\hat{h}_{ab};
\end{equation}

\nn
decomposing then the gauge measure in terms of diffeomorphism and Weyl variations $\D\delta\xi = \D \delta \omega\D\delta\sigma$ , equation \eqref{FP3} becomes

\begin{equation}
\begin{split}
\Delta_{FP}(\hat{h},\hat{\sigma})^{-1} &= n_R\int d^\m \delta m  \D \delta \omega\D\delta\sigma \delta \left[\delta m^\alpha de_{m_\alpha} \hat{h}_{ab}- 2(P\delta\sigma)_{ab}+(2\delta\omega-\Delta_c \delta\sigma^c)\hat{h}_{ab} \right ]\prod_i\delta(\delta\sigma(\hat{\sigma_i}))=\\&= n_R\int d^\m \delta m  \D \delta \omega\D\delta\sigma \\&\left [\int \D\beta d^kx e^{2\pi i \int d^2\sigma \sqrt{\hat{h}}\beta^{ab}(\delta m^\alpha  \de_{m_\alpha} \hat{h}_{ab}- 2(P\delta\sigma)_{ab}+(2\delta\omega-\Delta_c \delta\sigma^c)\hat{h}_{ab})+2\pi i\sum x_i \delta\sigma(\hat{\sigma_i}) } \right ],
\end{split}
\end{equation}

\nn
where we right delta functional in integral representation introducing a symmetric tensor field $\beta_{ab}$ and $k$ variables $x$.
Integrating now over $\delta\omega$, the delta functional forces $\beta$ to be traceless, denoting then by $\beta'$ a traceless symmetric tensor field, one finally has 

\begin{equation}
\Delta_{FP}(\hat{h},\hat{\sigma})^{-1}= n_R\int d^\m \delta m d^kx \D\delta\sigma\D\beta' e^{2\pi i \int d^2\sigma \beta'_{ab}(2(P\de \sigma)_{ab}-\delta m_{\alpha} \de_{m_\alpha}\hat{h}_{ab)}} e^{2\pi i \sum x_i \delta(\hat{\sigma_i})}.
\end{equation}

\nn
In order to invert the previous expression for obtaining Faddeev-Popov determinant one replaces all bosonic variables with Grassmanian variables:

\begin{equation}
\begin{split}
&\delta\sigma^a \rightarrow c^a\\&\beta'_{ab} \rightarrow b_{ab}\\& x_{ai} \rightarrow \eta_{ai} \\ &\delta m^\alpha \rightarrow \Xi^\alpha,
\end{split}
\end{equation}
\nn
one therefore has:

\begin{equation}
\begin{split}\Delta_{FP}(\hat{h},\hat{\sigma})&=\frac{1}{n_R}\int \D b\D c d^\m\Xi d^k\eta e^{-\frac{1}{4\pi}\int d^2\sigma b_{ab}[(2Pc)_{ab}-\Xi^\alpha\de_\alpha \hat{h}_{ab}]+\sum \eta_{ai}c^a(\hat{\sigma}_i)}\\&=\frac{1}{n_R}\int \D b\D ce^{-S_g} \prod_{\alpha=1}^\m \frac{1}{4\pi}\int d^2\sigma \sqrt{\hat{h}}b_{ab}\de_\alpha\hat{h}_{ab} \prod c^a(\hat{\sigma}_i).\end{split}
\label{FP4}
\end{equation}

\nn
where in last equality, the integration over the Grassmanian variables $\Xi$ and $\eta$ has been performed and $S_g$ is the ghost action defined in section \ref{PIQ}.
We are ready to write down the correct gauge fixed amplitude; putting \eqref{FP4} in \eqref{amplitude6}, one finally obtain:

\begin{equation}\begin{split}
\A_n =\sum_g e^{-\lambda\chi}&\int _{\M}\frac{d^\m m}{n_R} \D X\D b \D c e^{(-S_p-S_g)} \prod_{(a,i)\notin F}\int d^2\sigma_{i} \prod_{\alpha=1}^\m\frac{1}{4\pi}\int d^2 \sigma b_{ab}\de_\alpha\hat{h}^{ab}\times\\&\times\prod_{(a,i)\in F} c^a(\hat{\sigma}_i)\prod_{i=1}^n \sqrt{\hat{h}(\sigma_i)}\V_{j_i}(\sigma_i).\end{split}
\label{finalamplitude1}
\end{equation}

\nn
The previous expression can be further simplify performing the integral over ghosts. Expand ghosts field in a suitable complete basis 

\begin{equation}
c^a(\sigma)=c_jC^a_j(\sigma) \quad\quad \mbox{and} \quad\quad b_{ab}(\sigma)=b_{k} B^k_{ab} (\sigma).
\end{equation}

\nn
In order to perform the path integral we would like to turn it into gaussian diagonalizing the action, unfortunately the $P$ operator appearing in ghost action cannot diffeomorphic invariantly diagonalized; however the combinations $P^TP$ and $PP^T$ could be. Let then choose the basis such that

\begin{equation}
P^T P C^a_j = \lambda_j C^a_j \quad\quad \mbox{and} \quad\quad PP^T B^k_{ab}=\Lambda_k B^k_{ab},
\end{equation}

\nn
Notice that $PC_j$ is an eigenfunction of $PP^T$ as well as $P^TB^k$ is for $P^TP$; this means that there exists a one-to-one correspondence between these eigenfunction except when $PC_j=0$ and $P^TB^k=0$, that is for zero eigenvalue modes. Comparing this two last equations with \eqref{moduli2} and \eqref{CKV1}, one see that $c$ field zero modes, denoted by $C_{0j}$, are just CKVs, and $b$ field zero modes $B^k_0$ are holomorphic quadratic differential; their number is then $k$ and $\m$ respectively. For non zero modes, due to the one-to-one correspondence between eigenfunction, the integral becomes Gaussian and it just provides the determinant of the $PP^T$ operator, for zero modes on the other way the action does not contribute and one just has to integrate the unexponentiated terms; after all integrations one finds:

\begin{equation}
\begin{split}
\int \D b \D c e^{-S_g}\prod_{(a,i)\notin F}\int d^2\sigma_{i} &\prod_{\alpha=1}^\m\frac{1}{4\pi}\int d^2 \sigma b_{ab}\de_\alpha\hat{h}^{ab}\prod_{(a,i)\in F} c^a(\hat{\sigma}_i)=\\&= det \left ( \frac{\int d^2\sigma B_{0k}\de_{k'}\hat{h}}{4\pi}\right )det C_{0j}^a (\hat{\sigma}_i)det' \left ( \frac{PP^T}{2\pi}\right )^{1/2}.\end{split}
\end{equation}

\nn
where the prime in the last determinant denote the omission of the zero eigenvalues.
Inserting this result in \eqref{finalamplitude1}, one finally has:

\begin{equation}\begin{split}
\A_n=&\sum_g e^{-\lambda\chi}  det \left ( \frac{\int d^2\sigma B_{0k}\de_{k'}\hat{h}}{4\pi}\right )det C_{0j}^a (\hat{\sigma}_i)det' \left ( \frac{PP^T}{2\pi}\right )^{1/2} \times \\&\times \int _{\M}\frac{d^\m m}{n_R} \D X  \prod_{(a,i)\notin F}\int d^2\sigma_{i} e^{-S_p} \prod_{i=1}^n \sqrt{\hat{h}(\sigma_i)}\V_{j_i}(\sigma_i).\end{split}
\label{finalamplitude2}
\end{equation}

\nn
The computation of the $PP^T$ operator determinant is in general a vary complicated task, see for instance \cite{Bolte:1988ge}. For this reason we will not focus on its computation and we will absorb its value in an overall constant.\\

\section{Tree level amplitude}
As we saw earlier the tree level amplitude in string scattering corresponds to the integration over genus zero worlsheet. Consider the worldsheet of an $n$ closed string scattering, it is conformally equivalent to a sphere with $n$ points removed, called punctures. Each puncture corresponds to the insertion of a vertex operator. On the other hand, for open string scattering, the zero genus wordlsheet is conformally equivalent to a disk, with $n$ punctures along the boundary.

\begin{equation}
\Sigma_n^{0\,closed} \simeq S_2\backslash \{p_1..p_n\}  \simeq \hat{\C} \backslash \{p_1..p_n\} \quad\quad \Sigma_0^{open} \simeq D_2 \backslash \{p_1..p_n\} \simeq  \hat{\Ha} \backslash \{p_1..p_n\}
\end{equation}

 \begin{figure}[h!]
\centering
\begin{tikzpicture}
\amplitudetwo{(0,0)}
\end{tikzpicture}
\caption{Conformal map for four strings scattering.}
\label{amplitudepicture2}
 \end{figure}
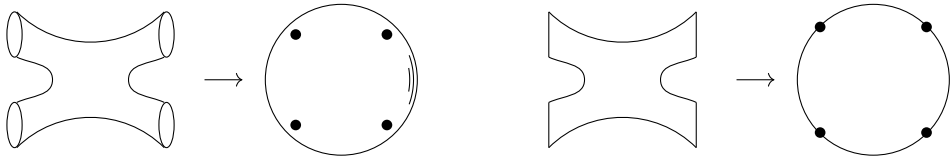

\nn
Consider the sphere $S_2$, equipped by the complex coordinate metric 

\begin{equation}
ds^2 = \frac{4dzd\z}{(1-z\z)^2}.
\end{equation}

\nn
According to our previous discussion, we now that, thanks to Riemann-Roch theorem, the sphere as no moduli and six real CKVs. That is, all spheres are conformally equivalent to $S_2$, and there exist six independent transformations leaving $S_2$ unchanged, for instance three of them are obviously rotations. In order to see this explicitly, considerer the differential equations \eqref{modulickv}:

\begin{equation}
\de_z \Delta h_{\z\z}= \de_{\z} \Delta h_{zz} =0\quad\quad \mbox{and} \quad\quad \de_z \delta \z=\de_{\z} \delta z=0,
\end{equation}

\nn
whose solutions, as we saw, provide the number of moduli and conformal killing vectors respectively. Such a solutions have to be defined on the whole sphere; in order to study the behaviour at infinity, consider the coordinate change $z\rightarrow u=1/z$, mapping the origin to infinity. Under such a coordinate transformation one has 
\begin{equation}
\Delta h_{zz}\rightarrow \Delta h_{uu}=z^4 h_{zz} \quad\quad \mbox{and} \quad\quad \delta z \rightarrow \delta u =-\frac{1}{z^2}\delta z.
\end{equation}

\nn
Focus on the first equation. In order for $\Delta_zz$ to be holomorphic at infinity ($u=0$) it should decrease at least as $z^{-4}$, but if it was it would not be holomorphic in $z=0$; we can therefore conclude that exists no solution of the first equation on the sphere: no moduli. As far as regard the second equation, on the other hand, one sees that $\delta z$ is holomorphic in $u=0$ provided it grows at most as $z^2$. The general CKV then is

\begin{equation}
\begin{split}
&\delta z = a_0 +a_1 z + a_2 z^2 \\& \delta \z =  a_0* + a_1*\z + a_2*\z^2.
\end{split}
\end{equation}

\nn
Therefore there are six real parameters or, if one prefers, three complex ones. Exponentiating these infinitesimal transformations one obtain the global transformation 

\begin{equation}
z'= \frac{az +b}{cz+d},
\end{equation}

\nn
with $a,b,c,d \in \C$. This transformations define the $SL(2,\C)$ group, also known as M\"{o}bius group. Notice that rescaling all parameters by the same factor the transformation remains unchanged, this allows one to impose the determinant of the transformation to be one: we have finally found 

\begin{equation}
CKG_{S_2}=\frac{SL(2,\C)}{Z}\equiv PSL(2,\C).
\end{equation}

\nn
Proceeding analogously for the disk $D_2$, or simply deducing it by topological arguments (the disk is just half a sphere), one finds that its moduli space is also trivial, then both sphere and disk have no moduli, and this implies that for both closed and open strings scattering amplitude \eqref{finalamplitude1}the integration over moduli space disappear, as well as $b$ ghost contribution. The only ghost contribution comes from $c$ ghost and it depends on the dimension $k$ of CKG. The general tree level amplitude for $n$ strings scattering turns then to be:

\begin{equation}
\A^{tree}_{j_1...j_n}(k_1...k_n)=g_c^{n_c+2n_o} \int \D X \D c e^{(-S_p-S_g)} \prod_{i=1}^k c(\hat{\sigma}_i) \prod_{(a,i)\notin F}\int d^2\sigma_{i}\prod_{i=1}^n \sqrt{\hat{h}(\sigma_i)}\V_{j_i}(\sigma_i).
\label{amplitude7}
\end{equation} 

\nn
As we saw in the previous section the integration over ghosts can be written in terms of determinants; equation \eqref{amplitude7} can be then finally expressed as 

\begin{equation}
\A^{tree}_{j_1...j_n}(k_1...k_n)=g_c^{n_c+2n_o} det C_{0j}^a(\hat{\sigma_i})det'\left (\frac{P^TP}{2\pi} \right )^{1/2}\int \D X \prod_{(a,i)\notin F}\int d^2\sigma_{i} e^{-S_p}\prod_{i=1}^n \sqrt{\hat{h}(\sigma_i)}\V_{j_i}(\sigma_i).
\label{amplitude8}
\end{equation}

\subsection{Closed Tachyon amplitude}
In this section we are going to specialize the result obtained above in the case of $n$ closed string scattering and we will compute explicitly the simplest case of $n$ closed tachyons, emphasising the four point case, known as Virasoro-Shapiro amplitude. 
Our starting point is the expression \eqref{amplitude8}, and we want to compute it for a spherical worldsheet.  Lets start computing the determinant of the $c$ ghost zero modes. As we said they correspond to CKVs and we computed them in previous section in the case of the sphere, thus we have

\begin{equation}
det C_{0j}^a= \begin{pmatrix}1&z_1&z_1^2&0&0&0\\1&z_2&z_2^2&0&0&0\\1&z_3&z_3^2&0&0&0\\0&0&0&1&\z_1&\z_1^2\\0&0&0&1&\z_2&\z_2^2\\ 0&0&0&1&\z_3&\z_3^2\end{pmatrix}= |z_1-z_2|^2|z_1-z_3|^2|z_2-z_3|^2.
\label{ghostdeterminant}
\end{equation}

\nn
Consider now the scattering of $n$ (closed) tachyons; that represents the most easier case due to the simple form of tachyon vertex operator.
We have 

\begin{equation}
\begin{split}
\A' &\equiv \int \D X e^{-S_p} \prod_{i=1}^n \V_{j_i}(k_i, \sigma_i)= \int \D X e^{-\frac{1}{4\pi\alpha'}\int d^2\sigma X\nabla^2 X+\sum_i ik_i X(\sigma_i)}= \\&=\int \D X e^{-\frac{1}{4\pi\alpha'}\int d^2\sigma \left [ X\nabla^2 X+i\sum_i \delta(\sigma-\sigma_i) k_i X(\sigma_i)\right ]}.\end{split}
\end{equation}

\nn
In order to perform these integrals one proceeds as usual: recognizing that they all are Gaussian, except for the zero mode one $X_0$, for which a delta function is given (momentum conservation), one can complete the square and integrate, the final result, in critical dimension, is 

\begin{equation}
\A' =  i (X_0)^{-26} det'\left ( \frac{-\nabla^2}{4\pi^2\alpha'}\right )^{-12}(2\pi)^{26} \delta^{26}(\sum_i k_i)e^{-\sum_{i<j=1}^n k_ik_j G(\sigma_i,\sigma_j)},
\label{Amplitudetreeclosed1}
\end{equation}

\nn
where $G(\sigma_i,\sigma_j)=(\nabla^2)^{-1}$ is the solution of the differential equation \footnote{Actually if one computes it carefully, the correct Green's function to be considered should be the one deprived by its zero mode, however terms due to it will drop out in the final answer, so we are going to completely ignore them.}

\begin{equation}
-\frac{1}{2\pi\alpha'}\nabla^2 G(\sigma_i,\sigma_j)=\frac{1}{\sqrt{h}}\delta^2(\sigma_i-\sigma_j).
\label{green1}
\end{equation}

\nn
In complex coordinate on the sphere, using the relation 

\begin{equation}
\de\bar{\de}ln|z|=2\pi \delta(z),
\end{equation}

\nn
equation \eqref{green1} has solution\footnote{Actually the full solution contains extra terms, coming from the zero mode, but they will drop out in the path integral and the turn to be irrelevant.}

\begin{equation}
G(z_i,\z_i,z_j,\z_j)= -\frac{\alpha'}{2} ln |z-z'|.
\label{Green2}
\end{equation}

\nn
Inserting \eqref{Green2} into \eqref{Amplitudetreeclosed1} one has 

\begin{equation}
\A'= i (X_0)^{-26} det'\left ( \frac{-\nabla^2}{4\pi^2\alpha'}\right )^{-12}(2\pi)^{26} \delta^{26}(\sum_i k_i) \prod_{i<j=1}^n |z_{ij}|^{\alpha' k_ik_j},
\label{Amplitudetreeclosed2}
\end{equation}

\nn
where we have defined $z_{ij}=(z_i-z_j)$. Finally, putting together \eqref{Amplitudetreeclosed2} and \eqref{ghostdeterminant}, after computing the correct normalization factors, one obtains for the tree closed tachyon amplitude (tct)

\begin{equation}
\A^{tct}(k_1..k_n) = \frac{8\pi i}{\alpha'}g_s^{n-2} (2\pi)^{26}\delta^{26}(k_i)|z_{12}|^2 |z_{13}|^2z_{23}|^2 \int \prod_{i=4}^n d^2z_i \prod_{i<j=1}^n |z_{ij}|^{\alpha'k_ik_j} 
\label{Amplitudetreeclosed3}
\end{equation}

\nn
For $n=1$ and $n=2$ one immediately see that \eqref{Amplitudetreeclosed3} is zero, because the last product is. This is due to the fact that no enough vertex operators are inserted in order to compensate gauge fixing, on the other hand, one can see that the first non zero ghost correlator on the sphere is the three point function.
For $n=3$ all vertex operator positions are fixed by $PSL(2,\C)$ and no integration variables is left, thus one has

\begin{equation}
\A^{tct}(k_1,k_2,k_3)=\frac{8\pi i}{\alpha'}g_s (2\pi)^{26}\delta^{26}(k_i)|z_{12}|^{2+\alpha'k_1k_2} |z_{13}|^{2+\alpha'k_1k_3}z_{23}|^{2+\alpha'k_2k_3}, 
\label{threepointtachyon}
\end{equation}

\nn
Using momentum conservation and reminding the tachyon mass is $m^2= -4/\alpha'$, one can write

\begin{equation}
2k_1k_2 = (k_1+k_1)^2-k_1^2-k_2^2 = k_3^2-k_1^2-k_2^2 = - \frac{4}{\alpha'},
\end{equation}

\nn
as well for other momentum scalar products. replace this result in \eqref{threepointtachyon} all exponents become zero, and then 

\begin{equation}
\A^{tct}(k_1,k_2,k_3)=\frac{8\pi i}{\alpha'}g_s (2\pi)^{26}\delta^{26}(k_i).
\end{equation}

\nn
Notice that this amplitude is well defined thanks to mass shell condition imposed on tachyons; if we had tried to compute it off-shell, the $z_{ij}$ terms would have not disappeared and the result would have depended on the choice of the vertex operator position, i.e. it would have been gauge dependent and thus inconsistent.
Consider now the four closed tachyon scattering, for $n=4$ expression \eqref{Amplitudetreeclosed3} becomes

\begin{equation}
\begin{split}
\A^{tct}(k_2,k_2,k_3,k_4)&=\frac{8\pi i}{\alpha'}g_s^{2} (2\pi)^{26}\delta^{26}(k_i)|z_{12}|^2 |z_{13}|^2|z_{23}|^2 \int d^2z_4 \prod_{i<j=1}^4 |z_{ij}|^{\alpha'k_ik_j}=\\&=\frac{8\pi i}{\alpha'}g_s^{2} (2\pi)^{26}\delta^{26}(k_i)|z_{12}|^{2+\alpha'k_1k_2} |z_{13}|^{2+\alpha'k_1k_3}z_{23}|^{2+\alpha'k_2k_3} \times \\&\quad \quad\quad\quad\times \int d^2z_4 |z_{14}|^{\alpha'k_1k_4}|z_{24}|^{\alpha'k_2k_4}|z_{34}|^{\alpha'k_3k_4} \end{split} 
\label{fourclosedtachyon}
\end{equation}

\nn
We have now to choose some specific values for fixed point, sure that the $z_i$-dependent factors grant the invariance of the result. A common choice is 

\begin{equation}
z_1 = 0 \quad\quad z_2=1\quad\quad z_3 = {\infty},
\end{equation}

\nn
and one has

\begin{equation}
\frac{8\pi i}{\alpha'}g_2^{2} (2\pi)^{26}\delta^{26}(k_i)\int d^2z |z|^{\alpha'k_1k_4}|1-z|^{\alpha'k_2k_4};
\end{equation}

\nn
where we rename $z_4\equiv z$.\\
As usual for four point scattering, it often convenient to express the amplitude in terms of Mandelstam variables

\begin{equation}
s=-(k_1+k_2)^2 \quad\quad t=-(k_2+k_3)^2 \quad\quad u=-(k_1+k_3)^2, 
\end{equation}

\nn
with

\begin{equation}
s+t+u= \sum m_i^2 = -\frac{16}{\alpha'}, 
\end{equation}

\nn
in these variables the amplitude becomes:

\begin{equation}
\A^{tct}(s,t,u)=\frac{8\pi i}{\alpha'}g_s^{2} (2\pi)^{26}\delta^{26}(k_i)\int d^2z |z|^{-\frac{1}{2}\alpha' t-4}|1-z|^{-\frac{1}{2}\alpha' u-4}.
\end{equation}

\nn

\subsection{Open tachyon amplitude}\label{opensection}
We said open string scattering worldsheet is conformally equivalent to the upper half complex plane $\hat{\Ha}_n$ with $n$ punctures along the real axis. Actually this is not strictly correct due the following subtlety. When one deals with close strings, in mapping wordlsheet into sphere the original reciprocal vertex operators positions turn to be completely irrelevant, because ordering of points has no sense in a sphere, in other words all worldsheets corresponding to different permutations of "legs" are conformally equivalent to each other and to the sphere with appropriate punctures. However in the open string case this is no longer true: punctures along the boundary of the disk have an ordering, and two different orderings may not be equivalent; actually for $n$ punctures and $k$ CKVs, the inequivalent configurations are the cycles of the symmetric group $S_k$. Therefore the amplitude we wrote for open string actually represents only the partial amplitude corresponding to a specific cyclical order of strings, the full amplitude is a sum over inequivalent permutations. With this in mind, for the sake of notation clarity, we will proceed for now without repeating "partial" every time nor explicitly writing the sum over configurations, we will restore it when needed.\\
In order to compute the ghost correlator we could proceed just like in the previous section, solving conformal killing equation on the disk, however we can do it in a cleverer way noticing that upper-half plane can be obtained from the whole complex plane by $z\sim \bar{z}$ identification. This means that we can take quantities computed on the sphere, and obtain the corresponding one on the disk by simply replacing $z$ by $\z$. 
From \eqref{ghostdeterminant}, after imposing vertex operator positions to be real, it immediately follows:
 
 \begin{equation}
 det C_{0j}^a (\sigma_i)|_{D}= |x_{12}x_{13}x_{23}|.
 \end{equation}

\nn
where we have renamed $z_i\equiv x_i$ to underline they belong to the real axis.\\
We can use the same trick for the tachyon path integral, but a little more care: when we computed \eqref{Green2} we omitted some term, that would have been irrelevant in that case. The solution of the differential equation \eqref{green1} on the sphere and with Neumann boundary conditions is

\begin{equation}
G(z_i,\z_i,z_j,\z_j)_{S_2}= -\frac{\alpha'}{2}ln |z_i-z_j|^2- -\frac{\alpha'}{2}ln |z_i-\z_j|^2,
\end{equation}

\nn
now we can identify $z\sim \z$ and we find the correct propagator on the disk:

\begin{equation}
G(z_i,z_j)_{D_2}= -\alpha' ln |z_i-z_j|^2= -\alpha' ln |x_{ij}|^2,
\end{equation}

\nn
Inserting these results into expression \eqref{finalamplitude2} and computing the constant factors, the tree open tachyon amplitude turns to be:

\begin{equation}
\A^{tot}(k_1..k_n)= \frac{2i}{\alpha'}g_s^{n-2}\delta^{26}(k_i) |x_{12}||x_{13}||x_{23}| \prod_{i=4}^n \int d x_i\prod_{i<j=1}^n |x_{ij}|^{2\alpha'k_ik_j}
\label{amplitudeopenfinal}
\end{equation}

\nn
Comparing \eqref{amplitudeopenfinal} and \eqref{Amplitudetreeclosed3} one sees the main differences between open and closed tachyon string amplitude, a part from normalization and renormalization factors,  are the range of integration, that in the first case is performed over the real axis while in second case on the complex plane, and the exponent of the integrand, that for open strings is twice the one for closed strings.\\
For $n=1$ and $n=2$ the amplitude \eqref{amplitudeopenfinal} is evidently zero, for the vary same reason concerning the closed case. For $n=3$, again no integration is left and momentum conservation grants the invariance from $x_{ij}$, one trivially obtains

\begin{equation}
\A^{tot}(k_1,k_2,k_3)=\frac{2 i g_s}{\alpha'}(2\pi)^{26}\delta^{26}(k_i).
\end{equation}

\nn
Consider now the case $n=4$, one has:

\begin{equation}
\A^{tot}(k_1,k_2,k_3,k_4)= \frac{2i}{\alpha'}g_s^{2}(2\pi)^{26}\delta^{26}(k_i) |x_{12}||x_{13}||x_{23}|\int d x \prod_{i<j=1}^4 |x_{ij}|^{2\alpha'k_ik_j},
\label{Veneziano1}
\end{equation}

\nn
where we renamed $x_4\equiv x$. 
We have  now to choose specific values for $x_1$,$x_2$ and $x_3$, in order to obtain the full amplitude we have to take into consideration all possible inequivalent cycles of $S_3$, that is $(123)$ and $(132)$, corresponding to the two possible choice of orientation of the boundary. Fixing
both  $(x_1,x_2,x_3 )$ and $(x_1,x_3,x_2)$ to $(0,1,{\infty})$ and summing the two partial amplitudes, after imposing momentum conservation and  rewriting it in terms of Mandelstam variables, one finally finds:

\begin{equation}
\A^{tot}(k_1,k_2,k_3,k_4)= \frac{2i}{\alpha'}g_s^{2}(2\pi)^{26}\delta^{26}(k_i) \int d x |x|^{-\alpha'u-2}|1-z|^{2\alpha'k_2k_4}+(u\leftrightarrow t).
\label{Veneziano2}
\end{equation}

\nn
Each integral along the real axis splits into three parts; as shown in figure \ref{orderpicture2} different ranges of integration correspond to different permutations of vertex operator positions. 

 \begin{figure}[h!]
\centering
\begin{tikzpicture}
\integralordering{(0,0)}
\end{tikzpicture}
\caption{Correspondence among integral range and vertex operator positions ordering}
\label{orderpicture2}
 \end{figure}

\nn
Thanks to $PLS(2,\R)$ invariance, this contributions can be mapped one into another, so they are equal and the final result can be written as

\begin{equation}
\A^{tot}(k_1,k_2,k_3,k_4)= \frac{4i}{\alpha'}g_s^{2}(2\pi)^{26}\delta^{26}(k_i) [A(s,t)+A(s,u)+A(t,u)],
\label{Veneziano3}
\end{equation}

\nn
where
\begin{equation}
A(s,t)=\int_0^1 dx x^{-\alpha's-2}(1-x)^{-\alpha't-2}.
\end{equation}

\nn
Defining $\alpha(y)=\alpha' y+1$, the previous integral in nothing but a beta function:

\begin{equation}
A(s,t)= \beta (-\alpha(s),-\alpha(t))=\frac{\Gamma(-\alpha(s))\Gamma(-\alpha(t))}{\Gamma(-\alpha(s)-\alpha(t))}.
\end{equation} 

\nn
This is the famous \textit{Veneziano Amplitude}. It represents the first calculation in string theory, although it was performed vary before ST was developed, introduced in \cite{Veneziano:1968yb} by Veneziano to describe hadrons. The more beautiful and manifest property of \eqref{Veneziano3} is the symmetry under Mandelstam invariants interchange.\\
In conclusion of this section, although we really do not need them for our discussion, it is dutiful to briefly mentioning Chan-Paton indices. Open strings come with two free extrema (Neumann boundary conditions), i.e. with two more degrees of freedom; then one is allowed and even prodded to attribute them quantum numbers. Suppose each endpoint may have $N$ different states, then open vertex operators carry two more indices running from $1$ to $N$, let's encode them in a factor $\lambda^\alpha_{ij}$, for the $\alpha-$-th vertex operator. Although these new indices have no role in worldsheet dynamic, because energy momentum tensor has no dependence on them, in computing string amplitudes they provided an extra normalization factor: writing them in exponential form as usual and performing the Gaussian path integral, the normalization determinant obtained can be written in terms of a trace \footnote{For a generic matrix $A$ the relation $|exp\{A\}|=exp\{trA\}$ holds.}, and one is finally left with a factor 

\begin{equation}
tr (\lambda^{\alpha_1}\lambda^{\alpha_2} ..\lambda^{\alpha_n}),
\end{equation}

\nn
for any partial amplitude with vertex operator permutation $(\alpha_1,\alpha_2,...\alpha_n)$. On one hand, thanks to trace cyclic property, permutations belonging to the same cycle come with the same factor, coherently to what we said a the beginning of this section; however, on the other hand, different partial amplitudes come with a different weight: Chan-Paton factors brake channel symmetry.
The matrices $\lambda^\alpha_{ij}$ turn to be representations of $U(N)$, and this is the vary reason why they were introduced, in order to equip the theory with colour quantum number; that is why sometimes one refer to partial amplitudes as color-ordered amplitude.  

\section{Excited states amplitude}
In this section we briefly discuss the tree $n$ amplitude for both closed and open excited string states. The only difference respect to tachyon amplitude comes from vertex operator. Consider, for instance, the open vector vertex operator

\begin{equation}
\V_p(\e, k, X )= i \e_\m \de X^{\m} e^{ikX},
\end{equation}

\nn
where the subscript stands for photon; the key idea to write down the $n$ vector amplitude is to absorb kinematic factor in the exponent by writing

\begin{equation}
\V(\e, k, X)= e^{ikX(x_i)+\e_{\m} \de X^{\m}}\Big |_{linear},
\end{equation}

\nn
when we mean the expansion must be truncated at terms linear in polarizations. Now the vertex operator is written as an exponential, and the product of $n$ operator also is, then one can compute the path integral in the same fashion as usual, cutting out nonmultilinear terms at the and of the calculation; then the amplitude of $n$ photons, a part from normalization factors, is:

\begin{equation}
\begin{split}
\A_{\m_1..\m_n}\e_1^{\m_1}..\e_n^{\m_n} &\sim  \int \D X \prod_{i=4}^{n}d x_i e^{-S_p}e^{i\sum_i (\e_i^\m \de X_{\m_i}(x_i)+k_i X(x_i)) }\Big |_{linear}=\\=&  \int\prod_{i=4}^{n}d x_i \prod_{i<j=1}^n |x_{ij}|^{2\alpha' k_ik_j}e^{\sum_{i<j} \frac{\e_i\e_j}{x_{ij}^2}-\sum_{i \neq j} \frac{\e_i k_j}{x_{ij}}}\Big |_{linear}=\\&=\int\prod_{i=4}^{n}d x_i \prod_{i<j=1}^n |x_{ij}|^{2\alpha'k_ik_j}\left [\sum_{i<j} \frac{\e_i\e_j}{x_{ij}^2}-\sum_{i \neq j} \frac{\e_i k_j}{x_{ij}}\right ]=\\&\equiv \int\prod_{i=4}^{n}d x_i \prod_{i<j=1}^n |x_{ij}|^{2\alpha'k_ik_j} F_p(x_{ij},\e, k),
\label{vectoramplitude}
\end{split}
\end{equation}

\nn
where in the last line we have define the function $F_p$, depending on polarizations, momenta and vertex position distances. Our final result shows an important propriety: the tree $n$ vector amplitude integrand may be written as the $n$ tachyon one times a rational function containing all informations about polarizations, moreover this means that the integrands of \eqref{vectoramplitude} differ from the tachyon case just for integer powers of $x_{ij}$, i.e. four string amplitudes can be written in terms of gamma functions, no matter what their states are!\\
Applying the same procedure to higher excited states one finds that this structure is common to all tree open amplitudes:

\begin{equation}
\begin{split}
\A^{open} (k_1..k_n, \e_1...\e_n) &\sim \int\prod_{i=4}^{n}d x_i \prod_{i<j=1}^n |x_{ij}|^{2\alpha'k_ik_j} F(x_{ij},\e, k),
\label{generalopenamplitude}
\end{split}
\end{equation}

\nn
where all the specific information about the states are encoded in the function $F$.\\
Let now proceed analogously for the closed string case; consider the tensor vertex operator 

\begin{equation}
\V_t(\e,k) = i \e_{\m\n} \de X^\m \bar{\de} X^\n e^{ikX},
\end{equation}

\nn
where $\e_{\m\n}$ is, as we saw, symmetric for gravitons and antisymmetric for $B$ field. Again we can write it as an exponential using the same trick plus the introduction of some vectors $\e_\m$ with the prescription that 

\begin{equation}
\e_\m \e_\n \rightarrow \e_{\m\n}
\end{equation}

\nn
at the and of the computation, then one has:

\nn
\begin{equation}
\V_t (\e,k)= e^{ikX+i\e_\m\de X^\m+i\bar{\e}_\n \bar{\de}X^\n}\Big |_{\substack{linear\\ \e_\m \e_\n \rightarrow \e_{\m\n}}}.
\end{equation}

\nn
Inserting the previous expression in the closed tree amplitude path integral and proceeding in the vary same way of the open case, of finds that the $n$ tree amplitude for tensor closed states, up to normalization, is: 

\begin{equation}
\begin{split}
\A (k_1..k_n)_{\m_1\n_1...\m_n\n_n} \e^{\m_1\n_1}..\e^{\m_n\n_n}\sim& \int \prod_{i=4}^n d z_i \prod_{i<j=1}^n (z_{ij})^{\frac{1}{2}\alpha'k_ik_j}\left [\sum_{i<j} \frac{\e_i\e_j}{z_{ij}^2}-\sum_{i \neq j} \frac{\e_i k_j}{z_{ij}}\right ]\times\\&\times \prod_{i=4}^n d \z_i \prod_{i<j=1}^n (\z_{ij})^{\frac{1}{2}\alpha'k_ik_j}\left [\sum_{i<j} \frac{\bar{\e}_i\bar{\e}_j}{\z_{ij}^2}-\sum_{i \neq j} \frac{\bar{\e}_i k_j}{\z_{ij}}\right ]=\\&=\int \prod_{i=4}^n d^2 z_i \prod_{i<j=1}^n |z_{ij}|^{\alpha'k_ik_j} F_t(z_{ij},\e,k)F_t(\z_{ij},\bar{\e},k),
\label{generalclosedamplitude}
\end{split}
\end{equation}

\nn
where in the last line we have defined the function $F_t$. Generalizing this expression for higher closed excited states one finds, just as in the open case, that the amplitude has the same structure for all states and all informations about them appear only trough the function $F$:

\begin{equation}
\begin{split}
\A^{closed} (k_1..k_n)\sim\int \prod_{i=4}^n d^2 z_i \prod_{i<j=1}^n |z_{ij}|^{\alpha'k_ik_j} F(z_{ij},\e,k)F(\z_{ij},\bar{\e},k).
\label{generalclosedamplitude2}
\end{split}
\end{equation}

\nn
Again, the function $F$ is rational in $z_{ij}$ and the integrands differs from that of $n$ tachyons just by integer powers of $z_{ij}$.\\
The factor 

\begin{equation}
\prod_{i<j=1}^n |z_{ij}|^{k_ik_j},
\label{KBfactor}
\end{equation}

\nn
appears both in open and string tree amplitude between any state, then is enough important to be named: it is called \textit{Koba-Nielsen factor}.
Comparing \eqref{generalclosedamplitude2}and \eqref{generalopenamplitude} we see closed and open string tree amplitude have a vary similar structure, in order to emphasize the analogy and write them in even close form, useful for the next section, it's convenient to manipulate them a little further. Let start by noticing that, although we have not underlying it before, the slopes $\alpha'$ appearing in closed and string amplitudes have not in principal to be equal, actually we can choose them to be:

\begin{equation}
\alpha'_{closed}=4\alpha'_{open}\equiv 2 \alpha.
\end{equation}

\nn
Moreover, in order to take account explicitly of ghost normalization factors we can write, for open strings

\begin{equation}
\begin{split}
|x_{12}||x_{13}||x_{23}|\int \prod_{i=4}^n dx_i \prod_{i<j=1}^n |x_{ij}|^{k_ik_j}&=\int \prod_{i=1}^n dx_i \left ( \frac{|x_{12}||x_{13}||x_{23}|}{dx_1dx_2dx_3} \right ) \prod_{i<j=1}^n |x_{ij}|^{k_ik_j}=\\&\equiv \int \frac{1}{SL(2,\R)}\prod_{i=1}^n dx_i \prod_{i<j=1}^n |x_{ij}|^{k_ik_j}.
\label{kbfactor1}
\end{split}
\end{equation}

\nn
and analogously for closed amplitudes

\begin{equation}
\begin{split}
|z_{12}|^2|z_{13}|^2|z_{23}|^2\int \prod_{i=4}^n d^2z_i \prod_{i<j=1}^n |z_{ij}|^{k_ik_j}&=\int \prod_{i=1}^n d^2z_i \left ( \frac{|z_{12}|^2|z_{13}|^2|z_{23}|^2}{d^2z_1d^2z_2d^2z_3} \right ) \prod_{i<j=1}^n |z_{ij}|^{k_ik_j}=\\&\equiv \int \frac{1}{SL(2,\C)}\prod_{i=1}^n d^2z_i \prod_{i<j=1}^n |z_{ij}|^{k_ik_j},
\label{kbfactor2}
\end{split}
\end{equation}

\nn
where in the last line of the two formulae we have formally defined the quotient respect to the appropriate $CKG$ as the quantities in round brackets.\footnote{The factors in round brackets in \eqref{kbfactor1} and \eqref{kbfactor2}  were originally introduced by Koba and Nielsen, and called Koba-Nielson factors; however nowadays this denomination is used to indicate \eqref{KBfactor}, and this is the use we will do.} 
Putting all together, with momentum conservation and $F$ dependence understood and up to constants, one can finally write:

\begin{equation}
\begin{split}
&\A^{open}(k_1..k_n)= \int \frac{1}{SL(2,\R)}\prod_{i=1}^n dx_i \prod_{i<j=1}^n |x_{ij}|^{\alpha k_ik_j} F(x),\\
&\A^{closed}(k_1..k_n)= \int \frac{1}{SL(2,\C)}\prod_{i=1}^n d^2 z_i \prod_{i<j=1}^n |z_{ij}|^{2\alpha k_ik_j} F(z)F(\bar{z}).
\end{split}
\label{closedopenfinalamplitude}
\end{equation}

\nn
It is important to underline here that the structure of tree amplitudes for open and closed bosonic strings, obtained in this chapter, is the same one obtains considering supersymmetric string theory; this fact is not true for generic genus, but at tree level the contribution due to fermionic variables factorize out and it just contribute to an overall normalization.

\chapter{String Amplitudes and Intersection Theory}\label{five}
This last chapter is dedicated to the interpretation of string amplitudes in the language of twisted De Rham theory. We will discuss KLT relations and we will see how a suitable choice of the dual co-homology space allows one to interpret closed string amplitudes as twisted cocycle-cocycle intersection and open string amplitudes as twisted cycle-cocycle intersection, showing how KLT relations naturally emerge as a twisted Riemann period relation. We will then discuss a diagrammatic method to compute multidimensional intersection number between twisted cycles in the moduli space of the punctured Riemann sphere, explicitly using it to KLT decompose a five closed string amplitudes into partial open string amplitudes. We implement a Mathematica script to compute intersection number with Mizera formula and we will apply it for computing intersection number between two-variables Parke-Taylor forms, analytically verifying the agreement with the result obtained by the recursive algorithm. Using then the latter, we will compute the projections onto the Perke-Taylor basis, obtaining the explicit Master integral decomposition of a five partial open tachyon amplitudes.

\section{Kawai-Lewellen-Tye relations}
In \cite{KAWAI19861} Kawai, Lewellen and Tye proved that any closed string tree amplitude can be expressed in terms of a sum of the products of appropriate open string amplitudes. These relations, known as KLT relations, were originally found by means of holomorphic properties, in this section we will prove they can be recovered using twisted de Rham Theory, and we will show they are nothing but twisted Riemann period relations.\\
In order to introduce such relations and to understand their structure, let first consider the simplest case, i.e. for $n=3$. In this case all vertices positions are fixed and no integration variable is left, then closed tree amplitude \eqref{closedopenfinalamplitude} trivially factorizes into

\begin{equation}
\begin{split}
\A^{closed}(k_1,k_2,k_3)&\sim  |z_{12}|^2 |z_{13}|^2 |z_{23}|^2\prod_{i<j=1}^3 |z_{ij}|^{2k_ik_j} F(z_{ij},\e,k)F(\z_{ij},\bar{\e},k)=\\&= \left (|z_{12}| |z_{13}| |z_{23}|\prod_{i<j=1}^3 (z_{ij})^{k_ik_j} F(z_{ij},\e,k)\right )\left (|z_{12}| |z_{13}| |z_{23}|\prod_{i<j=1}^3 (\z_{ij})^{k_ik_j} F(\z_{ij},\bar{\e},k) \right ) =\\&\sim \A^{open} (k_1,k_2,k_3)\bar{\A}^{open} (k_1,k_2,k_3).
\end{split}
\label{KLTn3}
\end{equation}

\nn
We therefore see that tree amplitude of three closed strings can be decomposed into two tree amplitudes of open strings.
 
In order to see the connection between KLT and twisted Riemann period relations, a suitable choice for the dual system is required: in section \ref{constrlocsys} we chose the dual local system to be the one associated to $u^{-1}$ $\Lod=\mathcal{L}_{-\omega}$, and that is actually the most convenient choice in the majority of cases, however this is not the case; for our purpose here it is appropriate to choose the local system related to the complex conjugate $\bar{u}$: $\Lod = \mathcal{L}_{\bar{\omega}}$. Consider the bilinear non degenerate pairing 

\begin{equation}
 H^n (\M, \Lo) \times H_c^n (\M,\mathcal{L}_{\bar{\omega}}) \rightarrow \C,
\end{equation}

\nn
for $\langle \f | \in H^n (\M, \Lo)$ and $|\bar{\f} \rangle \in H^n (\M,\mathcal{L}_{\bar{\omega}})$ their intersection number (see \cite{Mimachi}) is given by:

\begin{equation}
    \langle \f |\bar{\f} \rangle = \int |u|^2 reg_c(\f)\wedge \bar{\f},
\end{equation}

\nn
where $reg_c$ is the regularization map introduced in \eqref{regularization_map}. Although we are not going to use it here, its useful to point out that, with this choice of the dual local system, also the intersection number between cycles carries a $|u|^2$ extra factor, coming from monodromy factors.

Let $\f(z)= F(z)\prod dz_i$ and $\overline{\f(\z)}= F(\z) \prod d\z_i$ and select the modulus of Koba Nielsen factor as\footnote{The phase factor appearing in \cite{KAWAI19861} is hidden in this definition, we have choosen it in such a way that vertex operator positions lie on the real section of the sphere. } twist, then the closed string amplitude may be interpreted as the intersection number: 

\begin{equation}
    \A_n^{closed} = \int |u(z)|^2 \f(z)\overline{\f(\z)}= \langle \f(z)| \overline{\f(\z)}\rangle
\end{equation}

\nn
now, let $\{\Delta_\alpha\}=\{\Delta(\alpha)\otimes u\}$ be a basis for $H_n(\M_{0,n}, \Lo)$ and $\{\overline{\Delta_\alpha}\}=\{\overline{\Delta(\alpha)}\otimes \bar{u}\}$ a basis for $H_n(\M_{0,n},\mathcal{L}_{\bar{\omega}})$, then applying twisted Riemann relations \eqref{TwistedRiemannperiod}, one has:

\begin{equation}
    \langle \f(z)| \overline{\f(\z)}\rangle = \sum_{\alpha\beta} \langle \f(z)|\Delta_\alpha] (H^{-1})_{\alpha\beta} [\overline{\Delta_\beta} |\overline{\f(\z)}\rangle = \sum_{\alpha\beta} \left (\int_{\Delta(\alpha)} u(z) \f(z) \right )  (H^{-1})_{\alpha\beta} \left (\int_{\overline\Delta(\beta)} \bar{u}(z) \overline{\f(\z)} \right ).
    \label{KLTtwisted}
\end{equation}

\begin{figure}[h!]
  \centering
\begin{tikzpicture}
\sfera{(0,0)}
\end{tikzpicture}
\caption{Twisted cycle on $\M_{0,4}\simeq S^2_3$.}
\label{sferapunturata}
 \end{figure}
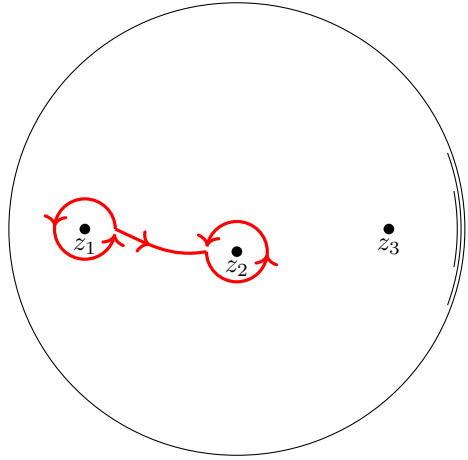

\nn
Now, in order to recognize partial open string amplitudes in the lhs one has to suitably choose $\Delta(\alpha)$ and $\overline{\Delta(\alpha)}$.
Consider the simplest case of $\M_{0,4}$, shown in figure \ref{sferapunturata}. As we saw the Moduli space of the sphere with four punctures is a sphere with three punctures $\M_{0,4}\simeq S^2_{3}$; fixing punctures on the real section of $S^2_3$ (the equator), cycles surrounding two punctures are equivalent to regularized paths between punctures on the boundary on a disk; i.e. precisely the integration contour of open string amplitudes.

\nn
Different twisted cycles, as discussed in \ref{opensection}, coincide to different partial amplitudes; moreover notice that in choosing real positions for punctures we can pick any possible order, being points on the sphere unordered, thus all partial amplitudes are recovered. In order to emphasize this fact we can label the index $\alpha$ of twisted cycles as permutation of the $n$ punctures.
For $n\geq 4$ the same result is obtained applying the above reasoning fiber by fiber to the Moduli space fibration 

\begin{equation}
    \M_{0,n} \rightarrow S^2_{n-1} \rightarrow S^2_{n-1} \rightarrow ...\rightarrow S^2_3.
\end{equation}

\nn
In the context of string amplitude the inverse of the intersection matrix appearing in \eqref{KLTtwisted} is called KLT kernel. Let now explicitly compute KLT kernel, first considering the two simplest cases of three and four strings scattering.
For $n=3$, twisted homology and cohomology groups are trivial and so is the intersection matrix, then \eqref{KLTtwisted} immediately reduces to \eqref{KLTn3}. Consider now the $n=4$ case. The twisted homology group $H_1(\M_{0,4},\Lo)$ has dimension one, then intersection matrix reduces to intersection number between a twisted cycle and a dual twisted cycle.
 twisted cycle defined in \eqref{twistedcycle} 
 
 Let 

\begin{equation}
    \Delta_\alpha \equiv \Delta_{1234} = \frac{1}{e^{2\pi\alpha_1}-1}S^1_\e (0) \otimes U_{S^1_\e(0)}+\langle \e,1-\e\rangle \otimes u_{(0,1)}- \frac{1}{e^{2\pi\alpha_2}-1}S^1_\e (1)\otimes u_{S^1_\e(1)}, 
\end{equation}

\nn
if we choose $\overline{\Delta(\alpha)}=\Delta(\alpha)$, the intersection number is given by the selfintersection number \eqref{cycleselfintersection}: 

\begin{equation}
    [\Delta_{1234}|\Delta_{1234}]= \frac{i}{2}\left ( \frac{1}{\tan{ \pi \alpha'k_1k_4}}+\frac{1}{\tan{\pi\alpha' k_2k_4}}\right);
\end{equation}

\nn
and KLT relation then becomes:

\begin{equation}
    \A^{closed}(k_1,k_2,k_3,k_4)= -2i\A^{open}(k_1,k_2,k_3,k_4)\left ( \frac{1}{\tanh{\pi \alpha'k_1k_4}}+\frac{1}{\tanh{ \pi k_2k_4}}\right)^{-1}\A^{open}(k_1,k_2,k_3,k_4).
\end{equation}

\nn
Picking up any another pair of twisted cycles as bases for the twisted cohomologies groups, one finds different decompositions of the closed string amplitude in terms of different partial open amplitude. For instance, the result found by Kawai and Lewellen in \cite{KAWAI19861}, corresponds to the choice $\Delta_\alpha= \Delta_{1234}$ and

\begin{equation}
    \overline{\Delta_\alpha}\equiv\Delta_{1324}=(1,\infty)\otimes u,
\end{equation}

\nn
in fact 

\begin{equation}
    [\Delta_{1234}|\Delta_{1324}]=\frac{i}{2}\frac{1}{\sin{(\pi k_1k_3)}},
\end{equation}

\nn
and then 

\begin{equation}
    \A^{closed}(k_1,k_2,k_3,k_4)= -2i \sin{(\pi k_1k_3)}\A^{open}(k_1,k_2,k_3,k_4)\A^{open}(k_1,k_3,k_2,k_4).
\end{equation}

\nn
In order to discuss $n\geq 5$ scattering amplitude, we first need to study how to compute intersection numbers for higher dimensional cycles.

 \section{(Co)Homology of the punctured sphere Moduli space}
 
Because of the vanishing theorem \ref{vanishing} the only untrivial homology and cohomology groups $H_k(\M_{0,n})$ and $H^k(\M_{0,n})$ are the ones for $k=\dim{(\M_{0,n})}=n-3$. Their dimension can be related to the Euler characteristic of $\M_{0,n}$ by

\begin{equation}
     \nu= \dim{H_{n-3}(\M_{0,n})}= \dim{H^{n-3}(\M_{0,n})}=(-1)^{3-n}\chi (\M_{0,n});
\end{equation}

\nn
the Euler characteristic of the $n$ punctured sphere Moduli space is known to be

\begin{equation}
    \chi (\M_{0,n} = (-1)^{n-3} (n-3)!, 
\end{equation}

\nn
therefore one simply finds 

\begin{equation}
    \nu = (n-3)!.
\end{equation}

\nn
After fixing the position of three punctures\footnote{In the previous chapter we fixed $(z_1,z_2,z_3)$, we made here a different choice to adopt the same convention used in literature.} to $(z_1,z_{n-1},z_n)=(0,1,\infty)$, regions of integration of open strings tree amplitudes are $(n-3)$- simplices labelled by a permutation $\beta\in S_n$: 

\begin{equation}
    \Delta(\beta)\equiv \overline{\{0<z_{\beta(2)}<z_{\beta(3)}....<z_{\beta(n-2)}<1\}}.
\end{equation}

\nn
We then define twisted cycle on $\M_{0,n}$, labelled by a permutation $\beta$, by

\begin{equation}
    \Delta_\beta \equiv \Delta(\beta)\otimes SL_\beta [u(z)],
\end{equation}

\nn
where the branch

\begin{equation}
    SL_{\beta}[u(z)]=\prod_{i<j} (z_{\beta(j)}-z_{\beta(i)})^{\alpha'k_{\beta(j)}k_{\beta(i)}}
\end{equation}

\nn
is chosen according the so called standard loading. Because only $(n-3)!$ twisted cycles are independent we can choose a basis to be the set 
\begin{equation} 
\{\Delta(\beta)|\beta=(1\sigma(2,3...n-2),n-1,n)\}, \quad \mbox{with} \quad \sigma\in S_{n-3}. 
\end{equation}

\nn
Twisted cohomology $H_{n-3}$ also is $(n-3)!$ dimensional, a convenient basis for this space is given by the so called Parke-Taylor forms\cite{Parke:1986gb}:
 \begin{equation}
     PT(\beta)=\frac{\bigwedge_{k=4}^n dz_n}{z_{\beta(2)}(z_{\beta(2)}-z_{\beta(3)})...(z_{\beta(n-2)}-1)},
 \end{equation}
 
 \nn
 where $\beta=(1,\sigma(2,3...n-2),n-1,n)$, with $\sigma\in S_{n-3}$.
 For instance for $n=5$ we have 
 
 \begin{equation}\begin{split}
     &PT(12345)= \frac{dz_2\wedge dz_3}{z_2(z_2-z_3)(z_3-1)},\\&PT(13245)= \frac{dz_2\wedge dz_3}{z_3(z_3-z_2)(z_2-1)}. \end{split}\label{PTforn4}
 \end{equation}

\nn
It can be shown Parke-Taylor forms can be written as a logarithmic form. 
Of course, one can obtain a basis for the twisted cohomology picking up any $(n-3)!$ Parke-Taylor forms, labelled by different permutations, as well as a basis for the twisted homology choosing $(n-3)!$ twisted cycles corresponding to different permutaions.

 \section{Higher dimensional intersection number for cycles}\label{Appendice1}
As pointed out in section \ref{Intc} the computation of higher dimensional twisted cycles is a quite involving topic of algebraic geometry, because one has to take into account contributions due to all intersection of lower dimensions. 
Actually, in the specific case of a Koba-Nielsen twist in the Moduli space of the puncture sphere, it is possible to compute them by a diagrammatic method. \\
Consider a twist having the form 

\begin{equation}
    u= \prod (z_i-z_j)^{s_{ij}},
\end{equation}

\nn
where $s_{ij}=\alpha'k_ik_j$.

 Suppose we want to compute the intersection number between $n-$ distinct and intersecting cycles $\Delta(\alpha)$ and $\Delta(\beta)$, with  $\alpha \neq \beta \in S_{n+3}$. In \cite{Mizera:2016jhj}, Mizera shows it can be written as  

\begin{equation}
[\Delta (\alpha)|\Delta(\beta)]= \left (\frac{i}{2}\right )^{n-3} m(\alpha|\beta),
\end{equation}

\nn
where $m(\alpha|\beta)$, known as bi-adjont scalar, is precisely the quantity we are going to compute digrammatically by means of the method proposed by Cachazo, He and Yuan in \cite{Cachazo:2013iea}.

\begin{itemize}
    \item Draw points on the boundary of a disk according to the ordering expressed by the permutation $\alpha$, then connect points by segments in the order expressed by the permutation $\beta$. The polygon chain constructed in this way self-intersects in $k$ points (internal vertices), identifying $n-k$ polygons with at least one of the vertices on the boundary;

    \item Draw a dual diagram where each polygon is mapped into a bubble and each vertex to a line. Boundary vertices become external legs coming in to a bubble and internal vertices become legs connecting bubbles. Label each bubble with $l$ legs $\nu_1,\nu_2,..\n_l$ by $(\n_1,\n_2..\n_{l-1})$;

    \item To each bubble $(\n_1,\n_2,..,\n_{l-1})$ associate the self-intersection number of the $l-3$ cycle $\Delta(\n_1,\n_2,..,\n_{l-1})$. Notice that $l$ is at most equal to $n-1$, then the sub intersecting cycles are granted to be less dimensional then the original ones. For $l=3$, one just has the self intersection of a point, trivially equal to one and for $l=4$, self-intersection between $1-$ cycles is given by \eqref{cycleselfintersection}. For $l>4$ we will see later on how to compute it;
    
    \item To each internal leg connecting the bubbles $(\n_1,\n_2,..,\n_{l-1})$ and $\m_{1},\m_{2},..\m_{l'-1}$ associate the propagator 
    
    \begin{equation}
        \frac{1}{\sin{\pi s_{\n_1,\n_2,..\n_{l-1}}}}= \frac{1}{\sin{\pi s_{\m_1,\m_2,..\m_{l'-1}}}}
    \end{equation}
    
    \item Compute the overall sing of the diagram by \begin{equation}
 (-1)^{w(\alpha|\beta)+1},
\end{equation}

\nn
where $w(\alpha|\beta)$ is the winding number, digrammatically computed by the rule depicted in \ref{windingn}: dispose points on a circle according to the orientation expressed by $\alpha$ and draw consecutive arcs according to the orientation
given by $\beta$; the number of completed cycles is precisely the winding number.
    
     \begin{figure}[h!]
  \centering
\begin{tikzpicture}
\winding{(0,0)}
\end{tikzpicture}
\caption{Graphic rule for winding number.}
\label{windingn}
 \end{figure}
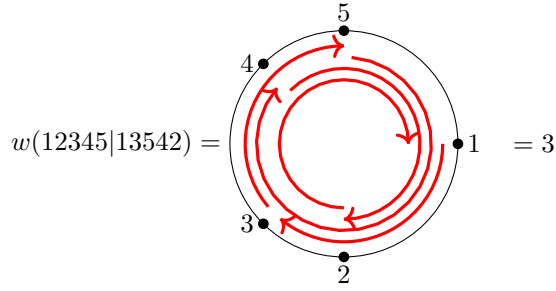

\end{itemize}

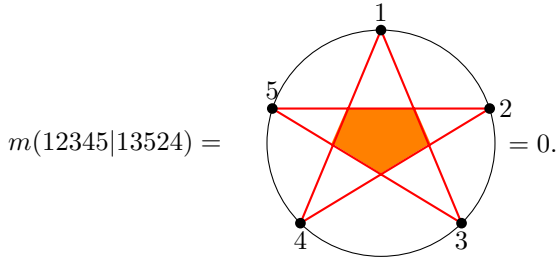
\begin{figure}[h!]
  \centering
\begin{tikzpicture}
\vanishingone{(0,0)}
\end{tikzpicture}
\caption{Example of vanishing intersection.}
\label{vanishingone}
 \end{figure}

\nn
Finally, if drawing the diagram, the polygon chain selfintersects such that a closed loop (orange region in figure \ref{vanishingone}) is enclosed, then the intersection vanishes
 
 \nn
Consider now some example; 
\begin{Ese}
Consider the $2-$ cycle intersection: 

\begin{equation}
[\Delta(12345)|\Delta(13425)]=-\frac{1}{4} m(12345|13425)].
\end{equation}

\nn
The diagram obtained applying the above rules is shown in figure \ref{ncicliintersection}.

  \begin{figure}[h!]
  \centering
\begin{tikzpicture}
\multinterese{(0,0)}
\end{tikzpicture}
\caption{Diagrammatic rule for a $2-$ cycles intersection.}
\label{ncicliintersection}
 \end{figure}
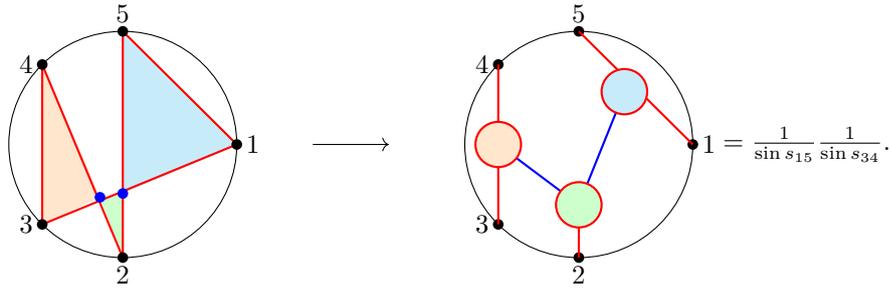
 
 \nn
 All bubbles are attached to three legs each, then they represent intersections between $0-$ cycles and their contribution is just $1$. There are two internal line, one attached to the orange $(34)$ bubble and one to the blue $(15)$ bubble, then there are two propagators with momentum $s_{45}$ and $s_{15}$ respectively.
 Finally the winding number is $w(12345|13425)=2$, then one has:
 
 \begin{equation}
     [\Delta(12345)|\Delta(13425)]=\frac{1}{4} \frac{1}{\sin{s_{15}}\sin{s_{34}}}.
 \end{equation}
 
 \nn
 \end{Ese}
 \begin{Ese}
 Consider now the $3-$ cycles intersection number 
 
 \begin{equation}
     [\Delta(123456)| \Delta(123645)]= -\frac{i}{8} m(123456|123645)
 \end{equation}
 
 \nn
 Again, green and orange bubbles represent trivial amplitudes, while the blue one $(123)$ now is the self intersection of the $1-$ cycle $\Delta(123)$. Internal lines are related to momenta $s_{45}$ and $s_{456}$, with $s_{i_1i_2..i_n}=\alpha'\sum_{1<i_1<..<i_n}k_{i_1}k_{i_n}$:
 	 
	 \begin{equation}
	     m(123456|123645)=\frac{1}{\sin{\pi s_{45}}} \frac{1}{\sin{\pi s_{456}}} m(123(456)|123(456))m(123|123)
	 \end{equation}
 
 \nn
 The winding number turns to be $w(123456|1236456)=2$, then one has:
 
 \begin{equation}
     [\Delta(123456)| \Delta(123645)]=\frac{i}{8}\frac{1}{\sin{\pi s_{45}}\sin{\pi s_{456}}} \left (\frac{1}{\tan \pi s_{12} }+\frac{1}{\tan \pi s_{23} } \right )
 \end{equation}
 
 \begin{figure}[h!]
  \centering
\begin{tikzpicture}
\multinteresedue{(0,0)}
\end{tikzpicture}
\caption{Diagrammatic rule for a $3-$ cycles intersection.}
\label{ncicliintersectiondue}
 \end{figure}
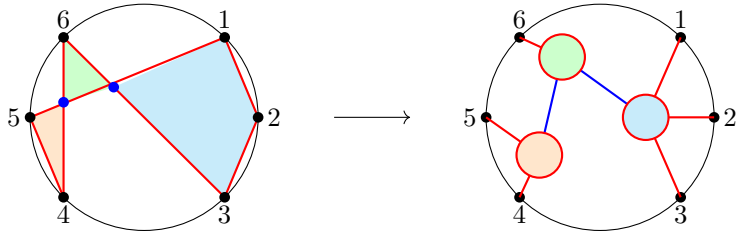
 
 \end{Ese}

\begin{figure}[h!]
  \centering
\begin{tikzpicture}
\multiselfinter{(0,0)}
\end{tikzpicture}
\caption{Diagrammatic rule for $n-$ cycles self-intersection.}
\label{multiselfinter}
 \end{figure}
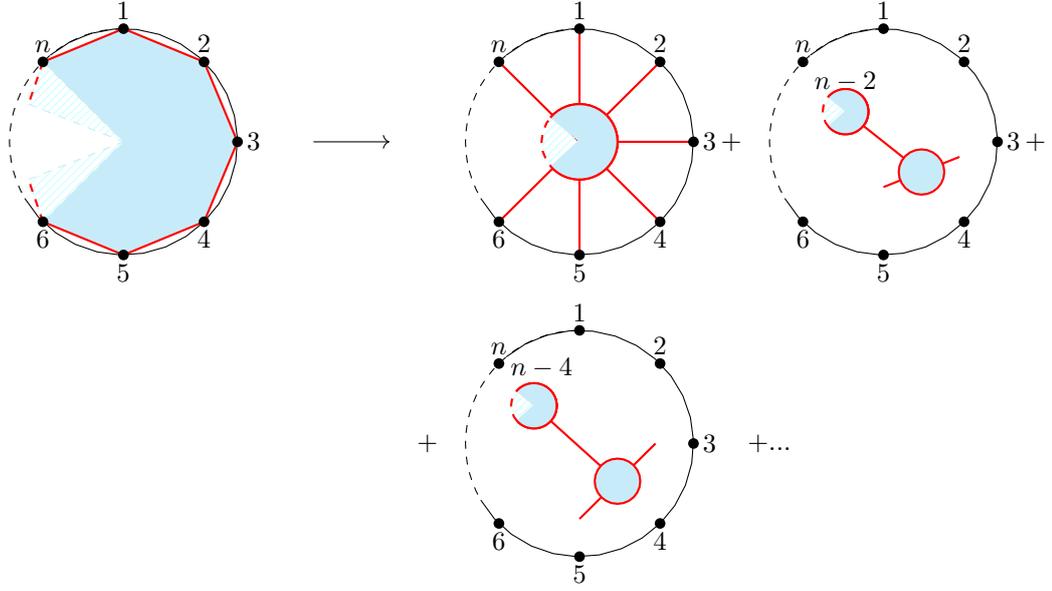

\nn
 Let now consider the self-intersection of an $n$ cycles. In this case, as shown in figure \ref{multiselfinter}, the diagrammatic prescription consists in drawing all possible diagrams involving amplitudes between $n-2$ cycles, $n-4$, and so on, preserving permutation orientation (only adjacent legs income in the same bubble and no legs intersect).

 \nn
now, the contribute of the $n-vertex$ diagram (the first) appearing only for odd $n$, is just a constant, that turn to be equal to $1,1,2,5$ for the first four odd numbers.
Then, in each subdiagram, associate to each internal line attached to a bubble $\nu_1\n_2,..\nu_l$ the propagator 

\begin{equation}
    \frac{1}{\tan \pi s_{\nu_1\n_2,..\nu_l}}.
\end{equation}

\nn
Finally notice that the winding number is always one, then no overall sign must be added.\\
Consider some example.\\
\begin{Ese}
The diagrammatic representation of the $1-$ cycle $\Delta(1234)$ self-intersection is:

  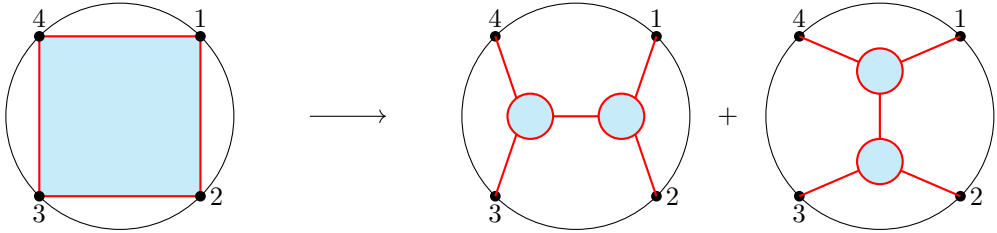
\begin{figure}[h!]
  \centering
\begin{tikzpicture}
\fourselfinter{(0,0)}
\end{tikzpicture}
\caption{Diagrammatic rule for $1$ cycles self-intersection.}
\label{1selfinter}
 \end{figure}

\nn
Then we have 

\begin{equation}
    [\Delta(1234)|\Delta(1234)]=\frac{i}{2} \left( \frac{1}{\tan \pi s_{12}}+ \frac{1}{\tan \pi s_{23}} \right ),
\end{equation}

\nn
the same result, of course, we obtained in \eqref{cycleselfintersection}.\\
\end{Ese}
\begin{Ese}

Consider now the $2-$cycle $\Delta(12345)$ we have 

  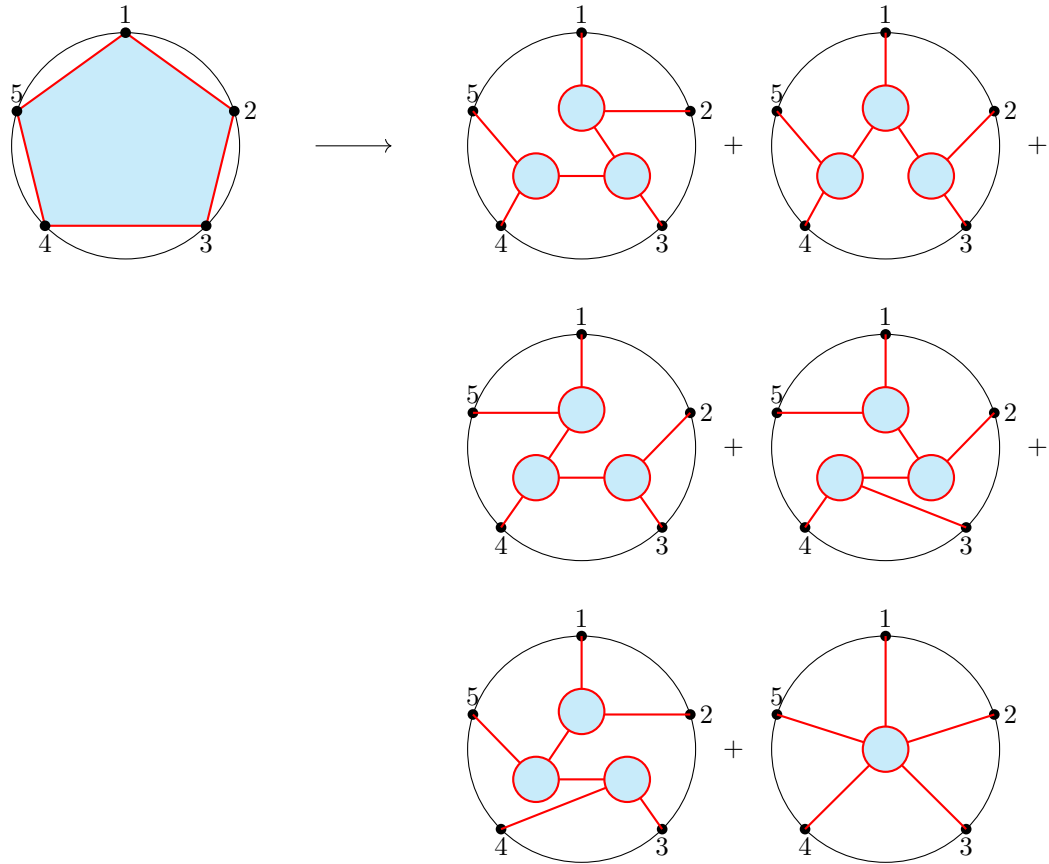
\begin{figure}[h!]
  \centering
\begin{tikzpicture}
\fiveselfinter{(0,0)}
\end{tikzpicture}
\caption{Diagrammatic rule for $1$ cycles self-intersection.}
\label{oneselfinter}
 \end{figure}

\begin{equation}
\begin{split}
    [\Delta(12345)|\Delta(12345)]&=-\frac{1}{4} \left( \frac{1}{\tan \pi s_{12}} \frac{1}{\tan \pi s_{45}} \right )+\left( \frac{1}{\tan \pi s_{23}} \frac{1}{\tan \pi s_{45}} \right )+\left( \frac{1}{\tan \pi s_{15}} \frac{1}{\tan \pi s_{23}} \right )+\\&+\left( \frac{1}{\tan \pi s_{15}} \frac{1}{\tan \pi s_{34}} \right )+\left( \frac{1}{\tan \pi s_{12}} \frac{1}{\tan \pi s_{34}} \right )+1,
    \end{split}
\end{equation}

\nn
where, for an easier visualization, each square bracket contains the contribute of each diagram listened in the same order.
\end{Ese}

 \section{Five Tachyon Amplitude}\label{lastone}
We finally have all tools needed to explicitly compute a KLT decomposition for five closed string amplitude. For notation clarity we will consider tachyon scattering, keeping in mind that for excited states the only difference lies in the function $F$, containing polarizations contributions, that plays no important role here. We have 

\begin{equation}\begin{split}
     \A_5^{closed}&=\int |z_2|^{2\alpha'k_1k_2}|z_3|^{2\alpha'k_1k_3}|z_2-1|^{2\alpha'k_2k_4}|z_3-1|^{2\alpha'k_3k_4}|z_2-z_3|^{2\alpha'k_2k_3}dz_2\wedge d z_3=\\&=\int |u(z_2,z_3)|^2 dz_2\wedge dz_3 = \langle 1 | 1\rangle.\end{split}, 
 \end{equation} 

\nn
equation \eqref{KLTtwisted} becomes

\begin{equation}
    \langle 1 | 1\rangle= \sum_{\alpha\beta} \langle 1 |\Delta(\alpha)] H_{\alpha\beta}^{-1}[\Delta(\beta)|1\rangle,
\end{equation}

\nn
Consider the basis for $H_{2}(\M_{0,5})$ given by $\Delta(12345)$ and $\Delta(13245)$, and for its dual given by $\Delta(21435)$ and $\Delta(31425)$, using the diagrammatic approach developed in previous section we can easily compute the intersection matrix $H_{\alpha\beta}$

\begin{equation}\begin{split}
H_{\alpha\beta}&=\begin{pmatrix} [\Delta(12345)|\Delta(21435)]&[\Delta(12345)|\Delta(31425)]\\ [\Delta(13245)|\Delta(21435)]&[ \Delta(13245)|\Delta(31425)] \end{pmatrix},\end{split}
\end{equation}
 
 \nn
evaluating the respective biadjoint scalars: 
 
 \begin{figure}[h!]
  \centering
\begin{tikzpicture}
\vanishingtwo{(0,0)}
\end{tikzpicture}
 \end{figure}
 
   \begin{figure}[h!]
  \centering
\begin{tikzpicture}
\multintereseone{(0,0)}
\end{tikzpicture}
 \end{figure}
  \begin{figure}[h!]
  \centering
\begin{tikzpicture}
\multinteresetwo{(0,0)}
\end{tikzpicture}
 \end{figure}
 
 \nn
thus one has:
 \begin{equation}\begin{split}
H_{\alpha\beta}&=-\frac{1}{4}\begin{pmatrix}\frac{1}{\sin (\pi \alpha'k_1k_2) \sin (\pi \alpha'k_3k_4)}&0\\0&\frac{1}{\sin (\pi\alpha' k_1k_3) \sin (\pi \alpha'k_2k_4)}\end{pmatrix}.\end{split}
\end{equation}

 \nn
and KLT relation reads:
 
 \begin{equation}\begin{split}
     \A_5^{closed}&= -4\A_5^{open}(12345)\bar{\A}_5^{open}(21435)\sin (\pi\alpha' k_1k_2) \sin (\pi \alpha'k_3k_4)+\\&-4\A_5^{open}(13245)\bar{\A}_5^{open}(31425)\sin (\pi \alpha'k_1k_3) \sin (\pi \alpha'k_2k_4).\end{split}\label{KLT5}
 \end{equation}

\nn
in fully agreement, up to normalization, with the original KLT result (eq.$(3.26)$. \cite{KAWAI19861}). 
 
 \nn
Expression \ref{KLT5} implies that the computation of a closed string scattering reduces to the evaluation of proper open partial amplitudes. Consider then the open amplitude
 \begin{equation}\begin{split}
     \A_5^{open}=\sum_{\beta} \A_5^{open}(\beta)&=\sum_{\beta}\int_{\Delta(\beta)} z_2^{\alpha'k_1k_2}z_3^{\alpha'k_1k_3}(z_2-1)^{\alpha'k_2k_4}(z_3-1)^{\alpha'k_3k_4}(z_2-z_3)^{\alpha'k_2k_3}dz_2\wedge d z_3\\&=\sum_{\beta}\int_{\Delta(\beta)} u(z_2,z_3) dz_2\wedge dz_3 = \sum_{\beta}\langle 1 | \Delta (\beta)\otimes u].\end{split}
 \end{equation}

 \nn
 applying the master decomposition formula we can write 
 
 \begin{equation}
\langle 1 | \Delta_\beta]= \sum_\alpha \langle 1|PT(\alpha)\rangle C_{\alpha\beta}^{-1} \langle PT(\beta)|\Delta_\beta].
    \label{5tachyon1}
 \end{equation}

 \nn
 where the PT basis of $H^2_\omega$ is given by \eqref{PTforn4}:

 \begin{equation}\begin{split}
     &PT(12345)= \frac{dz_2\wedge dz_3}{z_2(z_2-z_3)(z_3-1)},\\&PT(13245)= \frac{dz_2\wedge dz_3}{z_3(z_3-z_2)(z_2-1)}. \end{split}
 \end{equation}
 
 \nn
 The first intersection number appearing in \eqref{5tachyon1}, representing the projection into the PT basis, involves the non logarithmic form $\langle 1|$; it can therefore be computed by the recursive algorithm discussed in \ref{recursive}.  
Setting, for shorthand notation, $\alpha_1=\alpha'k_1k_2$, $\alpha_2=\alpha'k_1k_3$, $\alpha_3=\alpha'k_2k_4$, $\alpha_4=\alpha'k_3k_5$ and $\alpha_5=\alpha'k_2k_3$, we obtained:

\begin{equation}
\begin{split}
     &\langle 1 | PT(12345)\rangle = \frac{N_1(\alpha_i)}{D(\alpha_i)},\\& \langle 1 | PT(13245)\rangle = \frac{N_2(\alpha_i)}{D(\alpha_i)},
     \end{split}
\end{equation}

\nn
with

\begin{equation}\begin{split}
    &N_1(\alpha_i)=\alpha _2^2+\left(\alpha _3+\alpha _4+2 \alpha _5+2\right) \alpha _2+\left(\alpha _3+\alpha _5+1\right) \left(\alpha _3+\alpha _4+\alpha _5+1\right)+\alpha _1 \left(\alpha _2+\alpha _3+\alpha _4+\alpha
   _5+1\right),\\&N_2(\alpha_i)=\alpha _1^2+\left(\alpha _2+\alpha _3+\alpha _4+2 \alpha _5+2\right) \alpha _1+\left(\alpha _2+\alpha _4+\alpha _5+1\right) \left(\alpha _3+\alpha _4+\alpha _5+1\right),\\& D(\alpha_i)=\left(\alpha _1+\alpha _3+\alpha _5+1\right) \left(\alpha _2+\alpha _4+\alpha _5+1\right) \left(\alpha _1+\alpha _2+\alpha _3+\alpha _4+\alpha _5+1\right) \left(\alpha _1+\alpha _2+\alpha _3+\alpha _4+\alpha
   _5+2\right).\end{split}
\end{equation}

 \nn
The intersection matrix

 \begin{equation}
 C_{\alpha\beta}=\begin{pmatrix}\langle PT(12345)|PT(12345)\rangle &&\langle PT(12345)|PT(12345)\rangle \\\langle PT(13245)|PT(12345)\rangle && \langle PT(13245)|PT(13245)\rangle\end{pmatrix},
 \end{equation}
 
 \nn
 can be also computed by the simplified formula for logarithmic forms \ref{logamizera1}; we computed it with both methods and we numerically verified they agree. The final result is\footnote{The result showed here is the one obtained by the recursive method}:

 \begin{equation}
     \begin{split}
         &\langle PT(12345)|PT(12345)\rangle = \frac{N_{11}(\alpha_i)}{D_{11}(\alpha_i)},\\&\langle PT(13245)|PT(13245)\rangle =\frac{N_{22}(\alpha_i)}{D_{22}(\alpha_i)},\\&\langle PT(12345)|PT(13245)\rangle =\frac{N_{12}(\alpha_i)}{D_{12}},
     \end{split}
 \end{equation}

\nn
where 

\begin{equation}
    \begin{split}
&N_{11}=\left(\alpha _4+\alpha _5\right) \alpha _1^2+\left(\alpha _4+\alpha _5\right) \left(\alpha _2+\alpha _3+\alpha _4+2 \alpha _5\right) \alpha _1+\alpha _5 \left(\alpha _2+\alpha _4+\alpha _5\right) \left(\alpha _3+\alpha
   _4+\alpha _5\right),\\& D_{11}=\alpha _1 \alpha _4 \alpha _5 \left(\alpha _1+\alpha _2+\alpha _5\right) \left(\alpha _3+\alpha _4+\alpha _5\right),\\&N_{22}=\alpha _1 \alpha _2 \left(\alpha _3+\alpha _5\right)+\left(\alpha _2+\alpha _5\right) \left(\alpha _2+\alpha _3+\alpha _4+\alpha _5\right) \left(\alpha _3+\alpha _5\right)+\alpha _1 \alpha _5 \left(\alpha _3+\alpha _4+\alpha
   _5\right),\\&D_{22}=\alpha _2 \alpha _3 \alpha _5 \left(\alpha _1+\alpha _2+\alpha _5\right) \left(\alpha _3+\alpha _4+\alpha _5\right),\\&N_{12}=-\left (\alpha _1+\alpha _2+\alpha _3+\alpha _4+2\alpha _5\right),\\&D_{12}=\alpha _5(\alpha _1+\alpha _2+\alpha _5)(\alpha _3+\alpha _4+\alpha _5)
    \end{split}
\end{equation}

 \nn
 And the inverse intersection matrix is 
 
 \begin{equation}
     C^{-1}_{\alpha\beta}=\begin{pmatrix}\frac{\tilde{N}_{11}(\alpha_i)}{\tilde{D}_{11}(\alpha_i)}&\frac{\tilde{N}_{12}(\alpha_i)}{\tilde{D}_{12}(\alpha_i)}\\\frac{\tilde{N}_{12}(\alpha_i)}{\tilde{D}_{12}(\alpha_i)}&\frac{\tilde{N}_{22}(\alpha_i)}{\tilde{D}_{11}(\alpha_i)}\end{pmatrix},
 \end{equation}
 
 \nn
 where 
 \begin{equation}
    \begin{split}
&\tilde{N}_{11}=\alpha _1 \alpha _4 \left(\left(\alpha _2+\alpha _5\right) \left(\alpha _3+\alpha _5\right) \left(\alpha _2+\alpha _3+\alpha _4+\alpha _5\right)+\alpha _1 \left(\alpha _2 \left(\alpha _3+\alpha _5\right)+\alpha _5
   \left(\alpha _3+\alpha _4+\alpha _5\right)\right)\right),\\&\tilde{D}_{11}=\left(\alpha _1+\alpha _3+\alpha _5\right) \left(\alpha _2+\alpha _4+\alpha _5\right) \left(\alpha _1+\alpha _2+\alpha _3+\alpha _4+\alpha _5\right),\\&\tilde{N}_{22}=\alpha _2 \alpha _3 \left(\left(\alpha _4+\alpha _5\right) \alpha _1^2+\left(\alpha _4+\alpha _5\right) \left(\alpha _2+\alpha _3+\alpha _4+2 \alpha _5\right) \alpha _1+\alpha _5
   \left(\alpha _2+\alpha _4+\alpha _5\right) \left(\alpha _3+\alpha _4+\alpha _5\right)\right),
    \\&\tilde{N}_{12}=\alpha _1
   \alpha _2 \alpha _3 \alpha _4 \left(\alpha _1+\alpha _2+\alpha _3+\alpha _4+2 \alpha _5\right),\\&\tilde{D}_{12}=\left(\alpha _1+\alpha _3+\alpha _5\right) \left(\alpha _2+\alpha _4+\alpha _5\right) \left(\alpha _1+\alpha _2+\alpha _3+\alpha
   _4+\alpha _5\right).
    \end{split}
\end{equation}

\nn
Notice that all coefficients we computed should simplify under momentum conservation and on-shell conditions. 
The Master integral decomposition \eqref{5tachyon1} therefore reads

\begin{equation}\begin{split}
    \langle 1 | \Delta_\sigma]&=\left(\frac{N_1\tilde{N}_{11}}{D\tilde{D}_{11}}+\frac{N_2\tilde{N}_{12}}{D\tilde{D}_{12}}\right) \langle PT(12345) | \Delta_\sigma]+\left(\frac{N_1\tilde{N}_{12}}{D\tilde{D}_{12}}+\frac{N_2\tilde{N}_{22}}{D\tilde{D}_{22}}\right) \langle PT(13245) | \Delta_\sigma]=\\&\equiv \Lambda_1\langle PT(12345) | \Delta_\sigma] +\Lambda_2\langle PT(13245) | \Delta_\sigma].\end{split}
    \label{Masterdecompositiontachyon}
\end{equation}

\nn
Then, for instance 

\begin{equation}\begin{split}
    \A_5^{open}(12345)=&\int_0^1\int_0^1 z_2^{\alpha_1}z_3^{\alpha_2}(z_2-1)^{\alpha_3}(z_3-1)^{\alpha_4}(z_2-z_3)^{\alpha_5}dz_2\wedge dz_3  =\\&\Lambda_1 \int_0^1\int_0^1 z_2^{\alpha_1-1}z_3^{\alpha_2}(z_2-1)^{\alpha_3}(z_3-1)^{\alpha_4-1}(z_2-z_3)^{\alpha_5-1}dz_2\wedge dz_3 +\\&+\Lambda_2\int_0^1\int_0^1 z_2^{\alpha_1}z_3^{\alpha_2-1}(z_2-1)^{\alpha_3-1}(z_3-1)^{\alpha_4}(z_2-z_3)^{\alpha_5-1}dz_2\wedge dz_3 \end{split}
\end{equation}

\nn 
The validity of the previous decomposition has been analytically check in the case of equal exponents, ($\alpha_i=\beta$), in the case of only two different exponents ($\alpha_{i_1,i_2,i_3}=\beta_1$ and $\alpha_{i_4,i_5}=\beta_2$) and in the case of three different exponents ($\alpha_{i_1,i_2}=\beta_1$, $\alpha_{i_3,i_4}=\beta_2$ and $\alpha_{i_5}=\beta_3$). For the general case, due to the size of the expression, we checked it numerically for fixed values of exponents.
 Notice for excited state the master decomposition has the same structure of \eqref{Masterdecompositiontachyon}, with the only difference lying in the coefficients $\Lambda_i$, receiving contributions from the projections $\langle F(z)|PT(\alpha)\rangle$.
\addcontentsline{toc}{chapter}{Conclusions}
\chapter*{Conclusions and outlooks}

In this thesis, we have studied the properties of String theory amplitudes within the framework of Intersection Theory for twisted (co)homology,
which, as recently proposed, offered a novel approach to analyze the linear and quadratic relations between scattering amplitudes, in string theory as well as in Quantum Field Theory. As only recently pointed out, thanks to intersection theory, the analytic properties of scattering amplitudes can be related to the topological properties of the manifolds characterizing their integral representation.

We have presented an introduction to Morse theory, which provides a very useful mathematical method to obtain topological information of a manifold by only analytic tools, and we have seen how it can be applied to evaluate the dimension of (co)homology space.\\
We have studied homology and cohomology theory, with particular emphasis to De Rham cohomology of differential forms, discussing De Rham theorem and bilinear Riemann period relations.
We have explored (co)homology with local coefficients, focusing on the (co)homology with values in a Local System, called twisted (co)homology. We defined cycle-cycle and cocycle-cocycle intersection numbers as well as cycle-cocycle pairing, and we showed they enter both linear (contiguity) and quadratic (Riemann period) relations.
We presented different algorithms to evaluate univariate and multivariate intersection number between both logarithmic and non-logarithmic twisted cocycles.
We then explored a diagrammatic method for the computation of intersection numbers between twisted cycles of the moduli space of the $n-$ punctured Riemann sphere. \\
We have shown how a suitable choice of the dual local system allows to interpret closed strings tree amplitudes as intersection number between twisted forms and open strings tree amplitudes as pairings between a twisted form and a twisted cycle. We used intersection theory to rederive Kawai-Lewellen-Tye relations, expressing the decomposition of a closed string tree amplitudes into partial open strings tree amplitudes, and we showed in this framework they naturally appear as a twisted version of Riemann period relations.\\
We found two bases for two-dimensional twisted cycles and dual cycles diagonally intersecting and we explicitly computed their intersection matrix. We used this result to evaluate a five closed string tree amplitude Kawai-Lewellen-Tye decomposition, rederiving it in the form presented in the original paper.
We implemented a Mathematica code based on generalized residue theorem for multivariate logarithmic twisted cocycles intersection number, and we used it to explicitly determine the intersection matrix between the two-dimensional Parke-Taylor basis. We then used the recursive algorithm for generic $n$ forms to project five open tachyon tree amplitudes into the Parke-Taylor basis, obtaining its explicit Master decomposition.
We analytically checked the agreement of the recursive algorithm with the residue formula.\\
Remarkably, any open string tree-amplitude can be decomposed on the same basis of Master integrals appearing in the open tachyons 
tree-amplitudes, with coefficients that depend on the nature of the excited state, computed {\it via} intersection numbers.\\

The interest in studying string amplitudes also lies in the study of their field limit, that offers the possibility to explore quantum field theory amplitudes from a different perspective; this approach  allowed to prove the Bern-Carrasco-Johansson double-copy duality, for tree amplitudes, by Kawai-Lewellen-Tye relations; although the evidence is mounting with a series of papers showing double-copy behaviour for amplitudes at higher loop order, there is still no general proof of the double-copy formula. 
The application of intersection theory to higher genera string amplitudes, could provide new ideas useful to a better understanding of such duality;  this would probably require to investigate different (co)homologies and to study new methods for the evaluation of intersection numbers.
We applied twisted intersection theory to bosonic tree amplitudes, which share the same structure of the supersymmetric tree amplitudes. However, this is in general not true for higher genera: supersymmetric amplitudes become more involved than bosonic ones, because of the difficulty in splitting superspace onto bosonic and fermionic variables \cite{witten2013notes,witten2016superstring,donagi2013supermoduli}. \\

Intersection theory has turned out to be a very powerful mathematical language to interpret and evaluate scattering amplitudes, allowing a better understanding of their structure, and providing new methods for their evaluation. Because of the full generality of this approach, advances in this area would not be limited to particle physics or string theory, but could be broadly applied also in other contexts as condensed matter, statistical mechanics, gravitational-wave physics, as well as mathematics, owing to the ubiquity of Aomoto-Gel'fand integrals.

\section*{Acknowledgments}
This thesis has been written under the supervision of Prof. Pierpaolo Mastrolia, to whom I sincerely address my most deep gratitude, for giving me the opportunity 
to approach this fascinating world, for showing and transmitting me his great passion for research, for spending so much time on this work giving me his complete availability, and especially for his unlimited patience. 
Special thanks goes to Prof. Sergio L. Cacciatori, cu-supervisor of this work, for exciting discussions and precious teachings, and especially for having reminded me the main reason for studying physics is having fun.  
I am grateful to Angius Roberta, Chestnov Vsevolod, Frellesvig Hjalte, Gasparotto Federico and Mandal Manoj Kumar for very useful discussions, helpful comments and valuable exchanges. 

\begin{appendix}
\chapter{Appendix}

\section{Conformal Symmetry}
Let $g_{\mu \nu}$ be the metric in a d-dimensional space-time. Under a general coordinate transformation $x\rightarrow x'$, the metric tensor transforms as

\begin{equation}
g'_{\m\n}(x')= \de'_\m x^\alpha \de'_\n x^\beta g_{\alpha\beta}(x),
\end{equation}

\nn
with $\de'_\m = \frac{\de}{\de x'^{\m}}$. The set of these transformations leaving the metric invariant up to rescaling, namely  
\begin{equation}
g'_{\mu \nu} (x') = \Lambda (x) g_{\mu \nu} (x),
\end{equation}

\nn
forms a group, known as \textit{conformal group}. Conformal group contains all coordinate transformations preserving angles between vectors, and so, in flat space, it also includes Poincaré transformations, identified by $\Lambda(x)=1$.\\
In order to study the conformal group and to find its generators we proceed, as usual, considering infinitesimal transformations $x^\m \tu x'^\m= x^\m+ \epsilon^\m$, under which the metric changes, at first order, as  

\begin{equation}
\delta g_{\m\n}= g'_{\mu \nu}(x')-g_{\m\n}(x') = - \de_\m\epsilon^\alpha g_{\alpha\n}-\de_\n\epsilon^\beta g_{\beta\m}- \epsilon^\lambda\de_\lambda g_{\m\n},
\end{equation}

\nn
For this transformation to be conformal around the flat metric $\eta_{\m\n}$ (Euclidean or Minkowskian), the variation $\delta g_{\m\n}$ must be linear in the metric itself:

\begin{equation}
\partial_{\mu} \epsilon_{\nu} + \partial_{\nu} \epsilon_{\mu} = \Omega(x) \eta_{\mu \nu}.
\label{tfmcftinf}
\end{equation}

\nn
Taking the trace of equation \eqref{tfmcftinf}, we find $\Omega(x)$ in terms of $\de_\m\epsilon^\m \equiv  \de\epsilon$, and we can rewrite this relation as 

\begin{equation}
\de_\m\epsilon_\n+\de_\n\epsilon_\m = \frac{2}{d}(\de\epsilon)\eta_{\m\n}.
\end{equation}
\nn
After some algebrical manipulations we obtain the equation:

\begin{equation}
[\Box \eta_{\m\n}+(d-2)\de_\m\de_\n ]\de\epsilon =0.
\label{conformalconstrain}
\end{equation}

\nn
We see the constraints on the parameter depend on the space dimension $d$: in $d=1$ equation \eqref{conformalconstrain} is identically satisfied and no constraints are imposed on $\epsilon$,  $d=2$ evidently represents a special case and we will analyse it in detail later on, finally for $d>2$ \eqref{conformalconstrain} implies that $\de\epsilon$ can be at most linear in $x$, and so the parameter $\epsilon$ must be at most quadratic in the coordinates, we can therefore write it in the general form:

\begin{equation}
\epsilon_{\mu} = a_{\mu} + b_{\mu \nu} x^{\nu} + c_{\mu \nu \rho} x^{\nu} x^{\rho} \quad \quad \quad c_{\mu \nu \rho} = c_{\mu \rho \nu}.
\end{equation}

\nn
We now analyse this expression term by term. The easiest part is the constant term $\epsilon_\m = a_\m$ that we immediately recognize as corresponding to translations, as usual. The linear term can be decomposed into its symmetric and antisymmetric parts as:

\begin{equation}
\epsilon_\m = \omega_{\m\n}x^\n + \sigma_{\m\n}x^\n \quad\quad \mbox{with} \quad \omega_{\m\n}=-\omega_{\n\m}, \sigma_{\m\n}=\sigma_{\n\m},
\end{equation}

\nn
replacing it into equation \eqref{conformalconstrain} we find no more constraint are imposed on $\omega_{\m\n}$, that thus induces ordinary rotations; while the symmetric term $\sigma_{\m\n}$ turns to be constrained to be proportional to the metric tensor, providing the transformations

\begin{equation}
\epsilon_\m = \lambda x^\m,
\end{equation}

\nn
known as dilatations. Finally, the quadratic term, proceeding in the same way and after same manipulations turns to have the form:

\begin{equation}
\epsilon_\m = d_\m x^2 - 2x_\m(d\cdot x),
\end{equation}

\nn
with $d_\m$ a constant vector. These  are called special conformal transformations (SCT). From infinitesimal transformations we can immediately obtain the generators of the conformal group

\begin{center}
\begin{tabular}{|c l|}
\hline
Translation & $P_{\mu} = - i \partial_{\mu}$ \\
Dilatation & $D = -i x^{\mu} \partial_{\mu}$ \\
Rotation & $L_{\mu \nu} = i \left( x_{\mu} \partial_{\nu} - x_{\nu} \partial_{\mu} \right)$ \\
SCT & $K_{\mu} = -i \left( 2 x_{\mu} x^{\nu} \partial_{\nu} - x^2 \partial_{\mu} \right)$ \\
\hline
\end{tabular} 
\end{center}
satisfying the following commutation relations:
\begin{equation}
\begin{split}
& \left[ D , P_{\mu} \right] = i P_{\mu}, \\
& \left[ D , K_{\mu} \right]= - i K_{\mu}, \\
& \left[ K_{\mu} , P_{\nu} \right] = 2i \left( \eta_{\mu \nu} D - L_{\mu \nu} \right), \\
& \left[ K_{\rho} , L_{\mu \nu} \right] = i \left( \eta_{\rho \mu } K_{\nu} - \eta_{\rho \nu } K_{\mu} \right), \\
& \left[ P_{\rho}, L_{\mu \nu } \right] = i \left( \eta_{\rho \mu} P_{\nu} - \eta_{\rho \nu} P_{\mu} \right), \\
& \left[ L_{\mu \nu}, L_{\rho \sigma} \right] = i \left( \eta_{\nu \rho} L_{\mu \sigma} + \eta_{\mu \sigma} L_{\nu \rho} - \eta_{\mu \rho} L_{\nu \sigma} - \eta_{\nu \sigma} L_{\mu \rho} \right). \\
\end{split}
\label{algebracft}
\end{equation} 

\nn
If we redefine the generators as

\begin{equation}
\begin{split}
J_{\mu \nu}= L_{\mu \nu} ,\quad \quad & \quad \quad J_{ \mu d} = \frac{1}{2} \left(  K_{\mu}-P_{\mu} \right), \\
J_{d,d+1} = D, \quad \quad & \quad \quad J_{\mu,d+1} = \frac{1}{2} \left( P_{\mu} + K_{\mu} \right), \\
\end{split}
\end{equation}

\nn
we find the algebra 

\begin{equation}
\left[ J_{ab} , J_{cd} \right] = i \left( \eta_{ad} J_{bc } +\eta_{bc} J_{ad} - \eta_{ac} J_{bd} - \eta_{bd} J_{ac} \right),
\end{equation}

\nn
with latin indices running from $ 0$ to $d+1$.  In a space with signature $(p,q)$, this is the algebra of $so(p+1,q+1)$.\\
We will now come back to conformal transformations in the special case $d=2$; equation \eqref{conformalconstrain}, in Euclidean space ($\eta_{\mu \nu} = \delta_{\mu \nu}$) reduces to Cauchy-Riemann conditions:
\begin{equation}
\begin{split}
& (i) \quad \partial_1 \epsilon_1 = \partial_2 \epsilon_2,\\
& (ii) \quad \partial_1 \epsilon_2 = - \partial_2 \epsilon_1. \\
\end{split} 
\end{equation}

\nn
To simplified it further its convenient to use complex coordinates $z=x_1 +ix_2$ and $ \overline{z}= x_1 -i x_2$ and to introduce the complex parameters  $\epsilon (z) = \epsilon^1 +i \epsilon^2$ and $\overline{\epsilon} (\overline{z}) = \epsilon^1 -i \epsilon^2$, in this notations Cauchy-Riemann conditions become:

\begin{equation}
\de \bar{\epsilon}=0 \quad\quad \mbox{and}\quad\quad \bar{\de}\epsilon=0,
\end{equation}

\nn
where $\bar{\de}\equiv \de_{\bar{z}}$. This means $\epsilon$ can be any analytic function of $z$ independent of $\bar{z}$, and $\bar{\epsilon}$, vice versa, any analytic function of $\bar{z}$ independent of $z$; such a functions are respectively said \textit{holomorphic} and \textit{antiholomorphic}. Conformal transformations in two dimensions are therefore analytic coordinate transformations 

\begin{equation}
z \quad \rightarrow \quad f(z) \quad \quad \quad \quad \quad \overline{z} \quad \rightarrow \quad f( \overline{z}). 
\end{equation}

\nn
In order to obtain the generators of the group and its algebra, we proceed, just as before, considering infinitesimal transformations 

\begin{equation}
z \longrightarrow z'=z + \epsilon(z) \quad \quad \mbox{and}\quad \overline{z} \longrightarrow \overline{z}' = \overline{z}+ \overline{\epsilon}(\overline{z}),
\label{infdued}
\end{equation}

\nn
we can Laurent expand complex parameters as
\begin{equation}
\epsilon(z)= - \sum \epsilon_n z^{n+1} \quad \quad \mbox{and}\quad \quad \overline{\epsilon} (\overline{z}) = -\sum \overline{\epsilon}_n \overline{z}^{n+1},
\end{equation}

\nn
obtaining the generators:
\begin{equation}
l_n = - z^{n+1} \partial_z \quad \quad \quad \quad \overline{l}_n = - \overline{z}^{n+1} \partial_{\overline{z}}.
\label{generatori}
\end{equation}

\nn
Calculating commutators among them, we find the algebra:
\begin{equation}
\begin{split}
& \left[ l_m , l_n \right] = (m-n) l_{m+n} \\
& \left[ \overline{l}_m , \overline{l}_n \right] = (m-n) \overline{l}_{m+n} \\
& \left[ l_n , \overline{l}_m \right] =0.
\end{split}
\end{equation}

\nn
We first notice the conformal group in two dimensions is infinite dimensional and its algebra consists in the direct sum of two identical copies, known as Witt algebra. An other important feature to be notice, unique of the bidimensional case and more suitable to be detected, is that not all generators are globally well defined: if we imposed $\epsilon$ analyticity in $0$ and infinity we find that only $\epsilon_{n}$ for $n=-1,0,1$ are allowed to be not null, and consequently the only generators globally well defined are $l_{-1},l_0$ and $l_{1}$. Equivalent considerations are valid for barred quantities. \\
These three generators span an algebra isomorphic to $sl(2,\C)$ and they generate the finite transformations:

\begin{center}
\begin{tabular}{|c l|}
\hline
$l_{-1}: z \rightarrow z+\alpha $& Translation \\
$l_0\,: z \rightarrow \lambda z $ &  Scaling  \\
$l_{1}: z \rightarrow \frac{z}{1-\beta z}$& SCT\\
\hline
\end{tabular} 
\end{center}

\nn
rotations on the z-plane are generated by the linear combination $i(l_0-\bar{l}_0)$, while dilations by scaling in both $z$ and $\bar{z}$, i.e. $l_0+\bar{l}_0$. This transformations can be rearranged in a single expression:
\begin{equation}
\begin{split}
& z \quad \longrightarrow \quad f(z) = \frac{az+b}{cz+d} \\
\end{split}
\end{equation}
where $a,b, c,d \in \mathbb{C}$ such that $ad-bc =1$.This group, know as \textit{restricted conformal group} is the projective special linear group $PSL(2,\C)=SL(2, \mathbb{C}) / \mathbb{Z}_2 \sim SO(3,1)$, where $\Z_2$ fixes the freedom to change the sign to all parameters leaving the transformation invariant. \\

\end{appendix}

\printbibliography

@article{Mastrolia:2018uzb,
    author = "Mastrolia, Pierpaolo and Mizera, Sebastian",
    title = "{Feynman Integrals and Intersection Theory}",
    eprint = "1810.03818",
    archivePrefix = "arXiv",
    primaryClass = "hep-th",
    doi = "10.1007/JHEP02(2019)139",
    journal = "JHEP",
    volume = "02",
    pages = "139",
    year = "2019"}

@article{Frellesvig:2020qot,
    author = "Frellesvig, Hjalte and Gasparotto, Federico and Laporta, Stefano and Mandal, Manoj K. and Mastrolia, Pierpaolo and Mattiazzi, Luca and Mizera, Sebastian",
    title = "{Decomposition of Feynman Integrals by Multivariate Intersection Numbers}",
    eprint = "2008.04823",
    archivePrefix = "arXiv",
    primaryClass = "hep-th",
    doi = "10.1007/JHEP03(2021)027",
    journal = "JHEP",
    volume = "03",
    pages = "027",
    year = "2021"
}

@article{Frellesvig:2019uqt,
    author = "Frellesvig, Hjalte and Gasparotto, Federico and Mandal, Manoj K. and Mastrolia, Pierpaolo and Mattiazzi, Luca and Mizera, Sebastian",
    title = "{Vector Space of Feynman Integrals and Multivariate Intersection Numbers}",
    eprint = "1907.02000",
    archivePrefix = "arXiv",
    primaryClass = "hep-th",
    doi = "10.1103/PhysRevLett.123.201602",
    journal = "Phys. Rev. Lett.",
    volume = "123",
    number = "20",
    pages = "201602",
    year = "2019"
}

@article{Frellesvig:2019kgj,
    author = "Frellesvig, Hjalte and Gasparotto, Federico and Laporta, Stefano and Mandal, Manoj K. and Mastrolia, Pierpaolo and Mattiazzi, Luca and Mizera, Sebastian",
    title = "{Decomposition of Feynman Integrals on the Maximal Cut by Intersection Numbers}",
    eprint = "1901.11510",
    archivePrefix = "arXiv",
    primaryClass = "hep-ph",
    doi = "10.1007/JHEP05(2019)153",
    journal = "JHEP",
    volume = "05",
    pages = "153",
    year = "2019"
}

@article{Weinzierl:2020xyy,
    author = "Weinzierl, Stefan",
    title = "{On the computation of intersection numbers for twisted cocycles}",
    eprint = "2002.01930",
    archivePrefix = "arXiv",
    primaryClass = "math-ph",
    doi = "10.1063/5.0054292",
    journal = "J. Math. Phys.",
    volume = "62",
    number = "7",
    pages = "072301",
    year = "2021"
}

@article{Mizera:2017cqs,
    author = "Mizera, Sebastian",
    title = "{Combinatorics and Topology of Kawai-Lewellen-Tye Relations}",
    eprint = "1706.08527",
    archivePrefix = "arXiv",
    primaryClass = "hep-th",
    doi = "10.1007/JHEP08(2017)097",
    journal = "JHEP",
    volume = "08",
    pages = "097",
    year = "2017"
}

@article{Mizera:2017rqa,
    author = "Mizera, Sebastian",
    title = "{Scattering Amplitudes from Intersection Theory}",
    eprint = "1711.00469",
    archivePrefix = "arXiv",
    primaryClass = "hep-th",
    doi = "10.1103/PhysRevLett.120.141602",
    journal = "Phys. Rev. Lett.",
    volume = "120",
    number = "14",
    pages = "141602",
    year = "2018"
}

@phdthesis{Mizera:2019gea,
    author = "Mizera, Sebastian",
    title = "{Aspects of Scattering Amplitudes and Moduli Space Localization}",
    eprint = "1906.02099",
    archivePrefix = "arXiv",
    primaryClass = "hep-th",
    doi = "10.1007/978-3-030-53010-5",
    school = "Princeton, Inst. Advanced Study",
    year = "2020"
}

@article{Mimachi:2004ez,
    author = "Mimachi, K. and Yoshida, M.",
    title = "{Intersection numbers of twisted cycles associated with the Selberg integral and an application to the conformal field theory}",
    doi = "10.1007/s00220-004-1138-z",
    journal = "Commun. Math. Phys.",
    volume = "250",
    pages = "23--45",
    year = "2004"
}

@article{Mimachi:2002gi,
    author = "Mimachi, Katsuhisa and Yoshida, Masaaki",
    title = "{Intersection numbers of twisted cycles and the correlation functions of the conformal field theory. 2.}",
    eprint = "math/0208097",
    archivePrefix = "arXiv",
    doi = "10.1007/s00220-002-0766-4",
    journal = "Commun. Math. Phys.",
    volume = "234",
    pages = "339--358",
    year = "2003"
}

@article{Lee:2013hzt,
    author = "Lee, Roman N. and Pomeransky, Andrei A.",
    title = "{Critical points and number of master integrals}",
    eprint = "1308.6676",
    archivePrefix = "arXiv",
    primaryClass = "hep-ph",
    doi = "10.1007/JHEP11(2013)165",
    journal = "JHEP",
    volume = "11",
    pages = "165",
    year = "2013"
}

@article{Bolte:1988ge,
    author = "Bolte, J. and Steiner, F.",
    title = "{Determinants of Laplace Like Operators on Riemann Surfaces}",
    reportNumber = "DESY-88-189",
    doi = "10.1007/BF02096935",
    journal = "Commun. Math. Phys.",
    volume = "130",
    pages = "581--598",
    year = "1990"
}

@article{Veneziano:1968yb,
    author = "Veneziano, G.",
    title = "{Construction of a crossing - symmetric, Regge behaved amplitude for linearly rising trajectories}",
    doi = "10.1007/BF02824451",
    journal = "Nuovo Cim. A",
    volume = "57",
    pages = "190--197",
    year = "1968"
}

@inproceedings{Staessens:2010vi,
    author = "Staessens, Wieland and Vercnocke, Bert",
    title = "{Lectures on Scattering Amplitudes in String Theory}",
    booktitle = "{5th Modave Summer School in Mathematical Physics}",
    eprint = "1011.0456",
    archivePrefix = "arXiv",
    primaryClass = "hep-th",
    month = "11",
    year = "2010"
}

@book{matsumoto2002introduction,
  title={An Introduction to Morse Theory},
  author={Matsumoto, Y.},
  series={Iwanami series in modern mathematics},
  url={https://books.google.it/books?id=RDIEuAEACAAJ},
  year={2002},
  publisher={American Mathematical Society}
}

@inproceedings{1997HypergeometricFM,
  title={Hypergeometric functions, my love : modular interpretations of configuration spaces},
  author={Yoshida, M.},
  year={1997}
}

@book{Aomoto,
author = {Aomoto, Kazuhiko and Kita, Michitake},
year = {2011},
month = {01},
pages = {},
title = {Theory of Hypergeometric Functions},
doi = {10.1007/978-4-431-53938-4}
}

@article{10.2307/1969099,
 ISSN = {0003486X},
 URL = {http://www.jstor.org/stable/1969099},
 author = {N. E. Steenrod},
 journal = {Annals of Mathematics},
 number = {4},
 pages = {610--627},
 publisher = {Annals of Mathematics},
 title = {Homology With Local Coefficients},
 volume = {44},
 year = {1943}}

@book{Polchinski:1998rq,
    author = "Polchinski, J.",
    title = "{String theory. Vol. 1: An introduction to the bosonic string}",
    doi = "10.1017/CBO9780511816079",
    publisher = "Cambridge University Press",
    series = "Cambridge Monographs on Mathematical Physics",
    month = "12",
    year = "2007"
}

@book{Green:2012oqa,
    author = "Green, Michael B. and Schwarz, John H. and Witten, Edward",
    title = "{Superstring Theory Vol. 1}: {25th Anniversary Edition}",
    doi = "10.1017/CBO9781139248563",
    publisher = "Cambridge University Press",
    series = "Cambridge Monographs on Mathematical Physics",
    month = "11",
    year = "2012"
}

@book{Green:2012pqa,
    author = "Green, Michael B. and Schwarz, John H. and Witten, Edward",
    title = "{Superstring Theory Vol. 2}: {25th Anniversary Edition}",
    doi = "10.1017/CBO9781139248570",
    publisher = "Cambridge University Press",
    series = "Cambridge Monographs on Mathematical Physics",
    month = "11",
    year = "2012"
}

@article{KAWAI19861,
title = {A relation between tree amplitudes of closed and open strings},
journal = {Nuclear Physics B},
volume = {269},
number = {1},
pages = {1-23},
year = {1986},
issn = {0550-3213},
doi = {https://doi.org/10.1016/0550-3213(86)90362-7},
url = {https://www.sciencedirect.com/science/article/pii/0550321386903627},
author = {H. Kawai and D.C. Lewellen and S.-H.H. Tye}}

@book{kiritsis:in2p3-00714916,
  TITLE = {{String theory in a nutshell}},
  AUTHOR = {Kiritsis, E.},
  URL = {http://hal.in2p3.fr/in2p3-00714916},
  PUBLISHER = {{Princeton University Press}},
  YEAR = {2007},
  HAL_ID = {in2p3-00714916},
  HAL_VERSION = {v1},
}

@book{Milnor,
  TITLE = {{Morse Theory}},
  AUTHOR = {Milnor, J.},
  PUBLISHER = {{Princeton University Press}},
  YEAR = {1963},
}

@article{cho_matsumoto_1995, title={Intersection theory for twisted cohomologies and twisted Riemann’s period relations I}, volume={139}, DOI={10.1017/S0027763000005304}, journal={Nagoya Mathematical Journal}, publisher={Cambridge University Press}, author={Cho, Koji and Matsumoto, Keiji}, year={1995}, pages={67–86}}

@article{Matsumoto1998IntersectionNF,
  title={Intersection numbers for logarithmic k-forms},
  author={Keiji Matsumoto},
  journal={Osaka Journal of Mathematics},
  year={1998},
  volume={35},
  pages={873-893}
}

@article{Tong:2009np,
    author = "Tong, David",
    title = "{String Theory}",
    eprint = "0908.0333",
    archivePrefix = "arXiv",
    primaryClass = "hep-th",
    month = "1",
    year = "2009"
}

@book{zwiebach_2004, place={Cambridge}, title={A First Course in String Theory}, DOI={10.1017/CBO9780511841682}, publisher={Cambridge University Press}, author={Zwiebach, Barton}, year={2004}}

@article{Mimachi,
author = {Mimachi, Katsuhisa and Ohara, Katsuyoshi and Yoshida, Masaaki},
year = {2004},
month = {12},
pages = {},
title = {Intersection numbers for loaded cycles associated with Selberg-type integrals},
volume = {56},
journal = {Tohoku Mathematical Journal},
doi = {10.2748/tmj/1113246749}
}

@article{Aomoto1975OnVO,
  title={On vanishing of cohomology attached to certain many valued meromorphic functions},
  author={Kazuhiko Aomoto},
  journal={Journal of The Mathematical Society of Japan},
  year={1975},
  volume={27},
  pages={248-255}
}

@article{Aomoto2,
  title={Un théorème du type de Matsushima-Murakami concernant l'intégrale des fonctions multiformes},
  author={Kazuhiko Aomoto},
  journal={J. Math. pures et appl.},
  year={1973},
  volume={52},
  pages={1-11}
}

@article{Mizera:2016jhj,
    author = "Mizera, Sebastian",
    title = "{Inverse of the String Theory KLT Kernel}",
    eprint = "1610.04230",
    archivePrefix = "arXiv",
    primaryClass = "hep-th",
    doi = "10.1007/JHEP06(2017)084",
    journal = "JHEP",
    volume = "06",
    pages = "084",
    year = "2017"
}

@article{Cachazo:2013iea,
    author = "Cachazo, Freddy and He, Song and Yuan, Ellis Ye",
    title = "{Scattering of Massless Particles: Scalars, Gluons and Gravitons}",
    eprint = "1309.0885",
    archivePrefix = "arXiv",
    primaryClass = "hep-th",
    doi = "10.1007/JHEP07(2014)033",
    journal = "JHEP",
    volume = "07",
    pages = "033",
    year = "2014"
}

@article{Parke:1986gb,
    author = "Parke, Stephen J. and Taylor, T. R.",
    title = "{An Amplitude for $n$ Gluon Scattering}",
    reportNumber = "FERMILAB-PUB-86-042-T",
    doi = "10.1103/PhysRevLett.56.2459",
    journal = "Phys. Rev. Lett.",
    volume = "56",
    pages = "2459",
    year = "1986"
}

@article{2000, volume={15},
   url={http://dx.doi.org/10.1016/S0217-751X(00)00215-7},
   journal={International Journal of Modern Physics A},
   publisher={World Scientific Pub Co Pte Lt},
   author={Laporta},
   year={2000}
}

@article{BCJ,
author = {Bern, Zvi and Carrasco, John Joseph and Johansson, Henrik},
year = {2010},
title = {Perturbative Quantum Gravity as a Double Copy of Gauge Theory},
volume = {105},
url={http://dx.doi.org/10.1103/PhysRevLett.105.061602},
journal = {Physical review letters}
}

@article{2008,
   title={New relations for gauge-theory amplitudes},
   volume={78},
   url={http://dx.doi.org/10.1103/PhysRevD.78.085011},
   number={8},
   journal={Physical Review D},
   publisher={American Physical Society (APS)},
   author={Bern, Z. and Carrasco, J. J. M. and Johansson, H.},
   year={2008}
}

@article{TKACHOV198165,
title = {A theorem on analytical calculability of 4-loop renormalization group functions},
journal = {Physics Letters B},
volume = {100},
number = {1},
pages = {65-68},
year = {1981},
url = {https://www.sciencedirect.com/science/article/pii/0370269381902884},
author = {F.V. Tkachov}
}

@article{Chetyrkin1981IntegrationBP,
  title={Integration by parts: The algorithm to calculate $\beta$-functions in 4 loops},
  author={K. Chetyrkin and Fyodor V. Tkachov},
  journal={Nuclear Physics},
  year={1981},
  volume={192},
  pages={159-204}
}

@misc{friedman2021elementary,
      title={An elementary illustrated introduction to simplicial sets}, 
      author={Greg Friedman},
      year={2021},
      eprint={0809.4221},
      archivePrefix={arXiv},
      primaryClass={math.AT}
}

@InProceedings{10.1007/978-3-0346-0209-9_5,
author="Zein, Fouad El
and Snoussi, Jawad",
editor="El Zein, Fouad
and Suciu, Alexandru I.
and Tosun, Meral
and Uluda{\u{g}}, A. Muhammed
and Yuzvinsky, Sergey",
title="Local Systems and Constructible Sheaves",
booktitle="Arrangements, Local Systems and Singularities",
year="2010",
publisher="Birkh{\"a}user Basel",
}

@article{Cacciatori:2021nli,
    author = "Cacciatori, Sergio Luigi and Conti, Maria and Trevisan, Simone",
    title = "{Co-homology of Differential Forms and Feynman diagrams}",
    eprint = "2107.14721",
    archivePrefix = "arXiv",
    primaryClass = "hep-th",
    year = "2021"
}

@article{tHooft:1972tcz,
    author = "'t Hooft, Gerard and Veltman, M. J. G.",
    title = "{Regularization and Renormalization of Gauge Fields}",
    journal = "Nucl. Phys. B",
    volume = "44",
    pages = "189--213",
    year = "1972"
}

@article{Kita1994IntersectionTF,
  title={Intersection Theory for Twisted Cycles},
  author={Michitake Kita and Masaaki Yoshida},
  journal={Mathematische Nachrichten},
  year={1994},
  volume={166},
  pages={287-304}
}

@article{Kita2,
author = {Kita, Michitake and Yoshida, Masaaki},
year = {2006},
month = {11},
pages = {171 - 190},
title = {Intersection Theory for Twisted Cycles II ‐ Degenerate Arrangements},
volume = {168},
journal = {Mathematische Nachrichten},
}

@article{Kita3,
author = {Yoshida, Masaaki},
year = {2000},
month = {06},
pages = {173-185},
title = {Intersection Theory for Twisted Cycles III — Determinant Formulae},
volume = {214},
journal = {Mathematische Nachrichten - MATH NACHR},
}

@misc{witten2016superstring,
      title={More On Superstring Perturbation Theory: An Overview Of Superstring Perturbation Theory Via Super Riemann Surfaces}, 
      author={Edward Witten},
      year={2016},
      eprint={1304.2832},
      archivePrefix={arXiv},
      primaryClass={hep-th}}

@misc{witten2013notes,
      title={Notes On Holomorphic String And Superstring Theory Measures Of Low Genus}, 
      author={Edward Witten},
      year={2013},
      eprint={1306.3621},
      archivePrefix={arXiv},
      primaryClass={hep-th}}

@misc{donagi2013supermoduli,
      title={Supermoduli Space Is Not Projected}, 
      author={Ron Donagi and Edward Witten},
      year={2013},
      eprint={1304.7798},
      archivePrefix={arXiv},
      primaryClass={hep-th}}

@book{Bott,
  title={Differential Forms in Algebraic Topology},
  author={Raoul Bott and Loring W. Tu},
  year={1982},
  publisher={Springer, New York, NY}
}
\end{document}